\documentclass[botnum,fleqn,final]{unmpandathesis}

\pdfoutput=1

\usepackage[sort&compress]{natbib}
\usepackage[pdftex,colorlinks=true]{hyperref}
\usepackage{amsmath}
\usepackage{graphicx}
\usepackage{amsthm}
\usepackage{amssymb}
\usepackage{color}

\definecolor{gray}{gray}{.5}

{\theoremstyle{definition} \newtheorem{definition}{Definition}}
\newtheorem{theorem}{Theorem}
\newtheorem{corollary}{Corollary}
\newtheorem{lemma}{Lemma}
\newtheorem{result}{Result}

\newcommand{\nghb}{{\mathcal N}}
\newcommand{\eqmod}[1]{=_{\scriptscriptstyle{#1}}}
\newcommand{\tr}[1]{\mbox{Tr}\left[ #1 \right]}
\newcommand{\sign}[1]{\mbox{sign}( #1 )}
\newcommand{\leftexp}[2]{{\vphantom{#2}}^{#1}\!{#2}}
\newcommand{\CZ}{\leftexp{C}{Z}}
\newcommand{\CX}{\leftexp{C}{X}}
\newcommand{\C}[1]{\leftexp{C}{#1}}
\newcommand{\set}{\mathcal}
\newcommand{\hhbar}{\mathcal{H}_l\cap\overline{\mathcal{H}}_r}
\newcommand{\hbarh}{\overline{\mathcal{H}}_l\cap\mathcal{H}_r}
\newcommand{\avg}[1]{\langle #1 \rangle}
\newcommand{\bigavg}[1]{\Bigl \langle #1 \Bigr \rangle}
\newcommand{\bra}[1]{\langle #1 |}
\newcommand{\ket}[1]{| #1 \rangle}
\newcommand{\setH}{\set{H}}
\newcommand{\setS}{\set{S}}
\newcommand{\setZ}{\set{Z}}
\newcommand{\setM}{\set{M}}
\newcommand{\setMS}{\set{M}_{\mbox{\scriptsize{S}}}}

\newcommand{\setMH}{\set{M}_{\mbox{\scriptsize{H}}}}

\newcommand{\setMSE}{\set{M}_{\mbox{\scriptsize{SE}}}}
\newcommand{\smallsetMSE}{\set{M}_{\mbox{\tiny{SE}}}}
\newcommand{\hyph}[2]{#1-#2}

\begin{document}

\frontmatter

\title{Stabilizer states and \\ local realism}

\author{Matthew B. Elliott}

\degreesubject{Ph.D., Physics}

\degree{Doctor of Philosophy \\ Physics}

\documenttype{Dissertation}

\previousdegrees{B.A., Mathematics \& Physics, Carleton College, 2003}

\date{July, 2008}

\maketitle

\begin{acknowledgments}

\renewcommand{\baselinestretch}{1.45}\selectfont

There is an exceptional body of researchers at UNM I would like to thank, including my advisor Carl Caves, close collaborator Bryan Eastin, as well as collaborators Steve Flammia, Andrew Landahl, and Seth Merkel. Thanks is also deserved by others here at UNM, Sergio Boixo, Animesh Datta, Ivan Deutsch, and Anil Shaji.

I also have publications involving researchers not at UNM. My gratitude goes out to Jon Barrett, Dan Browne, Akimasa Miyake, Stefano Pironio, and Tony Short. I should also mention those not at UNM with whom I have had interesting physics discussions, namely Joonwoo Bae, Howard Barnum, Jaewan Kim, Alex Monras, and Patrick Rice.

The quantum circuits in this document were drawn using a wonderful \LaTeX package called Q-circuit~\cite{eastin:qcircuit}.

\end{acknowledgments}

\maketitleabstract

\begin{abstract}

The central focus of this work is to make progress towards understanding entanglement as a resource for computation by examining the quantum correlations that can be extracted from stabilizer states. As such, we focus on the stabilizer formalism, local realism, and the convergence of the two.

We contribute to the understanding and manipulation of stabilizer states by providing a simple and compact graphical description of them. Using our \hyph{stabilizer}{state} graphs we demonstrate how Clifford operations and Pauli measurements on stabilizer states can be illustrated graphically. Moreover, we demonstrate how our graphical description of stabilizer states can actually be extended to encompass all stabilizer codes.

The inability to explain quantum measurement results in local realistic terms is the most striking divergence of quantum mechanics from classical physics. We consider measurements on spatially separated quantum systems and learn when the measurement correlations are consistent with local realism. If they are, we provide an algorithm that generates a local realistic model for them. When the measurement correlations are not so describable, our algorithm generates a Bell inequality violated by them. Unfortunately, this program becomes computationally infeasible when the number of measurements becomes large. For this reason, we also explicitly give a large class of Bell inequalities for an arbitrary number of systems and dichotomic observables. We also examine local realism through probabilistic theories, which include local realistic models, quantum mechanics, and more.

Finally, we study the \hyph{stabilizer}{state} violation of local realism. We begin by examining local realistic models for Pauli measurements on stabilizer states which satisfy a single \hyph{highly}{motivated} property. Beginning with these models, we then examine how much stronger quantum correlations are than classical ones. This is done by allowing some classical communication. Under this paradigm, we introduce a simple communication model that reproduces all global quantum correlations. Studying the successes and failures of this simple model uncovers two properties essential for any successful communication model. Furthermore, we construct a universal model with these properties that reproduces all quantum measurement results for Pauli measurements on stabilizer states.

\clearpage 
\end{abstract}

\tableofcontents
\listoffigures
\listoftables

\mainmatter


\chapter{Preliminaries}
\label{chap:preliminaries}

\section{Introduction and overview}

The field of quantum information has taught us that \hyph{information}{processing} tasks are more appropriately discussed in the context of the physical apparatus implementing them. We cannot separate the physical system from the computational task. Moreover, our physical reality seems to be most fundamentally quantum mechanical, so it is reasonable to approach information theory assuming a physical system governed by the laws of quantum mechanics. Moreover, processing information encoded in states of quantum systems seems to be advantageous~\cite{shor:factor,feynman:simulate,grover:search,aharonov:jones,gisin:cryptography}, a fact that has even changed our notion of computational complexity~\cite{watrous:complexity,brassard:complexity}. However, despite these results of quantum information, there still remains the deep question of why quantum mechanics makes such feats possible.

From the perspective of \hyph{measurement}{based} quantum computation~\cite{raussendorf:oneway} the full power of quantum computation can be consigned to quantum correlations. This means studying the power of quantum computation can be done by studying outcomes of measurements on an appropriate quantum state. This remarkable discovery says that studying objects such as Bell inequalities~\cite{bell:epr} is akin to studying the power of quantum information itself. Most research in this direction concerns itself with the amount by which quantum states can violate the local realistic bound of a Bell inequality. However, one interesting related perspective~\cite{pironio:bell}, which was first explored in Ref.~\cite{maudlin:communication}, is to learn how much quantum correlations violate local realism by giving the parties access to local realistic correlations and asking how much additional information, in the form of classical communication, is needed to reproduce quantum correlations. Notice that the original shared randomness is essential, since, without it, results diverge considerably~\cite{massar:simulate}. This is the approach we choose to follow, meaning we study the power inherent in quantum information by examining local realistic models assisted with classical communication.

We do not study arbitrary measurements on arbitrary quantum states. Instead we restrict our attention to Pauli measurements on stabilizer states, which is actually not as restrictive as it initially seems. Most quantum states which are used as resources for \hyph{measurement}{based} quantum computation are examples of stabilizer states. More than this, the set of measurements usually includes Pauli measurements plus only one additional \hyph{non}{Pauli} measurement. Thus the situation we consider is shy of full \hyph{measurement}{based} quantum computation by only a single \hyph{non}{Pauli} measurement. Admittedly, Pauli measurements on stabilizer states can be efficiently simulated and are by themselves probably not even universal for classical computation~\cite{aaronson:simulate}. On the other hand, extending our results for Pauli measurements beyond stabilizer states opens the possibility of universal quantum computation as well~\cite{danos:mbqc,bravyi:clifford}. Hopefully, the progress made here will inspire results beyond stabilizer states, or on stabilizer states with this extra measurement included.

We break up the material that follows into three main areas of study. First, in Chapter~\ref{chap:stabilizer}, we review the stabilizer formalism in great detail. This material is essential since we want to eventually study quantum correlations resulting from outcomes of Pauli measurements on stabilizer states. Even readers familiar with the stabilizer formalism should read the background material in Chapter~\ref{chap:stabilizer}, as much of it cannot be found in standard texts. This chapter culminates with the presentation of \hyph{stabilizer}{state} graphs, a novel and useful graphical representation of the stabilizer formalism introduced in Ref.~\cite{elliott:graphs}. These graphs are inspired by graph states, which are a class of stabilizer states privileged with a natural graphical representation, and the fact that any stabilizer state is \hyph{local}{Clifford} equivalent to some graph state. In fact, a stabilizer state is usually \hyph{local}{Clifford} equivalent to several graph states. Representing these local Clifford operations by decorations on the nodes of the graph, then, produces a graphical representation of all stabilizer states. Since a given stabilizer state is typically \hyph{local}{Clifford} equivalent to several graph states, there are generally several graphs that represent the state. However, we show that only two graph transformations are sufficient to generate all the graphs representing a stabilizer state. As a result, these graph transformations also yield all graph states that are \hyph{local}{Clifford} equivalent to a given stabilizer state.

Certain transformations, namely Clifford operations and Pauli measurements, take stabilizer states to other stabilizer states. Since both of these stabilizer states can be represented by a graph, it follows that the transformation can be represented by a graph transformation. Aside from decoration changes, it turns out that all local Clifford operations are generated by a single kind of graphical transformation called local complementation. With regard to Pauli measurements, the \hyph{post}{measurement} state is described by a sequence of steps which take the initial stabilizer graph to the \hyph{post}{measurement} stabilizer graph. In the case of individual Pauli measurements, meaning those that act on only one node of the graph, the transformation always results in that node being disconnected from the graph, possibly preceded by a graph transformation.

The next main subject matter comes in Chapter~\ref{chap:localrealism}. This chapter introduces the concepts of locality and realism, which say that all measurement results have \hyph{pre}{established} values that do not change once particles are spatially separated. Quantum mechanical correlations violate local realism. Local realistic correlations are exactly those correlations achievable classically and, in our case, they serve as the basis for communication strategies. Communication is then used to quantify the difference between classical and quantum correlations.

Chapter~\ref{chap:localrealism} contains some novel results in the study of local realism, particularly in the area of Bell inequalities. These are correlation inequalities that must be satisfied by any local and realistic model. In an experimental setup to test local realism, particles are put into an entangled quantum state and then projective measurements are made on the them. Over many trials, the measurements are found to be correlated in such a way that cannot be reproduced under the conditions of locality and realism. Given a set of correlations, thought of as being obtained from an experimental setup, we describe a linear program that accomplishes one of two tasks. First, if the correlations can be described in a local and realistic model, the program outputs such a model. Second, if the correlations violate local realism, the program outputs a Bell inequality which the correlations violate.

Unfortunately, the runtime of the aforementioned linear program is exponential in the number of measurements used in the experiment. We use this as motivation to derive a novel technique for generating a large class of Bell inequalities applicable to arbitrary numbers of parties and measurements. These Bell inequalities result from the unique perspective that local realistic models can be mocked up by using random variables. We can derive the allowed correlations for random variables, and Bell inequalities arise from testing if the correlations are within this allowed range.

The third and last subject area we study, in Chapter~\ref{chap:cclhvmodels}, is a melding of the previous two chapters to study the stabilizer state violation of local realism. First, we study local realistic descriptions of Pauli measurements on stabilizer states. We do not want to include all local realistic descriptions of Pauli measurements, but we also do not want to limit our study to those local realistic descriptions that reproduce the correlations of quantum mechanics. Therefore, we end up looking at \hyph{probability}{preserving} local realistic descriptions of Pauli measurements on stabilizer states. These are local realistic descriptions which yield correlations that are the same as quantum mechanics, up to a sign. The idea is then that these tables can be fixed rather simply. We introduce a standard method for constructing \hyph{probability}{preserving} descriptions and show that any \hyph{probability}{preserving} local realistic model can be constructed in the standard way.

With \hyph{probability}{preserving} local realistic descriptions serving as a basis, we can then supplement them with classical communication. Following the pioneering work of Toner and Bacon~\cite{toner:bell}, and Tessier~\textit{et al.\/}~\cite{tessier:ghzmodel}, the intent is to use classical communication to reproduce the observed correlations present in Pauli measurements on stabilizer states exactly. The simplest kind of communication model we consider restricts communication between neighbors in the \hyph{stabilizer}{state} graph and insists that nodes in identical local environments behave identically. The first restriction is called \hyph{nearest}{neighbor} communication while the second is called site invariance. Surprisingly, such a model is capable of reproducing all global measurement correlations, global measurement correlations being those correlations involving all the individual Pauli measurements present in a given Pauli product. It is interesting to note that in the ubiquitous two party scenario, global measurement correlations are the only correlations considered. Therefore, it seems that the interesting features of quantum correlations come from submeasurements of a global measurement. These are correlations arising from the consideration of restricted sets of individual Pauli operators in the global Pauli product.

To reproduce all submeasurement correlations, it is necessary to relieve the constraints we placed on our model. In particular, we look at what happens to models for two special classes of graph states, \hyph{one}{dimensional} and \hyph{two}{dimensional} cluster states, when these restrictions are lifted. Cluster states are simply graph states where the underlying graph is a square lattice. We find that \hyph{nearest}{neighbor} communication is too strong a restriction to reproduce submeasurement correlations for either \hyph{one}{dimensional} or \hyph{two}{dimensional} cluster states. Site invariance, on the other hand, seems to be more revealing. Site invariance is too restrictive for \hyph{two}{dimensional} cluster states, but we explicitly provide a \hyph{site}{invariant} model for the \hyph{one}{dimensional} cluster state. This is interesting because \hyph{two}{dimensional} cluster states are universal for \hyph{measurement}{based} quantum computation while \hyph{one}{dimensional} cluster states are not.

In general, we show that the communication distance, defined as the number of successive edges through which communication can be sent, must scale linearly with the girth of the graph state in order for a communication model to reproduce all submeasurement correlations on all graph states. With this in hand, we know a model that successfully reproduces all submeasurement correlations must not be site invariant and must let communication span the entire graph. Indeed, under these conditions, we do generate a successful model.

As previously mentioned, much of the material in Chapter~\ref{chap:stabilizer} preceding the discussion of stabilizer graphs cannot be found in standard texts or lecture notes. Also, the results regarding \hyph{stabilizer}{state} graphs is entirely new. Stabilizer graphs, and the action of Clifford operations on them, are introduced in Ref.~\cite{elliott:graphs}. The material on the graphical description of Pauli measurements is presented in Ref.~\cite{elliott:measure}. Material involving the graphical representation of stabilizer codes has not yet been written up for publication. In addition, both the linear program and \hyph{random}{variable} Bell inequalities covered in Chapter~\ref{chap:localrealism} comprise novel unpublished work. The unique approach to generalized probabilistic theories in that chapter is unpublished as well. Chapter~\ref{chap:cclhvmodels} contains entirely new results. The material covering LHV tables for Pauli measurements on stabilizer states has not been written up for publication. However, much of the later content in the chapter is an expanded exposition of the published material in Ref.~\cite{barrett:graphmodel}.

In the rest of this chapter we talk about some conventions and notation we use, as well as briefly reviewing some of the central background in linear algebra and quantum mechanics. After all, understanding of the content in this thesis requires first an understanding of vectors and matrices, and then an understanding of basic quantum theory. Those readers already familiar with linear algebra and quantum mechanics can safely disregard most of the rest of this chapter, excepting the next section on some conventions that we choose to follow. Another purpose of this chapter is to introduce some notation we use later, since much of the notation, although standard, may be unfamiliar to some. The next section, below, includes all of the notation we use which is not considered standard practice. After that, the rest of this chapter is designed to highlight aspects of basic quantum theory central to the chapters that follow. It is therefore advised that even readers experienced in quantum information skim the material to make sure it is fresh.

\section{Conventions}
\label{sec:conventions}

Our first point of business is to discuss the conventions we use throughout this work. All of these conventions have been set up to help clarify the exposition of the material.

First, it is our hope that readers are able to go through each chapter without getting bogged down by overly complicated or tangential results. For this reason complicated proofs, or proofs of tangential topics, are put into an appendix. This allows us to state results and focus on their application rather than dwelling on their truth value.

We also find that sometimes the result of the calculations in a section is obscured by the calculations themselves. When this is the case, the result itself will be highlighted as such.

\noindent \textbf{Result.} This is the result of our calculations.

\noindent This way, the purpose of all our calculations is absolutely clear. Other important results are in the form of theorems, lemmas, and corollaries and so are already highlighted.

Now, a word on some notation we use which, though not standard, is both useful and natural for presenting the material herein. We find ourselves often performing algebra using modular arithmetic. There are even times when modular arithmetic using two different values appears in the same equation. To help clarify what sort of modular arithmetic is being used we use a few techniques.

First of all, the symbol ``$=$'' means that quantities are identical without using modular arithmetic. When we do use modular arithmetic, the two types we encounter are modulo $2$ arithmetic and modulo $4$ arithmetic. When an equation is meant to hold modulo $2$ we write the symbol `` $\eqmod{2}$''. In other words,
\begin{equation}
x = y \pmod{2} \;\; \Leftrightarrow \;\; x \eqmod{2} y.
\end{equation}
Similarly for equations that hold modulo $4$ we have
\begin{equation}
x = y \pmod{4} \;\; \Leftrightarrow \;\; x \eqmod{4} y.
\end{equation}
For example, we have that $0 \eqmod{2} 2$ and $3 \eqmod{4} -1$. The final situation we encounter is that modulo $2$ quantities sometimes appear in equations that involve arithmetic that is otherwise modulo $4$. Our notation when this situation arises is thus: any expression with an overbar is meant to be taken modulo $2$. For example we have that $\overline{1+1+5}+2 \eqmod{4} -1$, since $1+1+5 \pmod{2} = 1$ and $1+2 = 3 \eqmod{4} -1$. In contrast, notice that $1 + 1 + 5 + 2 \ne_{\scriptscriptstyle{4}} -1$.

Finally, a word on vector conventions. Much of the relevant modular arithmetic is formulated in terms of binary vectors, i.e., vectors whose entries are either $1$ or $0$. We would like for these vectors to not be confused with ordinary real valued vectors, which we also use. To make things worse, quantum states are also typically represented by vectors. To keep these vectors straight, we employ specific notation in each case. A binary vector, $a$, is represented by an unadorned letter. To make the distinction even more clear, binary vectors are always written as a row vector; in order to get the column vector, one must write $a^T$. Normal real valued vectors are taken by default to be column vectors, and we write them with an arrow over the letter such as $\vec{a}$. Finally vectors representing quantum states are also column vectors, and for these vectors we use Dirac notation. Table~\ref{tab:vectors} summarizes the different notation for vectors in these three contexts. 
\begin{table}
\center
\begin{tabular}{|l|c|c|} \hline
& Column Vector & Row Vector \\ \hline
Binary Vector & $\phantom{|} a^T$ & $\phantom{\langle} a^{\phantom{T}}$ \\ \hline
Real Valued Vector & $\phantom{|} \vec{a}^{\phantom{T}}$ & $\phantom{\langle} \vec{a}^T$ \\ \hline
Pure Quantum State & $\ket{a}$ & $\bra{a}$ \\ \hline
\end{tabular}
\caption[Vector notation]{There are three contexts in which vectors are used and so different vector notation is used for each context. Binary vectors are row vectors and they are represented by an unadorned letter. Regular real valued vectors are given the standard column vector notation with an arrow over the letter. Quantum states are column vectors written in Dirac notation, otherwise known as bra and ket notation. \label{tab:vectors}}
\end{table}

\section{Quantum theory of qubits}

The quantum states we consider are those of \emph{qubits\/}, meaning \hyph{two}{dimensional} quantum systems. \hyph{Two}{dimensional} quantum states encompass a very special class of quantum systems, yet there is enough richness in these states to occupy this entire thesis and more. Quantum information is also often expressed in terms of qubits because of the analogy with classical computers which operate using \hyph{two}{dimensional} classical systems called bits. Indeed, the term qubit derives from a shortened form of quantum bit.

We review only the basics of qubits below; it is assumed that any reader is familiar with both quantum mechanics and linear algebra. For those interested, a more thorough overview of the material here can be found in Ref.~\cite{nielsen:book}. The purpose of the following subsections is only to highlight and review those aspects of the quantum theory of qubits particularly relevant to the following chapters of this thesis.

\subsection{Single qubits}

If a qubit is in a pure state, then we can represent its state with a \hyph{two}{dimensional} vector with complex entries. Because of the connection to classical bits, we write our basis for the \hyph{two}{dimensional} vector space as
\begin{equation}
\begin{array}{ccc}
\left( \begin{array}{c} 1 \\ 0 \end{array} \right) \leftrightarrow \ket{0} & \mbox{and} & \left( \begin{array}{c} 0 \\ 1 \end{array} \right) \leftrightarrow \ket{1}.
\end{array}
\end{equation}
Any pure state of a qubit can be written as a column vector, $\ket{\psi}$, which must be a linear combination of our basis vectors, i.e.,
\begin{equation}
\ket{\psi} = \left( \begin{array}{c} \alpha \\ \beta \end{array} \right) = \alpha \ket{0} + \beta \ket{1}.
\end{equation}
The state of a qubit is really just an object that tells us what outcomes are produced when a measurement is made on the system. We also encounter other ways to represent states, with the purpose of all these representations being to predict measurement outcomes on the system.

A projective measurement of any $2 \times 2$ Hermitian operator, $A$, can be made on this state. According to the postulates of quantum mechanics, the measurement outcome is one of two possibilities corresponding to the two real eigenvalues of $A$. Once the measurement is carried out, the \hyph{post}{measurement} state is one of the eigenvectors of $A$, which one depending on the observed measurement outcome. Throughout this thesis, we take the nondegenerate measurement outcomes to be $+1$ and $-1$. If this seems strange, just think of multiplying $A$ by a constant and adding to it a multiple of the identity, $I$, so as to make the eigenvalues equal to $\pm 1$. This does not change the \hyph{post}{measurement} state, meaning the eigenvectors of $A$, and so it can be thought of as simply relabeling the two different measurement outcomes as $\pm 1$.

The probability, $p_{\pm}$, for a measurement on a system in the state $\ket{\psi}$ to produce an outcome $\pm 1$ can be inferred from the average value of the measurement, $\bra{\psi} A \ket{\psi} = \avg{A}$, since $p_{\pm} = (1/2) (1 \pm \avg{A})$. This is how the state gives rise to measurement outcome probabilities. If we do not know the state exactly but only know that it is some state $\ket{\psi_j}$ with probability $p_j$, then the average value is $\bra{\psi_j} A \ket{\psi_j}$ with probability $p_j$. Thus,
\begin{equation}
\avg{A} = \sum_j p_j \bra{\psi_j} A \ket{\psi_j} = \tr{A \sum_j p_j \ket{\psi_j} \bra{\psi_j}} = \tr{A \rho}.
\end{equation}
The $2 \times 2$ matrix $\rho$ allows us to assign a state to systems about which we have incomplete knowledge, assigning probabilities, $p_{\pm} = (1/2) (1 \pm \tr{A \rho})$, to measurement outcomes as best they can be.

Finally, the state must assign probabilities such that all the measurement outcome probabilities sum to $1$. The outcome probabilities summing to $1$ can be expressed as $\avg{I} = 1$, since both outcomes in this case are $1$ and hence $\avg{I} = p_+ + p_-$. For a pure state vector this means that $\langle \psi | \psi \rangle = 1$, so that
\begin{equation}
\langle \psi | \psi \rangle = \left( \begin{array}{cc} \alpha^* & \beta^* \end{array} \right) \left( \begin{array}{c} \alpha \\ \beta \end{array} \right) = |\alpha|^2+|\beta|^2 = 1,
\end{equation}
and more generally we also have that $\tr{I \rho} = \tr{\rho} = 1$.

\subsection{Projective measurements}

The measurements we talked about in the previous subsection, which are actually the only kind of measurements we ever consider, are projective measurements. They are called projective measurements because the state is projected onto an eigenstate of the observable, $A$. Thus the \hyph{post}{measurement} state is an eigenstate of $A$, and so any further measurement of $A$ must yield the same outcome previously obtained.

A projection operator $P$, which in general can be of any finite dimension, is such that one application may change the state, but subsequent applications do not, i.e., $P^2 = P$. By rewriting this condition as ${P(P-I)=0}$ we see that projection operators can only have eigenvalues $1$ and $0$. The number of entries with a $1$ is the number of linearly independent vectors unaffected by the projection operator. These vectors form a basis for the subspace onto which $P$ projects, since $P \ket{\psi}$ is in the subspace for any $\ket{\psi}$. The dimension of this subspace is simply calculated by taking $\tr{P}$, as this sums up all the eigenvalues of $P$.

If a projective measurement of $A$ gives an outcome of $\pm 1$ then it means the \hyph{post}{measurement} state is given by
\begin{equation}
\frac{P_{\pm} \ket{\psi}}{\sqrt{\bra{\psi} P_{\pm} \ket{\psi}}} = \frac{1}{\sqrt{\bra{\psi} \frac{1}{2} (I \pm A) \ket{\psi}}} \cdot \frac{1}{2} (I \pm A) \ket{\psi},
\end{equation}
where $P_{\pm} = (1/2) (I \pm A)$ is a \hyph{two}{dimensional} projection operator. If we want to measure another operator, $U^{\dagger} A U$, unitarily equivalent to $A$, then we get a \hyph{post}{measurement} state
\begin{equation}
\frac{1}{\sqrt{\bra{\psi} \frac{1}{2} (I \pm U^{\dagger} A U) \ket{\psi}}} \cdot \frac{1}{2} (I \pm U^{\dagger} A U) \ket{\psi} =U^{\dagger} \cdot \frac{1}{\sqrt{\bra{\phi} \frac{1}{2} (I \pm A) \ket{\phi}}} \cdot \frac{1}{2} (I \pm A) \ket{\phi},
\end{equation}
where $\ket{\phi} = U \ket{\psi}$. This is a fundamental result. In words, measuring $U^{\dagger} A U$ on $\ket{\psi}$ is equivalent to measuring $A$ on $U \ket{\psi}$ and then transforming the \hyph{post}{measurement} state by $U^{\dagger}$.

\subsection{Pauli matrices}

Any $2 \times 2$ matrix can be written as a linear combination of the identity and the \emph{Pauli matrices\/}. These are written in the $\ket{0}$, $\ket{1}$ basis as,
\begin{equation}
\begin{array}{cccc}
X = \left( \begin{array}{cc} 0 & 1 \\ 1 & 0 \end{array} \right),
& Y = \left( \begin{array}{cc} 0 & -i \\ i & 0 \end{array} \right), & \mbox{and}
& Z = \left( \begin{array}{cc} 1 & 0 \\ 0 & -1 \end{array} \right).
\end{array}
\end{equation}
The Pauli matrices are traceless, Hermitian, and unitary. They are sometimes written as $X = \sigma_1$, $Y = \sigma_2$, $Z = \sigma_3$, and may include $I = \sigma_0$. The three \hyph{non}{identity} Pauli matrices are sometimes combined into a column vector, $\vec{\sigma} = \left( \begin{array}{ccc} \sigma_1 & \sigma_2 & \sigma_3 \end{array} \right)^T$, whose entries are operators. There are many important properties of the Pauli matrices, for instance see Ref.~\cite{tannoudji:qmbook}, but we restrict our discussion to those aspects relevant to our studies.

The Pauli matrices are exceptionally useful in the study of qubits. The fact that Pauli matrices form a basis for $2 \times 2$ matrices means that quantum states and operations on quantum states can both be expressed in terms of them. For a quantum state $\rho$ we have
\begin{equation}
\rho = \frac{1}{2} (I + \vec{a} \cdot \vec{\sigma}),
\end{equation}
with $|\vec{a}| \leq 1$, and where the vector $\vec{a}$ is the Bloch vector defining the state. Pure states are those with $|\vec{a}| = 1$. Likewise, a Hermitian operator, $A$, with eigenvalues $\pm 1$ can be written as
\begin{equation}
A = \vec{a} \cdot \vec{\sigma},
\end{equation}
where this time $|\vec{a}| = 1$. In both cases, Hermiticity ensures that $\vec{a}$ has real entries.

\subsection{Multiple qubits} \label{subsec:multiplequbits}

Interesting features arise when we consider a system consisting of multiple, say $n$, \hyph{two}{dimensional} subsystems. The state of the overall system, if in a pure state, is represented by a vector with complex entries. Since each subsystem can be in one of two orthogonal states there are at very least $2^n$ orthogonal states of the overall system, one for each of the subsystem configurations. Thus the overall system can be taken to be $2^n$ dimensional. The general state of a system of multiple qubits is a linear combination of these orthogonal basis states.

Given this, the way to combine the vector spaces of multiple systems mathematically is quite intuitive. The operation that combines the vector spaces, called the tensor product, is denoted by the symbol ``$\otimes$.'' It combines vector spaces of dimensions $d_A$ and $d_B$ to yield a space of dimension $d_A d_B$. The essential property it must satisfy is that for subsystem operations $A$ and $B$ and subsystem states $\ket{\psi}$ and $\ket{\phi}$, it should not matter whether $A$ and $B$ act before or after we combine the systems, i.e.,
\begin{equation}
(A \otimes B) (\ket{\psi} \otimes \ket{\phi}) = A \ket{\psi} \otimes B \ket{\phi}.
\end{equation}
Given this requirement, there are multiple ways of representing the tensor product of two matrices, of which vectors are a special case. Our choice is to multiply each entry of $A$ by all the entries of $B$. Thus we have,
\begin{align}
\begin{split}
A \otimes B &= \left( \begin{array}{cc} a_{11} B & a_{12} B \\ a_{21} B & a_{22} B \end{array} \right) \\
&= \left( \begin{array}{cc} a_{11} \left( \begin{array}{cc} b_{11} & b_{12} \\ b_{21} & b_{22} \end{array} \right) & a_{12}\left( \begin{array}{cc} b_{11} & b_{12} \\ b_{21} & b_{22} \end{array} \right)  \\ a_{21} \left( \begin{array}{cc} b_{11} & b_{12} \\ b_{21} & b_{22} \end{array} \right) & a_{22} \left( \begin{array}{cc} b_{11} & b_{12} \\ b_{21} & b_{22} \end{array} \right) \end{array} \right) \\
&= \left( \begin{array}{cccc} a_{11} b_{11} & a_{11} b_{12} & a_{12} b_{11} & a_{12} b_{12} \\ a_{11} b_{21} & a_{11} b_{22} & a_{12} b_{21} & a_{12} b_{22} \\ a_{21} b_{11} & a_{21} b_{12} & a_{22} b_{11} & a_{22} b_{12} \\ a_{21} b_{21} & a_{21} b_{22} & a_{22} b_{21} & a_{22} b_{22} \end{array} \right).
\end{split}
\end{align}
Repeated application of this rule can be used to combine an arbitrary number of systems.

We use various notation for the tensor product. A subscript on a matrix, such as $A_j$, generally means the tensor product of all identity matrices, $I$, with a single $A$ matrix in the $j$th spot. Mathematically, we have $A_j = I \otimes \cdots \otimes I \otimes A \otimes I \otimes \cdots \otimes I$. For instance we can write $A \otimes B = A_1 B_2$. It is clear when this notation is not being used, as terms will appear such as $A_1 \otimes B_2$. When this happens, we know that $A_1$ does not mean $A \otimes I$. There is some special notation for tensor products of vectors as well. For the standard basis, we almost never write the \hyph{tensor}{product} symbol preferring the shorthand $\ket{00} = \ket{0} \otimes \ket{0}$. Lastly, we indicate $n$ copies of a state $\ket{\psi}$ by writing $\ket{\psi}^{\otimes n} = \ket{\psi} \otimes \cdots \otimes \ket{\psi}$, and likewise for matrices.

\subsection{The Pauli group}

The \hyph{multi}{qubit} extension of Pauli matrices leads to a very important class of unitary operators known as the Pauli group, which is as useful for states of multiple qubits as Pauli matrices are for single qubits. More specifically, we talk about the Pauli group on $n$ qubits and denote it as $\mathcal{P}_n$. This group is of central importance to the stabilizer formalism, and therefore merits an in depth discussion here. The Pauli group on $n$ qubits is defined as all \hyph{$n$}{fold} tensor products of the Pauli matrices $X$, $Y$, and $Z$, along with the identity $I$ and phases $\pm 1$, $\pm i$. An element of $\mathcal{P}_5$ is, for example, $X \otimes iI \otimes -Z \otimes iY\otimes -iI$. Of course, one can always pull off the overall phase and write $-i X \otimes I \otimes Z \otimes Y\otimes I$ instead.

The Pauli group is, first of all, a \emph{group\/} which means it is a set of elements, along with an associative multiplication operation, that satisfies three properties. The first is that it contains the identity operator. The second is that the product of any two elements of the group is also a element of the group. Finally the third property is that there must exist, for each element of the group, an inverse within the group. For any group there exists a subset of elements that \emph{generate\/} the group. That is, any group element can be written as a product of these generators. Moreover, elements in the generating set can be chosen to be \emph{independent\/}, in that no element in the generating set is the product of other elements in the set. This is simply because a dependent element in the generating set can be thrown out without losing the ability to generate the whole group. A \emph{subgroup\/} of a group is a subset of group elements that by themselves constitute a group.

We will now prove that the Pauli group is, indeed, a group. First of all it contains $I \otimes \cdots \otimes I$ by definition, which proves the first group property. As for closure under multiplication, two elements of $\mathcal{P}_n$, $g = g_1 \otimes \cdots \otimes g_n$ and $h = h_1 \otimes \cdots \otimes h_n$ multiply to $gh = g_1 h_1 \otimes \cdots \otimes g_n h_n$. The fact that this is also in the Pauli group then follows from from the case $n=1$ which is easily verified. Finally, given $g = g_1 \otimes \cdots \otimes g_n \in \mathcal{P}_n$, it follows, again from the $n=1$ case, that $g^{\dagger} = g_1^{\dagger} \otimes \cdots \otimes g_n^{\dagger} \in \mathcal{P}_n$. Moreover, $g^{\dagger} g = g_1^{\dagger} g_1 \otimes \cdots \otimes g_n^{\dagger} g_n = I$ also follows from the $n=1$ case and so there exists an inverse in the group for all elements.

First, we mention some basic properties satisfied by members of the Pauli group. Since Pauli matrices are traceless, this implies that all Pauli group elements are also traceless. From the unitarity of Pauli matrices, we also conclude that all members of the Pauli group are unitary as well. Elements of the Pauli group that have overall phase $\pm 1$ are Hermitian, while those with overall phase $\pm i$ are anti Hermitian. As a result, the Pauli group splits into two parts. Those elements, $g \in \mathcal{P}_n$, with overall phase $\pm 1$ satisfy $g^2 = I$ and those with overall phase $\pm i$ satisfy $g^2 = -I$. One consequence is that elements with a $\pm 1$ overall phase have eigenvalues $\pm 1$, while the other Pauli group elements have eigenvalues $\pm i$.

A remarkably useful property of the Pauli group is that any two elements either commute or anticommute. Verification of the $n=1$ case is left to the reader. Once this is established we can define the following function that takes elements of $\mathcal{P}_1$ as input.
\begin{equation} \label{eq:comm1}
\begin{array}{cc}
f(g_j,h_j)= \left\{
\begin{array}{ll}
0 & \mbox{if $g_j$ and $h_j$ commute} \\
1 & \mbox{if $g_j$ and $h_j$ anticommute.} 
\end{array} \right.
\end{array}
\end{equation}
This function allows us to reverse the order of multiplication through the identity $g_j h_j = (-1)^{f(g_j,h_j)} h_j g_j$. Now we consider two elements of $\mathcal{P}_n$, $g = g_{1} \otimes \cdots \otimes g_n$ and $h = h_{1} \otimes \cdots \otimes h_n$. We calculate,
\begin{equation} \label{eq:comm}
\begin{array}{l}
[g,h] = gh-hg = hg \left( (-1)^{\sum_{j=1}^n f(g_j,h_j)} -1 \right) \mbox{ and} \\
\{g,h\} = gh+hg = hg \left( (-1)^{\sum_{j=1}^n f(g_j,h_j)} +1 \right)
\end{array}
\end{equation}
We now see that $g$ and $h$ either commute or anticommute and they commute if and only if $\sum_{j=1}^n f(g_j,h_j)$ is even, which is to say they contain an even number of $\mathcal{P}_1$ elements that anticommute.


\chapter{The stabilizer formalism}
\label{chap:stabilizer}

Representing quantum states as a linear combination of basis states can quickly become futile as the dimension or number of systems grows. However, if we limited ourselves to studying states that can be written as a linear combination of just a few basis vectors, we would miss most quantum states and therefore many of the interesting features of quantum mechanics. Many methods have been developed to study more complicated quantum systems \cite{schollwock:dmrg,perez:mps,metropolis:montecarlo}, with one very important example being the stabilizer formalism \cite{gottesman:thesis}. The stabilizer formalism has found uses in quantum error correction \cite{gottesman:thesis}, \hyph{measurement}{based} quantum computation \cite{raussendorf:oneway}, the study of quantum entanglement \cite{fattal:entanglement,wang:entanglement,vandennest:resources}, as well as in other areas \cite{vandennest:ising,cleve:secret}.

We review the stabilizer formalism with special emphasis on those topics relevant to our needs. Much of what follows is focused on developing mathematical tools used to augment the stabilizer formalism to better understand its power. Probably the most significant result in this regard is the development of stabilizer graphs in Sec.~\ref{sec:stabilizergraphs}. Much of the work before this section is simply background on the stabilizer formalism, with a few novel results interspersed.

\section{Stabilizer states \label{sec:stabilizer}}

If a state $\ket{\psi}$ is a $+1$ eigenstate of some operator $A$, i.e., $A \ket{\psi} = \ket{\psi}$, we say that $\ket{\psi}$ is \emph{stabilized\/} by $A$. For example, $\ket{+} = (1/\sqrt{2}) (\ket{0}+\ket{1})$ is stabilized by the $X$ operator. In fact, $\ket{+}$ is the only state stabilized by $X$. If we consider a tensor product of two Pauli matrices, then we find that two linearly independent states are stabilized by the operator. As an example, we find that $X \otimes X$ stabilizes both $(1/\sqrt{2}) (\ket{00}+\ket{11})$ and $(1/\sqrt{2}) (\ket{01}+\ket{10})$. Stabilizer states are so called because they are states that are stabilized by tensor products of Pauli matrices. Thus we need to explore a bit the stabilizing properties of these operators.

We can generalize the previous examples to an \hyph{$n$}{fold} tensor product of Pauli matrices, $g$, by considering $(1/2)(I+g)$. This operator has the property that,
\begin{equation}
\begin{array}{c}
\left(\frac{1}{2}(I+g)\right)^2 = \frac{1}{4}(I+2g+g^2) = \frac{1}{4}(I+2g+I) = \frac{1}{2}(I+g),
\end{array}
\end{equation}
and so is a projection operator. The eigenvectors of $(1/2)(I+g)$ are the same as those of $g$, with $\frac{1}{2}(I+g) \ket{\psi} = \ket{\psi}$ for those $\ket{\psi}$ stabilized by $g$ and $\frac{1}{2}(I+g) \ket{\psi} = 0$ for the other eigenvectors of $g$. These properties together imply that $(1/2)(I+g)$ is the projector onto the states stabilized by $g$. Using this projector we determine the dimension of the subspace stabilized by $g$ to be $\tr{(1/2)(I+g)}=2^{n-1}$, which follows from $\tr{g} = 0$. Thus an \hyph{$n$}{fold} tensor product of Pauli matrices stabilizes $2^{n-1}$ linearly independent states.

Let us consider once again the states stabilized by $X \otimes X$. If we also impose the condition that the state be stabilized by $Z \otimes Z$, then the only state we are left with is $(1/\sqrt{2}) (\ket{00}+\ket{11})$. This is the unique state stabilized by both $X \otimes X$ and $Z \otimes Z$. Generally, let $g$ and $h$ be two commuting \hyph{$n$}{fold} tensor products of Pauli matrices. They, of course, need to commute in order to have common eigenvectors. As before, we can determine the number of linearly independent states stabilized by $g$ and $h$ by considering the projector onto the subspace stabilized by both operators. The trace of the projector yields the subspace dimension,
\begin{equation}
\tr{\frac{1}{2}(I+g)\frac{1}{2}(I+h)} = \tr{\frac{1}{4}(I+g+h+gh)} = 2^{n-2},
\end{equation}
since $gh$ is traceless as long as $g \ne h$. Thus $2^{n-2}$ states are stabilized by both $g$ and $h$. We see that this same projection argument will give that $k$ independent, commuting \hyph{$n$}{fold} tensor products of Pauli matrices stabilize $2^{n-k}$ linearly independent states, with each new independent operator cutting the dimension in half. The independence property means that no generator can be written as a product of other generators, up to a sign, and this is equivalent to products of operators being traceless. Thus the subspace dimension ends up being $(1/2^k) \tr{I} = 2^{n-k}$.

Everything still works the same way if we choose our operators to be \hyph{$n$}{fold} tensor products of Pauli matrices with overall phases $\pm 1$. For example, the state $(1/\sqrt{2}) (\ket{00}-\ket{11})$ is the unique state stabilized by $-X \otimes X$ and $Z \otimes Z$. Other elements of $\mathcal{P}_n$, however, are not valid as stabilizer operators because an overall phase of $\pm i$ means that there is no $+1$ eigenvalue of the operator. On a more fundamental level, a product of Pauli matrices with overall phase $\pm 1$ is Hermitian, and is therefore an observable, while those with phase $\pm i$ are not. The property that stabilizer operators are also observables is crucial for many applications of the stabilizer formalism. That being said, a subspace stabilized by tensor products of Pauli matrices, with overall phases $\pm 1$, is called a \emph{stabilizer subspace\/}. The case where the number of operators equals the number of qubits is special.

\begin{definition}
A \emph{stabilizer state\/} on $n$ qubits is the simultaneous $+1$ eigenstate of $n$ independent, commuting elements of the Pauli group.
\end{definition}

It is now appropriate to introduce some terminology related to stabilizer states. These terms arise from a connection between stabilizer operators and group theory. The key observation is that for two Hermitian Pauli group elements, $g$ and $h$, which stabilize some state, the product $gh$ also stabilizes the same state. This is simply the closure under multiplication property of groups. Clearly $I$ stabilizes all states, and also Hermitian Pauli group elements are their own inverses. Thus the Pauli group elements that stabilize a stabilizer state form a group. Once $n$ independent, commuting Pauli group elements that stabilize a state are specified, products of those operators generate the group. For this reason these $n$ operators are called \emph{stabilizer generators\/}. The whole group is called the \emph{stabilizer group\/}, and an element of the stabilizer group is called a \emph{stabilizer element\/}, \emph{stabilizer operator\/}, or simply \emph{stabilizer\/}. This is an abelian group by construction, meaning all elements in the group mutually commute.

Let us consider an example of a stabilizer state. We stated that $(1/\sqrt{2}) (\ket{00}-\ket{11})$ is the simultaneous $+1$ eigenstate of $-X \otimes X$ and $Z \otimes Z$. It is simple to verify that $-X \otimes X$ and $Z \otimes Z$ commute, since an even number of individual Pauli matrices in the tensor products anticommute. Thus $(1/\sqrt{2}) (\ket{00}-\ket{11})$ is a stabilizer state with stabilizer generators $-X \otimes X$ and $Z \otimes Z$. The stabilizer group for this state consists of the operators $I \otimes I$, $-X \otimes X$, $Y \otimes Y$, and $Z \otimes Z$, all of which stabilize the state. The operator $Y \otimes Y$ is a stabilizer element, but not one of the listed stabilizer generators.

We generally specify a stabilizer state by listing $n$ stabilizer generators for its stabilizer group. This set is not unique, however, since given two generators, $g$ and $h$, replacing $g$ by $gh$ generates the same group. It is possible to define a canonical set of generators for any stabilizer state, although doing so requires a more powerful description of the stabilizer formalism. The rest of this chapter is devoted to developing the mathematical machinery used to better understand and manipulate this important class of states.

\subsection{Binary representation \label{subsec:binaryrep}}

The binary representation of Pauli matrices associates a \hyph{two}{dimensional} binary vector $r(\sigma_j)$ with each Pauli matrix $\sigma_j$, where $r(I) = \left( \begin{array}{c|c} 0 & 0 \end{array} \right)$, $r(X) = \left( \begin{array}{c|c} 1 & 0 \end{array} \right)$, $r(Y) =  \left( \begin{array}{c|c} 1 & 1 \end{array} \right)$, and $r(Z) = \left( \begin{array}{c|c} 0 & 1 \end{array} \right)$. The vertical line is simply a means to help visualize the two halves of the vector, a schematic that will be helpful later. The first element of $r(\sigma_j)$ can be thought of as specifying the $X$ part of $\sigma_j$, since it is $1$ only for $X$ and $Y = iXZ$. Likewise, the second element specifies the $Z$ part, since it is $1$ only for $Z$ and $Y = iXZ$. In the sense that $Y=iXZ$, $Y$ has both an $X$ part and a $Z$ part, while $I$ has neither. By taking $r_1(\sigma_j)$ and $r_2(\sigma_j)$ to mean the first and second entries of $r(\sigma_j)$, we can reconstruct the Pauli matrix from the binary vector,
\begin{equation}
\sigma_j = i^{r_1(\sigma_j)r_2(\sigma_j)} X^{r_1(\sigma_j)} Z^{r_2(\sigma_j)}.
\end{equation}

The binary representation of an element of $\mathcal{P}_1$ specifies that element up to the overall phase of $\pm1$ or $\pm i$. We will worry about this phase later, but ignoring the phase does have a few benefits. For example let us consider the product of two Pauli matrices, using the binary vector representation.
\begin{equation} \label{eq:prod21}
\begin{array}{lll}
\sigma_j \sigma_k &=& i^{r_1(\sigma_j)r_2(\sigma_j)+r_1(\sigma_k)r_2(\sigma_k)} X^{r_1(\sigma_j)} Z^{r_2(\sigma_j)} X^{r_1(\sigma_k)} Z^{r_2(\sigma_k)} \\
&=& i^{r_1(\sigma_j)r_2(\sigma_j)+r_1(\sigma_k)r_2(\sigma_k)}(-1)^{r_1(\sigma_k)r_2(\sigma_j)} X^{r_1(\sigma_j)} X^{r_1(\sigma_k)} Z^{r_2(\sigma_j)} Z^{r_2(\sigma_k)} \\
&=& i^{r_1(\sigma_j)r_2(\sigma_j)+r_1(\sigma_k)r_2(\sigma_k) + 2 r_1(\sigma_k)r_2(\sigma_j)} X^{r_1(\sigma_j)+r_1(\sigma_k)} Z^{r_2(\sigma_j)+r_2(\sigma_k)}
\end{array}
\end{equation}
Since the binary representation ignores the overall phase, we find from Eq.~(\ref{eq:prod21}) that
\begin{equation} \label{eq:add1}
r(\sigma_j \sigma_k) \eqmod{2} r(\sigma_j)+r(\sigma_k).
\end{equation}
Multiple applications of this formula yields a similar formula for the product of an arbitrary number of Pauli matrices.

The binary representation is sometimes called the symplectic representation for the following reason. Define a symmetric $2n \times 2n$ matrix, $\Lambda$, as
\begin{equation} \label{eq:lambda}
\Lambda = \left( \begin{array}{c|c} 0 & I \\ \hline I & 0 \end{array} \right),
\end{equation}
where, as before, the lines split the matrix in half. The symplectic inner product of two binary vectors is determined by the $n=1$ version of this matrix as,
\begin{equation} \label{eq:symplectic1}
r(\sigma_j) \Lambda r^T(\sigma_k) = r_1(\sigma_j) r_2(\sigma_k) + r_2(\sigma_j) r_1(\sigma_k) \eqmod{2} \left\{ \begin{array}{ll} 0 & \mbox{if $[\sigma_j,\sigma_k]=0$} \\ 1 & \mbox{if $\{\sigma_j,\sigma_k\}=0$.} \end{array} \right.
\end{equation}
The last equality can be checked simply by plugging in all the binary vectors. Thus the symplectic inner product is a natural one for binary vectors representing Pauli matrices because it evaluates whether the matrices commute or anticommute. Notice that $r(\sigma_j) \Lambda r^T(\sigma_k)$ is simply the function $f(\sigma_j,\sigma_k)$ introduced in Eq.~(\ref{eq:comm1}).

The binary representation of Pauli matrices is generalized to an arbitrary \hyph{$n$}{fold} tensor product, $g$, as follows. Since we can represent each Pauli matrix in the tensor product as a \hyph{two}{dimensional} binary vector, we can represent the entire \hyph{$n$}{fold} tensor product as a \hyph{$2n$}{dimensional} binary vector, $r(g)$. It is conceptually helpful to break up $r(g)$ into its first $n$ components and last $n$ components so that $r(g) = \left( \begin{array}{c|c} r_1(g) & r_2(g) \end{array} \right)$. The binary vectors $r_1(g)$ and $r_2(g)$ generalize the notion of $r_1(\sigma_j)$ and $r_2(\sigma_j)$ for Pauli matrices. The $j$th component of $r_1(g)$ is $1$ only if the $j$th Pauli matrix in $g$ is $X$ or $Y$, while the $j$th component of $r_2(g)$ is $1$ only if the $j$th Pauli matrix in $g$ is $Z$ or $Y$. In other words, the $j$th and $(n+j)$th entries of $r(g)$ give the binary representation of the $j$th Pauli matrix in the \hyph{tensor}{product} decomposition of $g$.

Let us consider a few examples. For two qubits, the binary vector has dimension $2 \times 2 = 4$. Some examples in this case are $r(X \otimes X) = \left( \begin{array}{cc|cc} 1 & 1 & 0 & 0 \end{array} \right)$, $r(X \otimes Y) = \left( \begin{array}{cc|cc} 1 & 1 & 0 & 1 \end{array} \right)$, and $r(X \otimes Z) = \left( \begin{array}{cc|cc} 1 & 0 & 0 & 1 \end{array} \right)$. Since in these examples the first Pauli matrix is always $X$, that means the first entry is always $1$ and the third entry is always $0$. In the \hyph{$n$}{qubit} case, we have $r(I)$ as the zero vector, $r(X^{\otimes n}) = \left( \begin{array}{ccc|ccc} 1 & \cdots & 1 & 0 & \cdots & 0 \end{array} \right)$, $r(Z^{\otimes n}) = \left( \begin{array}{ccc|ccc} 0 & \cdots & 0 & 1 & \cdots & 1 \end{array} \right)$, and $r(Y^{\otimes n})$ as a vector of all $1$'s.

The way we construct $r(g)$ may, at first, seem a bit peculiar. Perhaps the obvious way to generalize the binary vector for a tensor product of Pauli matrices would have been to simply place the binary vector for each Pauli matrix side by side. The reason for our construction, however, becomes more obvious when we consider the symplectic inner product of $r(g)$ and $r(h)$,
\begin{align}
\begin{split}
r(g) \Lambda r^T(h) &= r_1(g) r^T_2(h) + r_2(g) r^T_1(h) \\
&= \sum_{j=1}^n \Bigl( [r_1(g)]_j [r_2(h)]_j + [r_2(g)]_j [r_1(h)]_j \Bigr).
\end{split}
\end{align}
From Eq.~(\ref{eq:symplectic1}), we see that the quantity $[r_1(g)]_j [r_2(h)]_j + [r_2(g)]_j [r_1(h)]_j$ is $0$ or $1$, modulo $2$, depending on whether the $j$th Pauli matrices of $g$ and $h$ commute or not. Therefore, following the discussion after Eq.~(\ref{eq:comm}), we see that $g$ and $h$ commute if $r(g) \Lambda r^T(h)$ is even, and anticommute otherwise. This gives the following result.
\begin{result}
\emph{For two Pauli group elements, $g$ and $h$, we have that}
\begin{equation} \label{eq:symplectic}
r(g) \Lambda r^T(h) \eqmod{2} \left\{ \begin{array}{ll} 0 & \mbox{\emph{if} $[g,h]=0$} \\ 1 & \mbox{\emph{if} $\{g,h\}=0$.} \end{array} \right.
\end{equation}
\end{result}

Eq.~(\ref{eq:add1}) also holds for arbitrary \hyph{$n$}{fold} tensor products of Pauli matrices $g$ and $h$. To see this we write,
\begin{align} \label{eq:paulisfromr}
\begin{split}
&g = i^{\sum_{j=1}^n [r_1(g)]_j [r_2(g)]_j} \bigotimes_{j=1}^n X^{[r_1(g)]_j} Z^{[r_2(g)]_j} \\
&h = i^{\sum_{j=1}^n [r_1(h)]_j [r_2(h)]_j} \bigotimes_{j=1}^n X^{[r_1(h)]_j} Z^{[r_2(h)]_j}.
\end{split}
\end{align}
Now we can generalize Eq.~(\ref{eq:add1}) by considering the product, $gh$,
\begin{equation} \label{eq:add}
\begin{array}{ll}
& gh = i^{\sum_{j=1}^n [r_1(g)]_j [r_2(g)]_j} i^{\sum_{j=1}^n [r_1(h)]_j [r_2(h)]_j} \\
& \phantom{gh = i^{\sum_{j=1}^n [r_1(g)]_j [r_2(g)]_j}}\times \bigotimes_{j=1}^n X^{[r_1(g)]_j} Z^{[r_2(g)]_j} X^{[r_1(h)]_j} Z^{[r_2(h)]_j} \\
\Rightarrow & gh = i^{\sum_{j=1}^n \{ [r_1(g)]_j [r_2(g)]_j + [r_1(h)]_j [r_2(h)]_j + 2 [r_2(g)]_j [r_1(h)]_j \}} \\
& \phantom{gh = i^{\sum_{j=1}^n [r_1(g)]_j [r_2(g)]_j}}\times \bigotimes_{j=1}^n X^{[r_1(g)]_j+[r_1(h)]_j} Z^{[r_2(g)]_j+[r_2(h)]_j}
\end{array}
\end{equation}
This leads us to our next result about binary vectors.
\begin{result}
\emph{For two Pauli group elements, $g$ and $h$, we have that}
\begin{equation}
r(gh) \eqmod{2} r(g)+r(h).
\end{equation}
\end{result}
\noindent This formula is important to the discussion of stabilizer states that follows.

\subsection{Generator matrix \label{subsec:genmatrix}}

So far we have discussed how to represent an \hyph{$n$}{fold} tensor product of Pauli matrices as a $2n$ dimensional binary vector. Our task now is to use this formalism to algebraically describe stabilizer states. Since a stabilizer state is specified by listing $n$ stabilizer generators, we can describe the stabilizer state by writing the $2n$ dimensional binary vector for each generator and then using these vectors as the rows of a $n \times 2n$ matrix. This matrix is called the \emph{generator matrix\/} for the stabilizer state, and is given the symbol $G$,
\begin{equation}
G = \left( \begin{array}{ccc} \mbox{\textemdash} & r(g_1) & \mbox{\textemdash} \\ & \vdots & \\ \mbox{\textemdash} & r(g_n) & \mbox{\textemdash} \end{array} \right),
\end{equation}
where we have labeled the $n$ stabilizer generators as $g_1, \ldots, g_n$. This matrix only specifies the generators up to an overall phase of $\pm 1$, a deficiency addressed in Subsec.~\ref{subsec:signs}.

Let us once again consider the stabilizer state $(1/\sqrt{2}) (\ket{00}-\ket{11})$, with stabilizer generators $g_1 = -X \otimes X$ and $g_2 = Z \otimes Z$. These generators have binary representations $r(g_1) = \left( \begin{array}{cc|cc} 1 & 1 & 0 & 0 \end{array} \right)$ and $r(g_2) = \left( \begin{array}{cc|cc} 0 & 0 & 1 & 1 \end{array} \right)$. Therefore a generator matrix for this state is,
\begin{equation} \label{eq:Gexample}
G = \left( \begin{array}{cc|cc} 1 & 1 & 0 & 0 \\ 0 & 0 & 1 & 1 \end{array} \right).
\end{equation}

Not every $n \times 2 n$ binary matrix is a valid generator matrix; there are two important properties generator matrices must satisfy. First, the stabilizer generators for a stabilizer state must be independent, meaning that no product of them can produce $\pm I$. Since Eq.~(\ref{eq:add}) tells us that taking a product of stabilizer generators corresponds to adding their binary vectors, no rows of $G$ can sum to zero and therefore $G$ must have full rank. The second important property comes from the fact that stabilizer generators must mutually commute. We can calculate commutativity by taking the symplectic product of stabilizer generators, giving us that
\begin{equation} \label{eq:genscommute}
G \Lambda G^T \eqmod{2} 0.
\end{equation}
Any $n \times 2n$ matrix satisfying Eq.~(\ref{eq:genscommute}) and having full rank is a valid generator matrix.

As previously noted, we are free to replace any stabilizer generator by a product of itself and another generator. This means that we are allowed to row reduce the generator matrix, performing addition modulo $2$, without changing the stabilizer state it represents. Furthermore, swapping qubits has the effect of swapping columns of the generator matrix; swapping qubits $j$ and $k$ is equivalent to swapping column $j$ with column $k$ and swapping column $n+j$ with column $n+k$. These operations are sufficient to put the generator matrix into \emph{standard form\/}.
\begin{result}
\emph{To any stabilizer state one can assign a generator matrix in standard form,}
\begin{equation} \label{eq:standardform}
G = \left( \begin{array}{c|c} \begin{array}{cc} I & A \\ 0 & 0 \end{array} & \begin{array}{cc} B & 0 \\ A^T & I \end{array} \end{array} \right).
\end{equation}
\end{result}
\noindent This standard form was obtained by performing row reduction, including swapping qubits, to the left half of $G$. The left half of $G$ need not have full rank, and so it is possible that this process terminates with rows of zeroes as shown. After this, we row reduced the right half of $G$ making the lower right corner the identity matrix, which is possible since $G$ has full rank. Finally, Eq.~(\ref{eq:genscommute}) implies that $A$ and $A^T$ appear as indicated and that $B = B^T$, i.e., $B$ is a symmetric matrix. As an example, we see that the generator matrix in Eq.~(\ref{eq:Gexample}) is already in standard form.

Any set of stabilizer generators that produce a generator matrix in standard form are called \emph{canonical\/} stabilizer generators. It turns out, and this is surprisingly fortunate, that the standard form of a generator matrix is not unique, and hence neither are canonical stabilizer generators. However, in Sec.~\ref{sec:stabilizergraphs} we will learn not only how to generate all sets of canonical stabilizer generators for a given stabilizer state, but also how to efficiently test whether two generator matrices in standard form correspond to the same stabilizer state. Before turning to this, though, we finish up the binary representation of stabilizer states by finally dealing with the sign ambiguity inherent in the formalism.

\subsection{Products of Pauli group elements \label{subsec:signs}}

The reason for this subsection is a bit subtle, so we begin with an example to illustrate the problem. Suppose we had been given the stabilizer generators for $(1/\sqrt{2}) (\ket{00} - \ket{11})$ as $g_1 = Y \otimes Y$ and $g_2 = Z \otimes Z$. The resulting generator matrix looks like,
\begin{equation} \label{eq:badGexample}
G = \left( \begin{array}{cc|cc} 1 & 1 & 1 & 1 \\ 0 & 0 & 1 & 1 \end{array} \right),
\end{equation}
and is not in standard form. This is not a problem because we can row reduce this matrix by adding the second row to the first. The result is the generator matrix in Eq.~(\ref{eq:Gexample}), which is in standard form. By converting the two rows of the matrix in Eq.~(\ref{eq:Gexample}) into tensor products of Pauli matrices, we get canonical generators $g_1' = X \otimes X$ and $g_2' = Z \otimes Z$. Of course, the correct canonical generators are $g_1' = - X \otimes X$ and $g_2' = Z \otimes Z$, which means we lost the overall phase on the first generator. The purpose of this subsection is therefore to determine the overall phase of the canonical generators $g'$ as a function of $r(g')$ and the given generator matrix $G$.

In order to determine the exact form of a canonical stabilizer generator, $g'$, all we need now is the overall phase of $g'$, henceforth called the \emph{sign} of $g'$ and denoted $\sign{g'}$. Since this sign is either $+1$ or $-1$ we will also sometimes find it convenient to write $\sign{g} = {(-1)}^{s_g}$, with the sign of generators being simplified to $\sign{g_j} = {(-1)}^{s_j}$. To determine the sign of $g'$ we will assume that during row reduction we kept track of which stabilizer generators were multiplied together to obtain $g'$. In other words, we know the values $a_j \in \{0,1\}$ such that $g' = g_1^{a_1} \cdots g_n^{a_n}$, where $g_1, \ldots, g_n$ are the given stabilizer generators. We know the canonical generator $g'$ must be of that form because $g_j^2 = I$ and the given generators commute. Thus we will determine sign$(g')$ in terms of the binary vector $a$ whose $j$th entry is $a_j$.

So let us assume we know the values of the $a_j$ such that $g' = g_1^{a_1} \cdots g_n^{a_n}$. In that case we want to generalize the product in Eq.~(\ref{eq:add}) to our situation. Let's begin by just writing down the product, recognizing the dot product that appears.
\begin{equation}
\begin{array}{lll}
g_1^{a_1} \cdots g_n^{a_n} &=& \left( i^{\sum_{j=1}^n [r_1(g_1)]_j [r_2(g_1)]_j} \bigotimes _{j=1}^n X^{[r_1(g_1)]_j} Z^{[r_2(g_1)]_j} \right) ^{a_1} \\
&& \phantom{gh} \cdots \left( i^{\sum_{j=1}^n [r_1(g_n)]_j [r_2(g_n)]_j} \bigotimes _{j=1}^n X^{[r_1(g_n)]_j} Z^{[r_2(g_n)]_j} \right) ^{a_n} \\
&=& i^{\sum_{j=1}^n \{ a_1 [r_1(g_1)]_j [r_2(g_1)]_j + \cdots + a_n [r_1(g_n)]_j [r_2(g_n)]_j \} } \\
&& \phantom{gh}\times \bigotimes_{j=1}^n X^{a_1 [r_1(g_1)]_j} Z^{a_1 [r_2(g_1)]_j} \cdots X^{a_n [r_1(g_n)]_j} Z^{a_n [r_2(g_n)]_j} \\
&=& i^{ a_1 r_1(g_1)r^T_2(g_1) + \cdots + a_n r_1(g_n) r^T_2(g_n) } \\
&& \phantom{gh}\times \bigotimes_{j=1}^n X^{a_1 [r_1(g_1)]_j} Z^{a_1 [r_2(g_1)]_j} \cdots X^{a_n [r_1(g_n)]_j} Z^{a_n [r_2(g_n)]_j} \\
&=& i^{\sum_{j=1}^n a_j r_1(g_j)r^T_2(g_j)} \\
&& \phantom{gh}\times \bigotimes_{j=1}^n X^{a_1 [r_1(g_1)]_j} Z^{a_1 [r_2(g_1)]_j} \cdots X^{a_n [r_1(g_n)]_j} Z^{a_n [r_2(g_n)]_j}.
\end{array}
\end{equation}
In the above formula we do not include a possible sign of the generators. However, if the generators do have a sign, we can simply put those signs in at the end. Now we want all the $X$'s together and $Z$'s together. It is clear that $X^{a_k [r_1(g_k)]_j}$ needs to commute through the $k-1$ $Z$ operators that come before it, giving us
\begin{equation} \label{eq:almostsign}
\begin{array}{lll}
g' &=& i^{\sum_{j=1}^n a_j r_1(g_j)r^T_2(g_j)} (-1)^{\sum_{j=1}^n \sum_{l<k} a_k a_l [r_1(g_k)]_j [r_2(g_l)]_j} \\
&& \phantom{gh}\times \bigotimes_{j=1}^n X^{a_1 [r_1(g_1)]_j+ \cdots +a_n [r_1(g_n)]_j} Z^{a_1 [r_2(g_1)]_j+ \cdots +a_n [r_2(g_n)]_j} \\
&=& i^{\sum_{j=1}^n a_j r_1(g_j)r^T_2(g_j) + 2 \sum_{k>l} a_k a_l r_1(g_k) r^T_2(g_l)} \bigotimes_{j=1}^n X^{r_1(g')} Z^{r_2(g')}.
\end{array}
\end{equation}
If we make the simple observation that
\begin{equation}
\label{eq:prodofpauli}
g' = \sign{g'} \times i^{r_1(g')r^T_2(g')} \bigotimes_{j=1}^n X^{r_1(g')} Z^{r_2(g')},
\end{equation}
then we can immediately pull off sign$(g')$ from Eq.~(\ref{eq:almostsign}) to get
\begin{equation} \label{eq:badsign}
\sign{g'} = i^{\sum_{j=1}^n a_j r_1(g_j)r^T_2(g_j) + 2 \sum_{k>l} a_k a_l r_1(g_k) r^T_2(g_l)} i^{-r_1(g')r^T_2(g')}.
\end{equation}
Ostensibly we are now done. However, Eq.~(\ref{eq:badsign}) can be simplified considerably with some basic, albeit tricky, algebra. Given the unsightly nature of Eq.~(\ref{eq:badsign}), we now carry out this simplification to obtain the final result of this subsection.

Using the notation, introduced in Sec.~\ref{sec:conventions}, that quantities underneath an overbar are understood to be evaluated modulo $2$, our first task is to simplify $2 \sum_{k>l} a_k a_l r_1(g_k)$ $r^T_2(g_l)$. This only needs to be known modulo $2$, and since $\overline{\sum_j f(j)} \eqmod{2} \overline{\sum_j \overline{f(j)}}$, it is sufficient to simplify the quantity $2 \sum_{k>l} a_k a_l \overline{r_1(g_k) r^T_2(g_l)}$. This allows us to use the fact that the generators commute, i.e., from Eq.~(\ref{eq:symplectic})
\begin{equation}
\overline{r_1(g_k) r^T_2(g_l)} = \overline{r_1(g_l) r^T_2(g_k)},
\end{equation}
to write,
\begin{equation} \label{eq:splitup}
\overline{r_1(g_k) r^T_2(g_l)} = \frac{1}{2} \left( \overline{r_1(g_k) r^T_2(g_l)} + \overline{r_1(g_l) r^T_2(g_k)} \right)
\end{equation}
Using Eq.~(\ref{eq:splitup}) we can simplify $2 \sum_{k>l} a_k a_l \overline{r_1(g_k) r^T_2(g_l)}$ because
\begin{equation} \label{eq:simplesum}
\begin{array}{ll}
2 \sum_{k>l} a_k a_l \overline{r_1(g_k) r^T_2(g_l)} &= \sum_{k>l} \{ a_k a_l \overline{r_1(g_k) r^T_2(g_l)} + a_k a_l \overline{r_1(g_l) r^T_2(g_k)} \} \\
&= \sum_{l,k=1}^n a_k a_l \overline{r_1(g_k) r^T_2(g_l)} - \sum_{k=1}^n a_k \overline{r_1(g_k) r^T_2(g_k)}.
\end{array}
\end{equation}
Plugging Eq.(\ref{eq:simplesum}) into Eq.~(\ref{eq:badsign}) yields the moderately nicer looking formula,
\begin{equation} \label{eq:nearlysign}
\sign{g'} =i^{\sum_{j=1}^n a_j \left\{ r_1(g_j) r^T_2(g_j) - \overline{r_1(g_j) r^T_2(g_j)} \right\}} i^{\sum_{j,k} a_j a_k \overline{r_1(g_j) r^T_2(g_k)}} i^{-r_1(g')r^T_2(g')}
\end{equation}
The best way to write this, though, is to recognize that we can split up the given generator matrix $G$ as
\begin{equation} \label{eq:twohalves}
G = \left( \begin{array}{c|c} G_1 & G_2 \end{array} \right).
\end{equation}
The combination $T = G_1 G^T_2$ is really the quantity of interest in these formulas for sign$(g')$. The reason for this is that
\begin{equation}
T_{jk} = \sum_{l=1}^n (G_1)_{jl} (G_2)_{kl} = \sum_{l=1}^n [r_1(g_j)]_l [r_2(g_k)]_l = r_1(g_j) r^T_2(g_k)
\end{equation}
So in terms of this $T$ matrix we have that
\begin{equation} \label{eq:sign}
\begin{array}{lll}
\sign{g'} &=& i^{\sum_{j=1}^n a_j ( T_{jj} - \overline{T}_{jj} )} i^{\sum_{j,k} a_j a_k \overline{T}_{jk}} i^{-r_1(g')r^T_2(g')} \\
&=& i^{\mbox{\scriptsize{Tr}}[T_a-\overline{T}_a+\overline{T}_a^2]} i^{-r_1(g')r^T_2(g')},
\end{array}
\end{equation}
where $T_a$ is the $T$ matrix restricted to those rows and columns corresponding to nonzero entries of $a$. In that last step we also used the fact that, for the binary matrix $\overline{T}_a$, we have $\sum_{jk} (\overline{T}_a)_{jk} = \tr{\overline{T}_a^2}$. To truly complete the formula, we should note that if the generators themselves have a sign then the signs of the generators needed to obtain $g'$ should be multiplied in front of Eq.~(\ref{eq:sign}). This is the final result of this subsection, and it is a quite useful result of which we often make use.

Just to make sure we did not get lost in all the algebra, let us use this formula for its originally intended purpose. The example we began with was the generator matrix given in Eq.~(\ref{eq:badGexample}). We wished to find the canonical generators for the stabilizer state specified by this generator matrix. After row reducing this matrix, we found that the canonical generator matrix for the state was the one given in Eq.~(\ref{eq:Gexample}), with canonical generators $g'_1 = g_1 g_2$ and $g'_2 = g_2$. In this case it is simple enough to multiply out $g_1 g_2$ to determine the first canonical generator. Since this is not always so simple, we could alternatively deduce this sign from Eq.~(\ref{eq:sign}). We note that $a = \left( \begin{array}{cc} 1 & 1 \end{array} \right)$ and we calculate
\begin{equation}
T_a = \left(\begin{array}{cc} 1 & 1 \\ 0 & 0 \end{array} \right) \left(\begin{array}{cc} 1 & 1 \\ 1 & 1 \end{array} \right) = \left(\begin{array}{cc} 2 & 2 \\ 0 & 0 \end{array} \right) \mbox{ and } \overline{T}_a = \left(\begin{array}{cc} 0 & 0 \\ 0 & 0 \end{array} \right).
\end{equation}
Hence we get, using $r_1(g'_1) r^T_2(g'_1) = 0$, that $\sign{g'_1} = i^{\mbox{\scriptsize{Tr}}[T_a-0+0]}i^{-0} = i^{2} = -1$.
While calculating the formula in this simple case was overkill, for more complicated cases it has to potential to save considerable effort.

\subsection{Canonical generator sets \label{subsec:canongen}}

Whenever we deal with stabilizer states we always describe them in terms of canonical generators, and the last few sections were devoted to showing how one can find canonical generators. This subsection will illustrate what one can do once canonical generators have been found.

Given a generator matrix, $G$, in canonical form, we define the canonical \emph{dual\/} generator matrix to be
\begin{equation}
H = \left( \begin{array}{cc|cc} 0 & 0 & I & 0 \\ 0 & I & 0 & 0 \end{array} \right).
\end{equation}
It is simple to check that the dual generator matrix $H$ satisfies the constraints in Subsec.~\ref{subsec:genmatrix} to be a generator matrix. It is also straightforward to check that $G \Lambda H^T \eqmod{2} I$. This means that a dual generator, $h_j$, commutes with all but one of the generators, $g_j$, with which it anticommutes. Dual generators with this property exist for any set of stabilizer generators, but for our canonical generators we have the dual generators taking a particularly simple form. The importance of dual generators is that they reveal the connection between $a$ and $r(g)$, where $g = g_1^{a_1} \cdots g_n^{a_n}$. Before, we assumed that we knew $a$ for a given $g$, but now with dual generators we can derive a simple relation.

The fact that $g = g_1^{a_1} \cdots g_n^{a_n}$ gives us $r(g)$ once we know $a$, since
\begin{equation}
r(g) \eqmod{2} \sum_{j=1}^n a_j r(g_j) = a G = \left( \begin{array}{cc} a & 0 \end{array} \right) \left( \begin{array}{cc} G \\ H \end{array} \right).
\end{equation}
This final equality turns out to be invertible, since from $G \Lambda H^T \eqmod{2} I$ we have
\begin{align} \label{eq:GHinverse}
\begin{split}
&\left( \begin{array}{c} G \\ H \end{array} \right) \Lambda \left( \begin{array}{cc} G^T & H^T \end{array} \right) \eqmod{2} \Lambda \\
\Leftrightarrow &\left( \begin{array}{c} G \\ H \end{array} \right)^{-1} \eqmod{2} \Lambda \left( \begin{array}{cc} G^T & H^T \end{array} \right) \Lambda,
\end{split}
\end{align}
where $\Lambda$ is the matrix in Eq.~(\ref{eq:lambda}). Hence we can find $a$ once we are given $r(g)$ by
\begin{align} \label{eq:finda}
\begin{split}
\left( \begin{array}{cc} a & 0 \end{array} \right) &\eqmod{2} r(g) \Lambda \left( \begin{array}{cc} G^T & H^T \end{array} \right) \Lambda,
\end{split}
\end{align}
producing the following result.
\begin{result}
\emph{For any stabilizer element $g = g_1^{a_1} \cdots g_n^{a_n}$, where $g_1, \ldots, g_n$ are canonical generators, we have that}
\begin{equation} \label{eq:a}
a = r(g) \Lambda H^T.
\end{equation}
\end{result}
\noindent We should note that this equation holds without using modulo $2$ arithmetic. The reason is because all entries of $r(g) \Lambda H^T$ are already either $1$ or $0$ due the form of $H$. The second condition from Eq.~(\ref{eq:finda}), that $r(g) \Lambda G^T \eqmod{2} 0$, simply states that $g$ must commute with all the stabilizer generators in order to be a stabilizer element.

Now the only missing piece of information about $g$ is its sign. However, this is remedied through the application of Eq.~(\ref{eq:sign}), with $\sign{g} = {(-1)}^{s_g}$ and $\sign{g_j} = {(-1)}^{s_j}$. We calculate that
\begin{equation}
T = \left( \begin{array}{cc} I & A \\ 0 & 0 \end{array} \right) \left( \begin{array}{cc} B & A \\ 0 & I \end{array} \right) = \left( \begin{array}{cc} B & 2 A \\ 0 & 0 \end{array} \right) \Rightarrow \overline{T} = \left( \begin{array}{cc} B & 0 \\ 0 & 0 \end{array} \right).
\end{equation}
This leads us to our next result.
\begin{result}
\label{res:sign}
\emph{Let $g = g_1^{a_1} \cdots g_n^{a_n}$, where $g_1, \ldots, g_n$ are canonical generators whose signs are specified by the entries of a binary vector $s$. Then,}
\begin{align} \label{eq:qmsign}
\begin{split}
2 s_g &\eqmod{4} 2 as^T + a \left( \begin{array}{cc} B & 0 \\ 0 & 0 \end{array} \right) a^T - r_1(g) r_2^T(g) \\
&\eqmod{4} 2 r(g) (s H \Lambda)^T + r_1(g) \left( \begin{array}{c|c} \begin{array}{cc} B & 0 \\ 0 & 0 \end{array} & -I \end{array} \right) r^T(g).
\end{split}
\end{align}
\end{result}
\noindent In the final equality we simply used Eq.~(\ref{eq:a}) to eliminate $a$ in favor of $r(g)$.

\section{Unitary operations and measurements}

Unitary operations and measurements are the typical actions on quantum states. Indeed any quantum operation be thought of as a unitary operation followed by a projective measurement whose outcome is ignored~\cite{peres:book}. For stabilizer states, we want to consider those operations and measurements that are natural in the context of the formalism we have developed. We begin this section by discussing the unitary evolution of stabilizer states and then we end the section by learning about projective measurements.

\subsection{Clifford operations}

This subsection concerns itself with the issue of unitary operations on stabilizer states. Given a stabilizer state $\ket{\psi}$ and a unitary operator $U$, the state resulting from the application of the unitary to the state, i.e., $U \ket{\psi}$, is generally not a stabilizer state. In other words, general unitary transformations take us out of the stabilizer formalism. In order to stay in the stabilizer formalism, we can only consider unitary transformations that take any stabilizer state to another stabilizer state. The group, and it is a group, of unitary operations that does this is called the \emph{Clifford group\/}. Elements of the Clifford group are called \emph{Clifford operations\/}, \emph{Clifford elements\/}, or \emph{Clifford unitaries\/}.

We now work towards a better understanding of the Clifford group. Since stabilizer states are described via stabilizer generators, we must know how stabilizer generators transform under Clifford operations. So let $\ket{\psi}$ be a stabilizer state, $U$ be a Clifford unitary, and $g_1, \ldots, g_n$ be stabilizer generators for $\ket{\psi}$. From these generators we can construct the stabilizer generators which stabilize $U \ket{\psi}$ by conjugating the stabilizer generators of $\ket{\psi}$, yielding the operators $Ug_1U^{\dagger}, \ldots, Ug_nU^{\dagger}$. It is simple to verify that these are stabilizers of the new state since $(U g_j U^{\dagger}) U \ket{\psi} = U g_j \ket{\psi} = U \ket{\psi}$. The generators $U g_j U^{\dagger}$ commute since $[U g_j U^{\dagger},U g_j U^{\dagger}] = U [g_j,g_j] U^{\dagger}=0$. They are also independent since if $\prod_{j=1}^n (U g_j U^{\dagger})^{a_j} = \pm I$ then it follows that $\prod_{j=1}^n g_j^{a_j} = \pm I$ as well. Thus we have indeed constructed a set of stabilizer generators for the new state.

Since we want $U \ket{\psi}$ to be a stabilizer state, we require that $U g_j U^{\dagger}$ be in the Pauli group. This means that the unitary operators $U$ map elements of the Pauli group, $g_j$, to elements of the Pauli group under conjugation. In group theory terms we say that the operators, $U$, \emph{normalize} the Pauli group. Indeed, the Clifford group is defined to be the normalizer of the Pauli group, meaning it consists of all unitaries that normalize the Pauli group. This group of unitary operators is a natural one to consider in the context of stabilizer states.

As an example, we point out that the Pauli group elements themselves are in the Clifford group. For instance, since two Pauli group elements $g$ and $h$ either commute or anticommute, we have that $g h g^{\dagger} = \pm h$. There are, not surprisingly, other elements of the Clifford group as explained below.

\subsection{Generators of the Clifford group}

The Clifford group is generated by the generators of the local Clifford group plus one additional \hyph{two}{qubit} Clifford operator~\cite{eastin:thesis}. Local Clifford operations are those elements of the Clifford group that act on a single qubit. Put another way, the local Clifford group is the normalizer of $\mathcal{P}_1$, and hence maps Pauli matrices to Pauli matrices under conjugation. Local Clifford operations are generated by,
\begin{equation}
\begin{array}{ccc}
H = \frac{1}{\sqrt{2}} \left( \begin{array}{cc} 1 & 1 \\ 1 & -1 \end{array} \right) & \mbox{and} & S = \left( \begin{array}{cc} 1 & 0 \\ 0 & i \end{array} \right),
\end{array}
\end{equation}
meaning that any local Clifford operation can be written as a product of $H$ and $S$ operators. We give these operators the names \emph{Hadamard\/} operator, for the $H$ matrix, and \emph{phase\/} operator, for the $S$ matrix.

The $H$ and $S$ operators are indeed elements of the local Clifford group. We can verify this by calculating,
\begin{equation} \label{eq:HSaction}
\begin{array}{llll}
HIH = I, & HXH = Z, & HYH = -Y, & HZH = X, \\
SIS^{\dagger} = I, & SXS^{\dagger} = Y, & SYS^{\dagger} = -X, & \mbox{and }SZS^{\dagger} = Z,
\end{array}
\end{equation}
where we use the fact that $H^{\dagger} = H$ to eliminate $H^{\dagger}$. In words, $H$ interchanges $X$ and $Z$, giving $Y$ a phase, and $S$ interchanges $X$ and $Y$, possibly adding a phase, and leaves $Z$ unchanged. In the binary representation, Eq.~(\ref{eq:HSaction}) becomes
\begin{equation} \label{eq:binaryHS}
\begin{array}{lll}
\left( \begin{array}{c|c} r_1(\sigma_j) & r_2(\sigma_j) \end{array} \right) & \underleftrightarrow{\mbox{\footnotesize{$H$}}} & \left( \begin{array}{c|c} r_2(\sigma_j) & r_1(\sigma_j) \end{array} \right) \\
\left( \begin{array}{c|c} r_1(\sigma_j) & r_2(\sigma_j) \end{array} \right) & \underleftrightarrow{\mbox{\footnotesize{$S$}}} & \left( \begin{array}{c|c} r_1(\sigma_j) & \overline{r_1(\sigma_j) + r_2(\sigma_j)} \end{array} \right),
\end{array}
\end{equation}
for any Pauli matrix $\sigma_j$ with $j=0,1,2,3$. Eqs.~(\ref{eq:HSaction}) and (\ref{eq:binaryHS}) verify that $H$ and $S$ are in the local Clifford group.

We will not prove that $H$ and $S$ generate the local Clifford group. For a proof of this fact the interested reader is referred to Refs.~\cite{eastin:thesis,gottesman:thesis,gottesman:tolerant}. The basic idea behind the proof is to show that the local Clifford group contains $24$ elements and that $H$ and $S$ generate them. The reason we omit the details is that there is nothing new this work would contribute to the proof; it would simply be a restatement of the referenced papers.

Refs.~\cite{eastin:thesis,gottesman:thesis,gottesman:tolerant} also prove that only one more unitary operation is needed to generate the entire Clifford group. There is some freedom in which extra unitary to add, but we choose this extra unitary to be the \hyph{two}{qubit} \emph{controlled-Z} operation,
\begin{equation}
\CZ = \left( \begin{array}{cc} I & 0 \\ 0 & Z \end{array} \right) = \left( \begin{array}{cccc} 1 & 0 & 0 & 0 \\ 0 & 1 & 0 & 0 \\ 0 & 0 & 1 & 0 \\ 0 & 0 & 0 & -1 \end{array} \right).
\end{equation}
The two qubits acted on by this transformation can be split up into a control qubit and a target qubit. If the control qubit is in the $\ket{0}$ state, $I$ is applied to the target qubit. Only when the control qubit is in the $\ket{1}$ state is the $Z$ operation applied to the target. General controlled operations act in the same way, with a controlled-$U$ operation being written as $\C{U}$. The controlled-$Z$ operation is special in that it happens to be symmetric in which qubit is the control and which is the target.

\subsection{Pauli measurements \label{subsec:measure}}

When a measurement described by a Hermitian matrix $M$ is made on a stabilizer state we want the \hyph{post}{measurement} state to also be a stabilizer state. We will see that the measurements satisfying this property correspond to tensor products of the identity and the Pauli matrices. Certainly, since the \hyph{post}{measurement} state is an eigenstate of the operator, the operator we measure must be a tensor product of the identity and the Pauli matrices in order for the \hyph{post}{measurement} state to be a stabilizer state. What we show below is that, in fact, the \hyph{post}{measurement} state is a stabilizer state every time such a Pauli product is measured. We could also consider the negatives of tensor products of the identity and the Pauli matrices, but we neglect this case because the results are the same as the case we consider only with the probabilities of the two outcomes interchanged.

So, suppose we want to measure a Hermitian Pauli group element, $M$, with sign $+1$. If we suppose that all the stabilizer generators for the initial state $\ket{\psi}$ commute with $M$ then $\pm M$ must be a stabilizer element, since adding $M$ to the list of generators yields $n+1$ generators for an \hyph{$n$}{qubit} state. We know that adding independent generators cuts the stabilized dimension in half, so $M$ cannot be independent of the existing generators. Thus $M = (-1)^m g_1^{m_1} \cdots g_n^{m_n}$, with $m, m_j \in \{0,1\}$ determining how the generators multiply to produce $M$. This gives us that $M \ket{\psi} = (-1)^m g_1^{m_1} \cdots g_n^{m_n} \ket{\psi} = (-1)^m \ket{\psi}$. A measurement of $M$ therefore gives result $(-1)^m$ with probability $\bra{\psi} (1/2) (I+(-1)^m M) \ket{\psi} = (1/2) (1+(-1)^m (-1)^m) = 1$, and the \hyph{post}{measurement} state is $(1/2) (I+(-1)^m M) \ket{\psi} = \ket{\psi}$.

The case we want to consider now is when $M$ anticommutes with at least one generator $g$. The first point of business is then to find an appropriate set of generators for the stabilizer state. For each given stabilizer generator, $g_j$, $M$ either commutes with $g_j$ or anticommutes with it. If $M$ anticommutes with a generator $g_j$, then $Mg_j g = -g_j M g = g_j g M$ and so $M$ commutes with the product $g_j g$. Therefore if we replace $g_j$ with $g_j g$ in our generator list, we end up replacing a generator that anticommutes with $M$ with a generator that commutes with $M$. In this way we can come up with a set of generators such that only one anticommutes with $M$.

Now we claim that these generators, $g_j \ne g$, that commute with $M$ are valid stabilizer generators of the \hyph{post}{measurement} state. For, consider the action of $g_j$ on the \hyph{un}{normalized} \hyph{post}{measurement} state, $g_j (1/2)(I + (-1)^m M) \ket{\psi} = (1/2)(I + (-1)^m M) g_j \ket{\psi} = (1/2)(I + (-1)^m M) \ket{\psi}$, where $(-1)^m$ is the measurement outcome. Now we only need one more independent stabilizer generator to completely specify the \hyph{post}{measurement} state. It should be clear now that $M$ performs this role. $M$ must certainly be independent of all the generators because it anticommutes with $g$. We also know that $(-1)^m M$ stabilizes the \hyph{post}{measurement} state because $(-1)^m M (1/2)(I + (-1)^m M) \ket{\psi} = (1/2)((-1)^m M + I) \ket{\psi} = (1/2)(I + (-1)^m M) \ket{\psi}$. Thus if we obtain the measurement result $(-1)^m$, we find independent, commuting stabilizer generators for the \hyph{post}{measurement} state to be $(-1)^m M$ and all $g_j \ne g$. Furthermore, it follows that $\bra{\psi} M \ket{\psi} = \bra{\psi} M g \ket{\psi} = \bra{\psi} -g M \ket{\psi} = - \bra{\psi} M \ket{\psi} \Rightarrow \bra{\psi} M \ket{\psi} = 0$, so the measurement outcomes are $\pm 1$ with equal probability. 

To summarize, the action of measuring a Hermitian Pauli group element, $M$, on a stabilizer state is as follows. Find a set of generators for the state such that at most one anticommutes with $M$. If all the stabilizer generators commute with $M$, then the outcome is certain and the state is unchanged. If one generator anticommutes with $M$, then replacing that generator by $\pm M$ gives a new set of generators that stabilize the \hyph{post}{measurement} state. In this case the measurement outcome is completely random.

An important special case is to consider the action of individual Pauli measurements on stabilizer states. These are precisely the kinds of measurements considered in Chapter~\ref{chap:cclhvmodels}. If we measure a Pauli operator $\sigma = X, Y, \mbox{or } Z$ on qubit $j$ the measurement operator looks like $M = I \otimes \cdots \otimes I \otimes \sigma \otimes I \otimes \cdots \otimes I$, with $\sigma$ occurring in the $j$th spot. This operator commutes with exactly those generators whose tensor product decomposition have either a $\sigma$ or an $I$ in the $j$th spot, and anticommutes with all the others. The generators for the \hyph{post}{measurement} state, except for $\pm M$, can all be chosen to have an $I$ in the $j$th spot.

\subsection{Measurement correlations}

The remarkable thing about quantum measurements is that they exhibit correlations that are stronger than those allowed classically. We will see why this is true in Chapter~\ref{chap:localrealism}, but for now we simply examine some basic properties of quantum measurement correlations.

We can only speak of correlations for two commuting Hermitian operators, $A$ and $B$. In this case the correlation between $A$ and $B$ comes from calculating the average of the product $AB$, which is Hermitian from our assumptions. The average value obtained by measuring the product $AB$ can be calculated by measuring $A$ and $B$ separately. This value is equal to measuring $A$ and measuring $B$, which can be done since $A$ and $B$ commute, and then multiplying together the results and averaging over many trials. This is not generally the same as multiplying the averages of $A$ and $B$, i.e., mathematically $\avg{AB} \ne \avg{A} \avg{B}$. In the case that they are the same, then $A$ and $B$ outcomes are independent of each other and so $A$ and $B$ are not correlated. The generic case allows for $A$ and $B$ to be correlated.

To illustrate this, we show how individual Pauli measurements can build up the statistics of any other measurement that is a tensor product of Pauli matrices. Consider, for example, a measurement of $\sigma \otimes \sigma'$ with $\sigma, \sigma' = X, Y, \mbox{or } Z$. The probability that the outcome is $+1$ is
\begin{equation}
\begin{array}{ll}
\bra{\psi} \frac{1}{2}\left( I \otimes I + \sigma \otimes \sigma' \right) \ket{\psi} &= \bra{\psi} \frac{1}{4}\left( I \otimes I + \sigma \otimes I \right) \left( I \otimes I + I \otimes \sigma' \right) \ket{\psi} \\
&+ \bra{\psi} \frac{1}{4}\left( I \otimes I - \sigma \otimes I \right) \left( I \otimes I - I \otimes \sigma' \right) \ket{\psi}.
\end{array}
\end{equation}
The quantity on the right is expressed in terms of various outcomes for measurements of the individual Pauli operators $\sigma \otimes I$ and $I \otimes \sigma'$. Specifically, we took the correlation between both outcomes being $+1$ and both outcomes being $-1$, since these are the cases in which their product is $+1$. Thus the statistics of $\sigma \otimes \sigma'$ can be obtained by looking at the average of the product of the individual Pauli measurements involved. However, we stress that while measuring $\sigma \otimes I$ and $I \otimes \sigma'$ gives the same statistics as measuring $\sigma \otimes \sigma'$, the \hyph{post}{measurement} states in these two scenarios are quite different.

To review, the essential point of this subsection is that one can look at the correlations between two commuting measurements. This is done by averaging over many trials the product of the measurement outcomes, rather than taking the product of the average of each measurement by itself. Mathematically, we account for correlations by taking the average value after multiplying the operators together, rather than multiplying together the average values of the individual operators. We also found that correlations between measurement outcomes are equivalent to correlations resulting from measuring a new operator which is the product of the two commuting operators whose outcome correlations are being studied.

\section{\hyph{Stabilizer}{state} graphs \label{sec:stabilizergraphs}}

The traditional representation of stabilizer states is to list $n$ commuting, independent stabilizer generators for the state. In this section we introduce an alternative description in which stabilizer states are represented by \emph{graphs\/} \cite{elliott:graphs}. We will see that this graphical representation of stabilizer states is immensely powerful, not only calling on results from the binary representation of stabilizer states but also drawing strength from the circuit model of quantum computation.

\subsection{Graph theory \label{subsec:graphtheory}}

\begin{figure}
\center
\includegraphics[width=12.5cm]{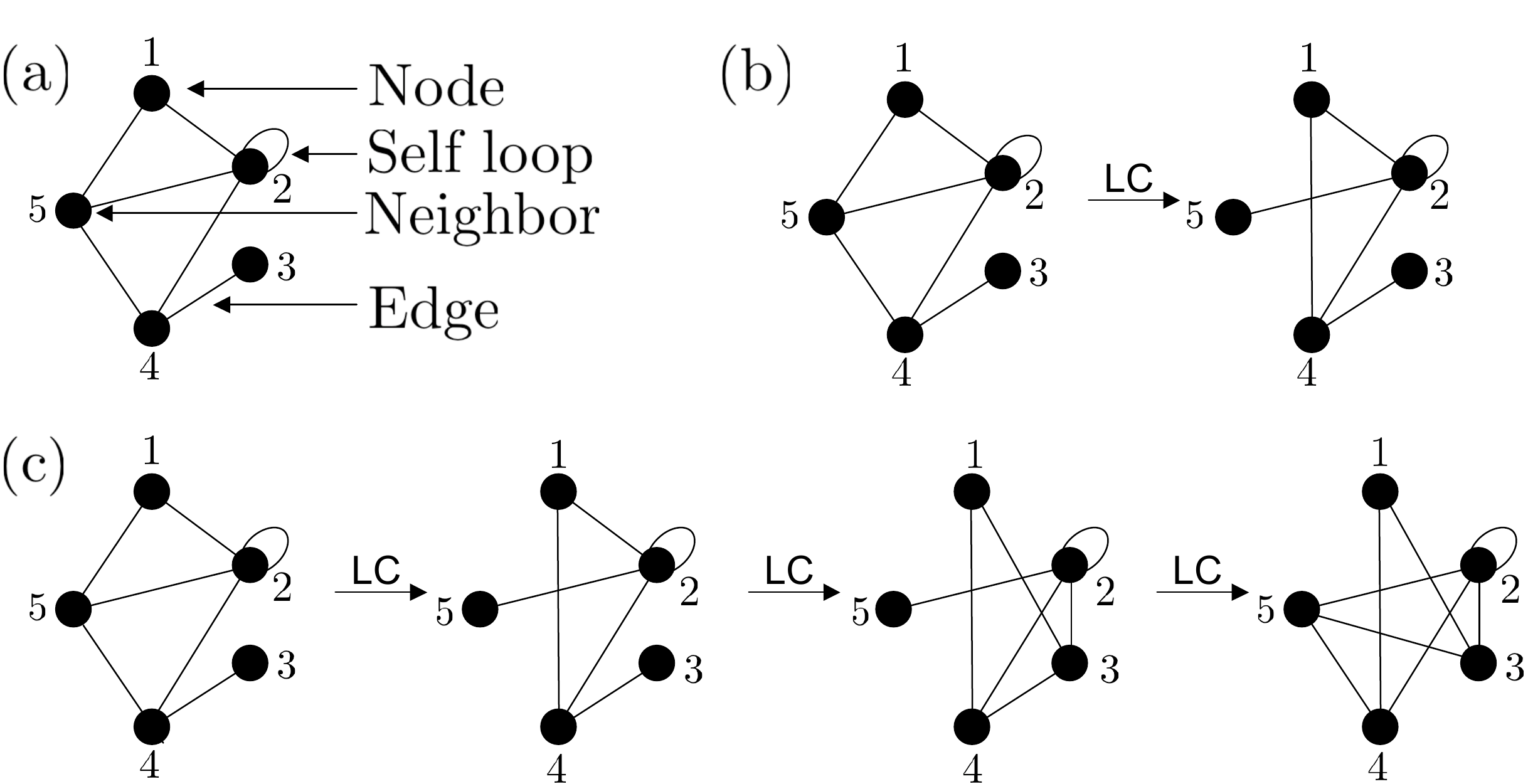}
\caption[Graph theory terminology]{The graph in (a) is used to illustrate the basic terms used to describe a graph. Each black dot is a node, with node number $2$ having a self loop. We indicate in (a) one of the neighbors of node $2$. Finally, we represent edges simply with a line that connects two nodes. In (b) we take the graph in (a) and perform local complementation on node $2$, the one with a loop. We do not consider node $2$ its own neighbor so it has only three neighbors, nodes $1$, $4$, and $5$. In local complementation we complement the edges between these neighbors. This means that the edges between pairs of nodes $\{1,5\},\{1,4\}$, and $\{4,5\}$ are complemented. The result is that the edges between nodes $1$ and $5$ and between nodes $4$ and $5$ are removed, since these nodes are connected in the graph in (a), and an edge is added between nodes $1$ and $4$ since they are not connected in (a). Finally, in (c), we take the graph in (a) and perform local complementation along the edge connecting nodes $2$ and $4$. This is done by first performing local complementation on node $2$, then performing local complementation on node $4$, and then performing local complementation on node $2$ again. If we had instead chosen to first perform local complementation on node $4$, then the intermediate stages would have looked different but the final graph would have been the same. \label{fig:graphterms}}
\end{figure}

There is a considerable collection of terminology associated with graph theory~\cite{diestel:graphtheory}. Fortunately, most of this terminology is intuitive, with only a few exceptions. All that aside, learning this terminology is central for being able to manipulate stabilizer states through their graphs. Hopefully Fig.~\ref{fig:graphterms} will help in this regard.

The simplest kind of graph is portrayed pictorially as points, called \emph{nodes\/}, connected by lines, called \emph{edges\/}. Two nodes that are connected by an edge are called \emph{neighbors\/}, while the \emph{neighborhood\/} of node $j$, denoted $\mathcal{N}(j)$, is a set including all of that node's neighbors. The \emph{adjacency matrix\/}, so called because its elements are determined by which nodes are neighbors, and therefore are adjacent to each other, is capable of completely describing any graph. The adjacency matrix, $\Gamma$, for an \hyph{$n$}{node} graph is an $n \times n$ binary matrix defined as
\begin{equation} \label{eq:adjacencydef}
\Gamma_{jk} = \left\{ \begin{array}{ll} 1 & \mbox{if nodes $j$ and $k$ are connected} \\ 0 & \mbox{otherwise.} \end{array} \right.
\end{equation}
A $1$ on the diagonal of an adjacency matrix is drawn as a \emph{self loop\/}, but in such cases we still do not consider that node as its own neighbor. Also note that adjacency matrices are symmetric, meaning $\Gamma^T = \Gamma$. It is clear that graphs and adjacency matrices contain the same amount of information.

There are a few important operations on graphs that we need to know about. The most basic operation is called \emph{complementation\/}. Complementing the edge between two nodes removes the edge if one is present and adds one otherwise. Note that an edge need not exist in order for it to be complemented. From this operation we construct two types of \emph{local complementation\/}, local complementation on a node and local complementation along an edge. Performing local complementation on a node complements the edges between all of the node's neighbors. Local complementation along an edge is equivalent to a sequence of local complementations at the nodes defining the edge. The sequence is as follows: first perform local complementation on one of the nodes, then local complement at the other node, and finally local complement at the first node again. Local complementation along an edge is symmetric in the two nodes defining the edge, so it does not matter at which node local complementation is first performed.

\subsection{Graph states \label{subsec:graphstates}}

Our scheme for drawing graphs corresponding to stabilizer states derives from a class of stabilizer states called \emph{graph states\/}, which are privileged with a natural graphical representation~\cite{hein:entanglement,hein:graphstates}. Consider a generator matrix $G$ in standard form, Eq.~(\ref{eq:standardform}). Let us call the rank of the left $n \times n$ matrix -- this is the $G_1$ matrix in Eq.~(\ref{eq:twohalves}) -- the \emph{left rank\/} of $G$. If the left rank of $G$ is $n$, then the canonical form of $G$ is
\begin{equation} \label{eq:graphmatrix}
G = \left( \begin{array}{c|c} I & B \end{array} \right),
\end{equation}
with the constraint that $B^T = B$. Since $B$ is a symmetric binary matrix, it is the adjacency matrix of some graph. Furthermore, if $B$ has all $0$'s on the diagonal the graph has no self loops. In the case where the generator matrix has full left rank, and where $B$ has all $0$'s on the diagonal, the generator matrix represents a stabilizer state called a graph state. We could just as well define a graph state to be any state whose generator matrix has full left rank, but for historical reasons graph states are defined not to have self loops. A graph state is completely determined by the graph with adjacency matrix $\Gamma = B$, and they are clearly a proper subset of all stabilizer states.

From Eqs.~(\ref{eq:adjacencydef}) and (\ref{eq:graphmatrix}), we can write the canonical stabilizer generators for a graph state as
\begin{equation} \label{eq:graphgens}
g_j = X_j \prod_{k \in \mathcal{N}(j)} Z_k.
\end{equation}
Recognizing that $\CZ (X \otimes I) \CZ = X \otimes Z$, we can see that graph states are the result of beginning with generators $g_j = X_j$ and then applying $\CZ$ between qubits corresponding to neighbors in the graph. The generators $g_j = X_j$ correspond to the state $\ket{+}^{\otimes n}$, so a graph state's graph can be regarded as a blueprint to build the graph state from a simple initial state.

\subsection{Graphs from generator matrices \label{subsec:gengraphs}}

We now have all the necessary tools needed to assign a graph to a general stabilizer state. 
Any stabilizer state with a generator matrix of canonical form can be converted by local Clifford operations to a state possessing a generator matrix with the form in Eq.~(\ref{eq:graphmatrix}). From Eq.~(\ref{eq:binaryHS}), applying Hadamard operations to the last $n-r$ qubits of the stabilizer state, where $r$ is the initial left rank of the generator matrix, exchanges columns $r+1$ through $n$ in the generator matrix with columns $n+r+1$ through $2n$, so that a generator matrix as in Eq.~(\ref{eq:standardform}) is transformed to
\begin{equation} \label{eq:graphform}
G = \left( \begin{array}{cc|cc} I & 0 & B & A \\ 0 & I & A^T & 0 \end{array} \right).
\end{equation}
The diagonal of $B$ in this generator matrix can then be stripped of $1$'s without otherwise changing the generator matrix by applying $S$ to offending qubits and using Eq.~(\ref{eq:binaryHS}). The resulting generator matrix has the form of a graph state and corresponds to a stabilizer state that differs from that represented by Eq.~(\ref{eq:standardform}) by at most a single $H$ or $S$ operation per qubit.

The close relationship between graph states and stabilizer states suggests the possibility of a graph like representation of stabilizer states. Our approach to such a representation is simply to transform the generator matrix of a stabilizer state into that of a graph state, draw the graph thereby obtained, and add decorations to each node indicating whether an $H$ or $S$ was applied to the corresponding qubit in the process. Motivated by the standard graph convention that a $1$ on the diagonal of an adjacency matrix denotes a self loop on that node, we choose to represent $S$ operations by self loops. Indicating the application of an $H$ is less clear and indeed there are several ways to do it~\cite{schlingemann:graphs,riera:graphs}. As there is no obvious choice, we employ the use of two different types of nodes. Normally we draw nodes as filled in dots, or as \emph{solid\/} nodes, so we indicate qubits to which an $H$ was applied as open circles, or \emph{hollow\/} nodes. We will sometimes use the term \emph{fill state\/} to reference the dichotomy in node types. 

\begin{figure}[h!]
\center
\includegraphics[width=11cm]{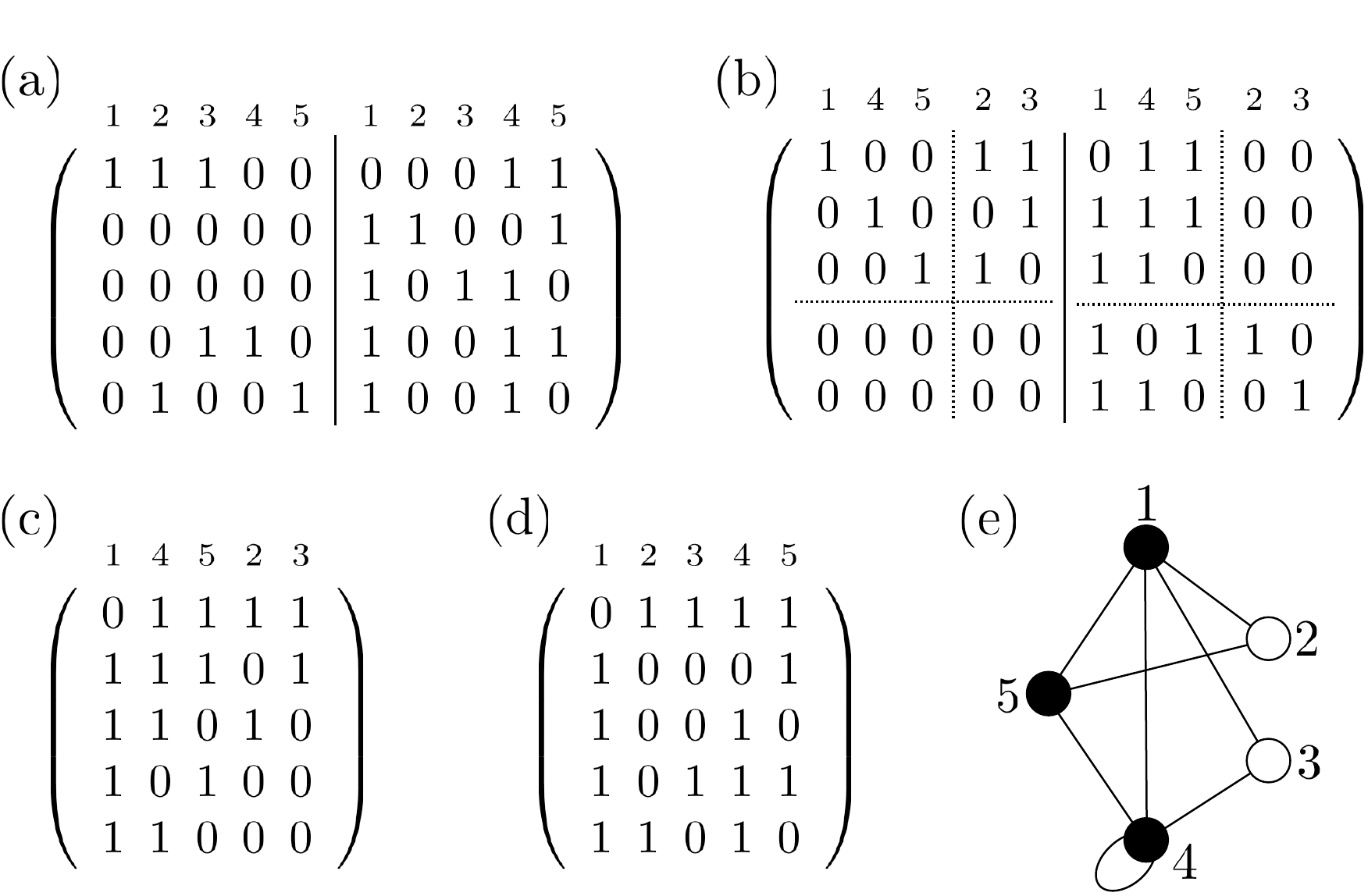}
\caption[Example of a generator matrix and an associated graph]{(a)~A generator matrix for a stabilizer state. (b)~A \hyph{canonical}{form} generator matrix obtained from (a) by row and qubit swapping. The dashed lines are to facilitate comparison to the standard form for generator matrices in Eq.~(\ref{eq:standardform}). (c)~The adjacency matrix indicated by (b).  (d)~The adjacency matrix of (c) with the qubit swaps undone.  (e)~The
stabilizer graph associated with (a). In parts (a)-(d) the columns have been labeled by the corresponding qubit.  In part (e) the nodes are labeled by the corresponding qubit. It is clear that the qubit swaps are not actually necessary, since we reverse them in the end. The generator matrix in (a) can be converted to graph form directly by exchanging columns $2$
and $3$ on the left with the matching columns on the right, an operation that corresponds to applying a Hadamard to qubits $2$ and $3$.  In graph form, the adjacency matrix is just the right half of the generator matrix. Loops arise from $1$'s on the diagonal of the adjacency matrix, and hollow nodes are used to indicate which columns were exchanged between the right and left halves of the generator matrix to get the adjacency matrix. Notice that there are no edges between hollow nodes, nor are there any loops on hollow nodes. \label{fig:redgraphex}}
\end{figure}

To summarize, we can graphically represent any stabilizer state with the following procedure. First, determine the canonical generators for the state, as explained in Subsecs.~\ref{subsec:genmatrix} and \ref{subsec:signs}, which yield the generator matrix in standard form, Eq.~(\ref{eq:standardform}). Second, draw the graph, including self loops, with adjacency matrix given by
\begin{equation} \label{eq:redadjacency}
\Gamma = \left( \begin{array}{cc} B & A \\ A^T & 0 \end{array} \right),
\end{equation}
making solid the nodes corresponding to the rows and columns of the submatrix $B$ and hollow the nodes corresponding to the rows and columns of the submatrix $0$. An example of a stabilizer state and an associated graph is given in Fig.~\ref{fig:redgraphex}.

There are many comments that should be made before proceeding further. For one, notice that this procedure does not associate every combination of edges, self loops, hollow nodes, and solid nodes with a stabilizer state. Because the submatrix $0$ in Eq.~(\ref{eq:redadjacency}) contains only $0$'s, hollow nodes never have loops, and there are no edges between hollow nodes. We should also point out, here, that a stabilizer state might be representable by more than one graph. This follows simply because a stabilizer state might be described by more than one set of canonical generators. We will learn to fully characterize all the stabilizer graphs representing the same state in Subsec.~\ref{subsec:equivrules}.

It is also important to note that the nodes in our graphs are labeled in that each node is associated with a particular qubit. Swapping two qubits is a physical operation that generally produces a different quantum state. In our graphs a SWAP gate can be described either by relabeling the corresponding nodes or by exchanging the nodes and all their decorations and connections while leaving the labeling the same. Since the process of bringing the generator matrix into canonical form can involve swapping qubits, we must keep track during this process of the correspondence between qubits and columns of the generator matrix and thus between these columns and the nodes of our graphs.

Possibly the biggest caveat of which to be aware is that our graphs, so far, represent stabilizer states in the same way that generator matrices represent stabilizer states; that is, they represent them up to an overall sign of the generators. This is a downside that will be remedied in the next section.

\subsection{Graphs from quantum circuits}

\begin{figure} \label{fig:quantumcircuit}
\center
\includegraphics[width=14cm]{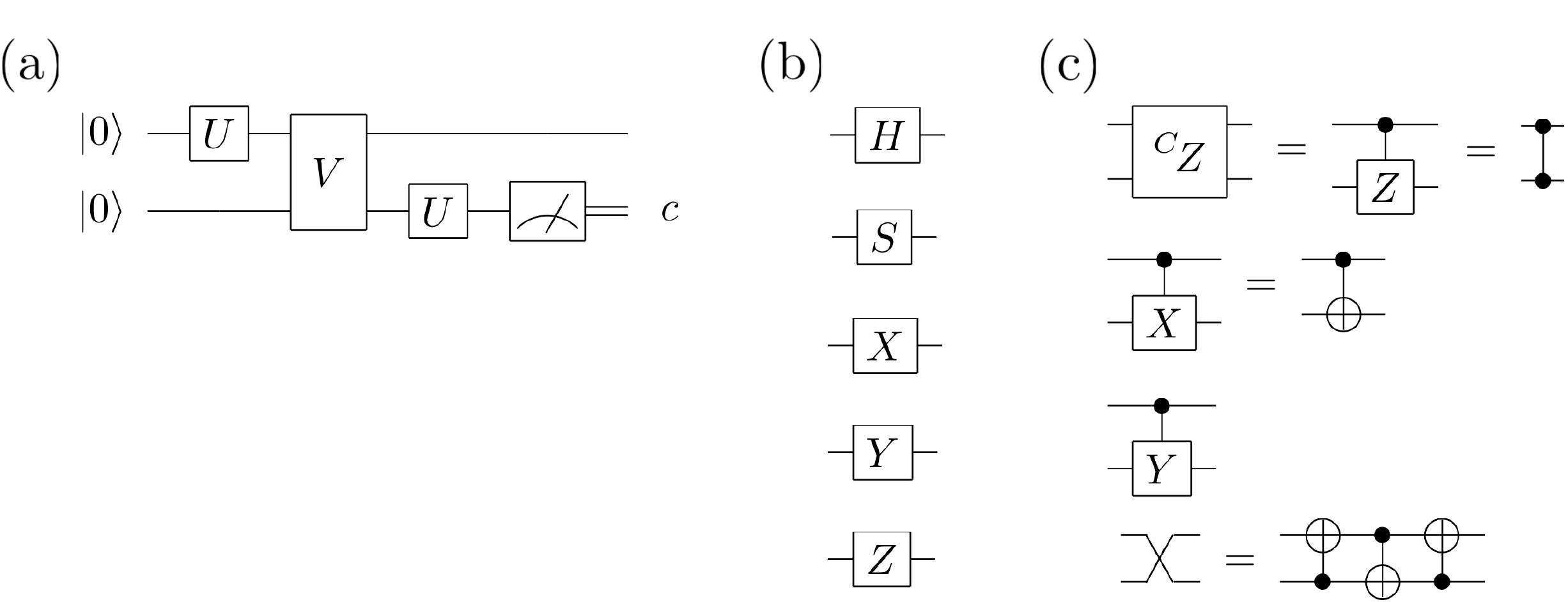}
\caption[Quantum circuits]{(a) A simple example of a quantum circuit. Each horizontal line tracks the evolution of a single qubit, with time moving forward as the circuit progresses towards the right. The state to the left of each horizontal line indicates the initial state of the corresponding qubit. A line incident on a box that means that an operation, which in the context of circuits is called a gate, is applied to that qubit. A \hyph{single}{qubit} gate $U$ corresponds to a box with a $U$ inside that only intersects one horizontal line. A \hyph{multi}{qubit} gate $V$ is also indicated by a box with a $V$ inside, except that the box intersects those lines corresponding to qubits on which it acts. The meter on the second qubit at the end of the circuit indicates a measurement. The output of the measurement is $c$, which is a classical bit whose evolution is indicated by a double horizontal line. Common \hyph{single}{qubit} Clifford gates are illustrated in (b), with the shorthand for common \hyph{two}{qubit} Clifford gates listed in (c). The gates in (c) are $\C{Z}$, $\C{X}$, and $\C{Y}$, with the last gate being the SWAP gate. The SWAP gate swaps two qubits and is equivalent to three $\C{X}$ gates as indicated.}
\end{figure}

An alternative perspective of our graphs for stabilizer states is that, in analogy with graphs states, they provide a blueprint for building the stabilizer state. Given a graph that represents a stabilizer state, if we remove all the decorations, that is we fill in all the nodes and remove all the self loops, we obtain a graph state which can be prepared as described in Subsec.~\ref{subsec:graphstates}. Then we can apply an $S$ to all qubits corresponding to nodes with a self loop and an $H$ to all qubits corresponding to hollow nodes, thereby producing the stabilizer state represented by the given graph. Thinking of our graphs for stabilizer states as a recipe for creating the state allows us to enrich our graphical formalism through a connection with \emph{quantum circuits}, a useful theoretical tool in the toolbox of quantum computation. An example quantum circuit is illustrated in Fig.~\ref{fig:quantumcircuit}(a), but for a more complete explanation see Ref.~\cite{nielsen:book}. Figs.~\ref{fig:quantumcircuit}(b) and~(c) list the quantum circuit version of some frequently used \hyph{single}{qubit} and \hyph{two}{qubit} Clifford operations, respectively.

Let us write the quantum circuit for the preparation procedure derived from a graph representing an \hyph{$n$}{qubit} stabilizer state. Such a quantum circuit consists of three layers of gates applied to $n$ qubits, each initially in the state $\ket{0}$. In the first layer, the Hadamard gate, $H$, is applied to each qubit. This prepares the qubits in the state, $\ket{+}^{\otimes n}$, required to create graph states using the method of Subsec.~\ref{subsec:graphstates}. In the second layer, controlled-$Z$ gates, $\CZ$, are applied between various pairs of qubits. Using only these first two layers, it possible to create any graph state. Finally, in the third layer, Hadamard gates and phase gates are applied to various subsets of qubits depending on whether the corresponding graph nodes are hollow or have a self loop. As a mnemonic, one can think of $H$ being applied to qubits corresponding to `H'ollow nodes, and $S$ to those corresponding to nodes with a `S'elf loop. In this final layer, there are two restrictions, arising from the restrictions on what graphs can represent stabilizer states. The first restriction is that a phase gate and Hadamard gate never occur on the same qubit. This corresponds to hollow nodes never having self loops. The second restriction is that two qubits with $H$'s in the third layer are never connected by a $\CZ$ gate in the second layer. This is the restriction that hollow nodes are never connected to each other in a graph. Quantum circuits consisting of these three layers are deemed to be in \emph{graph form\/}.

We mentioned earlier that quantum circuits would enable us to represent the signs of stabilizer generators, and we now demonstrate how this is done. Let us begin with a graph state, so that we have the generators in Eq.~(\ref{eq:graphgens}). In this case, the application of $Z_k$ to the graph state changes the sign of generator $g_k$ and leaves the other generators invariant. Using $\delta_{jk}$ which is $1$ if $j=k$ and $0$ otherwise, this can be written as
\begin{align}
\begin{split}
Z_k g_j Z_k &= Z_k \left( X_j \prod_{l \in \mathcal{N}(j)} Z_l \right) Z_k = Z_k X_j Z_k \prod_{l \in \mathcal{N}(j)} Z_l \\
&= (-1)^{\delta_{jk}} \left( X_j \prod_{l \in \mathcal{N}(j)} Z_l \right) = (-1)^{\delta_{jk}} g_j.
\end{split}
\end{align}
Thus we can dictate the sign of graph state generators by the presence of $Z$'s in the third layer of the quantum circuit; a $Z$ on qubit $j$ indicates that $\sign{g_j} = -1$. This procedure extends to all stabilizer states since a later application of $H$ or $S$ does not further change the sign of the generator,
\begin{align}
\begin{split}
&\sign{H_k g_j H_k} = \sign{X_j^{\delta_{jk}} Z_j^{\delta_{jk}} g_j} = \sign{g_j}, \\
&\sign{S_k g_j S_k^{\dagger}} = \sign{X_j^{\delta_{jk}} Y_j^{\delta_{jk}} g_j} = \sign{g_j}.
\end{split}
\end{align}
Just as the presence of an $H$ or $S$ in a quantum circuit is represented graphically as a hollow node or self loop, for obvious reasons we indicate a $Z$ in the quantum circuit as a minus sign in the corresponding node of the graph. This allows us to represent generator signs, and therefore completes our graphical description of stabilizer states.

Even though our graphs now provide a complete description of stabilizer states, a further generalization will prove beneficial in the context of Clifford operations and Pauli measurements. Our graphs for stabilizer states already suffer from the affliction that multiple graphs can represent the same state, so we generalize our graphical representation so that any graph consisting of solid and hollow nodes, with or without self loops and signs, connected by arbitrary edges represents some state. We call such a graph a \emph{\hyph{stabilizer}{state} graph\/}, or, more simply, a \emph{stabilizer graph\/}, while graphs with the property that hollow nodes do not have self loops and are not connected to each other are called \emph{reduced graphs\/}. Reduced graphs are capable of representing any stabilizer state and reduce to the usual notion of graphs for graph states. We will fully explore the connection between general stabilizer graphs and reduced graphs in Subsec.~\ref{subsec:equivrules}.

Now we explain how quantum circuits permit us to associate a stabilizer state with any stabilizer graph. To begin with, allowing arbitrary edges in a graph, so that hollow nodes can be connected to each other, corresponds to allowing arbitrary $\CZ$ gates in a \hyph{graph}{form} quantum circuit. Now it only remains to explain what is meant by a hollow node with a loop. We choose this to mean that, in the third layer of a \hyph{graph}{form} quantum circuit, an $S$ and an $H$ gate have been applied, in that order. The order is important since $SH\neq HS$. We have already chosen $Z$ gates to precede $H$ gates for good reason, so since $Z$ and $S$ commute it makes sense to have them both precede the $H$ gate.

\begin{figure}
\center
\includegraphics[width=10cm]{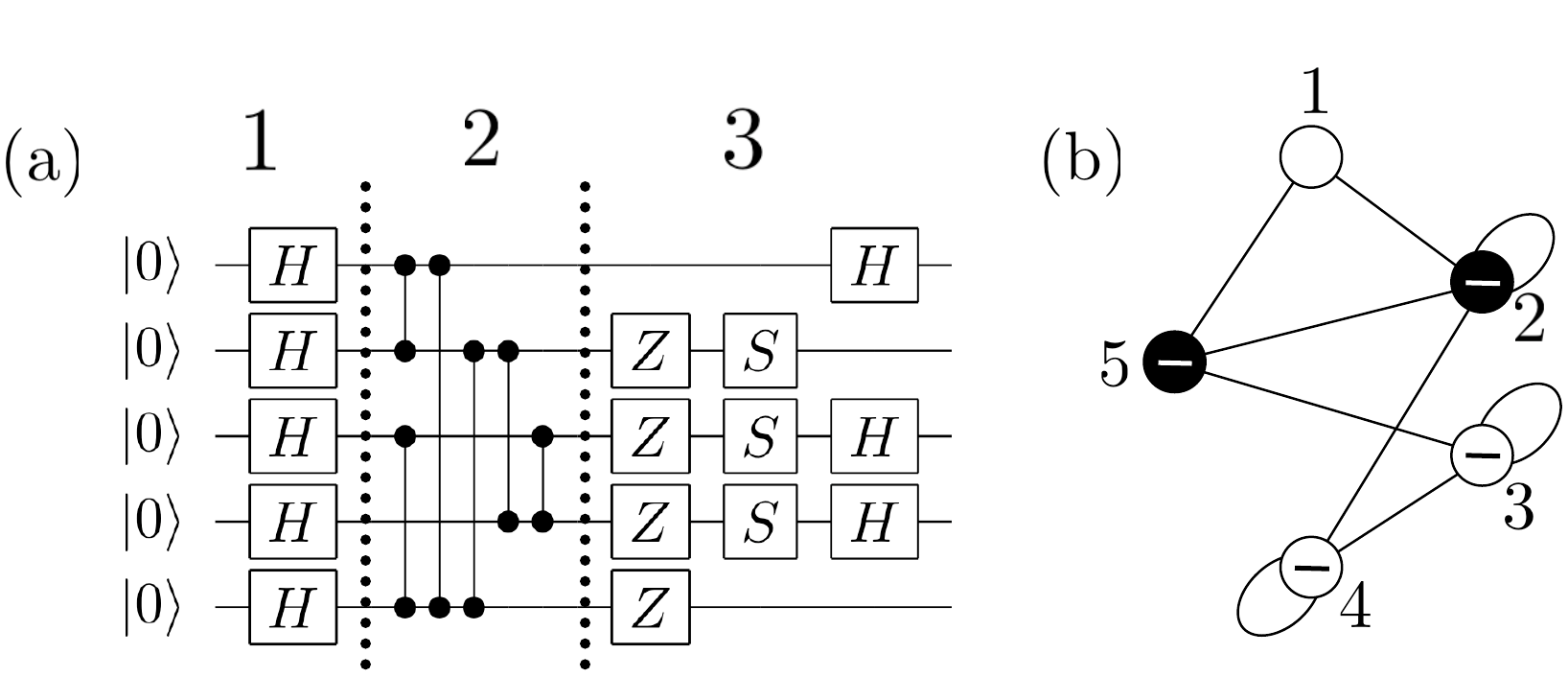}
\caption[Example of a \hyph{graph}{form} circuit and an associated graph]{(a)~A circuit in graph form, with the layers numbered and separated by dashed lines. (b)~A stabilizer graph
corresponding to the circuit in (a). $\protect\CZ$ gates between
qubits are transformed into links between nodes, terminating $Z$
gates become negative signs, terminating $S$ gates result in loops,
and terminating $H$ gates are denoted by hollow nodes. \label{fig:circgraphex}}
\end{figure}

To conclude this subsection, we concisely state the relationship between stabilizer graphs and \hyph{graph}{form} quantum circuits. \hyph{Stabilizer}{state} graphs and \hyph{graph}{form} quantum circuits are in one to one correspondence. In the second layer of a \hyph{graph}{form} quantum circuit, two qubits are linked by a $\CZ$ gate if and only if they are connected by an edge in the corresponding stabilizer graph. In the third layer of the quantum circuit, $Z$, $S$, and $H$ gates appear, in that order. A $Z$ gate appears if and only if the corresponding node has a minus sign, an $S$ gate appears if and only if it has a loop, and an $H$ gate appears if and only if the node is hollow. Both stabilizer graphs and \hyph{graph}{form} quantum circuits are in many to one correspondence with stabilizer states. Fig.~\ref{fig:circgraphex} has an example of a stabilizer graph along with its equivalent \hyph{graph}{form} quantum circuit.

\subsection{Graphical description of Clifford operations \label{subsec:transrules}}

Having explained our graphical representation of stabilizer states in some detail, now we get to see their utility in action. We will examine the usefulness of our graphs more in subsections to come, but an appropriate introductory demonstration of their applicability is to describe how the graph representing a stabilizer state changes when Clifford operations act on the state. Both stabilizer graphs and reduced stabilizer graphs have their relative merits, so we include an explanation of how Clifford operations effect both types of graphs. This section generalizes the results in Ref.~\cite{vandennest:complement}.

\begin{figure}
\center
\includegraphics[width=10cm]{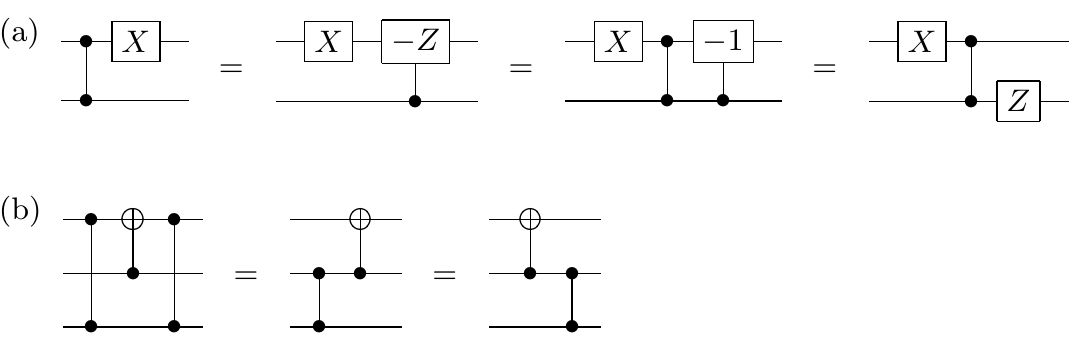}
\caption[Basic circuit identities]{Simple circuit identities, which we use to derive more
complex identities. Identity~(a) follows trivially from the steps
shown, while identity~(b) is easily verified in the standard
basis. \label{fig:basics}}
\end{figure}

Aside from the terminology in Subsec.~\ref{subsec:graphtheory}, there a few other terms we use to describe the action of Clifford operations. Two of these terms are \emph{flip\/} and \emph{advance\/}. Flip is used to describe the simple reversal of some binary property, such as the sign or the fill of a node. Advance refers specifically to an action on loops; advancing generates a loop on nodes where there was not previously one, and it removes the loop and flips the sign on nodes where there was a loop. Its action mirrors the application of the phase gate, since $S^2=Z$. The last bit of terminology is our sloppy use of language for the sake of brevity. That is, when we say that a gate acts on a node we mean that it acts on the qubit corresponding to that node. In fact, in what follows, qubit and node are used interchangeably.

We begin with the action of local Clifford operations on \hyph{stabilizer}{state} graphs. Since the local Clifford group is generated by $H$ and $S$, we need only describe the action of $H$ and $S$ on stabilizer states. This is accomplished by the following transformation rules.
\begin{enumerate}
\item[T1.] Applying $H$ to a node flips its fill.
\item[T2.] Applying $S$ to a solid node advances its loop.
\item[T3.] Applying $S$ to a hollow node without a loop performs local complementation on the node and advances the loops of its neighbors.

If the node has a negative sign, flip the signs of its neighbors as well.
\item[T4.] Applying $S$ to a hollow node with a loop flips its fill, removes its loop, performs local complementation on it, and advances the loops of its neighbors.

If the node does not have a negative sign, flip the signs of its neighbors as well.
\end{enumerate}

\begin{figure}[h!]
\center
\includegraphics[width=15cm]{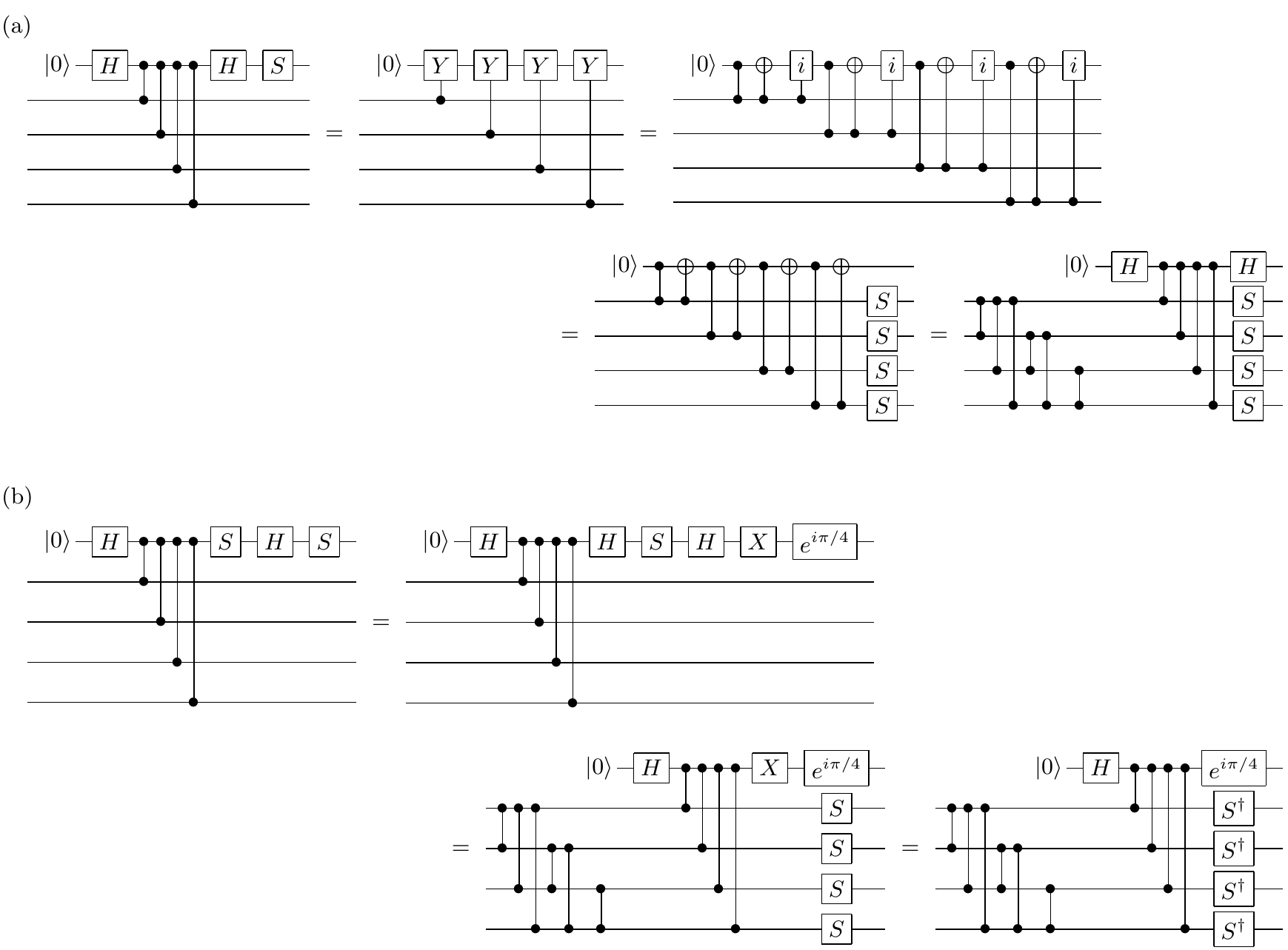}
\caption[Circuit identities for local Clifford transformations]{Circuit identities relevant to transforming a stabilizer graph under local Clifford operations. Both identities are illustrated for the case of four neighbors connected to a qubit of interest; equivalent expressions hold for other numbers of neighbors. The proof of identity~(a) relies on decomposing $\protect\C{Y}$ using $Y=iXZ$, the equivalence of $\protect\C{i}$ to $S$ on the control qubit, and the identity in Fig.~\ref{fig:basics}~(b). The first step in identity~(b) uses the fact that $I=e^{-i\pi/4}(HS)^3$, which gives $SHS=e^{i\pi/4}XHSH$; the second and third steps follow from the identities in~(a) and Fig.~\ref{fig:basics}~(a). The identity in Fig.~\ref{fig:basics}~(b) generates the thicket of controlled-$Z$ gates in the lower left of the final circuits in (a) and (b), thereby giving rise to local complementation in the corresponding stabilizer graphs. In~(b), the phase shift on the top qubit, $e^{i\pi/4}$, is a global phase shift and thus can be omitted. To complete the transformation rules, it is necessary to know what happens in both of these circuit identities when there is a $Z$ gate on the top qubit immediately after the $\protect\C{Z}$ gates. Ignoring overall phases, the effect on the third layer of the final circuit is to include an additional $Z$ on all qubits.\label{fig:localtrans}}
\end{figure}

These transformation rules can be derived from the circuit identities in Fig.~\ref{fig:localtrans}, which rely on the basic circuit identities given in Fig.~\ref{fig:basics}. Given an understanding of the relationship between circuits and graphs, transformation rules T$1$ and T$2$ are trivial. Transformation rules T$3$ and T$4$ derive from Figs.~\ref{fig:localtrans}(a) and (b), respectively.

The transformation rules T$1$--T$4$ do not generally take reduced stabilizer graphs to reduced stabilizer graphs. From Sec.~\ref{subsec:gengraphs}, however, we know that there exists a
reduced stabilizer graph corresponding to each stabilizer state, so it is always possible to represent the effect of a local Clifford operation as a mapping between reduced stabilizer graphs. The appropriate transformation rules for reduced stabilizer graphs are listed below.

\begin{enumerate}
\item[T(i).] Applying $H$ to a solid node without a loop, which is only connected to other solid nodes, flips the fill of that node.

\item[T(ii).] Applying $H$ to a solid node with a loop, which is only connected to other solid nodes, performs local complementation on the node and advances the loops of its neighbors.

Flip the node's sign, and if it now has a negative sign, flip the signs of its neighbors as well.
\item[T(iii).] Applying $H$ to a solid node without a loop, which is connected to a hollow node, flips the fill of the hollow node and performs local complementation along the edge connecting the nodes.

Flip the sign of nodes connected to both the solid and hollow nodes. If either of these two nodes has a negative sign, flip it and the signs of its current neighbors.

\item[T(iv).] Applying $H$ to a solid node with a loop, which is connected to a hollow node, performs local complementation on the solid node and then on the hollow node. Then it removes the loop from
the solid node, advances the loops of the solid node's current neighbors, and flips the fill of the hollow node.

Flip the signs of nodes that were originally connected to both the solid and hollow nodes. If the originally solid node initially had a negative sign, flip it and the signs of its current neighbors, and if the originally hollow node initially had a negative sign, flip the signs of its current neighbors.

\item[T(v).] Applying $H$ to a hollow node flips its fill.
\item[T(vi).] Applying $S$ to a solid node advances its loop.
\item[T(vii).] Applying $S$ to a hollow node performs local complementation on that node and advances the loops of its neighbors.

If the node has a negative sign, flip the signs of its neighbors as well.
\end{enumerate}

Of these transformation rules, T(i), T(v), and T(vi) are trivial, and T(vii) is a rewrite of T$3$. To prove the others requires results from Subsec.~\ref{subsec:equivrules}, in particular, equivalence rules E$1$ and E$2$. Specifically, T(ii) is obtained by applying equivalence rule~E$1$, which gives an equivalent, but unreduced graph, and then applying the Hadamard, via rule~T$1$, which leaves a reduced graph. For T(iii), one first applies the Hadamard, via rule~T$1$, and then uses equivalence rule~E$2$ to convert to a reduced graph. In the case of T(iv), one applies the Hadamard, using rule~T$1$, and then applies equivalence rule E$1$, first to the originally solid node and then to the hollow node. A key part of these transformations is the conversion of stabilizer graphs to reduced form. A section of Subsec.~\ref{subsec:equivrules} explains this process in more detail.

It is not hard to check that, in using rules T(iii) and T(iv), any other hollow nodes that are connected to the originally solid node do not become connected and do not acquire loops, in accordance with the need to end up with a reduced graph.

To complete the Clifford group, we need to present a graphical description of the action of controlled-$Z$ gates on stabilizer states. Since most of the interest in Clifford operations is in local Clifford operations \cite{hein:entanglement,vandennest:algorithm,ji:lulc,hostens:breeding}, and since the action of $\CZ$ generates all stabilizer states from a given initial one, we relegate the results relevant to $\CZ$ gates to Appendix~\ref{app:czgates}.

\subsection{Equivalent graphs \label{subsec:equivrules}}

In this subsection we explain how to test whether two graphs correspond to the same stabilizer state, and as a consequence we learn how to generate all graphs representing a given stabilizer state. Let us begin with an example of different \hyph{graph}{form} circuits corresponding to the same stabilizer state. Such an example can be found in Fig.~\ref{fig:localtrans}. Applying an additional $S^{\dagger}$ gate to the top qubit in Fig.~\ref{fig:localtrans}(b) makes the initial and final circuits both in graph form, but the two circuits lead to quite different graphs. These two graphs then represent the same stabilizer state and so are called \emph{equivalent} graphs. Two graphs are equivalent exactly when their corresponding \hyph{graph}{form} circuits are equal.

We now present two equivalence rules for stabilizer graphs. Applying either of the following two rules to a stabilizer graph yields an new one which represents the same stabilizer state. We will then use these rules to derive a similar set for reduced graphs. The proof that both sets of equivalence rules generate all graphs equivalent to the initial one is given in Appendix~\ref{app:equivproof}.

\begin{enumerate}
\item[E1.] Flip the fill of a node with a loop. Perform local complementation on the node, and advance the loops of its neighbors.

Flip the node's sign, and if the node now has a negative sign, flip the signs of its neighbors as well.

\item[E2.] Flip the fills of two connected nodes without loops, and local complement along the edge between them.

Flip the signs of nodes connected to both of the two original nodes. If either of the two original nodes has a negative sign, flip it and the signs of its current neighbors.
\end{enumerate}

\begin{figure}[h!]
\center
\includegraphics[width=15cm]{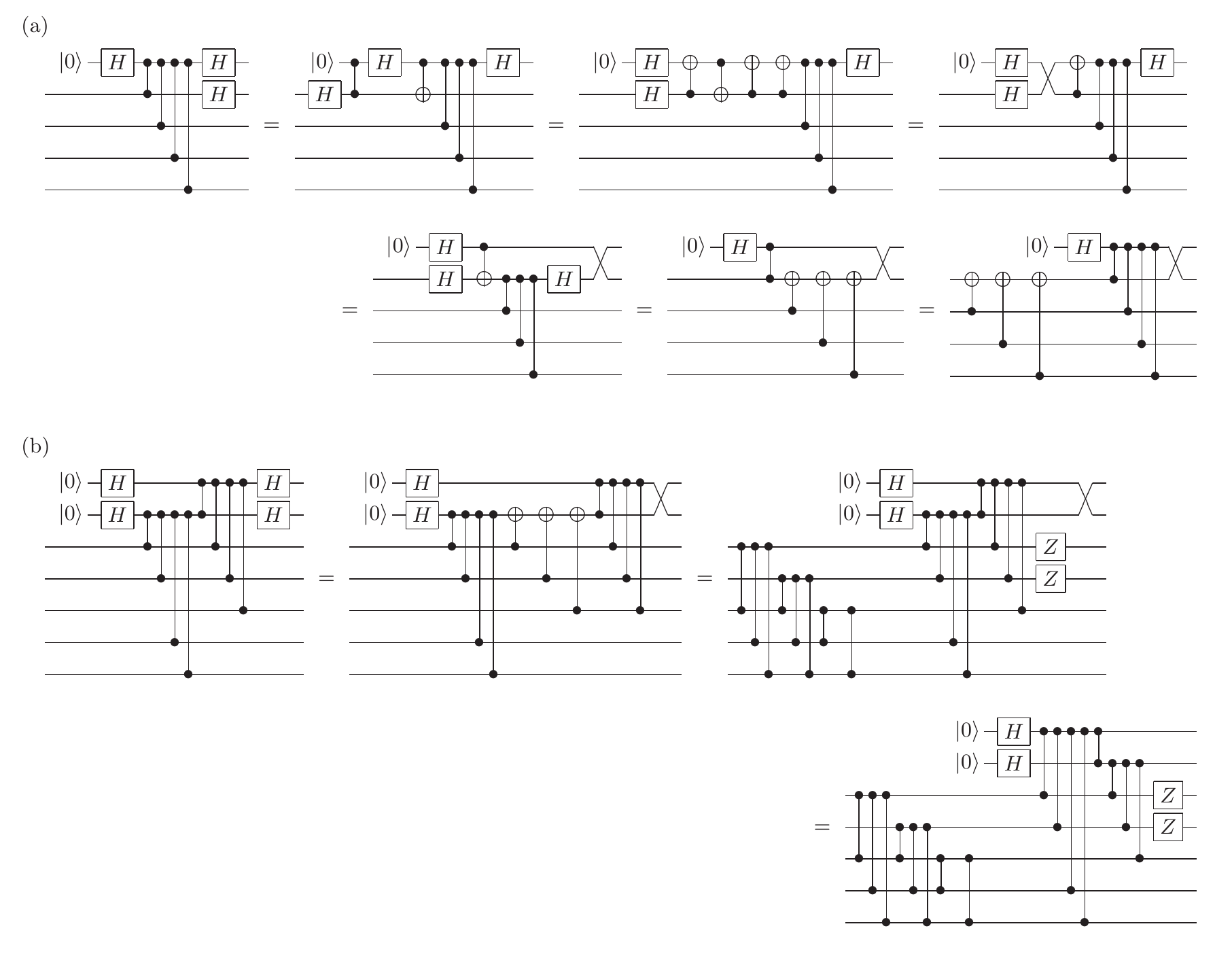}
\caption[Quantum circuit for equivalence rule E$2$]{(a) A circuit identity used in~(b). The identity is shown with four neighbors connected to the top qubit; equivalent expressions hold for other numbers of neighbors. The crucial trick here is to decompose \textit{SWAP} in terms of three $\protect\CX$ gates. (b) A circuit identity demonstrating transformation rule~E$2$. The selection of neighbors of the top two qubits is chosen to display all possibilities. The first equality in~(b) employs the identity from part~(a), while the subsequent ones follow from basic circuit identities. It is straightforward to verify that edges are changed according to local complementation along the edge connecting the top two qubits. Circuits illustrating E$2$ for other initial fill states can be obtained by applying additional terminal Hadamards to the circuits in~(b). The sign rule for E$2$ expresses the effect of a $Z$ gate on one of the top two qubits before the terminal Hadamard gate. Such a $Z$ gate can be pushed through the Hadamard, becoming an $X$ gate; in the final circuit, this $X$ gate can be processed using the identity in Fig.~\ref{fig:basics}(a).
\label{fig:E2circuit}}
\end{figure}

The first of these equivalence rules is our previous example, and can be obtained by applying an additional $S^{\dagger}$ gate to the top qubit in the identity of Fig.~\ref{fig:localtrans}(b). For the second rule, we need yet another circuit identity. Figure~\ref{fig:E2circuit}(a) shows how Hadamards can be removed from a pair of connected qubits without $S$ gates. Figure~\ref{fig:E2circuit}(b) extends this identity to a demonstration of rule~E2.

Equivalence rules E$1$ and E$2$ can be reworked to yield equivalence rules for reduced stabilizer graphs. These rules take reduced stabilizer graphs to equivalent reduced stabilizer graphs.

\begin{enumerate}
\item[E(i).]For a hollow node connected to a solid node with a loop, local complement on the solid node and then on the hollow node. Then remove the loop from the solid node, advance the loops of its
current neighbors, and flip the fills of both nodes.

As for signs, follow the sequence in transformation rule T(iv).
\item[E(ii).]For a hollow node connected to a solid node without a loop, local complement along the edge between them. Then flip the fills of both nodes.

As for signs, follow the sequence in equivalence rule~E$2$.
\end{enumerate}

There is a simple relationship between the two sets of equivalence rules. Equivalence rule E(ii) is identical to E2 for the case that the two connected nodes have opposite fill. Equivalence rule E(i) is simply rule E1 applied twice: first to the solid node with the loop and then once to the hollow node that has acquired a loop from the first application of E1. The second application of E1 is needed because the resulting graph is not reduced after one employment of the rule.

Both of these equivalence rules can also be derived by applying two Hadamards to a solid node, E(i) handling the case in which the solid node has a loop and E(ii) the case in which it does not. Thus equivalence rule~E(i) is simply transformation rule~T(iv) followed by use of rule~T(i) to apply a second Hadamard to the originally solid node. Likewise, E(ii) is rule~T(iii) followed by T(i) to apply a second Hadamard to the originally solid node. Notice that both rules preserve the number of hollow nodes.

It is worth summarizing here the results in Appendix~\ref{app:equivproof}, where the completeness of our equivalence rules is proved, because the ideas behind the proof are simple and can easily be missed when reading the details. In order to test whether two graphs represent the same state, the first step is to use equivalence rules E$1$ and E$2$ to put the graphs in reduced form. Then, if equivalence rules E(i) and E(ii) are used to put the hollow nodes in the same place, the two graphs represent the same state if and only if they are trivially identical. Given that we can test the equivalence of any two graphs using our equivalence rules, these rules must then generate all graphs equivalent to a given one. The most difficult part of the proof is showing that graphs represent the same state if and only if they are trivially identical, once the equivalence rules have been used. Because of this, an example illustrating the proof of this step is given in Fig.~\ref{fig:partthreeexample}. Fig.~\ref{fig:equivalenceexample} is also worth looking at since it demonstrates how to use the equivalence rules to efficiently test whether two graphs represent the same stabilizer state. Also, as a consequence of the proof in Appendix~\ref{app:equivproof}, we see that reduced graphs are special because they constitute all the graphs representing a state that use a minimal number of hollow nodes.

\subsection{Graphical description of \hyph{Pauli}{product} measurements \label{subsec:graphmeas}}

As another example of the usefulness of \hyph{stabilizer}{state} graphs, we now discuss how to graphically describe the effect of \hyph{Pauli}{product} measurements on stabilizer states~\cite{elliott:measure}. A \hyph{Pauli}{product} measurement is a measurement in the eigenbasis of an \hyph{$n$}{fold} tensor product of the identity, $I$, and the Pauli matrices, $X$, $Y$, and $Z$. In other words, a \hyph{Pauli}{product} measurement corresponds to a matrix $M$ such that
\begin{equation}
M = \bigotimes_{j=1}^n M_j\;,
\end{equation}
where $M_j = I, X, Y,$ or $Z$. If $M_j \ne I$, we call node~$j$ a \emph{measured\/} node; otherwise, if $M_j = I$, we say the node is not measured.

Given such a measurement operator, $M$, our task is two fold: first, to find the probability that a measurement of $M$ on a quantum system in the stabilizer state $\ket{\psi}$ gives an outcome ${(-1)}^m$, and, second, to determine the post-measurement quantum state of the system, i.e., a post-measurement \hyph{stabilizer}{state} graph. This section describes a general graphical rule, applicable to the graph that represents the stabilizer state, which accomplishes these tasks.

The difficulty of formulating a \hyph{Pauli}{product} measurement rule can be greatly reduced by using graph transformation and equivalence rules introduced in Subsecs.~\ref{subsec:transrules} and~\ref{subsec:equivrules}. The following three paragraphs describe a sequence of three simplifications that can be made to any measurement, thereby restricting its form to one more amenable to a measurement transformation rule. With this simplification carried out, we then present an applicable graphical measurement rule.

The first simplification relies on the fact that a measurement where $M_j = C_jZ_j C_j^{\dagger}$ is equivalent to a measurement where $M_j = Z_j$ preceded by application of the local Clifford operation $C_j^{\dagger}$ and followed by application of $C_j$ to the post-measurement state.  For an $X$ in the measured Pauli product $M$, $C=H$, and for a $Y$ in $M$, $C=SH$.  Thus, the first simplification is to transform the original graph, using rules~T1--T4, so that on the new graph the measurement becomes a product of $Z$s on the measured nodes.  This means that it suffices to determine the effect of $Z$-type measurements, that is, measurements with the property that $M_j = I$ or $Z$ for all $j$. The post-measurement state must be transformed by application of the appropriate local unitaries to the measured nodes, i.e., $C_j$ to measured node~$j$; in terms of graphs, this post-measurement transformation is handled by rules~T1--T4.

The second simplification is to reduce the graph, as explained in Subsec.~\ref{subsec:equivrules}. For the purposes of our measurement analysis, we are not required to reduce the entire graph, but only the measured nodes. After this second simplification, there are no loops on hollow measured nodes and no edges between hollow measured nodes.

The final simplification is to disconnect hollow measured nodes from unmeasured nodes using equivalence rules~E$1$ and~E$2$.  Suppose that a measured hollow node is connected to an unmeasured node. In the case that the unmeasured node does not have a loop, applying equivalence rule~E$2$ to the pair turns the measured node solid.  In the case that the unmeasured node has a loop, an application of E$1$ to the unmeasured node gives the measured hollow node a loop.  Now one can apply E$1$ to the measured hollow node to turn it solid.  One can verify that in both cases, application of the equivalence rules leaves the remaining measured hollow nodes loopless and unconnected to one another.  Thus, this last simplification terminates in a number of iterations no greater than the number of measured hollow nodes.

The end product of these simplifications is a \hyph{$Z$}{type} Pauli measurement on a graph in which measured hollow nodes are loopless and unconnected to one another and to unmeasured nodes. With the preceding simplifications carried out, we can now spell out the graphical description of the measurement. The proof of this description is given in Appendix~\ref{app:measproof}.

Our graphical description of \hyph{Pauli}{product} measurements is greatly facilitated by introducing special sets of nodes in the \hyph{stabilizer}{state} graph: $\setH$ is the set of hollow nodes, $\setS$ is the set of solid nodes, $\setZ$ is the set of nodes with a sign, $\setM=\{j\mid M_j\ne I\}$ is the set of measured nodes, $\setMS=\setM\backslash \setH$ is the set of measured solid nodes, $\setMH=\setM\cap\setH$ is the set of measured hollow nodes, and $\setMSE=\{j\in\setMS\mid|\setMH\cap\set{N}(j)|=\mbox{$0\pmod2$}\}$
is the set of measured solid nodes that have an even number of connections to measured hollow nodes.  Here $\set{A}\backslash\set{B}$ denotes the set of elements in $\set{A}$ that are not in $\set{B}$, $\set{A}\cap\set{B}$ denotes the intersection of $\set{A}$ and $\set{B}$, and $|\set{A}|$ denotes the
number of elements in $\set{A}$.

When a Pauli measurement is made on a stabilizer state, the outcome is either definite or random, with random outcomes giving equal probabilities of $1/2$ for the two outcomes.  Which case applies depends on $\setMSE$. The outcome in the deterministic case is specified by
\begin{equation}\label{eq:b}
b=|\setMH\cap\setZ|\;,
\end{equation}
the number of measured hollow nodes with a sign.

The result of a $Z$-type Pauli measurement is as follows.
\begin{enumerate}
\item If $\setMSE = \varnothing$, the measurement outcome is ${(-1)}^b$ with certainty, and the state is unchanged by the measurement.
\item If $\setMSE \ne \varnothing$, the measurement outcome, $(-1)^m$, is random, and a graph for the post-measurement state can be obtained according to steps~$1$--$4$ below.
\end{enumerate}

\begin{figure}
\begin{center}
  \begin{tabular}{c@{}c@{}c@{}c@{}c@{}c@{}c@{}c@{}}
\includegraphics[width=2.95cm]{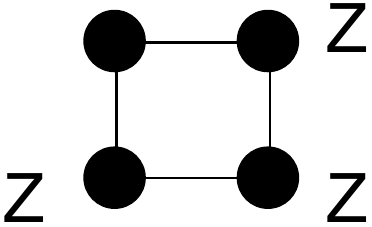} \hspace{.2em} & \raisebox{1.6em}{\Large $\stackrel{1}{\rightarrow}$} & \hspace{.3em} \raisebox{.15em}{\includegraphics[width=1.8cm]{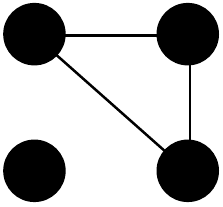}} \hspace{.3em} & \raisebox{1.6em}{\Large $\stackrel{2}{\rightarrow}$} & \hspace{.3em} \raisebox{.15em}{\includegraphics[width=1.8cm]{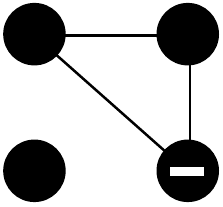}} \hspace{.3em} & \raisebox{1.6em}{\Large$\stackrel{3}{\rightarrow}$} & \hspace{.3em} \raisebox{.15em}{\includegraphics[width=1.8cm]{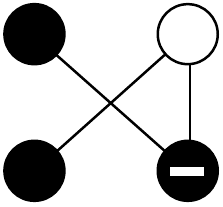}}
  \end{tabular}
  \end{center}
\caption[Example of a \hyph{Pauli}{product} measurement]{The steps that perform a \hyph{Pauli}{product} measurement $M = I \otimes Z \otimes Z \otimes Z$ on a \hyph{stabilizer}{state} graph for the four qubit cluster state.  The juxtaposition of Pauli operator and node indicates the presence of that operator in the intended measurement.  We label the nodes in the graphs clockwise starting from~$1$ in the upper left corner.  Thus, $\setM = \{ 2,3,4 \}$, $\setMS = \setMSE = \{ 2,3,4 \}$, and $\setMH = \varnothing$.  Since $\setMSE \ne \varnothing$, the outcome is random; for the sake of illustration, we take it to be $+1$. Node~$2$ is taken to be the chosen node, so the edges between its neighbors, nodes $\{ 1,3 \}$, and the unchosen nodes in $\setMSE$, nodes $\{ 3,4 \}$, are complemented in step~$1$. The chosen node does not have a sign, and $m \eqmod{2} b$ if we assume a $+1$ measurement outcome, so step~$2$ has the effect of giving a sign to the node in $\setMSE$ that is also a neighbor of node $2$, meaning node $3$. Step~3 removes all edges involving node~$2$ and then connects it to nodes~$3$ and~$4$ while making node~$2$ hollow.  Finally, step~$4$ has no effect since the chosen node does not have a loop.  If either node~$3$ or $4$ had been picked as the chosen node, the resulting graph could be transformed into this one using equivalence rule~E$2$.\label{fig:ZZZZexample}}
\end{figure}

To find the post-measurement state when $\setMSE \ne \varnothing$, it is necessary first to pick a node, which we call the \emph{chosen node}, from $\setMSE$.  The post-measurement state is then obtained by the following four steps.
\begin{enumerate}
\item For each neighbor of the chosen node, complement all of its edges to unchosen nodes in $\setMSE$.
\item If the chosen node does not have a sign, flip the signs of all its neighbors that are also in $\setMSE$; otherwise, if the chosen node has a sign, remove that sign, and flip the signs of all other nodes in $\setMSE$ that do not neighbor the chosen node. If $m \ne_{\scriptscriptstyle{2}} b$, flip the signs of the chosen node and all its neighbors.
\item Remove all edges involving the chosen node, and then connect the chosen node to all the other nodes in $\setMSE$.  Make the chosen node hollow.
\item If the chosen node has a loop, remove that loop, perform local complementation on the chosen node, advance the loops of its neighbors, and if $m \ne_{\scriptscriptstyle{2}} b$, flip the signs of the unchosen nodes in $\setMSE$.
\end{enumerate}
These steps constitute a complete graphical description for the effect of the measurement $M$ on the state.  Notice that, in step~$1$, an edge between two nodes that are in $\setMSE$ and are initially neighbors of the chosen node gets complemented twice, so it remains unchanged.  Figure~\ref{fig:ZZZZexample} illustrates the use of this measurement rule for the case of a three qubit measurement on a four qubit cluster state.

\subsection{Graphical description of \hyph{single}{qubit} Pauli measurements \label{subsec:singlemeas}}

\hyph{Single}{qubit} measurements are a straightforward but important special case~\cite{aaronson:simulate}, as illustrated, for example, by the use of such measurements in \hyph{measurement}{based} quantum computation~\cite{raussendorf:oneway}. While we could easily apply our measurement rules above to this special case, we instead formulate a simpler alternate description of \hyph{single}{qubit} measurements. This description does not generalize to other \hyph{Pauli}{product} measurements, but it is more simple than our general rule in this special case.

\begin{figure}
\center
\includegraphics[width=10cm]{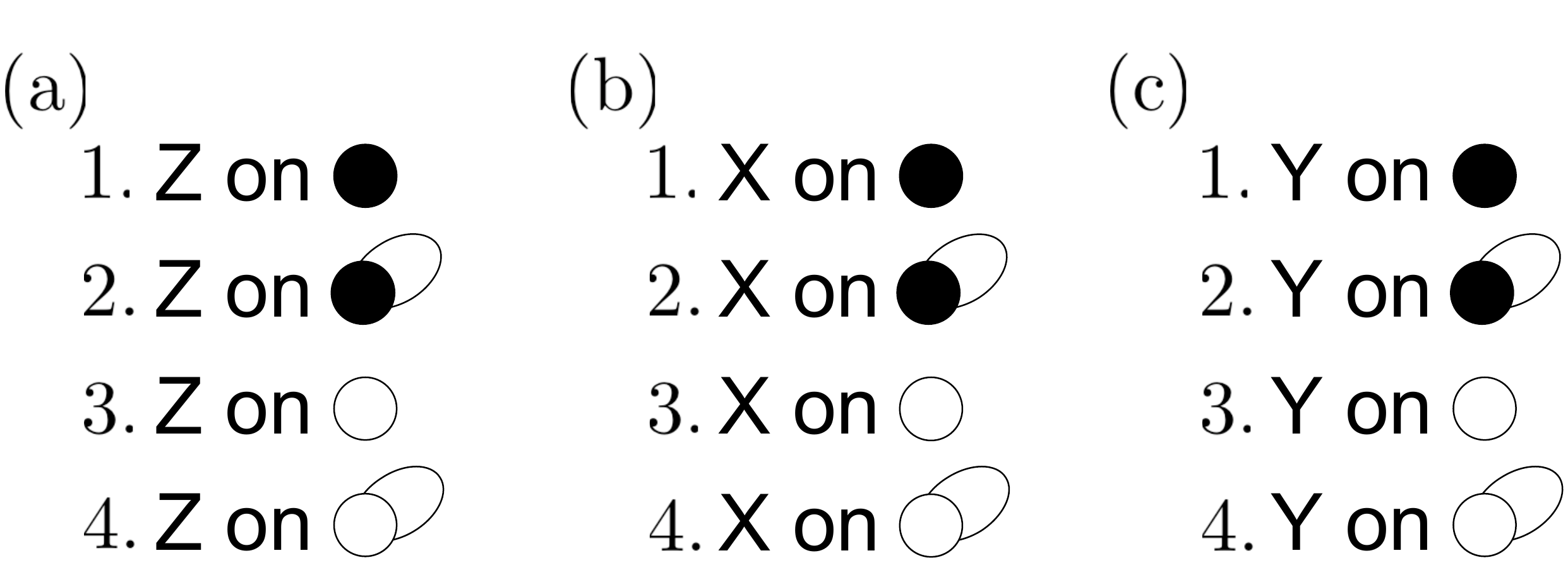}
\caption[Possible \hyph{single}{qubit} Pauli measurements]{The different possible Pauli measurements on the different possible node types. The sign of the node is irrelevant for classification purposes and is therefore omitted.
\label{fig:cases}}
\end{figure}

Since there are three possible measurements of Pauli matrices that can be made on four possible signless node types, there are $12$ possible measurement configurations to consider. These are illustrated in Fig.~\ref{fig:cases}. We first consider each case in complete generality, and then we finish this subsection with a specific example of \hyph{single}{qubit} Pauli measurements on a stabilizer state.

The simplest case to consider is that of a $Z$ measurement on a solid node without a loop, which is case $1$ in Fig.~\ref{fig:cases}(a). It is simple to determine that in this case the node disconnects from the graph~\cite{hein:entanglement}. This simply means that all edges to that qubit are removed. Since the measured qubit is in an eigenstate of $Z$ after the measurement, it is now represented by a hollow node with a sign determined by the measurement result.

Let us see why this is true. If the measurement is on node $j$, then the claim is that all generators except for $g_j$ have $Z$ or $I$ in the $j$th spot in the tensor product decomposition of the generator and $g_j$ has an $X$ in that spot. Thinking of the circuit that creates the stabilizer state, after the second layer we have a graph state and we know this is true from Eq.~(\ref{eq:graphgens}). Moreover, the only possible Clifford operator in the third layer on the $j$th qubit is a $Z$. Therefore it must also be true for the final stabilizer state as well. Now it is clear that only $g_j$ anticommutes with the measurement and so must be replaced by $g_j' = Z_j$ after the measurement. We can remove the $Z$ from all generators with a $Z$ in their $j$th spot through multiplication by $g_j'$. In this way we obtain the stabilizer generators for a stabilizer graph obtained from the original one by simply disconnecting qubit $j$.

We will now show the remarkable fact that a graphical description of all other measurements can be determined from this simple case by using equivalence rules E$1$ and E$2$ and transformation rules T$1-$T$4$. The plan is to use the equivalence rules to represent the stabilizer state by a graph which permits a simple graphical description of the given Pauli measurement.

We now determine the effect of a $Z$ measurement on a solid node with a self loop, case $2$ in Fig.~\ref{fig:cases}(a). The loop is a graphical representation of an $S$ gate being applied to a solid qubit without a loop. Now this application of $S$ followed by a $Z$ measurement is equivalent to measuring in the $SZS^{\dagger} = Z$ basis followed by an $S$ gate on the measured qubit. We know that a $Z$ measurement on a solid node without a loop disconnects that node, and therefore a $Z$ measurement on a solid node with a loop also disconnects the node from the graph. The \hyph{post}{measurement} qubit state is once again represented by a hollow node. This trick of using the node decorations to change the measurement basis is a useful one and is used repeatedly.

We now finish the last two $Z$ measurement possibilities. In order to graphically describe these measurement situations, we use equivalence rules E$1$ and E$2$. In the first situation, case $3$ in Fig.~\ref{fig:cases}(a), suppose a $Z$ measurement is made on a hollow node without a loop. We want to flip the fill of the measurement node. This can be done via E$2$ if the node is connected to another node without a loop, local complementing along the edge connecting the node and its loopless neighbor. If all of the neighbors of the node have a loop, then we must first use E$1$ on a neighbor to produce a loop on the measurement node. The resulting task is to describe a $Z$ measurement on a hollow node with a loop, which is case $4$ in Fig.~\ref{fig:cases}(a). This can be accomplished by applying E$1$ to the node to flip its fill and then disconnecting the node, since we are now measuring $Z$ on a solid node. The \hyph{post}{measurement} node is hollow.

The next cases involve $X$ measurements which are nearly analogous to $Z$ measurements. The difference is that $X$ measurements on hollow nodes have the effect of disconnecting those nodes from the graph. This follows from recognizing that a hollow node, with or without a loop, represents a solid node, with or without a loop, followed by an $H$ gate. This $H$ gate changes the measurement basis from $X$ to $HXH = Z$, and the result is equivalent to making a $Z$ measurement on a solid node, which we know disconnects that node. In order to make an $X$ measurement on a solid node, we need to first make the node hollow. This can be done via equivalence rules E$1$ and E$2$, analogously to cases $3$ and $4$ in Fig.~\ref{fig:cases}(a). After the $X$ measurement the disconnected node is solid. This covers cases $1-4$ in Fig.~\ref{fig:cases}(b).

Now the only cases that are left are the $Y$ measurements, cases $1-4$ in Fig.~\ref{fig:cases}(c). However, if we once again change the measurement basis by writing $Y = SXS^{\dagger}$, then we can replace the $Y$ measurement by an $X$ measurement preceded by an $S$ gate. The action of this $S$ gate is described graphically by transformation rules T$2$, T$3$, and T$4$. Now we are left with measuring $X$ on the transformed graph, which can be done as described for cases $1-4$ in Fig.~\ref{fig:cases}(b). The \hyph{post}{measurement} state of the measured qubit is represented by a solid node with a loop. This completes our graphical representation of Pauli measurements on stabilizer states.

\begin{figure}
\center
\includegraphics[width=13cm]{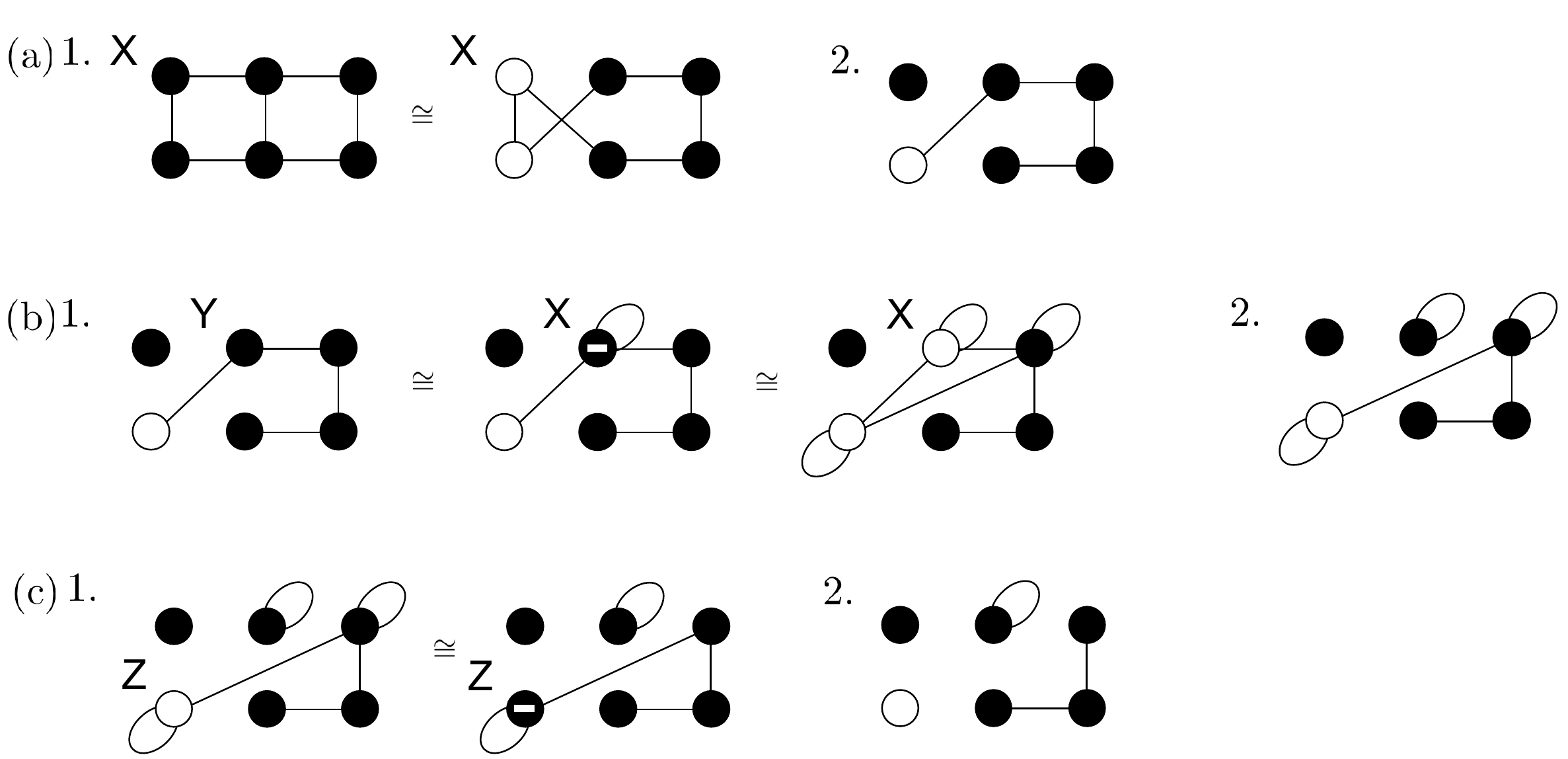}
\caption[Examples of \hyph{single}{qubit} Pauli measurements]{Examples of $X$, $Y$, and $Z$ measurements on various node types. See the text for a more detailed explanation.
\label{fig:examples}}
\end{figure}

To illustrate the utility of our graphical representation of Pauli measurements on stabilizer states we use the examples in Fig.~\ref{fig:examples}. We draw attention to the applicability of our results to \hyph{measurement}{based} quantum computation~\cite{raussendorf:oneway} by analyzing $X$, $Y$, and $Z$ measurements on a $2 \times 3$ cluster state, which is a graph state corresponding to a square lattice~\cite{briegel:ising}. We purposefully follow an inefficient measurement order for pedagogical reasons. We also always assume the measurement outcome $+1$.

We begin with the $X$ measurement illustrated in Fig.~\ref{fig:examples}~(a). Representing an $X$ measurement on the indicated note can be accomplished by applying equivalence rule E$2$, thereby making the node, along with one neighbor of the node, hollow. Local complementation along the edge defined by the two nodes yields the graph on the right side of Fig.~\ref{fig:examples}~$(a)1$. The two graphs in Fig.~\ref{fig:examples}~$(a)1$ are equivalent in that they both represent the $2 \times 3$ cluster state. The purpose of using the graph on the right, however, is that the \hyph{post}{measurement} state can be obtained by disconnecting the measured node from the graph, as illustrated in Fig.~\ref{fig:examples}~$(a)2$.

Next we consider Fig.~\ref{fig:examples}~(b) where a $Y$ measurement is made on a solid node of the graph. This $Y$ measurement has the same effect as an $X$ measurement on the middle graph of Fig.~\ref{fig:examples}~$(b)1$. This $X$ measurement can be carried out by first applying equivalence rule E$1$, which results in the far right graph in Fig.~\ref{fig:examples}~$(b)1$, and then disconnecting the node, as illustrated in Fig.~\ref{fig:examples}~$(b)2$. Since we changed the measurement basis from $Y$ to $X$, we must apply an $S$ gate to the measured node, giving it a loop.

Finally, Fig.~\ref{fig:examples}~(c) depicts a $Z$ measurement on a hollow node with a loop. Equivalence rule E$1$ can be used in this case to turn the node solid, resulting in the graph on the right of Fig.~\ref{fig:examples}~$(c)1$. Now that the node is solid, the $Z$ measurement can be made by simply disconnecting the node. We present the graph for the \hyph{post}{measurement} state in Fig.~\ref{fig:examples}~$(c)2$. Note that the unmeasured nodes are in a graph state.

\subsection{Graphical description of stabilizer codes}

The stabilizer formalism was originally developed for the purpose of using stabilizer codes in quantum error correction. Fortunately, it turns out that stabilizer codes, too, can be represented using our graphical formulation of stabilizer states. The method we use is an adaptation of the one in Ref.~\cite{schlingemann:graphs}, extended in Refs.~\cite{schlingemann:codes,schlingemann:cluster}, for codes using graph states. We note here that our description is easily adapted to subsystem stabilizer codes~\cite{kribs:oqec,poulin:stabilizeroqec}, and even holds hope for being extended to codeword stabilized codes~\cite{cross:cwscodes,chuang:cwscodes}.

A \emph{stabilizer code\/} that encodes $k$ logical qubits in $n$ physical ones is defined to be a $2^k$ dimensional stabilizer subspace of $n$ qubits. This means that all stabilizer codes are completely specified by listing $n-k$ stabilizer generators on $n$ qubits, $g_1, \ldots, g_{n-k}$. In order to understand how stabilizer codes encode information, consider the following process. Pick any state in the stabilizer subspace that happens to also be a stabilizer state and call it $\ket{\overline{0 \cdots 0}}$, where the overbar means this is an \hyph{$n$}{qubit} encoding of a $k$ qubit state. To define this state we need to list $k$ more stabilizer generators, $\overline{Z}_1, \ldots, \overline{Z}_k$, in addition to $g_1, \ldots, g_{n-k}$ which we already know stabilize the state. Given this, we can define a total of $2^k$ stabilizer states all of which are in the given stabilizer subspace. These states are written as $\ket{\overline{c_1 \cdots c_k}}$, for $c_j \in \{0,1\}$, and are defined to be stabilizer states stabilized by $g_1, \ldots, g_{n-k}, (-1)^{c_1} \overline{Z}_1, \ldots, (-1)^{c_k} \overline{Z}_k$. Hence we have defined our encoded basis states and our effective Pauli $Z$ operators which put negative signs in the appropriate way on the basis states. It only remains to prove our encoded basis states are orthogonal,
\begin{align}
\begin{split}
\Big\langle \overline{c'_1 \cdots c'_k} \Big| \overline{c\phantom{'}_1 \cdots c\phantom{'}_k} \Big\rangle &= \Big\langle \overline{c'_1 \cdots c'_k} \Big| (-1)^{c_j} \overline{Z}_j \Big| \overline{c\phantom{'}_1 \cdots c\phantom{'}_k} \Big\rangle \\
&= (-1)^{c_j} (-1)^{c'_j} \Big\langle \overline{c'_1 \cdots c'_k} \Big| \overline{c\phantom{'}_1 \cdots c\phantom{'}_k} \Big\rangle \\
& \Rightarrow c_j = c'_j \mbox{ or } \Big\langle \overline{c'_1 \cdots c'_k} \Big| \overline{c\phantom{'}_1 \cdots c\phantom{'}_k} \Big\rangle = 0,
\end{split}
\end{align}
and we have completely specified how the $k$ qubits of information are encoded.

Since stabilizer graphs are constructed using canonical generators, we begin by using the given stabilizer generators to construct a canonical set of generators for $\ket{\overline{0 \cdots 0}}$. Doing this, we obtain canonical stabilizer generators, $g'_j(0 \cdots 0)$, which can be written in terms of the original generators as
\begin{equation}
g'_j (0 \cdots 0) = g_1^{a_1(j)} \cdots g_{n-k}^{a_{n-k}(j)} \overline{Z}_1^{b_1(j)} \cdots \overline{Z}_k^{b_n(j)},
\end{equation}
with the binary vector $\left( \begin{array}{cc} a(j) & b(j) \end{array} \right)$ specifying how to obtain the $j$th canonical generator. This binary vector also permits the obtention of canonical generators for $\ket{\overline{c_1 \cdots c_k}}$, since 
\begin{equation} \label{eq:canoncodegens}
\begin{array}{lll}
g'_j (c_1 \cdots c_k) & = & g_1^{a_1(j)} \cdots g_{n-k}^{a_{n-k}(j)} \left( (-1)^{c_1} \overline{Z}_1 \right)^{b_1(j)} \cdots \left( (-1)^{c_k} \overline{Z}_k \right)^{b_{n}(j)} \\
& = & (-1)^{b(j) c^T} g'_j (0 \cdots 0).
\end{array}
\end{equation}
Thus all of the $2^k$ stabilizer states in our stabilizer code are specified by the same canonical generators, up to an overall sign. In other words, all of the states $\ket{\overline{c_1 \cdots c_k}}$ in our code are described by the same generator matrix.

Now we are ready to represent any stabilizer code by a graph. First, using the canonical generators $g'_j (0 \cdots 0)$ draw a reduced stabilizer graph on $n$ nodes that represents the state $\ket{\overline{0 \cdots 0}}$. Then draw $k$ more nodes, which we call \emph{input nodes\/}, giving a total of $n+k$ nodes. These $k$ additional input nodes correspond to the $k$ qubit state being encoded. Next, we want connect the input nodes to the stabilizer graph by connecting the $l$th input node to the $j$th node of the stabilizer graph if $b_l(j) = 1$, i.e., when $\overline{Z}_l$ was used to obtain the $j$th canonical generator, $g'_j(0 \cdots 0)$, for $\ket{\overline{0 \cdots 0}}$. The resulting graph provides a complete description of the basis states and logical operators for the given stabilizer code. 

Using this graph we can reconstruct the stabilizer generators for $\ket{\overline{c_1 \cdots c_k}}$ as follows. First of all, if $c_l = 1$ let us call the $l$th input node an \emph{active\/} input node. To determine the $j$th stabilizer generator, simply begin with $g'_j (0 \cdots 0)$ and multiply it by $-1$ for each edge connecting the $j$th node of the stabilizer graph to an active input node. Doing this we find that the generator is multiplied by $-1$ if and only if $b_l(j) = c_l = 1$. Hence we have,
\begin{equation}
g'_j (c_1 \cdots c_k) = (-1)^{\sum_{l=1}^k b_l(j) c_l} g'_j (0 \cdots 0) = (-1)^{b(j) c^T} g'_j (0 \cdots 0),
\end{equation}
in agreement with Eq.~(\ref{eq:canoncodegens}).

Describing in words how to construct graphs for stabilizer codes is difficult. Therefore, we choose to illustrate the procedure by constructing a graph for the well known \hyph{$5$}{qubit} code. This code encodes one qubit by using $5$ qubits, and this is the smallest code that can correct any \hyph{single}{qubit} error. According to the \hyph{$5$}{qubit} code, the stabilizer generators for $\ket{\overline{c}}$, with $c=0,1$, are
\begin{equation}
\begin{array}{l}
g_1 = X \otimes Z \otimes Z \otimes X \otimes I, \mbox{ } g_2 = I \otimes X \otimes Z \otimes Z \otimes X, \\
g_3 = X \otimes I \otimes X \otimes Z \otimes Z, \mbox{ } g_4 = Z \otimes X \otimes I \otimes X \otimes Z, \\
\mbox{and}\hspace{1em} (-1)^c \overline{Z} = (-1)^c Z \otimes Z \otimes Z \otimes Z \otimes Z.
\end{array}
\end{equation}
\begin{figure}
\center
\includegraphics[width=4cm]{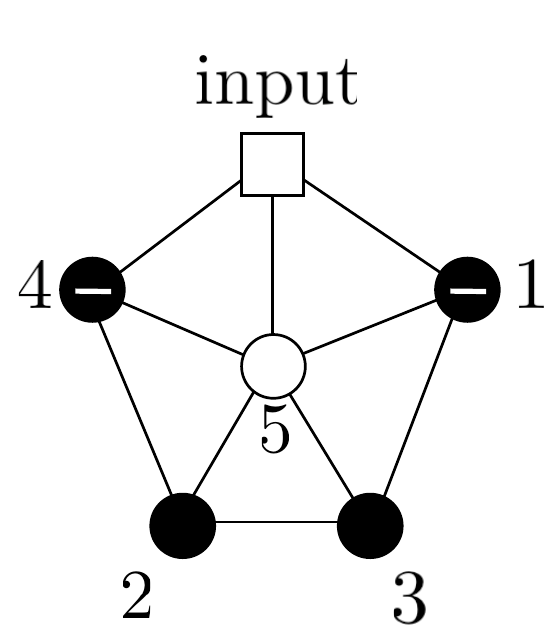}
\caption[Graph for the \hyph{$5$}{qubit} code]{A graph that represents all features of the \hyph{$5$}{qubit} code. This code encodes $1$ qubit using $5$ qubits, and so our graph is made of $1+5 = 6$ nodes. The round nodes, considered by themselves, represent the stabilizer state $\ket{\overline{0}}$. The square input node connects to nodes $1$, $4$, and $5$. This indicates that the canonical generators for $\ket{\overline{1}}$ are obtained from those of $\ket{\overline{0}}$ by switching the signs of generators $1$, $4$, and $5$. \label{fig:stabcodegraph}}
\end{figure}
By writing out the generator matrix for these generators and row reducing, we determine a set of canonical generators to be
\begin{equation}
\begin{array}{l}
g'_1 = (-1)^{c+1} X \otimes I \otimes Z \otimes I \otimes X, \mbox{ } g'_2 = I \otimes X \otimes Z \otimes Z \otimes X, \\
g'_3 = Z \otimes Z \otimes X \otimes I \otimes X, \mbox{ } g'_4 = (-1)^{c+1} I \otimes Z \otimes I \otimes X \otimes X, \\
\mbox{and}\hspace{1em} g'_5 = (-1)^c Z \otimes Z \otimes Z \otimes Z \otimes Z.
\end{array}
\end{equation}
As illustrated in Fig.~\ref{fig:stabcodegraph}, we obtain a graph for this stabilizer code by first drawing the stabilizer graph for $\ket{\overline{0}}$ and then drawing an edge from the input node to all those graph nodes whose corresponding canonical stabilizer generators contains a factor $(-1)^c$.

\begin{figure}
\center
\includegraphics[width=10cm]{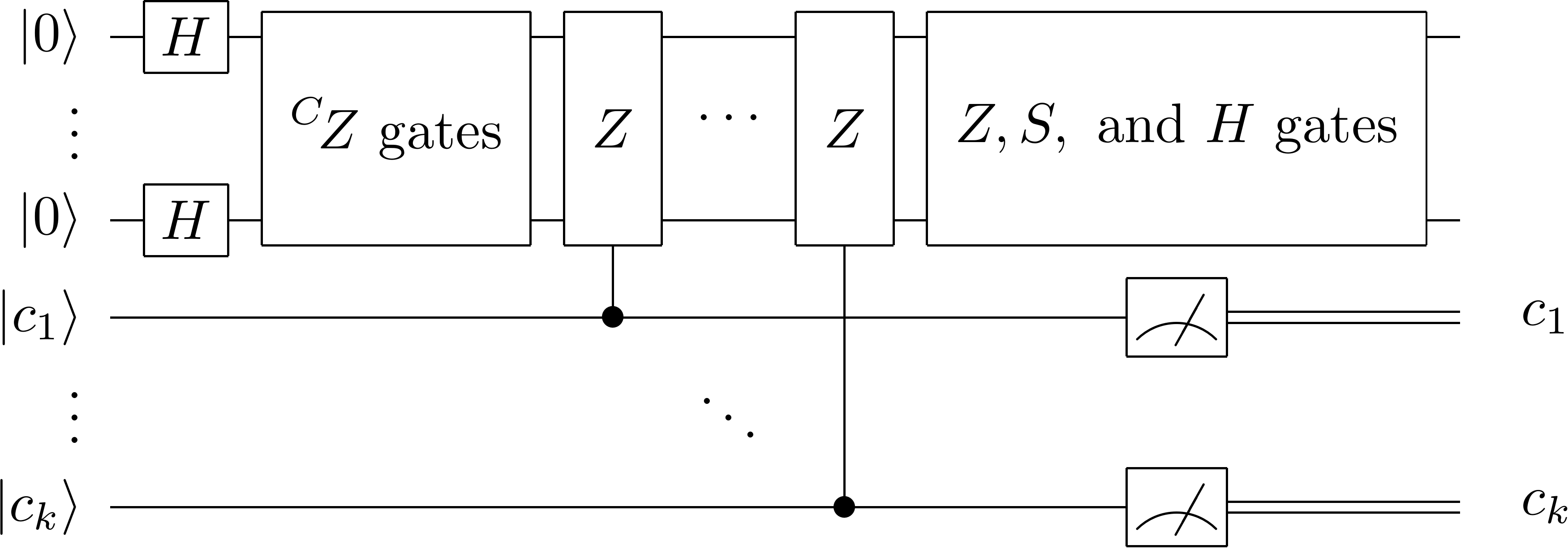}
\caption[Circuit encoding stabilizer code basis states]{A circuit that encodes the basis state $\ket{c_1 \cdots c_k}$. This circuit is directly obtained from the graph that describes the stabilizer code. The $\CZ$ gates, as well as the $Z, S,$ and $H$ gates, are determined by the stabilizer graph part of the code's graph. The final $k$ qubits correspond to input nodes, and their controlled-$Z$ gates act on those qubits connected to that input node in the graph. If $c_1, \ldots c_k = 0$, then the controlled-$Z$ gates do not act, and the top $n$ qubits are in state characterized by the stabilizer graph, i.e., the $\ket{\overline{0 \cdots 0}}$ state. If some $c_j$ are $1$, then the $\CZ$ gates from those qubits deposit a $Z$ onto qubits connected to them. These $Z$ gates flip generator signs so that the resulting state encodes $\ket{\overline{c_1 \cdots c_k}}$. Before the measurement, the final $n+k$ qubit state is $\ket{\overline{c_1 \cdots c_k}}\ket{c_1 \cdots c_k}$. \label{fig:codecircuit}}
\end{figure}

We often have occasion to think of \hyph{stabilizer}{state} graphs as a prescription for how to create the state. It turns out this is true for stabilizer codes as well, in the sense that the graph gives a set of instructions for how to encode a $k$ qubit state. Fig.~\ref{fig:codecircuit} shows a circuit corresponding to a graph that represents a stabilizer code, where edges to input nodes are also represented by $\CZ$ gates, followed by a measurement in the $Z$ basis on the $k$ input nodes. This circuit is capable of encoding any basis state, but fails to encode arbitrary unknown states. This failure of the circuit to encode arbitrary states fundamentally arises because quantum circuits are linear in the input, while our method of encoding is not linear in the input. However, a slight modification of the circuit in Fig.~\ref{fig:codecircuit} does give rise to a circuit that is capable of encoding arbitrary unknown quantum states.

To encode an arbitrary state, $\ket{\psi} = \sum_{c_1, \ldots c_k = 0}^1 \alpha_{c_1 \cdots c_k} \ket{c_1 \cdots c_k}$, we simply switch the measurement basis to the $X$ basis. From Fig.~\ref{fig:codecircuit} and linearity, we know that before the measurement the full $n+k$ qubit state is
\begin{equation}
\sum_{c_1 \cdots c_k = 0}^1 \alpha_{c_1 \cdots c_k} \ket{\overline{c_1 \cdots c_k}} \ket{c_1 \cdots c_k}.
\end{equation}
Now measure in the $X$ basis by applying a Hadamard and measuring in the $Z$ basis. Hadamards on the $k$ input qubits transform the state to
\begin{align}
\begin{split}
& \sum_{c_1 \cdots c_k = 0}^1 \alpha_{c_1 \cdots c_k} \ket{\overline{c_1 \cdots c_k}} \left( \frac{\ket{0}+(-1)^{c_1}\ket{1}}{\sqrt{2}} \right) \cdots \left( \frac{\ket{0}+(-1)^{c_k}\ket{1}}{\sqrt{2}} \right) \\
= &\; \frac{1}{2^{k/2}} \sum_{c_1 \cdots c_k = 0}^1 \sum_{d_1 \cdots d_k = 0}^1 (-1)^{c d^T} \alpha_{c_1 \cdots c_k} \ket{\overline{c_1 \cdots c_k}} \ket{d_1 \cdots d_k}.
\end{split}
\end{align}
Now we make the $Z$ measurements and obtain the results $x_1, \ldots, x_k \in \{0,1\}$, making the \hyph{post}{measurement} state of the first $n$ qubits
\begin{equation}
\sum_{c_1 \cdots c_k = 0}^1 (-1)^{c x^T} \alpha_{c_1 \cdots c_k} \ket{\overline{c_1 \cdots c_k}}.
\end{equation}
This can be made the correct encoded state by applying the logical $(\overline{Z}_j)^{x_j}$ operator for all $j$, since
\begin{align}
\begin{split}
\ket{\overline{\psi}} = & \prod_{j=1}^k (\overline{Z}_j)^{x_j} \sum_{c_1 \cdots c_k = 0}^1 (-1)^{c x^T} \alpha_{c_1 \cdots c_k} \ket{\overline{c_1 \cdots c_k}} \\
= & \sum_{c_1 \cdots c_k = 0}^1 (-1)^{c x^T} \alpha_{c_1 \cdots c_k} \prod_{j=1}^k (\overline{Z}_j)^{x_j} \ket{\overline{c_1 \cdots c_k}} \\
= & \sum_{c_1 \cdots c_k = 0}^1 (-1)^{c x^T} \alpha_{c_1 \cdots c_k} \prod_{j=1}^k (-1)^{c_j x_j} \ket{\overline{c_1 \cdots c_k}} \\
= & \sum_{c_1 \cdots c_k = 0}^1 (-1)^{c x^T} \alpha_{c_1 \cdots c_k} (-1)^{c x^T} \ket{\overline{c_1 \cdots c_k}} = \sum_{c_1 \cdots c_k = 0}^1 \alpha_{c_1 \cdots c_k} \ket{\overline{c_1 \cdots c_k}}.
\end{split}
\end{align}
Thus we see, in analogy with stabilizer states, graphs for stabilizer codes gives a circuit that is capable of encoding an arbitrary state.

\section{Concluding remarks}

This chapter explored the stabilizer formalism for qubits. Since stabilizer states are eigenstates of products of Pauli operators, representing Pauli operators as binary vectors gives a powerful description of stabilizer states. We described how to manipulate these binary vectors to accomplish various tasks, such as multiplying Pauli operators together, applying Clifford operations, and achieving a canonical form for the stabilizer generators of a state.

Inspired by this binary representation of the stabilizer formalism, we introduce \hyph{stabilizer}{state} graphs. These are simple graphs consisting of solid and hollow nodes possibly with loops and signs attached to them. Aided by the quantum circuit formalism, we prove that Clifford operations and Pauli measurements are simply represented using stabilizer graphs. Moreover, stabilizer graphs are even capable of representing stabilizer codes in quantum error correction. Because of the relationship to quantum circuits, graphs for stabilizer states and codes have the interpretation that they represent a set of instructions for how to construct or encode the state given a small set of elementary gates.

Applying our graphical representation of stabilizer codes to problems in quantum error correction is an appealing direction for future research. In particular, subsets of nodes in our graph correspond to stabilizer elements, or syndrome measurements in the language of quantum error correction. Knowing which syndrome measurements facilitate a simple and effective error decoding strategy is an important open problem in quantum error correction. Results in this area may be forthcoming using our graphical approach. It also seems that our graphical representation of stabilizer codes could be adapted to represent subsystem stabilizer codes and possibly even codeword stabilized quantum codes. These advances are on the frontier of known error correcting codes and a unifying graphical description that proves useful would be desirable.

Another interesting open problem is the question of which stabilizer states are local unitarily equivalent to a given stabilizer state. In the case of \hyph{local}{Clifford} equivalence, we already know the answer; two stabilizer states are \hyph{local}{Clifford} equivalent if and only if their graph edges are related by local complementation. However, recent progress suggests that \hyph{local}{Clifford} equivalence is different than general local unitary equivalence~\cite{ji:lulc}. Thus, it may be worthwhile to apply our circuit identities to a \hyph{non}{Clifford} local unitary operation, thereby extending the results here to general local unitary equivalence of stabilizer states.


\chapter{Local realism}
\label{chap:localrealism}

Much of the reason we learned about stabilizer states in such detail was so that we could examine their properties from the point of view of \emph{local realism\/}. Local realism was introduced in Ref.~\cite{einstein:epr} to expose an apparent inadequacy of quantum theory. The basic elements of that argument exposed a conflict between quantum mechanics and physical intuition, namely that physical reality conforms to \emph{locality\/} and \emph{realism\/}. This conflict remained untested in the laboratory until Bell's theorem \cite{bell:epr} provided a physical experiment that could be performed. When the experimental results were published \cite{aspect:bell,rowe:loophole,weihs:loophole}, the results strongly supported quantum mechanics and refuted local realism. This result puts quantum mechanics fundamentally at odds with concepts of classical physics, and truly sets it apart from all other accepted physical theories.

\begin{figure}
\center
\includegraphics[width=10cm]{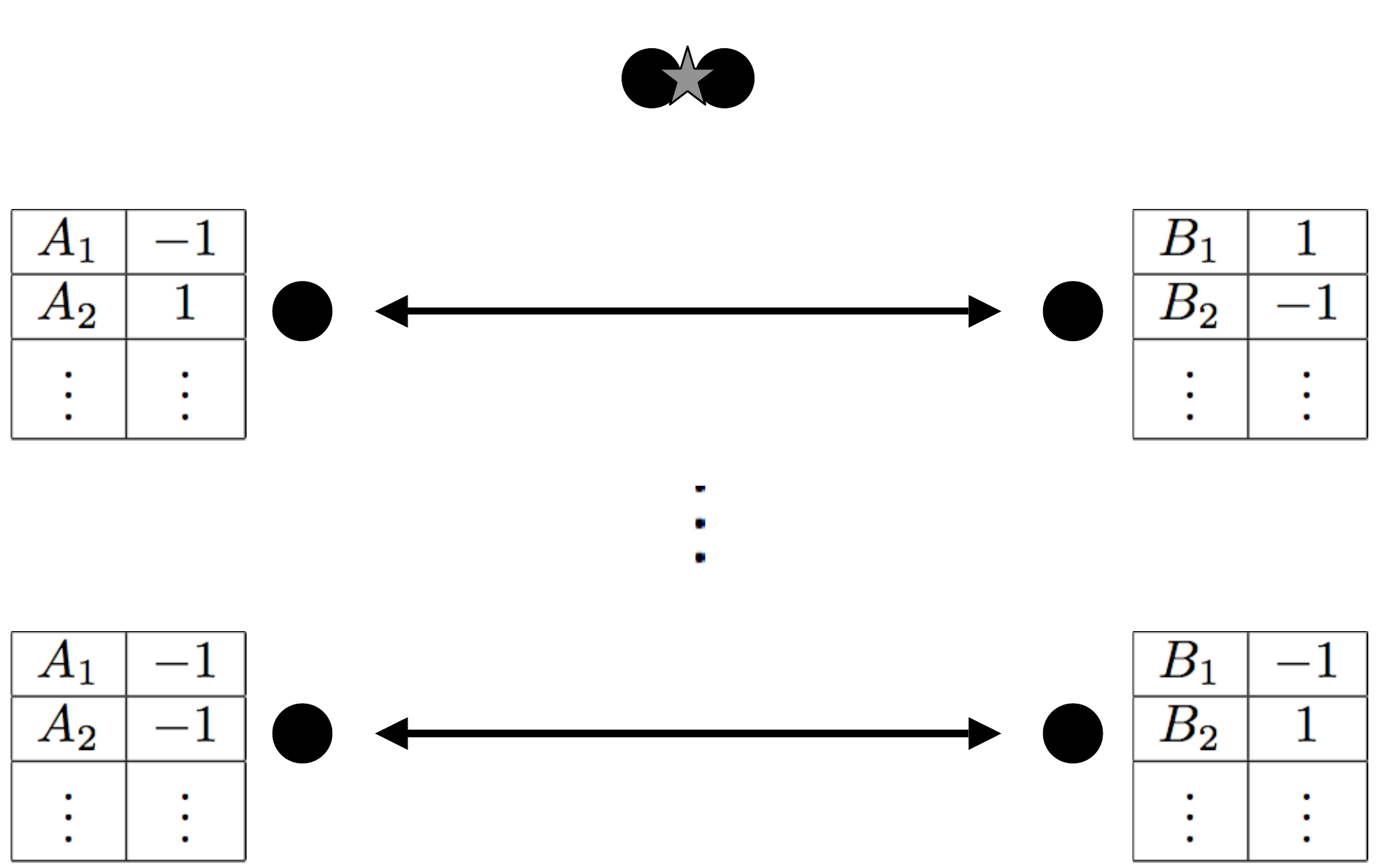}
\caption[Illustration of the concept of realism]{Realism can be thought of as asserting that there exist tables attached to each system which determine the outcomes of all possible measurements. Here we imagine that two particles, drawn as filled in circles, interact and then separate, bringing with them a table that determines their measurement outcomes. The left column of the table is the measurement $A_j$ or $B_k$, and the right column is the outcome, if one would decide to make that measurement. In this case outcomes are either $\pm 1$ in some appropriate units. Each time the experiment is repeated a new pair of particles and tables is generated, with the randomness of quantum mechanics implying that the tables change in some probabilistic fashion. \label{fig:realism}}
\end{figure}

Simply put, realism is the assumption that all possible measurements are assigned outcomes. For instance, quantum mechanics states that one cannot learn both the position and momentum of a particle, due to the Heisenberg uncertainty principle. Realism states, however, that even though we can only learn the value of position exclusive of learning momentum, both position and momentum have well defined values prior to the measurement giving us one of them. A nice conceptual tool is to think of a list of measurements attached to each system, as illustrated in Fig.~\ref{fig:realism}. This list of measurements has a value associated with each measurement which is the outcome that will be given if the measurement is made. Throughout this chapter, we always assume that there are two measurement outcomes $+1$ and $-1$.

The idea of locality is just that operations and measurements on one system cannot instantaneously affect a spatially separated second system. For our purposes, this implies that the tables in Fig.~\ref{fig:realism} cannot change once the particles become spatially separated.

So the task at hand is first to learn about the power inherent in local realistic models to capture \emph{correlations\/} in measurement outcomes. When we say correlation we mean the average of the product of measurement outcomes. We can then effectively compare local realistic models with quantum mechanics to find that quantum mechanics produces stronger correlations than local realism allows. Specifically, after reviewing some important historic work, we find a general approach to testing whether a set of experimental results could be reproduced by a local realistic theory. This method, however, is limited since the general case of this problem is believed to be very hard~\cite{pitowsky:np}. So, we derive next a large class of efficiently computable necessary conditions on the experimental results in order for them to be describable by a local realistic theory. Finally, we review a mathematical framework for learning about all probabilistic theories, which include local realistic models as well as quantum mechanics, to better understand the relationship between the two.

\section{LHV tables \label{sec:lhvtables}}

The list of measurements and outcomes associated with each separated system in Fig.~\ref{fig:realism} is called a \emph{local-hidden-variable table\/}, or LHV table for short. The term local hidden variable comes from the supposed hidden variables that determine the specific values in the table, so that randomness need not be invoked. If we assume local realism, then we assume these tables can be constructed and that observed measurement results are consistently described by such tables. In this section we learn how to mathematically construct LHV tables, of which there are two main methods, and then we will find that such tables are in conflict with quantum mechanics.

\subsection{Probability distributions} \label{subsec:probdist}

The most common, and historically relevant, method of LHV table construction is by using probability distributions. By looking at many instances of an LHV table, an instance meaning a particular set of values for the table entries, we can generate statistics for measurement results. Then we can speak of the probability that a particular set of outcomes is realized for a given set of measurements. These probabilities can be written as $P(A_1 \rightarrow a_1, A_2 \rightarrow a_2, \ldots ; B_1 \rightarrow b_1 , B_2 \rightarrow b_2, \ldots) = P(\{A_j \rightarrow a_j\};\{B_k \rightarrow b_k\})$ for two systems $A$ and $B$ and where $A_j$ and $B_k$ are the measurements and $a_j,b_k \in \{-1,1\}$ are the outcomes. These probabilities must be positive and normalized,
\begin{align} \label{eq:posandnorm}
\begin{split}
& \hspace{1em} P(\{A_j \rightarrow a_j\};\{B_k \rightarrow b_k\}) \geq 0, \\
\mbox{and} & \sum_{\{a_j\}=-1}^1 \sum_{\{b_k\}=-1}^1 P(\{A_j \rightarrow a_j\};\{B_k \rightarrow b_k\}) = 1.
\end{split}
\end{align}
Given this probability distribution we know that the LHV table assigning values of $a_j$ to $A_j$ and $b_k$ to $B_k$ occurs with probability $P(\{A_j \rightarrow a_j\};\{B_k \rightarrow b_k\})$.

\subsection{Random variables}

An alternative formulation of LHV tables that we give a bit more attention is their construction through random variables. Much work has been done analyzing local realism through probability distributions \cite{bell:epr,barrett:theories}, while this alternative perspective using random variables has been largely ignored. However, this is the perspective for much of our work.

Random variables are used to give measurement outcomes, and since we only consider measurement outcomes $\pm 1$, we only consider two valued random variables that produce values of $\pm 1$.
\begin{definition}
\label{def:vars}
A \emph{random variable\/}, $R(a)$, is such that
\[
R(a) = \left\{ \begin{array}{lll}
+1 & \mbox{with probability} & (1+a)/2 \\
-1 & \mbox{with probability} & (1-a)/2
\end{array}\right.
\]
for $a \in [-1,1]$.
\end{definition}
Note that the parameter $a$ is the expectation value of the random variable since $\avg{R(a)} = ((1+a)/2)(+1) + ((1-a)/2)(-1) = a$. To get a feeling for this definition of a random variable, let us consider a few cases. One special case is the random variable $R( \pm 1 )$ which takes the value $\pm 1$ with unit probability. We take the product of two random variables to mean a random variable whose outcome is the product of the outcomes for the two random variables being multiplied. With this understanding, we have $R(\pm 1)R(a)=\pm R(a)=R(\pm a)$. Also another case of interest is when $a=0$. In this case $R(0)$ is a $50$-$50$ random variable, like an unbiased coin flip.

The way LHV tables are constructed using random variables is to use an instance of $R(\avg{A_j})$ to determine the value of $a_j$, and likewise for $B_k$. This guarantees that the average value of $a_j$ agrees with the measurement statistics since $\bigavg{R(\avg{A_j})} = \avg{A_j}$. What is left is to make sure the individual random variables are correlated so as to reproduce all the correct measurement correlations, meaning that $\bigavg{R(\avg{A_j}) R(\avg{B_k})} = \avg{A_j B_k}$. If there are more than two systems, then we must find correlated random variables that also reproduce correlations involving the other systems. Once the proper random variables have been found, instances of LHV tables are produced through instances of the random variables.

It may or may not be obvious that random variables and probability distributions are equivalent methods of generating LHV tables. If this is not obvious, simply note that not only can probability distributions and random variables generate LHV tables, but also a probability distribution or a set of random variables could be constructed, in the asymptotic limit, given access to instances of an LHV table. Thus, these two approaches must be equivalent.

\subsection{Allowed correlations for random variables \label{subsec:varcorr}}

Now we back away from LHV tables and consider random variables in full abstractness. Our task is to figure out the possible correlations that can be found among random variables. We do this in two steps. First, we consider only two random variables and learn what correlations can exist between them. Then we can generalize this result to $n$ random variables inductively. In the next section, with this task completed, we use random variables to construct LHV tables.

A simple calculation yields the product of two uncorrelated random variables. $R(a)$ and $R(b)$ multiply to $+1$ when $R(a)$ and $R(b)$ are both $+1$ or both $-1$. The probability that this happens is
\begin{equation}
\left( \frac{1+a}{2} \right) \left( \frac{1+b}{2} \right) + \left( \frac{1-a}{2} \right) \left( \frac{1-b}{2} \right) = \frac{1+ a b}{2}
\end{equation}
which means that $R(a) R(b) = R(a b)$.

\begin{figure}
\center
\includegraphics[width=9cm]{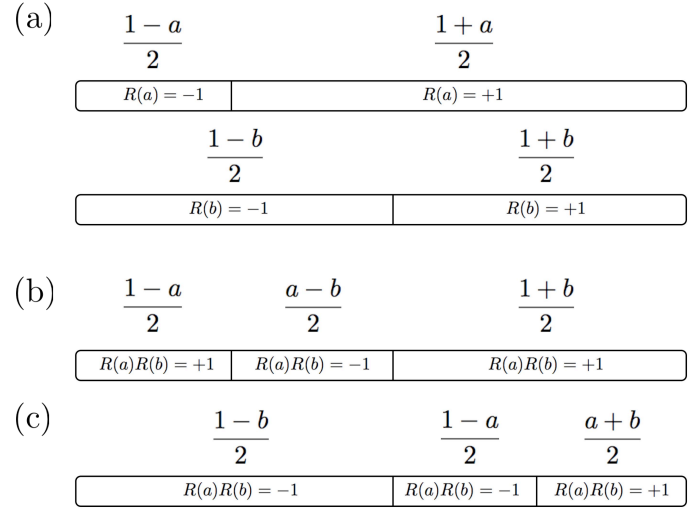}
\caption[Visual aid for random variables]{Visual aids useful for finding the possible values of two correlated random variables. The figure in (b) represents the case of maximally correlated random variables, while (c) represents the case of maximal anticorrelation.}
\label{fig:twocorr}
\end{figure}

For correlated random variables, $R(a) R(b) = R(p)$ and we wish to find the possible values of $p = \bigavg{R(a) R(b)}$, the average value of the product of the random variables. To do this, consider Fig.~\ref{fig:twocorr}(a). The bars in the figure have length $1$ and if you pick a point in the top bar, you will pick it on the left part with probability $(1-a)/2$ and on the right part with probability $(1+a)/2$. Therefore randomly picking a point on the line is equivalent to an instance of the random variable with the left and right sides corresponding to $-1$ and $+1$ outcomes, respectively.

First let's see how correlated our variables can be. We want to arrange things so that we pick both random variables to have the same value as often as we can. This is illustrated in Fig.~\ref{fig:twocorr}(b). Clearly every time $R(a) = -1$, that is $(1-a)/2$ of the time, we can also pick $R(b)=-1$ and every time $R(b)= +1$, which is $(1+b)/2$ of the time, we can simultaneously pick $R(a)=+1$. After this, though, we are stuck having $R(a)$ and $R(b)$ with opposite signs at least $(a-b)/2$ of the time. In this case the probability that the product of $R(a)$ and $R(b)$ is $+1$ is
\begin{equation}
\left( \frac{1-a}{2} \right) + \left( \frac{1+b}{2} \right)= \frac{1+(1-(a-b))}{2}
\end{equation}
which means that $R(a) R(b) = R(1-(a-b)) = R(1-|a-b|)$. Note that this argument only requires that $a \leq b$. However, since reversing $a$ and $b$ gives the same formula, it must be valid for all $a$ and $b$ as well. Random variables $R(a)$ and $R(b)$ satisfying $R(a) R(b) = R(1-|a-b|)$ are called \textit{maximally correlated}.

Now let's see how anticorrelated our random variables can be. In the figure $(1-a)/2 \leq (1+b)/2$, so we can pick $R(a)$ to be negative every time $R(b)$ is positive. Likewise we can pick $R(b)$ to be negative every time $R(a)$ is positive. As is illustrated in Fig.~\ref{fig:twocorr}(c), this means the probability of getting a negative result is
\begin{equation}
\left( \frac{1-a}{2} \right) + \left( \frac{1-b}{2} \right) = \frac{1-((a+b)-1)}{2}
\end{equation}
which means that $R(a) R(b) = R((a+b)-1) = R(|a+b|-1)$. The assumption here was that $a \geq -b$. But if $a \leq -b$, then $-a \geq - (-b)$ and the above result, being insensitive to replacing $a$ by $-a$ and $b$ by $-b$, gives us the maximal anticorrelation for $R(-a) R(-b)$ which is the same as for $R(a) R(b)$. Random variables $R(a)$ and $R(b)$ satisfying $R(a) R(b) = R(|a+b|-1)$ are called \textit{maximally anticorrelated}. Combining these two results and noting that these maximal correlations can always be achieved gives us the following result\footnote{In Ref.~\cite{groblacher:nonlocal} it was shown that $R(a) R(b) = R(p) \Rightarrow |a+b|-1 \leq p \leq 1-|a-b|$ for some nonlocal-hidden-variable theories.}.

\begin{lemma}
\label{lma:twovars}
Let $a,b \in [-1,1]$. Then $R(a) R(b) = R(p)$ is possible if and only if
\[
|a+b|-1 \leq p \leq 1-|a-b|.
\]
\end{lemma}

Let's test out this result on a few special cases. Suppose $b=\pm1$. Then Lemma~\ref{lma:twovars} gives us that $p=\pm a$ as expected. Also, we see that $p$ can take any value in $[-1,1]$ if and only if $a=b=0$, which is also not surprising. A special case that will permit a generalization to $n$ variables is when the random variables are more likely to be $+1$ than $-1$. For this case we get the following corollary to Lemma~{\ref{lma:twovars}}.

\begin{corollary}
\label{corr:twovars}
Let $a,b \in [0,1]$. Then $R(a) R(b) = R(p)$ is possible if and only if
\[
a+b-1 \leq p \leq 1-|a-b|.
\]
\end{corollary}

This corollary is the one we will be able to generalize to $n$ random variables. When we do so, we get the following theorem.

\begin{theorem}
\label{thm:nvars}
Let $a_1\geq a_2 \geq \cdots \geq a_n \in [0,1]$. Then $R(a_1) \cdots R(a_n) = R(p)$ is possible if and only if $p \in [-1,1]$ and
\[
(a_1 + \cdots + a_n) - (n-1) \leq p \leq (n-1) - (a_1 + \cdots + a_{n-1} -a_n).
\]
\end{theorem}

The proof of Theorem~\ref{thm:nvars} is a bit lengthy, and it is not terribly enlightening, so it is given in Appendix~\ref{app:proofnvars}. We can now use this result to prove the corresponding generalization of Lemma~\ref{lma:twovars}.

\begin{corollary}
\label{corr:nvars}
Let $|a_1| \geq |a_2| \geq \cdots \geq |a_n|$, with  $a_1, a_2,  \cdots, a_n \in [-1,1]$. Then $R(a_1) \cdots R(a_n) = R(p)$ is possible if and only if $p \in [-1,1]$ and
\[
(|a_1| + \cdots + |a_n|) - (n-1) \leq s_1 \cdots s_n p \leq (n-1) - (|a_1| + \cdots + |a_{n-1}| -|a_n|),
\]
where $s_j = a_j/|a_j|$ is the sign of $a_j$.
\end{corollary}

\begin{proof}
$R(a_1) \cdots R(a_n) = R(p) \Leftrightarrow R(|a_1|) \cdots R(|a_n|) = R(s_1 \cdots s_n p)$ and the result follows from Theorem~\ref{thm:nvars}.
\end{proof}

What we do now is consider for a moment the content in Theorem~\ref{thm:nvars}. First, notice that if any $a_j$ is $0$ then the interval is symmetric; this is why we do not need to define the sign of $a_j = 0$. Also, recall that for two random variables, both need to have $0$ expectation value in order for the average of their product to be anything in the full interval, $[-1,1]$. So, suppose that all random variables have the same expectation value $a \geq 0$. In this case, the upper bound in Theorem~\ref{thm:nvars} is always greater than or equal to $1$. For the lower bound to be $-1$ we need $a \leq (n-2)/n$. For $n=2$, $a$ of course needs to be $0$, but we now see that $a \rightarrow 1$ in the limit that $n \rightarrow \infty$. That is, for very large $n$, it is possible to find random variables all with expectation values close to 1 and yet with arbitrary correlations between them. With a bit of thought, though, it is not a surprising result that increasing the number of random variables also generally increases the possible correlations among them.

Also note that proving this theorem for a particular value of $n$ automatically proves it for all smaller $n$. For, inserting $a_1=1$ into the result of Theorem~\ref{thm:nvars} yields the same formula for one less random variable.

\section{Bell inequalities \label{sec:bellinequalities}}

We return now to considering a local realistic description for measurement results on quantum systems. It turns out that we can develop an experimentally testable quantity capable of revealing nature's nonconformity to local realism. The idea is to construct a Hermitian observable, $\mathcal{B}$, called a \emph{Bell operator\/}, whose expectation value under LHV tables is bounded by some value, $\avg{\mathcal{B}}_{\mbox{\scriptsize{LR}}}$, the maximal local realistic value, yet quantum mechanics predicts a possible expectation value exceeding this bound~\cite{braunstein:bellop}. Thus given a Bell operator, we have the \emph{Bell inequality\/}
\begin{equation}
\avg{\mathcal{B}} \leq \avg{\mathcal{B}}_{\mbox{\scriptsize{LR}}}.
\end{equation}
A predicted violation by quantum mechanics of any Bell inequality would prove that quantum mechanics is not both local and realistic.

Developing experimentally useful Bell inequalities, or even Bell inequalities at all, has been the subject of an immense body of research, not a fraction of which we could hope to cover here. As just a taste of what has been done see Refs.~\cite{clauser:bell,peres:bell,mermin:bell,gisin:bell,werner:bell,masanes:bell,braunstein:bell,guhne:bell,popescu:bell,zukowski:bell,cabello:bell,laskowski:bell,acin:werner}. Instead, we will review the most important early work on Bell inequalities before discussing some novel results and approaches.

\subsection{CHSH inequality}

The most well known, useful, and historically relevant Bell inequality was invented by Clauser, Horne, Shimony, and Holt and is therefore called the CHSH inequality~\cite{clauser:bell}. This Bell inequality emerges from the following thought experiment. Suppose that two qubits are allowed to interact for some time and then they are sent in opposite directions, one to Alice and the other to Bob. Alice performs one of two projective measurements, $A_1$ or $A_2$, while Bob measures either $B_1$ or $B_2$. This hypothetical experiment is performed many times so that statistics can be gathered on the measurement results, but with the chosen measurements being varied each time. With these statistics, one can calculate the average value of the CHSH Bell operator,
\begin{equation}
\mathcal{B}_{\mbox{\scriptsize{CHSH}}} = A_1 \otimes B_1 + A_1 \otimes B_2 + A_2 \otimes B_1 - A_2 \otimes B_2.
\end{equation}

If the results of the experiment are describable by a local realistic model, then we can bound the value of $\avg{\mathcal{B}_{\mbox{\scriptsize{CHSH}}}}$. Each time this experiment is run, the table expressing the measurement results has $a_j,b_k = \pm 1$, and in this run we can write the value of $\mathcal{B}_{\mbox{\scriptsize{CHSH}}}$ as $a_1 b_1 + a_1 b_2 + a_2 b_1 - a_2 b_2 = a_1 (b_1 + b_2) + a_2 (b_1 - b_2)$ which has the value $\pm 2 a_1$ or $\pm 2 a_2$ depending on whether $b_1 = b_2$ or $b_1 = -b_2$. Thus when averaging over all the trials we have that $\avg{\mathcal{B}_{\mbox{\scriptsize{CHSH}}}}_{\mbox{\scriptsize{LR}}} \leq 2$, giving the CHSH inequality,
\begin{equation} \label{eq:chsh}
\avg{\mathcal{B}_{\mbox{\scriptsize{CHSH}}}} = \avg{A_1 \otimes B_1 + A_1 \otimes B_2 + A_2 \otimes B_1 - A_2 \otimes B_2} \leq 2.
\end{equation}

We now demonstrate how quantum mechanics defies local realism by finding that this inequality can be violated. To see this, write the projective measurements, with assumed eigenvalues $\pm 1$, of Alice and Bob as $A_j = \vec{a}_j \cdot \vec{\sigma}$ and $B_k =\vec{b}_k \cdot \vec{\sigma}$, where $\vec{a}_j$ and $\vec{b}_k$ are unit vectors. This gives us a quantum mechanical upper bound,
\begin{equation}
\begin{array}{l}
\avg{A_1 \otimes B_1 + A_1 \otimes B_2 + A_2 \otimes B_1 - A_2 \otimes B_2} \\
\hspace{5em} = \avg{\vec{a}_1 \cdot \vec{\sigma} \otimes (\vec{b}_1 + \vec{b}_2) \cdot \vec{\sigma} + \vec{a}_2 \cdot \vec{\sigma} \otimes (\vec{b}_1 - \vec{b}_2) \cdot \vec{\sigma}} \\
\hspace{5em} = \avg{\vec{a}_1 \cdot \vec{\sigma} \otimes \vec{b}_+ \cdot \vec{\sigma}} |\vec{b}_1 + \vec{b}_2| + \avg{\vec{a}_2 \cdot \vec{\sigma} \otimes \vec{b}_- \cdot \vec{\sigma}} |\vec{b}_1 - \vec{b}_2| \\
\hspace{5em} \leq |\vec{b}_1 + \vec{b}_2| + |\vec{b}_1 - \vec{b}_2| \leq 2 \sqrt{2},
\end{array}
\end{equation}
where $\vec{b}_{\pm}$ is a unit vector along $\vec{b}_1 \pm \vec{b}_2$. The second to last inequality comes from the fact that $\vec{b}_{\pm} \cdot \vec{\sigma}$ is an observable with outcomes $\pm 1$. The last inequality comes from considering the angle $\theta$ between $\vec{b}_1$ and $\vec{b}_2$,
\begin{equation}
|\vec{b}_1 + \vec{b}_2| + |\vec{b}_1 - \vec{b}_2| = \sqrt{2} \left( \sqrt{1+\cos{\theta}} + \sqrt{1-\cos{\theta}} \right),
\end{equation}
and maximizing over $\theta$.

This quantum mechanical upper bound can furthermore be achieved by setting $A_1 = X$, $A_2 = Z$, $B_1 = (1/\sqrt{2})(X+Z)$, and $B_2 = (1/\sqrt{2})(X-Z)$, and considering the state $\ket{\psi} = (1/\sqrt{2})(\ket{00}+\ket{11})$. Since $2 \sqrt{2} > 2$, we then have that quantum mechanics makes predictions beyond the capabilities of any local realistic theory.

\subsection{Linear programming approach \label{subsec:lp}}

Eq.~(\ref{eq:chsh}) is a necessary condition for local realism to hold, meaning that its violation by a theory disqualifies it from being local and realistic, but that does not mean it is sufficient, meaning that its satisfaction guarantees that the theory violates no Bell inequality. The issue of whether the CHSH inequality is sufficient has been considered by others, with current progress suggesting that it is not~\cite{fine:bell,garg:bell,collins:bell,vertesi:werner}. This leads to the question of whether one could find a set of Bell inequalities that, when taken together, are sufficient for a local realistic description of measurement outcomes to exist. This prospect has been studied by Peres \cite{peres:bell} and he has indeed found such a complete set. He also found that determining all the Bell inequalities is an NP hard problem.

Peres' method for constructing a sufficient set of Bell inequalities answers the fundamental question of whether such a finite set exists and allows one to generate Bell inequalities for a given number of measurement settings. In a typical Bell inequality experiment, one of these Bell inequalities is chosen and then the experimental setup is designed to demonstrate a violation of that Bell inequality. In this subsection, we outline a much simpler approach to generating Bell inequalities than the one taken by Peres. Our approach is also particularly relevant to experimental violations of local realism because, given a robust experimental setup, our approach generates Bell inequalities that should be violated by the experimental results.

Suppose, for simplicity, that we have two separated systems on which measurements $A_j$, $1\leq j \leq m_A$, and $B_k$, $1\leq k \leq m_B$, are made. The generalization to more than two systems is straightforward. If we suppose a local realistic description of the measurement outcomes exists, then there must be, as per Subsec.~\ref{subsec:probdist}, a column vector $\vec{P}$, which if there are two outcomes per measurement has dimension $2^{m_A+m_B}$, whose entries are the probabilities that a certain table is realized, $P(\{A_j \rightarrow a_j\};\{B_j \rightarrow b_j\})$. For instance, in the case as before of two measurements we have,
\begin{equation}
\begin{array}{l}
\vec{P} = \left( \begin{array}{c} P(A_1 \rightarrow 1, A_2 \rightarrow 1; B_1 \rightarrow 1, B_2 \rightarrow 1), \end{array} \right. \\
\hspace{2.5em} \begin{array}{c} P(A_1 \rightarrow 1, A_2 \rightarrow 1; B_1 \rightarrow 1, B_2 \rightarrow -1), \end{array} \\
\left. \begin{array}{cc} \cdots &  P(A_1 \rightarrow -1, A_2 \rightarrow -1; B_1 \rightarrow -1, B_2 \rightarrow -1) \end{array} \right)^T.
\end{array}
\end{equation}
Note that this permits a generalization to measurements with any number of outcomes if desirable.

Any correlation can be written as a linear combination of entries in $\vec{P}$. Let us write the correlations that can be observed in some experiment, of which there will be $(m_A +1) (m_B +1)$ if we include the $I$ measurement, as entries in a vector, $\vec{C}$. Then this implies that a local realistic description of these correlations exists if and only if, for the appropriate matrix $M$, there exists a vector $\vec{P}$ with \hyph{non}{negative} entries such that $M \vec{P} = \vec{C}$. In the case that measurements have two outcomes $\pm 1$, $M$ will be an $(m_A~+~1) (m_B~+1)~\times~2^{m_A+m_B}$ matrix whose entries are $\pm 1$. In general, $M$ is completely determined by the number of measurements and the number of outcomes, as well as their allowed values. Thus the problem is, given $M$ and $\vec{C}$, to solve,
\begin{equation} \label{eq:solveprobdist}
\begin{array}{ccc}
M \vec{P} = \vec{C} & \mbox{subject to} & \vec{P} \geq 0.
\end{array}
\end{equation}
Notice we do not include the normalization constraint separately because it is contained in the equation $M \vec{P} = \vec{C}$. In fact, we assume that the first row of $M$ has all $1$'s and that the first entry in $\vec{C}$ is $1$, as this takes care of normalization. We also adopt the notation that $\vec{A} \geq \vec{B}$ when every entry of $\vec{A}$ is bigger than the corresponding entry of $\vec{B}$. Problems like these have been studied extensively and are called \emph{linear programs\/}.

The most general form of a linear program is, given $\vec{L}$, $M$, and $\vec{C}$, to solve
\begin{equation} \label{eq:lp}
\begin{array}{cccccc}
\mbox{minimize} & \vec{L}^T \vec{P} & \mbox{subject to} & M \vec{P} \geq \vec{C} & \mbox{and} & \vec{P} \geq 0.
\end{array}
\end{equation}
Just about any linear optimization problem involving linear constraints can be put into this form. We will see later how our problem can be put into this form, but first there is an important feature of linear programming needing discussion. For every linear program, Eq.~(\ref{eq:lp}), there is a related \emph{dual\/} linear program.
\begin{equation} \label{eq:duallp}
\begin{array}{cccccc}
\mbox{maximize} & \vec{C}^T \vec{Q} & \mbox{subject to} & M^T \vec{Q} \leq \vec{L} & \mbox{and} & \vec{Q} \geq 0.
\end{array}
\end{equation}
Notice that $\vec{P}$ appears nowhere in the dual program. This program seeks a completely different quantity $\vec{Q}$ that even generally has a different dimension than $\vec{P}$. Yet there is an important relationship between these problems. The strong duality theorem for linear programming states that the minimum value of $\vec{L}^T \vec{P}$ in Eq.~(\ref{eq:lp}) is exactly equal to the maximum value of $\vec{C}^T \vec{Q}$ in Eq.~(\ref{eq:duallp})~\cite{chvatal:book}. We will see that duality in linear programming reveals a duality between probability distributions and Bell inequalities.

Returning to Eq.~(\ref{eq:solveprobdist}), we try to rewrite our problem as a linear program,
\begin{equation}
\begin{array}{cccccc}
\mbox{minimize} & \vec{1}^{\phantom{.}T} (M \vec{P} - \vec{C}) & \mbox{subject to} & M \vec{P} \geq \vec{C} & \mbox{and} & \vec{P} \geq 0,
\end{array}
\end{equation}
where $\vec{1}$ is a vector of all $1$'s of appropriate dimension. A solution to Eq.~(\ref{eq:solveprobdist}) exists if and only if this minimum is $0$. The reason is that $M \vec{P} \geq \vec{C}$ assures that every entry of $M \vec{P} - \vec{C}$ is positive. Thus the sum of all the entries in $M \vec{P} - \vec{C}$ is $0$ if and only if each entry of $M \vec{P} - \vec{C}$ is zero, i.e $M \vec{P} = \vec{C}$. Now we easily turn this into a linear program.
\begin{equation} \label{eq:lpprobdist}
\begin{array}{cccccc}
\mbox{minimize} & (M^T \vec{1}\phantom{.})^T \vec{P} & \mbox{subject to} & M \vec{P} \geq \vec{C} & \mbox{and} & \vec{P} \geq 0.
\end{array}
\end{equation}
If the minimum value is $\vec{1}^{\phantom{.}T} \vec{C}$, then there exists a probability distribution, $\vec{P}$, producing the given measurement correlations, $\vec{C}$. Otherwise, no such probability distribution exists.

As a practical matter, this linear program gives us a method for determining whether a set of correlations violates local realism. In terms of theory, it gives a deep connection between probability distributions and Bell inequalities as we show below. The dual program of Eq.~(\ref{eq:lpprobdist}), which eventually produces Bell inequalities, is
\begin{equation} \label{eq:duallpbi}
\begin{array}{cccccc}
\mbox{maximize} & \vec{C}^T \vec{Q} & \mbox{subject to} & M^T \vec{Q} \leq M^T \vec{1} & \mbox{and} & \vec{Q} \geq 0.
\end{array}
\end{equation}
From the strong duality theorem we know that there exists a local realistic description of the correlations in $\vec{C}$ if and only if this maximum value is $\vec{1}^{\phantom{.}T} \vec{C}$. Since $\vec{Q} = \vec{1}$ satisfies the constraints and achieves this value, we know that generally the maximum value must exceed $\vec{1}^{\phantom{.}T} \vec{C}$.

From this we deduce that there exists a local realistic description of the correlations if and only if
\begin{equation} \label{eq:duallpbi}
\begin{array}{cccccc}
\mbox{maximize} & \vec{C}^T (\vec{Q}-\vec{1}) & \mbox{subject to} & M^T (\vec{Q}-\vec{1}) \leq 0 & \mbox{and} & (\vec{Q}-\vec{1}) \geq -\vec{1}
\end{array}
\end{equation}
is $0$. This gives us the set of Bell inequalities $\vec{C}^T (\vec{Q}-\vec{1}) \leq 0$ for all vectors $\vec{Q}-\vec{1}$ satisfying the constraints. This is a complete set of Bell inequalities in that they are both necessary and sufficient for a local realistic description to exist. They all have the form of a linear combination of correlations being less than $0$.

We can simplify these Bell inequalities in a few ways. The first thing is to recognize that the constraint $(\vec{Q}-\vec{1}) \geq -\vec{1}$ is not needed. For, suppose that we have a vector $\vec{Q}-\vec{1}$ satisfying $M^T (\vec{Q}-\vec{1}) \leq 0$, but such that $(\vec{Q}-\vec{1}) \geq -q \vec{1}$ for some $q > 1$. Then the vector $(\vec{Q}-\vec{1})/q$ satisfies all constraints and thus we get the Bell inequality $\vec{C}^T (\vec{Q}-\vec{1})/q \leq 0 \Leftrightarrow \vec{C}^T (\vec{Q}-\vec{1}) \leq 0$. So in fact we get a Bell inequality for all vectors $\vec{Q}-\vec{1}$ satisfying $M^T (\vec{Q}-\vec{1}) \leq 0$, regardless of whether they satisfy the second constraint.

A second simplification can also be used to deal with the constraint $M^T (\vec{Q}-\vec{1}) \leq 0$. Since we assume the first row of $M$ is all $1$'s, due to $\vec{P}$ needing to be normalized, we can break up $\vec{Q}-\vec{1}$ as $\vec{Q} - \vec{1} = \left( \begin{array}{cc} -q_0 & \vec{q} \end{array} \right)^T$, where $\vec{q}$ has one less dimension than $\vec{Q}$. If we also break up $\vec{C} = \left( \begin{array}{cc} 1 & \vec{c} \end{array} \right)^T$, then the Bell inequalities are $-q_0+\vec{c}^{\phantom{.}T} \vec{q} \leq 0$ for all $-q_0 \vec{1} + \tilde{M}^T \vec{q} \leq 0$, where $\tilde{M}$ is $M$ without the first row. To have a complete set of Bell inequalities it suffices to pick the smallest $q_0$ and so we have $q_0 = (\tilde{M}^T \vec{q})_{\mbox{\scriptsize{max}}}$, where max denotes the maximum entry of the vector, giving us all the Bell inequalities,
\begin{equation}
\begin{array}{ccc}
\vec{c}^{\phantom{.}T} \vec{q} \leq (\tilde{M}^T \vec{q})_{\mbox{\scriptsize{max}}} & \mbox{for all} & \vec{q}.
\end{array}
\end{equation}

In some sense this is an obvious result. It states that all the Bell inequalities are obtained by considering an arbitrary linear combination of correlations, and then saying that this must be less than the maximum possible value of those correlations under extremal LHV tables. However, this result is very important because it says we need only consider linear combinations of correlations, a fact that is far from obvious to begin with. It is also important because if a local realistic description of a set of correlations does not exist, it provides an algorithm that generates a Bell inequality which the correlations violate.

What we have shown here is a computational method that determines whether a set of correlations can be described within a local realistic framework. If the set is so describable, then the linear program gives a probability distribution resulting in those correlations. If the set is not describable by a local realistic model, then the linear program is capable of generating a Bell inequality that the correlations violate. Practically, this is a useful thing to be able to do. In terms of theory, linear programming gives a way to generate all the Bell inequalities and through the principle of duality in linear programming we see that probability distributions and Bell inequalities are also dual to each other.

\subsection{\hyph{Random}{variable} Bell inequalities \label{subsec:rvbi}}

\begin{figure}
\center
\includegraphics[width=6cm]{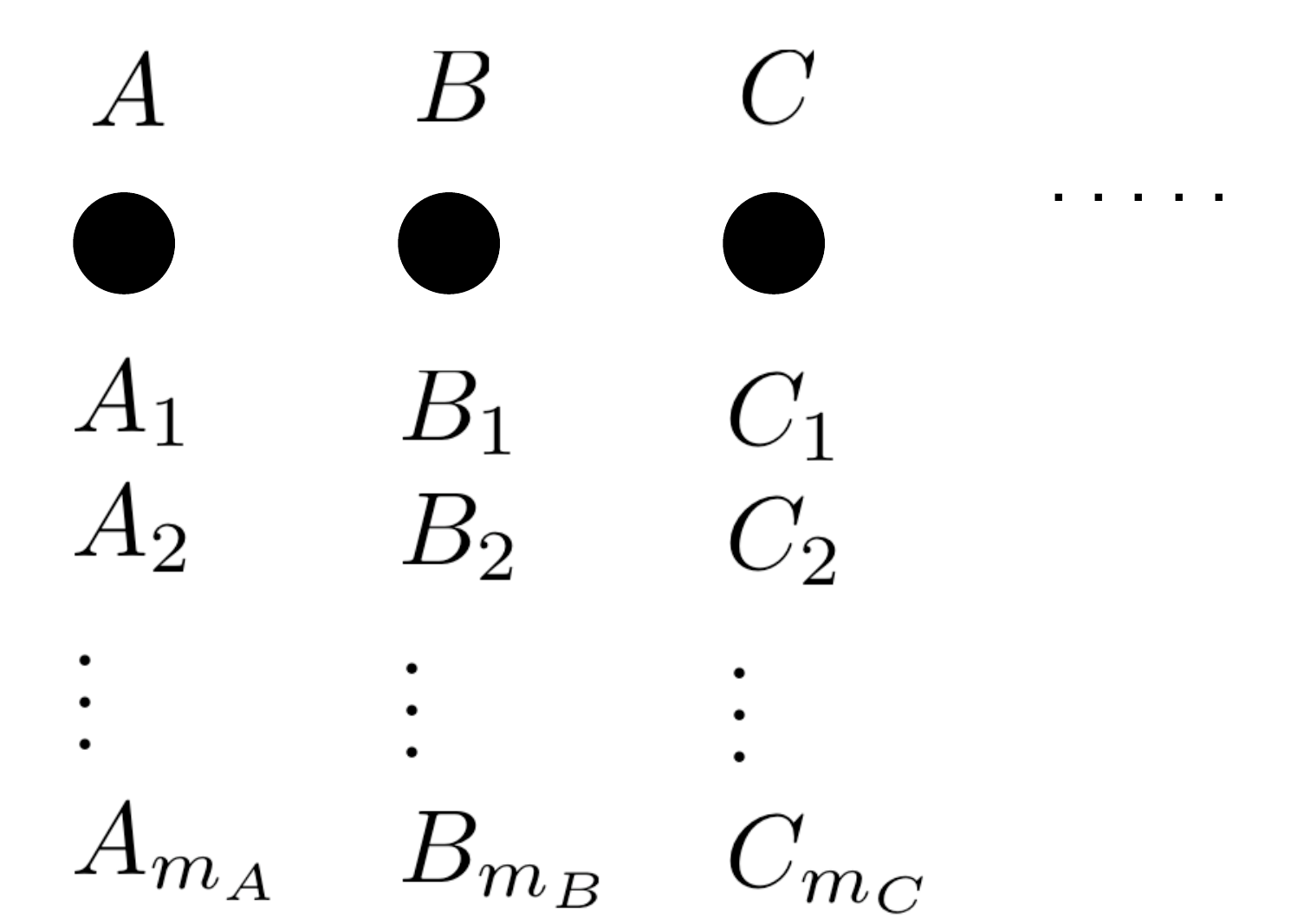}
\caption[Setting for \hyph{random}{variable} Bell inequalities]{To derive Bell inequalities from random variables we assume the following setting. We allow an arbitrary number of parties, listed here as party $A$, $B$, $C$, etc. Each party controls a subsystem of the total system, denoted as filled in circles. Each party can furthermore make an arbitrary number of measurements, although these measurements must give only two outcomes, $\pm 1$. This last restriction is the only lack of generality in the setup. \label{fig:situation}}
\end{figure}

It is typically computationally difficult to solve the linear program in Subsec.~\ref{subsec:lp} because the number of columns in the matrix $M$ grows exponentially with the number of measurements. Therefore, it is useful to use a different approach to derive explicit Bell inequalities, although such a set may not be complete. In order to come up with Bell inequalities for a wide range of configurations, we take a novel approach involving random variables. We derive Bell inequalities using random variables for the situation illustrated in Fig.~\ref{fig:situation}, which is indeed quite general.

Before considering the most general case, let us consider the specific example of two parties with two measurements each. The types of correlations that exist are then of the form $\avg{A_j B_k}$, giving us a list of four correlations that must be simultaneously satisfied in order for a local realistic description of the correlations to exist.
\begin{equation}
\begin{array}{l}
R(\avg{A_1}) R(\avg{B_1}) = R(\avg{A_1 B_1}) \\
R(\avg{A_1}) R(\avg{B_2}) = R(\avg{A_1 B_2}) \\
R(\avg{A_2}) R(\avg{B_1}) = R(\avg{A_2 B_1}) \\
R(\avg{A_2}) R(\avg{B_2}) = R(\avg{A_2 B_2})
\end{array}
\end{equation}
A necessary condition for these four equations to have a solution is that their product have a solution. Thus multiplying these four conditions together gives us the \emph{consistency condition\/}, $1 = R(\avg{A_1 B_1}) R(\avg{A_1 B_2}) R(\avg{A_2 B_1}) R(\avg{A_2 B_2})$. This condition can be turned into an algebraic quantity by using Corollary~\ref{corr:nvars} in Subsec.~\ref{subsec:varcorr}, giving us the Bell inequalities
\begin{align}
\begin{split}
& |\avg{A_1 B_1}|+|\avg{A_1 B_2}|+|\avg{A_2 B_1}|+|\avg{A_2 B_2}|-3 \leq s_{11}s_{12}s_{21}s_{22} \\
\mbox{and}\hspace{1em}&s_{11}s_{12}s_{21}s_{22} \leq 3- |\avg{A_1 B_1}|-|\avg{A_1 B_2}|-|\avg{A_2 B_1}|+|\avg{A_2 B_2}|,
\end{split}
\end{align}
where $s_{jk}$ is the sign of $\avg{A_j B_k}$. If $s_{11}s_{12}s_{21}s_{22} = 1$ then the first inequality is always satisfied, while if $s_{11}s_{12}s_{21}s_{22} = -1$ the second inequality is always satisfied. Because of this, we can combine both above inequalities into a single Bell inequality,
\begin{equation}
|\avg{A_1 B_1}|+|\avg{A_1 B_2}|+|\avg{A_2 B_1}|- s_{11}s_{12}s_{21}s_{22}|\avg{A_2 B_2}| \leq 2.
\end{equation}
After a little investigation, it becomes clear that this is equivalent to the familiar CHSH inequality.

The general case is a straightforward generalization of the previous example. The types of correlations that we must capture have the form $\avg{A_j^a B_k^b C_l^c \cdots}$, where $a,b,c,\ldots \in \{0,1\}$ determine whether a measurement is included in the correlation. This means the random variables that produce the measurement outcomes must be correlated so that
$\bigavg{[R(\avg{A_j})]^a [R(\avg{B_k})]^b [R(\avg{C_l})]^c \cdots} = \avg{A_j^a B_k^b C_l^c \cdots}$. This gives us a list of correlations that must be simultaneously satisfied in order for a local realistic description to exist. Multiplying together subsets of these equations and using Corollary~\ref{corr:nvars} gives us that
\begin{equation} \label{eq:genccineq}
\sum_{j=1}^n |\avg{M_j}|-(n-1) \leq \prod_{j=1}^n s_j \leq (n-1)-\sum_{j=1}^{n-1} |\avg{M_j}| + |\avg{M_n}|,
\end{equation}
where the $M_j$, of which there are $n$, are measurements in the subset that do not cancel and $s_j$ is the sign of $\avg{M_j}$. One thing we should point out is that each individual system measurement must appear an even number of times throughout all the $M_j$, which implies that all the $M_j$ must multiply to $\pm I$. This is a constraint that is somewhat difficult to understand unless one writes down a few examples along the lines of our random variable derivation of the CHSH inequality. Now, as before, one of the inequalities in Eq.~(\ref{eq:genccineq}) is always satisfied. This allows us to write the most general form of a \hyph{random}{variable} Bell inequality as
\begin{equation}
\label{eq:genbi}
\sum_{j=1}^{n-1} |\avg{M_j}|- \left( \prod_{j=1}^n s_j \right) |\avg{M_n}| \leq n-2.
\end{equation}
According to Theorem~\ref{thm:nvars} we should choose $|\avg{M_j}| \leq |\avg{M_n}|$ for all $j$, but in fact Eq.~(\ref{eq:genbi}) needs to hold for any ordering in order for a local realistic model to exist, so we define \hyph{random}{variable} Bell inequalities to include any choice for $M_n$. We can easily write down this Bell inequality after picking a set of correlations, $\avg{M_j}$, although this set of inequalities is typically not complete. For an example of a Bell inequality inequivalent to any in this set, see Ref.~\cite{laskowski:bell}.

\subsection{Two qubits}

In this subsection we analyze \hyph{random}{variable} Bell inequalities, Eq.~(\ref{eq:genbi}), in the special case of two parties, but an arbitrary number of measurements. In this case we prove that all \hyph{random}{variable} Bell inequalities are equivalent to the CHSH inequality. That is, if the CHSH inequality holds for all choices of measurement then so do all \hyph{random}{variable} Bell inequalities on two qubits. Conversely, since the CHSH inequality is a special case of a \hyph{random}{variable} Bell inequality, satisfaction of Eq.~(\ref{eq:genbi}) guarantees that the CHSH inequality holds.

If we consider the value of $n$ in Eq.~(\ref{eq:genbi}), which is the number of joint measurements, we can phrase the problem as follows. For $n=1$, Eq.~(\ref{eq:genbi}) says that $- |\avg{I \otimes I}| \leq -1$ which always holds. For $n=2$, we have that $|\avg{A \otimes B}| - |\avg{A \otimes B}| \leq 0$ which also always holds. Continuing with $n=3$, we get that $|\avg{A \otimes I}| + |\avg{I \otimes B}| - |\avg{A \otimes B}| \leq 1$, and sign variations thereof. Because of how \hyph{random}{variable} Bell inequalities are constructed, we know that $I$ must appear as indicated so that the three operators can multiply to the identity. This inequality must also be satisfied because it is equivalent to $R(\avg{A \otimes I}) R(\avg{I \otimes B}) = R(\avg{A \otimes B})$. One can find such random variables as follows. Prepare the quantum system in a lab and make measurements of $A \otimes B$. The outcomes of $A$ and $B$ determine $R(\avg{A \otimes I})$ and $R(\avg{I \otimes B})$, and these multiply to $R(\avg{A \otimes B})$. The first instance of Eq.~(\ref{eq:genbi}) that is not necessarily satisfied by quantum states is $n=4$, which gives rise to CHSH type inequalities. Our claim is that the $n=4$ case being satisfied is sufficient for all values of $n$ to be satisfied.

\begin{lemma}
For two qubits, all \hyph{random}{variable} Bell inequalities are satisfied if and only if the special case of $n=4$ is satisfied.
\end{lemma}

\begin{proof}
The forward direction is trivial. To prove the converse, suppose that
\begin{equation}
|\avg{A_1 \otimes B_1}|+|\avg{A_1 \otimes B_2}|+|\avg{A_2 \otimes B_1}|-s_{11}s_{12}s_{21}s_{22}|\avg{A_2 \otimes B_2}| \leq 2.
\end{equation}
To show that Eq.~(\ref{eq:genbi}) must now be satisfied we consider the possible forms of $M_n$.

The first case we consider is $M_n = I \otimes B_1$, with $B_1 \ne I$ since we can always reorder the $M_j$ to have $M_n \ne I \otimes I$ provided a \hyph{non}{identity} measurement exists. Since $B_1$ must occur in some other $M_j$, we choose it to appear in $M_{n-1} = A_1 \otimes B_1$. Now we add and subtract $\avg{M} = \avg{A_1 \otimes I}$ from Eq.~(\ref{eq:genbi}) and what we want to prove becomes
\begin{align}
\begin{split}
\sum_{j=1}^{n-2} |\avg{M_j}| &- (s_1 \cdots s_{n-2} s) |\avg{A_1 \otimes I}| + (s_1 \cdots s_{n-2} s) |\avg{A_1 \otimes I}| \\
&+ |\avg{A_1 \otimes B_1}| - (s_1 \cdots s_{n-2} s)(s_{n-1} s_n s) |\avg{I \otimes B_1}| \leq n-2,
\end{split}
\end{align}
where $s$ is the sign of $\avg{M}$. Now if we can prove that
\begin{equation} \label{eq:oneless}
\sum_{j=1}^{n-2} |\avg{M_j}| - (s_1 \cdots s_{n-2} s) |\avg{A_1 \otimes I}| \leq n-3,
\end{equation}
then combining this with the fact that $(s_1 \cdots s_{n-2} s) |\avg{A_1 \otimes I}| + |\avg{A_1 \otimes B_1}| - (s_1 \cdots s_{n-2} s)(s_{n-1} s_n s) |\avg{I \otimes B_1}| \leq 1$, which follows after some thought from the $n=3$ case, we will have proven Eq.~(\ref{eq:genbi}). Notice that Eq.~(\ref{eq:oneless}) is an instance of a \hyph{random}{variable} Bell inequality with one fewer measurement. This follows from the fact that we eliminated an even number of individual measurements, thereby leaving the equation with an even number of individual measurements. Thus we have used the $n=3$ case to reduce the value of $n$ by one.

This strategy of reducing the value of $n$ works with the other possible forms of $M_n$ as well. If $M_n = A_1 \otimes I$, then we can use the same strategy as before to reduce the value of $n$. The only remaining case is if $M_n = A_1 \otimes B_1$, with $A_1, B_1 \ne I$. In this case we choose $\avg{M_{n-1}} = \avg{A_2 \otimes B_1}$, which we know must be a term in the sum because $B_1$ must appear again. We then choose $M_{n-2} = A_1 \otimes B_2$, which also must exist because $A_1$ must appear again but cannot appear in $M_{n-1}$. Now we add and subtract $M = A_2 \otimes B_2$ from Eq.~(\ref{eq:genbi}) to get
\begin{align}
\begin{split}
\sum_{j=1}^{n-3} |\avg{M_j}| &- (s_1 \cdots s_{n-3} s) |\avg{A_2 \otimes B_2}| + (s_1 \cdots s_{n-3} s) |\avg{A_2 \otimes B_2}| \\
&\hspace{3em}+ |\avg{A_1 \otimes B_2}| + |\avg{A_2 \otimes B_1}|\\
&- (s_1 \cdots s_{n-3} s)(s_{n-2} s_{n-1} s_n s) |\avg{A_1 \otimes B_1}| \leq n-2.
\end{split}
\end{align}
As before, this equation follows provided we can prove that
\begin{equation} \label{eq:twoless}
\sum_{j=1}^{n-3} |\avg{M_j}| - (s_1 \cdots s_{n-3} s) |\avg{A_2 \otimes B_2}| \leq n-4,
\end{equation}
and
\begin{equation}
\begin{array}{l}
(s_1 \cdots s_{n-3} s) |\avg{A_2 \otimes B_2}| + |\avg{A_1 \otimes B_2}| + |\avg{A_2 \otimes B_1}| \\
\hspace{4em}- (s_1 \cdots s_{n-3} s)(s_{n-2} s_{n-1} s_n s) |\avg{A_1 \otimes B_1}| \leq 2.
\end{array}
\end{equation}
This last equation, again with some thought, follows from the $n=4$ case. The first equation, Eq.~(\ref{eq:twoless}), is simply a \hyph{random}{variable} Bell inequality with two fewer measurements because, as before, we removed an even number of individual measurements.

Thus we can always reduce the problem to proving a \hyph{random}{variable} Bell inequality involving either one or two fewer measurements. In this way we reduce the problem, by step sizes of one or two, to a \hyph{random}{variable} Bell inequality involving either three or four measurements. We conclude that all \hyph{random}{variable} Bell inequalities follow from the $n=4$ case.
\end{proof}

For more than two qubits a similar result likely holds, but the details for more parties has not yet been worked out. The conjecture would be that for $m$ qubits one only need consider $2^m$ joint measurements or fewer, with all larger number of measurements following as a consequence of these cases.

\section{Generalized probabilistic theories}

So far we have discussed the fact that measurement correlations in quantum mechanics cannot, in general, be described in local realistic terms. In order to better understand this violation of local realism, it helps to look at quantum mechanics in terms of a broader framework. By comparing quantum mechanics to other theories, many of which may also make predictions contrary to local realism, we can better understand the nature of quantum correlations and pose questions such as why do quantum correlations not violate local realism more.

Just to motivate why one should consider quantum mechanics as part of a larger framework, let us consider, once again, the CHSH Bell operator $\mathcal{B}_{\mbox{\scriptsize{CHSH}}} = A_1B_1 + A_1B_2 + A_2B_1 - A_2B_2$. Since we have that $-1 \leq \avg{A_jB_k} \leq 1$, the naive bound one would place on $\avg{\mathcal{B}_{\mbox{\scriptsize{CHSH}}}}$ would be $\avg{\mathcal{B}_{\mbox{\scriptsize{CHSH}}}} \leq 4$. This is the fundamental bound on the expectation value of the CHSH Bell operator. We found that $\avg{\mathcal{B}_{\mbox{\scriptsize{CHSH}}}}_{\mbox{\scriptsize{LR}}} = 2$ and $\avg{\mathcal{B}_{\mbox{\scriptsize{CHSH}}}}_{\mbox{\scriptsize{QM}}}=2 \sqrt{2}$, where $\avg{\mathcal{B}}_{\mbox{\scriptsize{QM}}}$ is the maximum value of $\avg{\mathcal{B}}$ over all quantum states~\cite{tsirelson:bound, wehner:tsirelson}. So neither local realism nor quantum mechanics achieves the CHSH fundamental limit.

Popescu and Rohrlich asked whether the fact that quantum mechanics does not achieve the CHSH fundamental limit is a consequence of a principle called \emph{no signaling\/}~\cite{popescu:prbox}. The \hyph{no}{signaling} principle states that a party, Alice, cannot do anything to her system that would affect the distribution of measurement outcomes on a spatially separated system controlled by Bob. Note that this is different than locality. Locality assumes that Alice cannot affect Bob's system in any way, while no signaling only states that Bob cannot measure the effects produced by Alice. In a realistic theory, the two concepts coincide. In the end, Popescu and Rohrlich found that there could indeed exist a \hyph{no}{signaling} theory in which the fundamental limit of $4$ for the CHSH Bell operator is achieved. To prove this, they constructed a fictitious device called a PR box, named after its inventors, which accomplishes this feat.

Thus the question arises of why quantum mechanics does not violate the CHSH inequality more than it does ~\cite{brassard:prbox,broadbent:prbox,buhrman:nonlocality,vandam:nonlocal,marcovitch:complexity}. While there has been some success arguing that no physical theory should be able to achieve $\avg{\mathcal{B}_{\mbox{\scriptsize{CHSH}}}} = 4$, no one has yet been able explain why quantum mechanics stops at $2 \sqrt{2}$. We will review the idea of a PR box, and also consider a framework for considering all probabilistic theories. Doing this will enable a comparison between the correlations in quantum mechanics and those in other \hyph{no}{signaling} theories. Perhaps, in the future, this will lead to a fundamental understanding of quantum correlations.


\subsection{PR box}

\begin{figure}
\center
\includegraphics[clip=true,width=8cm]{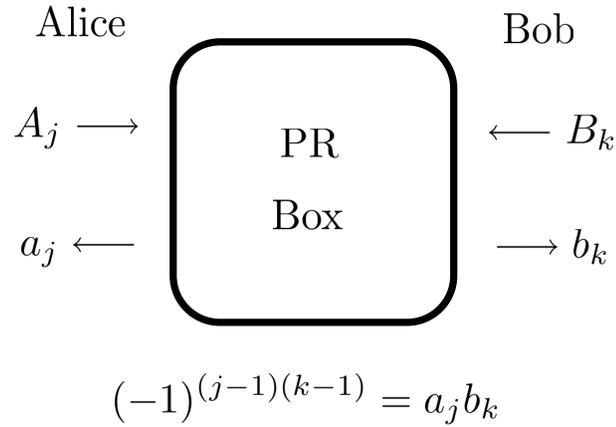}
\caption[PR box]{A visualization of a hypothetical PR box. The device has two ends, one at the location of Alice and the other at Bob. At her end, Alice can input what measurement she wants to make, and likewise for Bob. After Alice's measurement of $A_j$, with $j=1,2$, is input into the box, she immediately gets an output $a_j \in \{-1,1\}$. This measurement result, from the perspective of Alice, is completely random. Bob gets an output as well, and when Alice and Bob meet later to compare results, they find that their results are perfectly anticorrelated if $j=k=2$ and perfectly correlated if either $j=1$ or $k=1$. The box did this by giving one party a random result, and then giving the other party either the same result or its negative depending on Alice and Bob's measurement choices. \label{fig:prbox}}
\end{figure}

In this subsection we give an example of a theory, other than quantum mechanics, that makes predictions violating local realism. In this theory there exists a device called a PR box~\cite{popescu:prbox,barrett:prbox} which cannot be used to signal instantaneously between Alice and Bob, but which gives Alice and Bob access to correlations that achieve $\avg{\mathcal{B}_{\mbox{\scriptsize{CHSH}}}} = 4$.

The way a PR box works is explained in Fig.~\ref{fig:prbox}. The outputs of a PR box are perfectly correlated for $A_1B_1$, $A_1B_2$, and $A_2B_1$, and perfectly anticorrelated for $A_2B_2$. This gives $\avg{\mathcal{B}_{\mbox{\scriptsize{CHSH}}}} = 1+1+1-(-1) = 4$. This device furthermore cannot be used to signal between Alice and Bob, because regardless of one party's measurement choice the other party simply gets random measurement results. Thus this device preserves no signaling yet achieves the CHSH fundamental limit. Obviously no such device, then, could be constructed in the real world.

\subsection{General framework for a single system}

We now discuss a general framework in which local realistic theories, quantum mechanics, and PR boxes can all be discussed and compared~\cite{barrett:theories,barnum:broadcast,barnum:cloning,masanes:nonsignaling}. A basic property of these theories is that they all accept the existence of systems on which operations can be performed and measurements can be made. Measurements then have outcomes that occur with some probability, with determinism being recovered when probabilities are $1$ and $0$. Theories in this framework are therefore called \emph{generalized probabilistic theories\/}. Our first order of business is to discuss single systems in generalized probabilistic theories. Once we do this we can then understand how to deal with multiple systems, which are needed in order to discuss correlations and local realism.

If we have many independent copies of a system, then by making a measurement on those copies we can determine the probabilities of the measurement outcomes. In general we need many copies because a measurement might disturb the system. We can repeat this process until we have done all measurements, or just until we have done enough measurements to be able to predict measurement probabilities for all measurements. For example, in quantum mechanics, we know the probability for all measurement results given that we know $\avg{X}$, $\avg{Y}$, and $\avg{Z}$, so we can stop making measurements once we know the probabilities for the Pauli measurements. Then for any measurement $A=\vec{a} \cdot \vec{\sigma}$, $\avg{A}=\avg{\vec{a} \cdot \vec{\sigma}} = \sum_{j=1}^3 a_j \avg{\sigma_j}$. Thus for any generalized probabilistic theory we assume that there exists some number of \emph{fiducial\/} measurements, $F_j$, whose outcome probabilities determine the outcome probabilities for all measurements.

Once we know the set of fiducial measurements for a system, of which we assume there are $M$, we can then completely characterize the system by the fiducial measurement probabilities. The state of a system in a generalized probabilistic theory then is described by a vector, $\vec{P}$, whose entries are the probability of a fiducial measurement outcome,
\begin{equation} \label{eq:genstate}
\vec{P} = \left( \begin{array}{l} P(F_1 \rightarrow 1) \\ P(F_1 \rightarrow -1) \\ \hspace{2.5em} \vdots \\ P(F_M \rightarrow 1) \\ P(F_M \rightarrow -1) \end{array} \right).
\end{equation}
Note that this vector is different than the vector $\vec{P}$ used to generate LHV tables. In particular, we now have that the entries of $\vec{P}$ sum to $M$ rather than $1$. We assumed in Eq.~(\ref{eq:genstate}) that there are two measurement outcomes, $\pm 1$. The state vector $\vec{P}$ is easily generalized to more than two outcomes, but the two outcome scenario is sufficient for our purposes. In the case of a qubit where $M=3$ and the state is characterized by the Bloch vector $\vec{a}$, we have that
\begin{equation}
\vec{P} = \frac{1}{2} \left( \begin{array}{l} 1+a_1 \\ 1-a_1 \\ 1+a_2 \\ 1-a_2 \\ 1+a_3 \\ 1-a_3 \end{array} \right).
\end{equation}

Notice that $3$ parameters are sufficient to describe the state of a qubit, while $\vec{P}$ has $6$ entries. This redundancy comes from the fact that we have not yet enforced that the state be normalized. Thus we require that a valid state, Eq.~(\ref{eq:genstate}), be normalized in that the probabilities for each fiducial measurement sum to $1$. This requirement can be written as
\begin{equation} \label{eq:norm}
\begin{array}{lllll}
1 &=& P(F_j \rightarrow 1) + P(F_j \rightarrow -1) && \\
&=& \left( \begin{array}{cccc} 0 \cdots 0 & 1 & 1 & 0 \cdots 0 \end{array} \right) \left( \begin{array}{c} \vdots \\ P(F_j \rightarrow 1) \\ P(F_j \rightarrow -1) \\ \vdots \end{array} \right) &=& \vec{n}_j^T \vec{P},
\end{array}
\end{equation}
for $j=1,\ldots,M$ and where $\vec{n}_j^T= \vec{e}_j^{\phantom{.}T} \otimes \left( \begin{array}{cc} 1 & 1 \end{array} \right)^T$ with $\vec{e}_j$ being an $M$ dimensional vector with a $1$ in the $j$th spot and zeroes elsewhere. Our normalization constraint is now conveniently written as $\vec{n}_j^T \vec{P} = 1$, and so we call the vectors $\vec{n}_j$ \emph{normalization vectors\/}. Normalization vectors are orthogonal with
\begin{equation} \label{eq:orthn}
\vec{n}_j^T \vec{n}_k = \vec{e}_j^{\phantom{.}T} \vec{e}_k \cdot \left( \begin{array}{cc} 1 & 1 \end{array} \right) \left( \begin{array}{c} 1 \\ 1 \end{array} \right) = 2 \delta_{jk}.
\end{equation}

Since $\vec{P}$ is $2 M$ dimensional, and since we have $M$ orthogonal normalization vectors, we only need $M$ more vectors and we will have a convenient basis for all states. This role is filled by the \emph{mean vectors\/}, $\vec{m}_j$. These vectors are defined by $\vec{m}_j^T= \vec{e}_j^{\phantom{.}T} \otimes \left( \begin{array}{cc} 1 & -1 \end{array} \right)^T$, and their inner product with $\vec{P}$ gives the mean values of the fiducial measurements in that state,
\begin{equation} \label{eq:mean}
\begin{array}{lll}
\vec{m}_j^T \vec{P} &=& \left( \begin{array}{cccc} 0 \cdots 0 & 1 & -1 & 0 \cdots 0 \end{array} \right) \left( \begin{array}{c} \vdots \\ P(F_j \rightarrow 1) \\ P(F_j \rightarrow -1) \\ \vdots \end{array} \right) \\
&=& P(F_j \rightarrow 1) - P(F_j \rightarrow -1) = \avg{F_j}.
\end{array}
\end{equation}
The mean vectors are orthogonal to each other and also to the normalization vectors,
\begin{equation} \label{eq:orthogonal}
\begin{array}{l}
\vec{m}_j^T \vec{m}_k = \vec{e}_j^{\phantom{.}T} \vec{e}_k \cdot \left( \begin{array}{cc} 1 & -1 \end{array} \right) \left( \begin{array}{c} 1 \\ -1 \end{array} \right) = 2 \delta_{jk}, \mbox{ and} \vspace{1em} \\
\vec{m}_j^T \vec{n}_k = \vec{e}_j^{\phantom{.}T} \vec{e}_k \cdot \left( \begin{array}{cc} 1 & -1 \end{array} \right) \left( \begin{array}{c} 1 \\ 1 \end{array} \right) = 0 \cdot \delta_{jk} = 0.
\end{array}
\end{equation}

So now we have $2 M$ orthogonal vectors which serve as a basis for all states. This gives us that
\begin{equation}
\vec{P} = \frac{1}{2} \sum_{j=1}^M ( a_j \vec{n}_j + b_j \vec{m}_j),
\end{equation}
for some $a_j, b_j$, and with an overall factor of $(1/2)$ put in for convenience. There are still $6$ parameters here, but by using the normalization condition, Eq.~(\ref{eq:norm}), we eliminate $3$ of them,
\begin{equation}
\vec{n}_k^T \vec{P} = 1 \;\; \Rightarrow \;\; \frac{1}{2} \sum_{j=1}^M ( a_j \vec{n}_k^T \vec{n}_j + b_j \vec{n}_k^T \vec{m}_j) = a_k = 1,
\end{equation}
where we used Eqs.~(\ref{eq:orthn}) and~(\ref{eq:orthogonal}) to evaluate the inner products. So now we actually have,
\begin{equation} \label{eq:statedecomp}
\vec{P} = \frac{1}{2} \sum_{j=1}^M (\vec{n}_j + b_j \vec{m}_j).
\end{equation}
Finally, since the numbers $b_j$ represent the average value of the fiducial measurements, we have that $-1 \leq b_j \leq 1$. The vector $\vec{b}$ whose entries are $b_j$ is analogous to the Bloch vector for qubits, since the entries of the Bloch vector are the average values of Pauli measurement outcomes.

\begin{figure}
\center
\includegraphics[width=8cm]{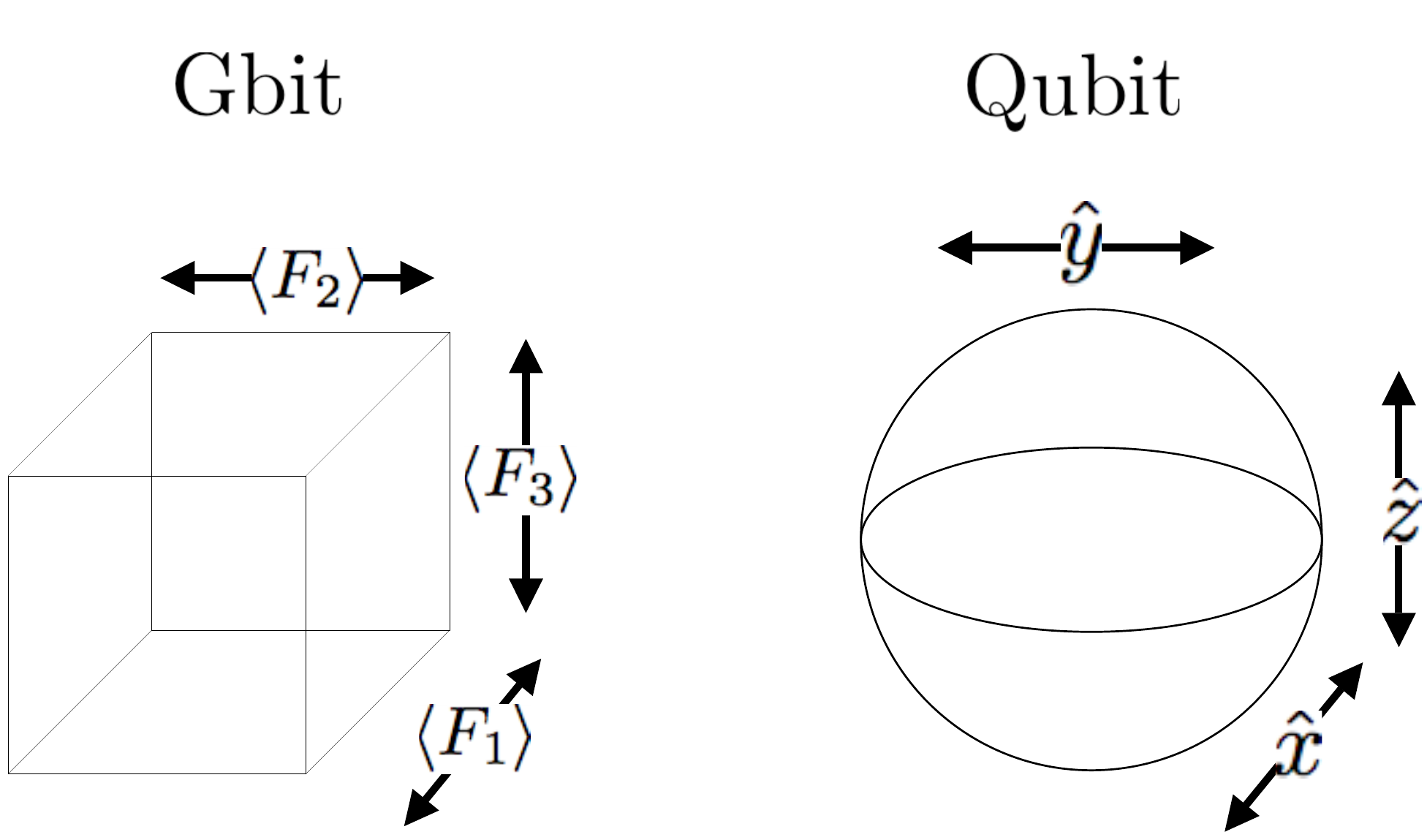}
\caption[Gbit versus qubit]{Comparison of gbits, for $M=3$, and qubits. On the left is a $2 \times 2 \times 2$ cube and on the right is a sphere of radius $1$. Vectors inside the cube give rise to states in GNST, or gbits, and vectors inside sphere give rise to states of qubits. The sphere, if moved on top of the cube, would fit perfectly inside it. This means that states of qubits are also states of gbits, but the converse not true. The corners of the cube, for instance, do not correspond to any quantum state. \label{fig:gbit}}
\end{figure}

Any generalized probabilistic theory must have systems whose states, $\vec{P}$, conform to Eq.~(\ref{eq:statedecomp}) since the states are normalized and assign positive probabilities to fiducial measurement outcomes. This does not mean that any vector $\vec{P}$ in Eq.~(\ref{eq:statedecomp}) with $-1 \leq b_j \leq 1$ represents a state in a given theory. To specify a particular theory, one must specify the set of allowed states. States in this set must be represented by normalized vectors, $\vec{P}$ with positive entries, but the set need not contain all allowed states. The theory which does allow all states is called GNST, or the \emph{generalized \hyph{no}{signaling} theory\/}~\cite{barrett:theories}. Quantum mechanics, for instance, allows only vectors $\vec{b}$ whose norm is no greater than $1$. This permits a representation of qubit states by vectors $\vec{b}$ lying inside a sphere. GNST, on the other hand, permits any vector whose entries are less than $1$ in magnitude. These vectors lie not inside a sphere, but inside a hypercube. Just as quantum bits are called qubits, we call states in GNST generalized bits, or gbits. Gbits and qubits are contrasted in Fig.~\ref{fig:gbit}.

\subsection{General framework for multiple systems}

Now that we have a complete description of single systems in generalized probabilistic theories, we move to tackle the case of multiple systems. For obvious reasons we insist that no generalized probabilistic theory allow instantaneous signaling, and this is the fundamental principle on which we base our study of multiple systems. The \hyph{no}{signaling} principle is implied by stating that for two systems, $A$ which is controlled by Alice and $B$ which is controlled by Bob, measurements and operations acting separately on $A$ and $B$ must commute. The reasoning is thus. Suppose all operators acting separately on $A$ and $B$ commute. Then results of measurements on $A$ are the same whether Bob does his operations before or after Alice. Since operations Bob does after the measurement by Alice cannot affect its result, neither can operations Bob performs before Alice's measurement have any effect. In what follows we assume only two systems for simplicity. The extension to more than two systems is fairly obvious.

If we assume that operations acting separately on $A$ and $B$ commute, then the probability $P(A \rightarrow a ; B \rightarrow b)$ is unambiguously defined. Now suppose for the moment that systems $A$ and $B$ are completely uncorrelated with each other. In this case, $P(A \rightarrow a ; B \rightarrow b) = P(A \rightarrow a) P(B \rightarrow b)$. Now we want to represent the state of these two systems with one state vector, using reasoning analogous to our motivation of the tensor product in Subsec.~\ref{subsec:multiplequbits}. Since the state of system $A$ is determined by the fiducial measurements $F_j^{\mbox{\scriptsize{A}}}$, and likewise for $B$, we conclude that the state of the entire system is the tensor product of the individual system states,
\begin{equation} \label{eq:multisys}
\begin{array}{lll}
\vec{P}^{\mbox{\scriptsize{AB}}} &=& \vec{P}^{\mbox{\scriptsize{A}}} \otimes \vec{P}^{\mbox{\scriptsize{B}}} = \left( \begin{array}{c}
P(F_1^{\mbox{\scriptsize{A}}}  \rightarrow 1) \\
\vdots \\
P(F_M^{\mbox{\scriptsize{A}}}  \rightarrow -1
\end{array} \right) \otimes \left( \begin{array}{c}
P(F_1^{\mbox{\scriptsize{B}}}  \rightarrow 1) \\
\vdots \\
P(F_M^{\mbox{\scriptsize{B}}}  \rightarrow -1
\end{array} \right) \vspace{1em} \\
&=& \left( \begin{array}{c}
P(F_1^{\mbox{\scriptsize{A}}}  \rightarrow 1) P(F_1^{\mbox{\scriptsize{B}}}  \rightarrow 1) \\
\vdots \\
P(F_M^{\mbox{\scriptsize{A}}}  \rightarrow -1) P(F_M^{\mbox{\scriptsize{B}}}  \rightarrow -1)
\end{array} \right) \vspace{1em} \\
&=& \left( \begin{array}{c}
P(F_1^{\mbox{\scriptsize{A}}}  \rightarrow 1;F_1^{\mbox{\scriptsize{B}}}  \rightarrow 1) \\
\vdots \\
P(F_M^{\mbox{\scriptsize{A}}}  \rightarrow -1;F_M^{\mbox{\scriptsize{B}}}  \rightarrow -1)
\end{array} \right),
\end{array}
\end{equation}
where we assumed $M$ fiducial measurements for both systems. Extending this to correlated systems leads us to the assumption that the state of multiple systems is a vector in the tensor product space of the individual system vector spaces. The overall state vector looks like the final equality in Eq.~(\ref{eq:multisys}).

Since we have a nice basis for a single system, involving normalization and mean vectors, we can just take a tensor product of these basis vectors to get a basis for multiple systems. Thus a state of two systems can be written as,
\begin{equation} \label{eq:almosttwosys}
\vec{P}^{\mbox{\scriptsize{AB}}} = \frac{1}{4} \sum_{j=1}^M \sum_{k=1}^M (\vec{n}_j \otimes \vec{n}_k + a_{jk} \vec{m}_j \otimes \vec{n}_k + b_{jk} \vec{n}_j \otimes \vec{m}_k + c_{jk} \vec{m}_j \otimes \vec{m}_k),
\end{equation}
where we know, as before, that the coefficient of $\vec{n}_j \otimes \vec{n}_k$ is $1$ because of normalization. We will see later that $a_{jk}$ and $b_{jk}$ determine the average values of fiducial measurement results at $A$ and $B$, and that the numbers $c_{jk}$ express correlations between fiducial measurement outcomes. We find below the restrictions on $a_{jk}$, $b_{jk}$, and $c_{jk}$.

To learn more about the states of multiple systems, let us consider the reduced states of Eq.~(\ref{eq:almosttwosys}). If we want to recover the state of Alice's system alone then we need to know the probabilities of her fiducial measurement outcomes. We can determine $P(F_j^{\mbox{\scriptsize{A}}} \rightarrow a_j)$ from $\vec{P}^{\mbox{\scriptsize{AB}}}$ by assuming that Bob makes a measurement of $F_k^{\mbox{\scriptsize{B}}}$ and summing over the outcomes. Mathematically,
\begin{equation}
P(F_j^{\mbox{\scriptsize{A}}} \rightarrow a_j) = P(F_j^{\mbox{\scriptsize{A}}}  \rightarrow a_j; F_k^{\mbox{\scriptsize{B}}}  \rightarrow 1) + P(F_j^{\mbox{\scriptsize{A}}}  \rightarrow a_j; F_k^{\mbox{\scriptsize{B}}}  \rightarrow -1).
\end{equation}
This is accomplished via normalization vectors by the action $(I \otimes \vec{n}_k^T) \vec{P}^{\mbox{\scriptsize{AB}}}$, which needs to obtain the same result regardless of $k$. Thus we obtain the reduced state for Alice's system,
\begin{equation}
\vec{P}^{\mbox{\scriptsize{A}}} = (I \otimes \vec{n}_k^T) \vec{P}^{\mbox{\scriptsize{AB}}},
\end{equation}
and similarly for Bob. Using this result on Eq.~(\ref{eq:almosttwosys}) gives us that
\begin{equation}
\vec{P}^{\mbox{\scriptsize{A}}} = \frac{1}{2} \sum_{j=1}^M (\vec{n}_j + a_{jk} \vec{m}_j),
\end{equation}
and similarly for Bob. Since this result needs to be independent of $k$, we get that $a_{jk}$ needs to be independent of $k$, i.e., $a_{jk} \rightarrow a_j$. A similar argument also gives that $b_{jk} \rightarrow b_k$. This gives us the state of two systems as,
\begin{equation} \label{eq:twosys}
\vec{P}^{\mbox{\scriptsize{AB}}} = \frac{1}{4} \sum_{j=1}^M \sum_{k=1}^M (\vec{n}_j \otimes \vec{n}_k + a_j \vec{m}_j \otimes \vec{n}_k + b_k \vec{n}_j \otimes \vec{m}_k + c_{jk} \vec{m}_j \otimes \vec{m}_k).
\end{equation}
Now we see that $a_j = \avg{F_j^{\mbox{\scriptsize{A}}}}$ and $b_k = \avg{F_k^{\mbox{\scriptsize{B}}}}$ determine the individual system results, and we have the restriction that $-1 \leq a_j,b_k \leq 1$.

Now we seek to find the allowed values of $c_{jk}$, given values of $a_j$ and $b_k$. The only constraint we have left is that the probabilities of outcomes need to be positive. Taking this view will lead us to the allowed values of $c_{jk}$. However, there is a simpler method. If we use that $a_j = \avg{F_j^{\mbox{\scriptsize{A}}}}$ and $b_k = \avg{F_k^{\mbox{\scriptsize{B}}}}$, then the outcomes at $A$ and $B$ can be thought of as instances of random variables $R(a_j)$ and $R(b_k)$. The average value $\avg{F_j F_k} = (\vec{m}_j \otimes \vec{m}_k)^T \vec{P}^{\mbox{\scriptsize{AB}}} = c_{jk}$ must be equal to $\avg{R(a_j)R(b_k)}$, giving us that
\begin{equation} \label{eq:cvals}
|a_j +b_k| - 1 \leq c_{jk} \leq 1 - |a_j - b_k|
\end{equation}
by Lemma~\ref{lma:twovars}. So we find that two system states in GNST have the form in Eq.~(\ref{eq:twosys}) with $-1 \leq a_j,b_k,c_{jk} \leq 1$ satisfying Eq.~(\ref{eq:cvals}). States of multiple systems in other theories are a subset of these states.

With this formalism of generalized probabilistic theories, we can now compare quantum mechanics to other theories which allow a greater violation of CHSH. We write, for instance, the state of a PR box as
\begin{align} \label{eq:prstate}
\begin{split}
\vec{P}_{\mbox{\scriptsize{PR box}}}^{\mbox{\scriptsize{AB}}} = \frac{1}{4} \Bigg( & \sum_{j=1}^M \sum_{k=1}^M \vec{n}_j \otimes \vec{n}_k \\
&+ \vec{m}_1 \otimes \vec{m}_1 + \vec{m}_1 \otimes \vec{m}_2 + \vec{m}_2 \otimes \vec{m}_1 - \vec{m}_2 \otimes \vec{m}_2 \Bigg),
\end{split}
\end{align}
which satisfies all the constraints on coefficients to be in GNST. We can compare this to a quantum state, which we call the Bell state, that maximally violates the CHSH inequality,
\begin{align} \label{eq:bellstate}
\begin{split}
\vec{P}_{\mbox{\scriptsize{Bell}}}^{\mbox{\scriptsize{AB}}} = \frac{1}{4} \Bigg( & \sum_{j=1}^M \sum_{k=1}^M \vec{n}_j \otimes \vec{n}_k \\
&+ \frac{1}{\sqrt{2}} ( \vec{m}_1 \otimes \vec{m}_1 + \vec{m}_1 \otimes \vec{m}_2 + \vec{m}_2 \otimes \vec{m}_1 - \vec{m}_2 \otimes \vec{m}_2 ) \Bigg),
\end{split}
\end{align}
which also satisfies the needed constraints. Now it is absolutely clear that the correlations in the quantum state are simply weaker than the PR box correlations by exactly a factor of $1/\sqrt{2}$.

There is considerably more to this framework than what is presented here. We have illustrated those aspects of generalized probabilistic theories which can be used to answer questions about local realism, which is the topic of interest for this work. Refs.~\cite{barrett:theories,barnum:broadcast,barnum:cloning,linden:prbox} contain a lot of results particularly regarding quantum information processing tasks in generalized probabilistic theories.

Using this general framework might lead us to a better understanding of quantum correlations, since now we can compare quantum mechanics to other theories that allow stronger correlations. For instance, an interesting question is to consider the theory which allows only locally quantum states and measurements but allows arbitrary correlations between systems. How does this theory differ from quantum mechanics? This formalism allows one to ask many interesting questions regarding the power of quantum correlations.

\section{Concluding remarks}

This chapter discusses aspects of local realism and the quantum mechanical violation of it. We introduce the random variable and probability distribution representations of local realism, and use both to demonstrate different results. Using probability distributions we come up with a linear program that solves the problem of whether a set of correlations, usually thought of as being generated by measurements on a quantum state, can be reproduced in a local and realistic way. If they can be, the program generates an appropriate probability distribution. Otherwise, the program generates a Bell inequality that is violated; a Bell inequality being a correlation inequality that all local and realistic theories must satisfy. Since this program scales exponentially in the number of measurements, we then use random variables to efficiently generate a set of Bell inequalities that apply to any number of parties and any number of dichotomic measurement settings.

We finish the chapter by introducing a general probabilistic model in which local realistic theories and quantum mechanics can be discussed on equal footing. We find that this model also permits other \hyph{non}{signaling} theories which violate Bell inequalities more than quantum mechanics. The question thus arises of why nature does not permit such super strong correlations.

Current research in Bell inequalities suggests that nonlinear Bell inequalities may be more powerful than linear ones. Usually, a single nonlinear Bell inequality is either capable of replacing many linear Bell inequalities~\cite{zukowski:bell}, or it is capable of demonstrating a violation of local realism using far fewer measurement settings~\cite{vertesi:werner}. It is interesting to note that \hyph{random}{variable} Bell inequalities are nonlinear. Therefore, future research that investigates nonlinear Bell inequalities seems appropriate. On a related subject, it would be exciting to know if one could supplement \hyph{random}{variable} Bell inequalities with either one additional Bell inequality, or possibly a small set of Bell inequalities, so that the new set of Bell inequalities is sufficient. Doing this even for the case of two qubits would be remarkable.

There are many open problems in generalized probabilistic theories. Perhaps the most interesting is to consider that theory which insists that states be locally quantum mechanical, but allows arbitrary correlations between local systems. By this we mean that the reduced state is that of a qubit, and furthermore that positive probabilities are assigned to all \hyph{joint}{measurement} outcomes. States in this theory can be represented by density matrices that, instead of having positive expectation values on all states, have positive expectation values on all product states. If this theory does not violate any Tsirelson bound~\cite{tsirelson:bound}, then quantum mechanics would essentially be as nonlocal as possible; if this theory does violate some Tsirelson bound, then quantum mechanics is not as nonlocal as it could be. Preliminary results suggesting the former statement is true for \hyph{two}{level} systems can be found in Ref.~\cite{barnum:locallyquantum}. Either way, an answer to this question would provide deep insight into quantum nonlocality.


\chapter{Stabilizer states and local realism}
\label{chap:cclhvmodels}

The last two chapters on stabilizer states and local realism have both been building up to this chapter, whose namesake is the title of this work. In this chapter we will study the correlations present in outcomes of Pauli measurements on stabilizer states. Pauli measurements are a good class of measurements to start with because their outcomes exhibit correlations that violate local realism~\cite{mermin:ghz,guhne:bell,scarani:nonlocality}, yet they are well understood within the stabilizer formalism~\cite{anders:simulate} as per Subsecs.~\ref{subsec:measure} and~\ref{subsec:graphmeas}.

The general problem we consider is as follows. Suppose that an \hyph{$n$}{qubit} stabilizer state, which we will assume corresponds to a connected graph, is distributed between $n$ parties so that each qubit is spatially separated from the rest. On each qubit we assume that either a measurement of $I$, $X$, $Y$, or $Z$ is made, with $I$ corresponding to no measurement at all. Once this is done, we imagine an independent investigator collecting a subset of the results and multiplying them together. This subset of results may be correlated in a particular way, and it is these correlations we wish to study.

The way we study the correlations present in Pauli measurements breaks up into two main ingredients that should be clearly distinguished. The first is that we begin with a local realistic description of all measurement results. This means that we attach an LHV table to each qubit which states what each measurement result, $X$, $Y$, and $Z$, will be. It turns out that an essentially complete description of all possible LHV tables is possible. However, these LHV tables cannot correctly reproduce all measurement correlations because no realistic description of all Pauli measurements generally exists. Thus we need to augment our LHV tables somehow in order to faithfully reproduce all quantum correlations.

We do this with our second ingredient, classical communication~\cite{toner:bell,pironio:bell, regev:communication,toner:thesis}. Specifically, after the measurements are made we allow some of the measurement choices to be communicated to other parties. With this new information parties are allowed to change their measurement outcomes so as to reproduce the correct correlations for the measurements that were made. By examining how many measurement outcomes need to be sent and to how many parties they need to be sent, we learn how difficult it is to reproduce quantum correlations. If LHV tables performed well, then we would only need a little communication. On the other hand if LHV tables necessarily get many correlations wrong, then it is reasonable to assume that a complicated communication strategy will be needed to fix them.

A model of measurement correlations that uses a local-hidden-variable table and supplements it with classical communication is called a~\emph{classical-communication-assisted local-hidden-variable model\/}, or more simply just a~\emph{communication model\/}. Classical-communication-assisted LHV models were pioneered by Toner and Bacon~\cite{toner:bell}, and Tessier~\textit{et al.\/}~\cite{tessier:ghzmodel}.

\section{LHV tables for stabilizer states \label{stabLHV}}

In this section we learn about constructing LHV tables for Pauli measurements on stabilizer states. Of course no table will reproduce all the predictions of quantum mechanics, as can be seen with the simple example in Ref.~\cite{mermin:ghz}. Suppose we have an LHV table that provides results of $X$ and $Y$ measurements on all the qubits of a GHZ state,
\begin{equation}
\ket{\mbox{\footnotesize{GHZ}}} = \frac{1}{\sqrt{2}} (\ket{000}+\ket{111}).
\end{equation}
Among the stabilizer elements of this state are $- X \otimes Y \otimes Y$, $- Y \otimes X \otimes Y$, $- Y \otimes X \otimes Y$, and $X \otimes X \otimes X$, which only involve the measurements under consideration. As in Subsec.~\ref{subsec:rvbi}, this gives four conditions on the random variables that generate instances of our LHV table,
\begin{equation}
\begin{array}{l}
R(\avg{X_1}) R(\avg{Y_2}) R(\avg{Y_3}) = R(\avg{X_1 Y_2 Y_3}) = R(-1) = -1 \\
R(\avg{Y_1}) R(\avg{X_2}) R(\avg{Y_3}) = R(\avg{Y_1 X_2 Y_3}) = R(-1) = -1 \\
R(\avg{Y_1}) R(\avg{Y_2}) R(\avg{X_3}) = R(\avg{Y_1 Y_2 X_3}) = R(-1) = -1 \\
R(\avg{X_1}) R(\avg{X_2}) R(\avg{X_3}) = R(\avg{X_1 X_2 X_3}) = R(1) = 1. \\
\end{array}
\end{equation}
Multiplying these four equations together yields $1 = -1$, and so a simultaneous solution to these equations does not exist. Thus any LHV table must get at least one of these wrong.

Even though we cannot hope to get all measurement results correct, LHV tables can capture much about the quantum correlations. This is because the correlations, even though stronger than local realism allows, are still rather simple. As discussed in Subsec.~\ref{subsec:measure}, a tensor product of Pauli matrices, $g$, can be classified as one of two types for a given stabilizer state. If $\pm g$ is in the stabilizer group, meaning that $\pm g = g_1^{a_1} \cdots g_n^{a_n}$ where $g_1,\ldots,g_n$ are canonical stabilizer generators, then the result of the measurement is $+1$ or $-1$ with unit probability. We will call these \emph{definite\/} measurements. In all other cases the measurement gives $\pm 1$ with equal probability; hence we call these \emph{random\/} measurements. For example, if the stabilizer state corresponds to a connected graph, then no individual Pauli measurement or its opposite is in the stabilizer group, and therefore all individual Pauli measurement outcomes are random.

It is useful here to reiterate the results of Chapter~\ref{chap:stabilizer} regarding products of Pauli group elements. The product of a set of Pauli group elements can always be written as in Eq.~(\ref{eq:prodofpauli}), where Eq.~(\ref{eq:badsign}) is the sign that sits in front of the Pauli product. For the measurement of a stabilizer element $g$, the result of the measurement is $+1$ with unit probability. For such stabilizer elements, we generally consider the definite measurement $\sign{g}g$, because this is a Pauli product with a plus sign in front. The outcome of a measurement of $\sign{g}g$ is $\sign{g} = {(-1)}^{s_g}$ with certainty. Throughout this chapter we will also be interested in the LHV prediction for a definite measurement $\sign{g}g$, called the \emph{value\/} of $g$, which we write as $(-1)^{v_g}$. We are careful to distinguish between LHV results and quantum results, giving the outcomes different symbols and different names. If $s_g \eqmod{2} v_g$ then the LHV table and quantum mechanics agree on the result of that measurement, while if $s_g \ne_{\scriptscriptstyle{2}} v_g$ then they disagree.

Our goal is to find LHV tables that, since they cannot capture all measurement correlations, assign values of $\pm 1$ to all definite measurements and assign a uniformly random variable $R(0)$ to all random measurements. This will ensure that all random measurements are correctly assigned random values. The way such a table could make a prediction contrary to quantum mechanics is either to give a value of $+1$ to definite measurements needing a $-1$ outcome, or to assign a $-1$ value to those needing to be $+1$. We say that tables with this property are \emph{probability preserving\/} in the sense that they assign outcomes with the right probability, either certain or random, but they may assign the wrong outcome with that probability. In what follows we explain a standard method for constructing \hyph{probability}{preserving} LHV tables for Pauli measurements on stabilizer states, and then we show that any \hyph{probability}{preserving} LHV table can be constructed in this way.

\subsection{Standard LHV table construction}

This method for constructing LHV tables is originally due to my advisor, Carlton M. Caves, and the material presented in this subsection is more or less a review of his document on the subject. Once we review how this construction works, we can then prove some novel results about the tables.

As usual we begin with random variables corresponding to individual Pauli measurement results, and then we derive how to correlate them appropriately. The way we write the individual Pauli results is
\begin{equation}
\label{eq:indresults}
\begin{array}{lll}
R(\avg{X_k}) = (-1)^{x_k}, & R(\avg{Z_k}) = (-1)^{z_k}, & R(\avg{X_k}) R(\avg{Y_k}) R(\avg{Z_k}) = (-1)^{c_k}.
\end{array}
\end{equation}
The notation above writes the outcomes as $-1$ to a power, with the power being $50$-$50$ random or definite, depending on the stabilizer state. Essentially, tables are created by assigning values to $X$ and $Z$, and then, instead of assigning a value to $Y$, by stating the fixed correlation of $X$, $Y$, and $Z$.

Using these values for the individual measurements, we can calculate the values assigned to definite measurements under the table. The first step is to find the values assigned to the definite measurements corresponding to our canonical stabilizer generators. This can be done by writing the part of the stabilizer generators that is a tensor product of Pauli matrices in binary notation,
\begin{equation}
\label{eq:generator}
\sign{g_j} g_j = i^{\sum_k [r_1(g_j)]_k [r_2(g_j)]_k} \bigotimes_{k=1}^N X^{[r_1(g_j)]_k} Z^{[r_2(g_j)]_k}.
\end{equation}
If we use the individual Pauli results in Eq.~(\ref{eq:indresults}) then the value assigned to this tensor product of Pauli matrices, which is labeled as $(-1)^{v_j}$, is
\begin{equation}
\label{eq:initval}
(-1)^{v_j} = (-1)^{\sum_k c_k [r_1(g_j)]_k [r_2(g_j)]_k}  \prod_{k=1}^N (-1)^{x_k [r_1(g_j)]_k} (-1)^{z_k [r_2(g_j)]_k}.
\end{equation}
At this point, if we wanted to get right the outcome of $\sign{g_j} g_j$, we would set this equal to $s_j$ modulo $2$. However, we want to give LHV tables the freedom to assign any value whatsoever to $\sign{g_j} g_j$.

The exponent of Eq.~(\ref{eq:initval}) gives the value of $v_j$.
\begin{align} \label{eq:vj}
\begin{split}
v_j &\eqmod{2} \sum_k \left( c_k [r_1(g_j)]_k [r_2(g_j)]_k + x_k [r_1(g_j)]_k + z_k [r_2(g_j)]_k \right) \\
&= \left( \sum_k c_k [r_1(g_j)]_k [r_2(g_j)]_k\right) + \left( \begin{array}{cc} x & z \end{array} \right) r^T(g_j),
\end{split}
\end{align}
where $x$ and $z$ are binary vectors whose entries are $x_k$ and $z_k$. Now we simplify this by introducing a binary vector, $V$, whose $j$th entry is $v_j + \sum_k c_k [r_1(g_j)]_k [r_2(g_j)]_k$. Note that $V$ depends on the generators as well as the values of $c_k$. What we get is that
\begin{equation} \label{eq:vrel}
V \eqmod{2} \left( \begin{array}{cc} x & z \end{array} \right) G^T,
\end{equation}
with $G$ being the generator matrix for the stabilizer state.

What we have done is to relate $V$, which involves the values assigned to the signless stabilizer generators as well as the correlation between $X$, $Y$, and $Z$, to $\left( \begin{array}{cc} x & z \end{array} \right)$ by using the generator matrix $G$. Now, if we could invert this equation to solve for $\left( \begin{array}{cc} x & z \end{array} \right)$, then by picking the values we want to assign to the signless generators as well as the values of $c_k$, we could actually solve for the required values of $x$ and $z$. These values of $x$ and $z$, along with the chosen $c_k$, would then determine a table that assigns the desired values $v_j$ to the signless generators. Unfortunately, solving for $\left( \begin{array}{cc} x & z \end{array} \right)$ in Eq.~(\ref{eq:vrel}) is problematic because $G$ does not have a unique inverse.

We solve this problem by using the dual generator matrix, $H$, introduced in Subsec.~\ref{subsec:canongen}. Specifically, since the dual generators, $h_j$, give rise to random measurements,  we assign them random values $(-1)^{r_j}$. This gives us a binary vector $R$ whose $j$th entry is $r_j + \sum_k c_k [r_1(h_j)]_k [r_2(h_j)]_k$, and we have the analogous version of Eq.~(\ref{eq:vrel})
\begin{equation} \label{eq:rrel}
R \eqmod{2} \left( \begin{array}{cc} x & z \end{array} \right) H^T.
\end{equation}
Combining this equation with Eq.~(\ref{eq:vrel}) gives us that
\begin{equation} \label{eq:rvrel}
\left( \begin{array}{cc} V & R \end{array} \right) \eqmod{2} \left( \begin{array}{cc} x & z \end{array} \right) \left( \begin{array}{cc} G^T & H^T \end{array} \right).
\end{equation}
Now we can use the inverse of $\left( \begin{array}{cc} G^T & H^T \end{array} \right)$ in Eq.~(\ref{eq:GHinverse}) to invert Eq.~(\ref{eq:rvrel}),
\begin{align} \label{eq:tablecons}
\begin{split}
\left( \begin{array}{cc} x & z \end{array} \right) \eqmod{2} \left( \begin{array}{cc} V & R \end{array} \right) \left( \begin{array}{cc} G^T & H^T \end{array} \right)^{-1} \eqmod{2} \left( \begin{array}{cc} V & R \end{array} \right) \Lambda \left( \begin{array}{c} G \\ H \end{array} \right) \Lambda.
\end{split}
\end{align}
This, of course, works the same for any set of generators since dual generators exist for any set of given stabilizer generators. However, we only consider, here, the case of canonical stabilizer generators.

With this result we can construct an LHV table once we are given three quantities. First, we need to know what values are to be assigned to tensor products of Pauli matrices corresponding to a set of canonical stabilizer generators, $g_j$, for the state under consideration. For instance, we can choose to get all the generators correct by setting $(-1)^{v_j} = \sign{g_j}$, but note that this does not guarantee $(-1)^{v_g} = \sign{g}$ for all stabilizer elements. Then we need to know the random values assigned to the dual generators $h_j$. Since these generators are not in the stabilizer group for the state, they must be assigned random outcomes. Finally, we need to know the value of $c_j$ in Eq.~(\ref{eq:indresults}) for all $j$. Once we know these values, we can construct an LHV table that respects them.

Standard LHV tables for Pauli measurements on stabilizer states are those tables constructed by picking values for $v_j$ and $c_j$ and using Eq.~(\ref{eq:tablecons}) to determine all individual results. We now prove the following result concerning such tables.
\begin{lemma} \label{lma:standpp}
Standard LHV tables for Pauli measurements on stabilizer states are probability preserving.
\end{lemma}

\begin{proof}
When we say that standard LHV tables are probability preserving, we mean that they assign random values to all random measurements and fixed values to all definite measurements.

First we observe that the $2 n$ rows of $\left( \begin{array}{c} G \\ H \end{array} \right)$ are linearly independent, where $n$ is the number of qubits in the stabilizer state for which we want to construct the table. As a consequence, any \hyph{$n$}{fold} tensor product of Pauli matrices can be written as a product of generators $g_j$ and $h_j$, possibly with a sign. Since only stabilizer elements correspond to definite measurements, if the decomposition of the measurement $g$ contains any $h_j$, quantum mechanics will predict its result to be random. Thus what we need to prove is that a measurement of a tensor product of Pauli matrices $g=\pm g_1^{a_1} \ldots g_n^{a_n} h_1^{b_1} \ldots h_n^{b_n}$ is random if and only if some $b_j$ is nonzero.

We can write down the value of $v_g$ for a measurement of $\sign{g} g$ in the same way we wrote down the value of $v_j$ in Eq.~(\ref{eq:vj}),
\begin{align}
\begin{split}
v_g  \eqmod{2} \sum_{k=1}^n c_k [r_1(g)]_k [r_2(g)]_k &+ \sum_{k=1}^n x_k [r_1(g)]_k + \sum_{k=1}^n z_k [r_2(g)]_k.
\end{split}
\end{align}
We want to write this in terms of the values of the generators, $v_j$ and $r_j$, because those are the quantities we know. So by writing the binary vector $r(g)$ in terms of $r(g_j)$ and $r(h_j)$ we get that
\begin{align}
\begin{split}
v_g \eqmod{2} \sum_{k=1}^n c_k [r_1(g)]_k [r_2(g)]_k &+ \sum_{k=1}^n x_k \left( \sum_{j=1}^n a_j [r_1(g_j)]_k + b_j [r_1(h_j)]_k \right) \\
&+ \sum_{k=1}^n z_k \left( \sum_{j=1}^n a_j [r_2(g_j)]_k + b_j [r_2(h_j)]_k \right) \\
\phantom{v_g} \eqmod{2} \sum_{k=1}^n c_k [r_1(g)]_k [r_2(g)]_k &+ \sum_{j=1}^n a_j \left( \sum_{k=1}^n x_k [r_1(g_j)]_k + \sum_{k=1}^n z_k [r_2(g_j)]_k \right) \\
&+ \sum_{j=1}^n b_j \left( \sum_{k=1}^n x_k [r_1(h_j)]_k + \sum_{k=1}^n z_k [r_2(h_j)]_k \right).
\end{split}
\end{align}
Now we can use Eq.~(\ref{eq:vj}) and its counterpart for $r_j$ to write $v_g$ in terms of the known quantities $v_j$, $r_j$, and $c_k$,
\begin{align}
\begin{split}
v_g \eqmod{2} \sum_{k=1}^n c_k [r_1(g)]_k [r_2(g)]_k &+ \sum_{j=1}^n a_j \left(v_j + \sum_{k=1}^n c_k [r_1(g_j)]_k [r_2(g_j)]_k \right) \\
&+ \sum_{j=1}^n b_j \left(r_j + \sum_{k=1}^n c_k [r_1(h_j)]_k [r_2(h_j)]_k \right) \\
\phantom{v_g} \eqmod{2} \left( \mbox{\begin{tabular}{c} Terms that do \\ not involve $r_j$s \end{tabular}} \right) &+ \sum_{j=1}^n b_j r_j,
\end{split}
\end{align}
Now it is easy to see that the result is random if and only if there exists a $b_j \ne 0$.
\end{proof}

To summarize, we can now construct LHV tables, which assign whatever values we want to generators, that produce definite results for all definite measurements and random results for all random measurements. What we will discover now is that our method for constructing LHV tables allows us to construct any \hyph{probability}{preserving} LHV table in existence.

\subsection{All \hyph{probability}{preserving} LHV tables}

Suppose we are given a \hyph{probability}{preserving} LHV table that assigns values to all individual Pauli measurements on some stabilizer state. In this subsection we will prove that a standard LHV table exists that reproduces all correlations of the given table. Thus standard LHV tables are capable of representing all \hyph{probability}{preserving} LHV tables.

\begin{lemma}
For any \hyph{probability}{preserving} LHV table describing Pauli measurements on a stabilizer state, there exists a standard table with the same correlations.
\end{lemma}

\begin{proof}
Since the given table assigns values to all individual measurements, we can consider one particular instance of the table and write, for that instance, the values of $X$, $Y$, and $Z$ measurements,
\begin{equation}
\begin{array}{lll}
R(\avg{X_k}) = (-1)^{x'_k}, & R(\avg{Z_k}) = (-1)^{z'_k}, & R(\avg{X_k}) R(\avg{Y_k}) R(\avg{Z_k}) = (-1)^{c'_k}.
\end{array}
\end{equation}
where $x'_j,z'_j,c'_j \in \{0,1\}$. Although all three of these variables might be random in the LHV table, primed values are fixed because they correspond to one particular instance of the given LHV table. The reason I consider one instance is that, in order to construct a standard table, all I need to know is the values of the outcomes for definite measurements corresponding to stabilizer generators, and these values are fixed for all instances of the table. Therefore the value of $v_j$ for this instance must be the value of $v_j$ for all instances, and so we get that
\begin{equation}
v_j \eqmod{2} \sum_k \left( c'_k [r_1(g_j)]_k [r_2(g_j)]_k + x'_k [r_1(g_j)]_k + z'_k [r_2(g_j)]_k \right).
\end{equation}
The standard LHV table we want to construct uses this value for $v_j$, and has $c_k = c'_k$. From this it follows that, for all $j$,
\begin{equation}
\label{eq:genvals}
\sum_k x_k [r_1(g_j)]_k + z_k [r_2(g_j)]_k \eqmod{2} \sum_k x'_k [r_1(g_j)]_k + z'_k [r_2(g_j)]_k,
\end{equation}
It only remains to prove that this standard table assigns the same values to all definite measurements as the given one, and we have proven what we wanted.

So consider the value $v'_g$ of a definite measurement under the given table, once again decomposing its binary vector into a linear combination of binary vectors for stabilizer generators,
\begin{align}
\begin{split}
v'_g \eqmod{2} \sum_{k=1}^n c'_k [r_1(g)]_k [r_2(g)]_k &+ \sum_{k=1}^n (x'_k [r_1(g)]_k + z'_k [r_2(g)]_k) \\
\phantom{v'_g} \eqmod{2} \sum_{k=1}^n c'_k [r_1(g)]_k [r_2(g)]_k &+ \sum_{k=1}^n x'_k \left( \sum_{j=1}^n a_j [r_1(g_j)]_k \right) \\
&+ \sum_{k=1}^n z'_k \left( \sum_j a_j [r_2(g_j)]_k \right) \\
\phantom{v'_g} = \sum_{k=1}^n c'_k [r_1(g)]_k [r_2(g)]_k &+ \sum_{j=1}^n a_j \left( \sum_{k=1}^n x'_k [r_1(g_j)]_k + z'_k [r_2(g_j)]_k \right).
\end{split}
\end{align}
Now we can use Eq.~(\ref{eq:genvals}) along with $c_k = c'_k$ to write $v_g$ in terms of standard LHV table quantities,
\begin{align} \label{eq:vg}
\begin{split}
v'_g \eqmod{2} \sum_{k=1}^n c_k [r_1(g)]_k [r_2(g)]_k &+ \sum_{j=1}^n a_j \left( \sum_{k=1}^n x_k [r_1(g_j)]_k + z_k [r_2(g_j)]_k \right) \\
\phantom{v'_g} = \sum_{k=1}^n c_k [r_1(g)]_k [r_2(g)]_k &+ \sum_{k=1}^n x_k \left( \sum_{j=1}^n a_j [r_1(g_j)]_k \right) \\
&+ \sum_{k=1}^n z_k \left( \sum_{j=1}^n a_j [r_2(g_j)]_k \right) \\
\phantom{v'_g} \eqmod{2} \sum_{k=1}^n c_k [r_1(g)]_k [r_2(g)]_k &+ \sum_{k=1}^n (x_k [r_1(g)]_k + z_k [r_2(g)]_k) \eqmod{2} v_g.
\end{split}
\end{align}
This says that $v'_g \eqmod{2} v_g$, so the standard table assigns the same values to all definite measurements as does the given table. Since the standard table automatically gives random results to random measurements, this completes our proof.
\end{proof}

The essence of the above proof is that all definite measurement values are determined by the signless generator values, along with the values of $c_k$. Since these are variables in creating a standard table, standard tables are capable of representing any \hyph{probability}{preserving} LHV table.

\subsection{One additional measurement}

For universal \hyph{measurement}{based} quantum computation using stabilizer states, it is crucial that one additional measurement, other than a Pauli measurement, be included. Since our investigation in this chapter is based on \hyph{probability}{preserving} LHV tables, the existence of \hyph{probability}{preserving} tables for \hyph{non}{Pauli} measurements deserves some attention. Measurements involving \hyph{non}{Pauli} operators are no longer either definite or uniformly random, so we first make explicit what is meant by probability preserving in this case. \hyph{Probability}{preserving} tables for Pauli measurements are defined so that $\avg{M}_{\mbox{\scriptsize{LHV}}} = \pm \avg{M}_{\mbox{\scriptsize{QM}}}$, and so we take this as the general definition of a \hyph{probability}{preserving} LHV table for non-Pauli measurements as well. Unfortunately, we prove in this subsection that such tables do not exist, in general, when a \hyph{non}{Pauli} measurement is allowed. Therefore, extending our results to include a \hyph{non}{Pauli} measurement seems likely to require more complicated probabilistic communication strategies. Whether \hyph{probability}{preserving} tables exist for Pauli measurements on more general states~\cite{danos:mbqc} is still open.

We now work through an example where the construction of \hyph{probability}{preserving} LHV tables fails when any \hyph{non}{Pauli} measurement is allowed. In this vein, consider the stabilizer state
\begin{equation}
\ket{\psi} = \frac{1}{\sqrt{2}} \left( \ket{00} + \ket{11} \right)
\end{equation}
and an arbitrary \hyph{single}{qubit} measurement $T(\theta,\phi)=\vec{u}(\theta,\phi) \cdot \vec{\sigma}$, for some unit vector $\vec{u}(\theta,\phi)$. The angles $\theta$ and $\phi$ used to describe the vector are not necessarily the usual spherical polar angles. To define $\theta$ and $\phi$, we need to first define new axes $\vec{x}\,'$, $\vec{y}\,'$, and $\vec{z}\,'$ as follows. First write $\vec{u}$ in terms of the unit vectors $\pm \vec{x}$, $\pm \vec{y}$, and $\pm \vec{z}$, such that the coefficients of $\vec{u}$ are not negative. For example, if $\vec{u} = - \vec{x}$, then $\vec{u} = +1 (- \vec{x})$. Once this is done, we choose $\vec{z}\,'$ as the axis with the smallest positive coefficient and $\vec{x}\,'$ as the axis with the largest positive coefficient. Notice that a primed axis, $\vec{a}\,'$, need not be $\pm \vec{a}$ and that the resulting axes need not form a \hyph{right}{handed} coordinate system. In the example that $\vec{u} = +1 (- \vec{x})$, we must choose $\vec{x}\,' = - \vec{x}$, but we can choose $\vec{z}\,'$ to be either $\pm \vec{y}$ or $\pm \vec{z}$. If we choose $\vec{z}\,' = \vec{y}$ and $\vec{y}\,' = - \vec{z}$, then the resulting coordinate system is not a \hyph{right}{handed} one.

\begin{figure}
\center
\includegraphics[height=6cm]{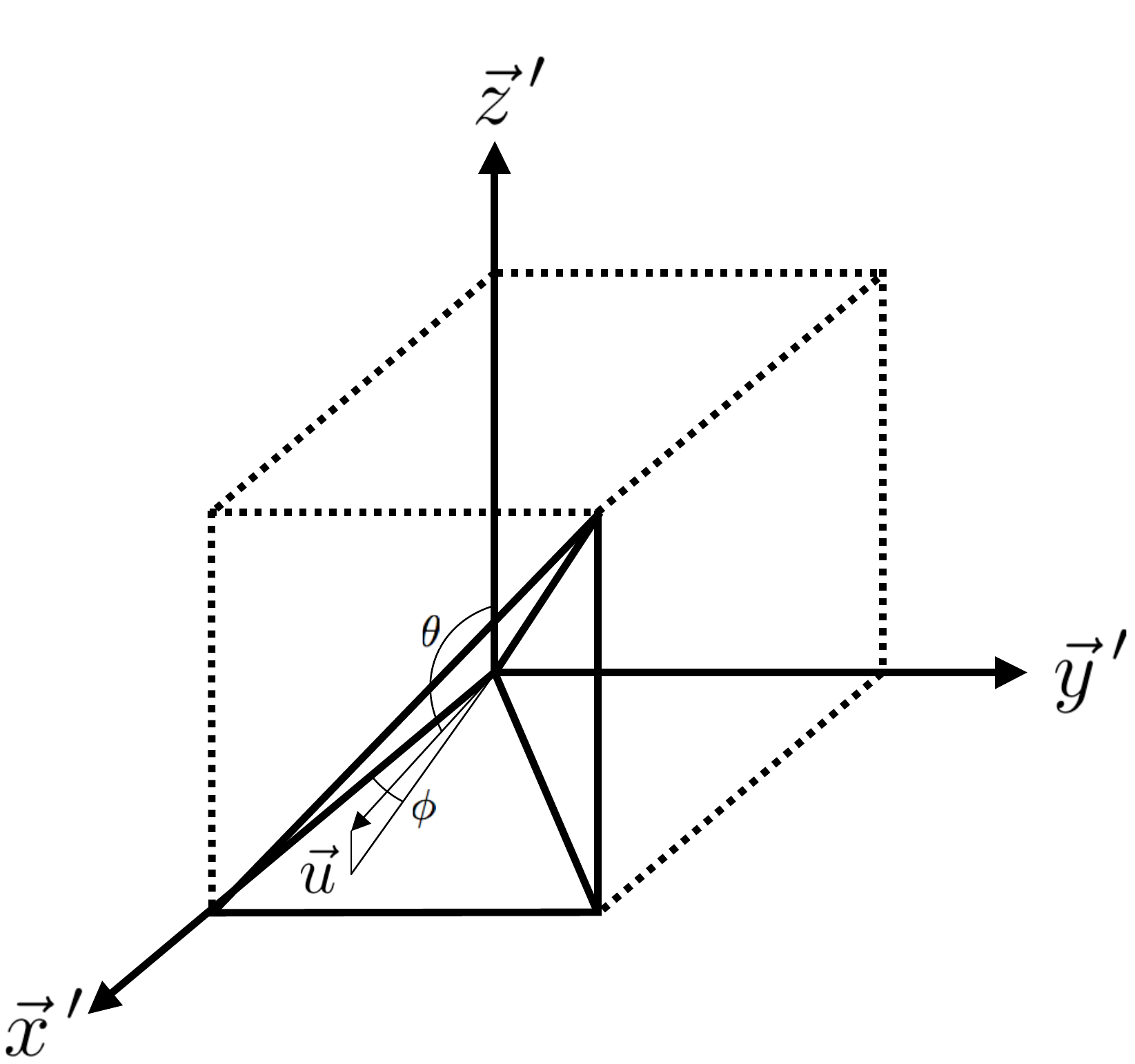}
\caption[Definition of angles $\theta$ and $\phi$]{Definitions of $\theta$ and $\phi$ once the axes $\vec{x}\,'$, $\vec{y}\,'$, and $\vec{z}\,'$ have been identified. The primed axes shown result in a \hyph{right}{handed} coordinate system, but this need not be the case in general. Since $\vec{u}$ is furthest from $\vec{z}\,'$ and closest to $\vec{x}\,'$, it must lie in the tetrahedron outlined in the figure.}
\label{fig:angles}
\end{figure}

The angles $\theta$ and $\phi$ are illustrated in Fig.~\ref{fig:angles} and are simply spherical polar angles with respect to the primed axes. Specifically, $\theta$ is the polar angle from the $\vec{z}\,'$ axis and $\phi$ is the azimuthal angle about the $\vec{z}\,'$ axis, measured in a \hyph{right}{handed} way for a \hyph{right}{handed} coordinate system and in a \hyph{left}{handed} way for a \hyph{left}{handed} coordinate system. The primed axes are defined so that $\vec{u}$ lies in the tetrahedron in the figure. The fact that the vector $\vec{u}$ lies in the tetrahedron constrains the angles so that
\begin{equation}
\label{eq:allowedangles}
0 \leq \phi \leq \frac{\pi}{4}\mbox{  and  } \theta_\phi \leq \theta \leq \frac{\pi}{2},
\end{equation}
where $\theta_\phi$ satisfies $\cos\theta_\phi = \sin\theta_\phi \sin\phi$. The upper face of the tetrahedron lies along the plane $z' = y' \Rightarrow \cos\theta = \sin\theta \sin\phi$, hence $\theta_\phi$ is the smallest allowed value for $\theta$.

Now consider measurements on the state $\ket{\psi}$ of $X' \otimes X'$, $Y' \otimes X'$, $Y' \otimes T(\theta,\phi)$, and $X' \otimes T(\theta,\phi)$, where $X'$ and $Y'$ correspond to measurements along $\vec{x}\,'$ and $\vec{y}\,'$, respectively. We calculate that
\begin{equation}
\label{eq:examplecorrelations}
\begin{array}{ccc}
\avg{X' \otimes X'} = \pm 1 &\hspace{1em}& \avg{Y' \otimes X'} = 0 \\
\avg{Y' \otimes T(\theta,\phi)} = \pm \sin\theta \sin\phi && \avg{X' \otimes T(\theta,\phi)} = \pm \sin\theta \cos\phi.
\end{array}
\end{equation}
We want to show that it is impossible to preserve these quantum probabilities with an LHV table unless $T(\theta,\phi)$ is proportional to a Pauli matrix.

Following Subsec.~\ref{subsec:rvbi} we want, for a \hyph{probability}{preserving} LHV table, random variables so that
\begin{align}
\begin{split}
R(\avg{X'_1}) R(\avg{X'_2}) &= \pm 1 \\
R(\avg{Y'_1}) R(\avg{X'_2}) &= R(0) \\
R(\avg{Y'_1}) R(\avg{T(\theta)}) &= \pm R(\sin\theta \sin\phi) \\
R(\avg{X'_1}) R(\avg{T(\theta)}) &= \pm R(\sin\theta \cos\phi).
\end{split}
\end{align}
Multiplying these four equations together, and absorbing all signs into $R(0)$, gives us the following consistency condition.
\begin{equation}
R(\sin\theta \cos\phi) R(\sin\theta \sin\phi) = R(0)
\end{equation}
Using Theorem~\ref{thm:nvars}, what we want to know is the values of $\theta$ and $\phi$ satisfying the inequality,
\begin{equation}
\label{eq:nolhv}
|\sin\theta \cos\phi + \sin\theta \sin\phi| - 1 \leq 0 \leq 1- |\sin\theta \cos\phi - \sin\theta \sin\phi|.
\end{equation}
For the allowed values of $\theta$ and $\phi$ given in Eq.~(\ref{eq:allowedangles}), our claim is that $|\sin\theta \cos\phi + \sin\theta \sin\phi| - 1 \leq 0$ is true if and only if $\theta = \phi = 0$ which is the case that $T(\theta,\phi)$ is proportional to a Pauli matrix.

First, we rewrite the left hand side of Eq.~(\ref{eq:nolhv}) as $|\cos\phi + \sin\phi| \leq 1 / {\sin\theta}$, where we use that $\sin\theta > 0$ supposing that $\theta \ne 0$. Now from Eq.~(\ref{eq:allowedangles}), we get that $1/{\sin\theta} \leq 1/{\sin\theta_\phi} = \sqrt{1+\sin^2\phi}$. So let us compare $|\cos\phi + \sin\phi|$ and $\sqrt{1+\sin^2\phi}$. Squaring them both gives $|\cos\phi + \sin\phi|^2 = 1 + ( 2 \cos\phi ) \sin \phi$ and $(\sqrt{1+\sin^2\phi})^2 = 1+( \sin\phi ) \sin\phi$. Supposing that $\phi \ne 0$ and given that $0 \leq \phi \leq \pi /4$, it is clear that
\begin{equation}
2 \cos\phi > \sin \phi \Rightarrow |\cos\phi + \sin\phi| > \sqrt{1+\sin^2\phi} \geq 1 / {\sin\theta}.
\end{equation}
Hence we have a contradiction with Eq.~(\ref{eq:nolhv}), and we conclude that $|\sin\theta \cos\phi + \sin\theta \sin\phi| - 1 \leq 0$ holds true only for $\phi = 0 \Rightarrow \theta = 0$. Therefore we have shown that the probabilities associated with the correlations in Eq.~(\ref{eq:examplecorrelations}) can be preserved only if $T(\theta,\phi)$ is proportional to a Pauli matrix.

\subsection{Comparing LHV tables and quantum mechanics}

It is an interesting question to ask how well our standard \hyph{probability}{preserving} tables for Pauli measurements perform, a task that can be accomplished, for example, by calculating the probability of failure. However, in order to do this, we need to know the exact prediction of all definite measurements under the table. We can then compare these predictions to the quantum mechanical predictions and see when they agree or disagree.

We learned in the beginning of this section that quantum mechanics predicts the result of a definite measurement, $M_g = \sign{g}g$, to be $\sign{g} = {(-1)}^{s_g}$. So, we actually already know the quantum mechanical prediction of definite measurements from Subsec.~\ref{subsec:signs}. We are interested in the sign of products of canonical generators $g_1^{a_1} \cdots g_n^{a_n}$ given in Eq.~(\ref{eq:qmsign}) of Result~\ref{res:sign}. What we will now find is the corresponding formula for standard LHV tables so that we can compare measurement outcomes.

To determine the result, ${(-1)}^{v_g}$, assigned to a definite measurement under a standard LHV table, insert the generator values in Eq.~(\ref{eq:vj}) into the value of $v_g$ in Eq.~(\ref{eq:vg}), thereby eliminating $x_k$ and $z_k$. Doing this gives us that
\begin{align}
\begin{split}
v_g &\eqmod{2} \sum_{l=1}^n c_l [r_1(g)]_l [r_2(g)]_l + \sum_{j=1}^n a_j \left( v_j + \sum_{l=1}^n c_l [r_1(g_j)]_l [r_2(g_j)]_l \right) \\
&\eqmod{2} av^T + \sum_{l=1}^n c_l [r_1(g)]_l [r_2(g)]_l + \sum_{j,l=1}^n a_j c_l [r_1(g_j)]_l [r_2(g_j)]_l.
\end{split}
\end{align}
Now we want to write the binary vector for $g$ as a linear combination of the binary vectors for $g_j$. When we do this, the quantity $\mathcal{C}_{jk} = \left( \sum_{l=1}^n c_l [r_1(g_j)]_l [r_2(g_k)]_l \right)$ appears several times.
\begin{align} \label{eq:origvg}
\begin{split}
v_g \eqmod{2} av^T &+ \sum_{l=1}^n c_l \left( \sum_{j=1}^n a_j [r_1(g_j)]_l \right) \left( \sum_{k=1}^n a_k [r_2(g_k)]_l \right) \\
&+ \sum_{j,l=1}^n a_j c_l [r_1(g_j)]_l [r_2(g_j)]_l \\
= av^T &+  \sum_{j,k=1}^n a_j a_k \mathcal{C}_{jk} + \sum_{j=1}^n a_j \mathcal{C}_{jj} \eqmod{2} av^T +  \sum_{j \ne k} a_j a_k \mathcal{C}_{jk}.
\end{split}
\end{align}

What we have here is that the value under the standard LHV table of a definite measurement of $g$ is the sum of two quantities. The first, rather expectedly, is the product of the values for definite measurements corresponding to the generators needed to obtain $g$, ${(-1)}^{av^T}$. The other quantity involves looking at the generators involved in obtaining $g$ and carefully taking into account the value of $c_k$ for those generators.

To compare this with the quantum mechanical result, we should write the LHV measurement outcome in terms of $r(g)$ by replacing the binary vector $a$ with $r(g)$ via Eq.~(\ref{eq:a}). The result is most simply written in terms of $\tilde{\mathcal{C}}$, which is $\mathcal{C}$ with its diagonal entries set to zero. In that case,
\begin{align}
\begin{split}
v_g \eqmod{2} a v^T + a \tilde{\mathcal{C}} a^T = r(g) (vH \Lambda)^T + r(g) \left(\Lambda H^T \right) \tilde{\mathcal{C}} \left( H \Lambda \right) r^T(g).
\end{split}
\end{align}
To carry out the calculation of this quantity, let us write $\mathcal{C} = G_1 C G_2^T$, where $G = \left( \begin{array}{c|c} G_1 & G_2 \end{array} \right)$ and $C$ is a diagonal matrix whose entries are the values $c_k$, i.e $C_{jk} = c_k \delta_{jk}$. Using the canonical generator matrix, we have
\begin{align}
\begin{split}
\mathcal{C} = G_1 C G_2^T &= \left( \begin{array}{cc} I & A \\ 0 & 0 \end{array} \right) \left( \begin{array}{cc} C^{(r)} & 0 \\ 0 & C^{(n-r)} \end{array} \right) \left( \begin{array}{cc} B & A \\ 0 & I \end{array} \right) \\
&= \left( \begin{array}{cc} C^{(r)} B & C^{(r)} A + A C^{(n-r)} \\ 0 & 0 \end{array} \right),
\end{split}
\end{align}
where $r$ is the left rank of the generator matrix $G$, and we split up the matrix $C$ into its first $r$ and last $n-r$ rows and columns. Now we get $\tilde{\mathcal{C}}$ by making the diagonal entries zero,
\begin{equation} \label{eq:stabC}
\tilde{\mathcal{C}} = \left( \begin{array}{cc} C^{(r)} \tilde{B} & C^{(r)} A + A C^{(n-r)} \\ 0 & 0 \end{array} \right),
\end{equation}
where $\tilde{B}$ is obtained from $B$ by setting its diagonal entries to zero. Using this matrix, we finally get the outcome of an LHV table.
\begin{result}
\emph{Let $M_g = \sign{g} g$ be a tensor product of Pauli matrices corresponding to a stabilizer element $g = g_1^{a_1} \cdots g_n^{a_n}$. A standard LHV table, defined by a binary vector $v$ and a matrix $C$ with $C_{jk} = c_k \delta_{jk}$, assigns to this measurement an outcome ${(-1)}^{v_g}$, with\/}
\begin{align} \label{eq:lhvvalue}
\begin{split}
v_g &\eqmod{2} a v^T + a^{(r)} \left( \begin{array}{cc} C^{(r)} \tilde{B} & C^{(r)} A + A C^{(n-r)}\end{array} \right) a^T \\
&= r(g) (v H \Lambda)^T + r_1(g) \left( \begin{array}{cc|cc} C^{(r)} \tilde{B} & 0 & 0 & C^{(r)} A + A C^{(N-r)} \end{array} \right) r^T(g).
\end{split}
\end{align}
\end{result}
\noindent The superscript $r$ or $n-r$ indicates whether to take the first $r$ or last $n-r$ entries, and we have $\tilde{B}_{jk} = B_{jk} (1-\delta_{jk})$.

Now, in order for an LHV table to predict all the same results as quantum mechanics, one needs to have that $v_g \eqmod{2} s_g$ for all $g$. For every $g$ for which this is not true, that is a measurement the LHV table gets wrong.

It would be nice to know the best LHV tables for Pauli measurements on stabilizer states. However, the answer to this question is not known for the general case. There are some preliminary results in this regard. For instance, permutations of $C$ corresponding to a symmetry of the graph yields a table that gets the same number of outcomes correct. It is also known that given a table, another table that assigns opposite results to all measurements does not exist. Results like these lead towards understanding how many correct results a table can predict, but since we do not have a general answer we omit the proof of these claims.

\subsection{Equivalent standard LHV tables}

Every \hyph{probability}{preserving} stabilizer LHV table has a standard counterpart, however it can happen that two apparently different standard LHV tables actually yield identical correlations. We now focus on when this happens using the tools we just developed to calculate LHV measurement outcomes.

Clearly for each of the $2^n$ ways to assign canonical generator values we get a unique table. For a given choice of the $v_j$, there are also $2^n$ choices for the $c_j$, however it is possible that two different choices of $c_j$ yield equivalent tables. If we just consider the special case that the stabilizer state is a graph state, it turns out that flipping all the $c_j$ yields an equivalent table, and we can prove that this leads to exactly a factor of $2$ redundancy.

\begin{lemma}
For an \hyph{$n$}{qubit} graph state corresponding to a connected graph, there are $2^n 2^{n-1} = 2^{2n - 1}$ unique \hyph{probability}{preserving} LHV tables.
\end{lemma}

\begin{proof}
First of all, specializing the value of $v_g$ in Eq.~(\ref{eq:lhvvalue}) to graphs states gives us
\begin{equation} \label{eq:graphsign}
v_g \eqmod{2} av^T + aCBa^T = av^T + aBCa^T,
\end{equation}
since in this case $r = n$ and $G_1 C G^T_2 = I C B^T = CB$ already has all zeroes on the diagonal.

So now we begin with two tables, one table that assigns values to the stabilizer generators according to $v$ and values to the product of $X$, $Y$, and $Z$ measurements according to $c$ and the other table that assigns values according to $w$ and $d$. The idea is to show that the two tables assign the same values to all stabilizer elements if and only if $w = v$ and either $d=c$ or $d \eqmod{2} c + 1$, where $c+1$ adds $1$ to each element of $c$.

First, suppose $w = v$ and $d = c+1$. Then using Eq.~(\ref{eq:graphsign}) with $D_{jk} = d_k \delta_{jk}$, we see that
\begin{equation}
a w^T + a B D a^T = a v^T + a B (C+I) a^T = a v^T + a B C a^T + a B a^T \eqmod{2} a v^T + a B C a^T,
\end{equation}
where $B_{jk} = B_{kj}$ was used in determining that $a B a^T$ is even, and therefore equal to $0$ modulo $2$. So these two tables assign the same outcomes to all measurements.

Now suppose that the two tables assign identical values to all stabilizer elements. If we write $w \eqmod{2} v+u$ and $D \eqmod{2} C + P$, then this condition becomes that, for all $a$,
\begin{equation}
a u^T + a B P a^T \eqmod{2} 0.
\end{equation}
This is the condition that the table specified by $u$ and $P$ assign $+1$ to all stabilizer elements. If we let $a = e_j$, i.e., the measurement is of stabilizer generator $g_j$, then this becomes
\begin{equation}
0 \eqmod{2} u_j + \sum_{k,l,p = 1}^n \delta_{jk} B_{kl} P_{ll} \delta_{lp} \delta_{pj} = u_j + B_{jj} P_{jj} = u_j.
\end{equation}
So we get that $u = 0 \Rightarrow w = v$, which is the first part of what we needed to show. What is left is to find all possible $P$ such that $a B P a^T \eqmod{2} 0$ for all $a$. The claim is that either $P = 0$ or $P= I$. To prove this, suppose that there were two nodes such that $P_{jj} = 0$ and $P_{kk} = 1$. Now we can assume that $B_{jk} = 1$ because the graph is connected, and so there must be two connected nodes that have different diagonal values of $P$. Otherwise all nodes would have either $P_{jj} = 0$ or $P_{jj} = 1$, contradicting our assumption. So we find that for $a = e_j + e_k$,
\begin{align}
\begin{split}
0 & \eqmod{2} a B P a^T = \sum_{l,p,q = 1}^n ( \delta_{jl} + \delta_{kl} ) B_{lp} P_{pp} \delta_{pq} ( \delta_{qj} + \delta_{qk} ) \\
   & = B_{jj} P_{jj} + B_{jk} P_{kk} + B_{kj} P_{jj} + B_{kk} P_{kk} = 1.
\end{split}
\end{align}
Hence the contradiction, and we conclude that $P = 0$ or $P= I \Rightarrow D = C$ or $D \eqmod{2} C+I$, as desired.
\end{proof}

We also have the following corollary.

\begin{corollary}
\label{corr:stabilizeruniqueness}
For an \hyph{$n$}{qubit} stabilizer state corresponding to a connected graph, there are $2^{2n-1}$ unique \hyph{probability}{preserving} LHV tables.
\end{corollary}

\begin{proof}
First assume that two tables assign the same values to all definite measurements. From Eq.~(\ref{eq:lhvvalue}), and following previous notation, we get that
\begin{align} \label{eq:defval}
\begin{split}
&v_g \eqmod{2} a v^T + a^{(r)} C^{(r)} \tilde{B} (a^{(r)})^T + a^{(r)} (C^{(r)} A + A C^{(n-r)}) (a^{(n-r)})^T \\
&\Rightarrow u a^T + a^{(r)} P^{(r)} \tilde{B} (a^{(r)})^T + a^{(r)} (P^{(r)} A + A P^{(N-r)}) (a^{(N-r)})^T \eqmod{2} 0.
\end{split}
\end{align}
As before, setting $a = e_j$ and $a = e_j + e_k$ for some neighbors $j,k \leq r$ we get $u=0$ and $P^{(r)} = 0$ or $P^{(r)} = I$. The resulting constraint is now
\begin{equation}
0 \eqmod{2} a^{(r)} (P^{(r)} A + A P^{(N-r)}) (a^{(N-r)})^T = \sum_{j = 1}^r \sum_{k = r+1}^n a_j A_{jk} a_k (P_{jj} + P_{kk})
\end{equation}
for all $a$. But by connectivity, there is some $A_{jk} = 1$ so for that $j$ and $k$ we get $P_{jj} = P_{kk}$. Now since all the $P_{jj}$ are equal, we conclude that all the $P_{kk}$ are also equal, which gives us that $P = 0$ or $P = I$.

Similar to before we can check that such choices for $P$ generate equivalent tables, and so we get the same redundancy factor of $2$ that we had for graph states.
\end{proof}
We also note that the \hyph{local}{Clifford} equivalence of stabilizer states to graph states also gives us Corollary \ref{corr:stabilizeruniqueness}.

This concludes a very thorough examination of LHV tables for Pauli measurements on stabilizer states. We now move on to learning how classical communication can correct these tables when they make incorrect predictions.

\section{Simple \hyph{classical}{communication} models \label{sec:simplemodels}}

In a communication model we assume that Pauli measurements are made on a stabilizer state and that the results are obtained from an LHV table. We furthermore choose this table to be the \hyph{probability}{preserving} tables of the previous section so that the decision to change LHV results need not be probabilistic. Measurement choices are then communicated between nodes in the graph. The party at each node can then, based on this new information, choose to either output their LHV result or output its negative. The decision of which choice to make depends on the details of the particular model.

We begin by discussing the simplest kinds of communication models. These models share two properties that make them simple. The first is that communication take place only between neighbors in the underlying stabilizer graph. This restriction to communication only with neighbors in the graph makes intuitive sense if we think of a graph as a recipe for constructing the corresponding stabilizer state. In that case, nodes that are connected have interacted in the past and therefore occupy a privileged position with regard to exchange of  information. We call a model that restricts communication to neighbors a \emph{\hyph{nearest}{neighbor} communication model\/}.

Another simplification we make is to consider \emph{\hyph{site}{invariant}\/} models. Site invariance is the property that nodes in symmetric situations perform the same action. This property is natural because identical particles that have identical histories should behave identically. \hyph{site}{invariant} \hyph{nearest}{neighbor} communication models, i.e., models that obey both of these simplifications, are explored in Refs.~\cite{tessier:ghzmodel,barrett:graphmodel}.

Finally, in what follows, instead of considering stabilizer states in general we only consider Pauli measurements on graph states whose stabilizer generators all have a sign of $+1$. Fortunately, we do not lose any generality by doing this because any communication strategy for stabilizer states can be translated into a communication model for a graph state. This follows from the \hyph{local}{Clifford} equivalence of stabilizer states to graph states and the \hyph{local}{Clifford} manipulation of LHV tables described in Refs.~\cite{tessier:thesis,tessier:ghzmodel}. In addition, the underlying graph of a stabilizer state describes the same interaction pattern as for the local unitarily equivalent graph state, so communication in the graph model will still be confined to be along edges between neighbors in the graph. However, we do point out that in general this adds the complication of having to communicate node type as well, so that each party knows which local Clifford operations were applied to the parties with which they are communicating.

It turns out that these two properties that make models so simple also assure their failure. In this section and the next, we consider the effects of each of these properties in some detail and show that if a model has either property it necessarily fails to capture some quantum correlations. However, it is still helpful to consider these models because they give us minimal conditions for other models to work and because they can still capture many quantum correlations as explained below. 

\subsection{Definite measurements and definite submeasurements}

Before getting to a simple communication model, we first review quantum mechanical results concerning Pauli measurements on graph states. When these measurements are made, the product of all measurement outcomes is the same as the value of measurement of the corresponding Pauli product. This tensor product of Pauli matrices, $M$, is called a \emph{global measurement\/} because it results from considering all the measurement outcomes that were made. A \emph{submeasurement\/} of a global Pauli product~$M$ is a Pauli product $\tilde M$ such that the \hyph{non}{identity} elements of $\tilde M$ all appear in $M$, i.e., $\tilde M_j=M_j$ or $\tilde M_j=I$ for all~$j$. LHV models implicitly predict the result of measuring such a subset of the Pauli operators of a global measurement, the measurement of an identity operator being simply the omission of the corresponding LHV entry.  A proper \hyph{communication}{assisted} LHV model for graph states should not only reproduce the predictions of quantum mechanics for global measurements but also for all possible submeasurements. LHV tables by themselves are capable of reproducing neither type of measurement.

If we consider a global measurement $M$, which may be random or definite, of a tensor product of Pauli matrices then a submeasurement of $M$ can be specified by a binary vector $e$ such that $e_j = 1$ if $M_j$ is included in the submeasurement and $e_j = 0$ if it is not. Given the vector $e$ that specifies the submeasurement, we can use $[r_1(M)]_j e_j$ and $[r_2(M)]_j e_j$ to write the binary vector for the submeasurement. In this way we force all measurements with $e_j = 0$ to be $I$, and leave the rest unchanged.

Our models will automatically predict correctly the quantum result of random global measurements and random submeasurements because the underlying LHV table assigns random results to those measurements. Thus in all of what follows, we concern ourselves only with definite measurements and definite submeasurements. Recall that any definite measurement or submeasurement $M_g$ corresponds to a stabilizer element $g$, with $M_g = \sign{g} g$.

The condition to be a definite measurement is simple to calculate. If $M_g = \sign{g} g$, then it must commute with all the generators since $M_g =\sign{g} g_1^{a_1} \cdots g_n^{a_n} \Rightarrow [M_g,g_j] = 0$. Whether $M_g$ commutes with the generators $g_j$ can be determined by the symplectic product in Eq.~(\ref{eq:symplectic}), so this tells us that
\begin{align} \label{eq:definite}
\begin{split}
r(M_g) \Lambda G^T \eqmod{2} 0 &\Leftrightarrow \left( \begin{array}{cc} r_1(M_g) & r_2(M_g) \end{array} \right) \left( \begin{array}{cc} 0 & I \\ I & 0 \end{array} \right) \left( \begin{array}{cc} I \\ \Gamma \end{array} \right) \eqmod{2} 0 \\
&\Leftrightarrow r_2(M_g) \eqmod{2} r_1(M_g) \Gamma
\end{split}
\end{align}
where $\Gamma$ is the adjacency matrix that defines the graph state. We can use this result to determine whether a submeasurement of $M$, corresponding to binary vector $e$, is random or definite. We can then recover the above formula in the case that $e=1$. So, if we take the binary vector for a submeasurement and plug it in to the above condition, we get the condition that the submeasurement be definite. This is an important result.
\begin{result}
\emph{The condition for a submeasurement of $M$ corresponding to $e$ to be definite is that\/}
\begin{equation} \label{eq:defsubcon}
[r_2(M)]_j e_j \eqmod{2} \sum_{k=1}^n \Gamma_{jk} [r_1(M)]_k e_k,
\end{equation}
\end{result}
\noindent for all $j = 1, \ldots, n$.

Now we turn to the quantum mechanical prediction of a definite measurement or submeasurement. The quantum mechanical prediction for the outcome of a definite submeasurement of a global measurement $M$ which is specified by the binary vector $e$ can be determined by plugging in the binary vector of the submeasurement into Eq.~(\ref{eq:qmsign}) of Result~\ref{res:sign}. We find this result to be $i$ to the power,
\begin{equation}
\sum_{j,k=1}^n [r_1(M)]_j e_j \Gamma_{jk} [r_1(M)]_k e_k - \sum_{j=1}^n [r_1(M)]_j e_j [r_2(M)]_j e_j
\end{equation}
Now if we use Eq.~(\ref{eq:defsubcon}), denoting quantities that should be taken modulo $2$ with an overbar, we get the quantum mechanical result of a definite submeasurement.
\begin{result}
\emph{The quantum prediction for a definite submeasurement of $M$ corresponding to $e$ is $i$ to the power\/}
\begin{equation} \label{eq:qmsubsign}
\sum_{j=1}^n [r_1(M)]_j e_j \left( \sum_{k=1}^n \Gamma_{jk} [r_1(M)]_k e_k - \overline{\sum_{k=1}^n \Gamma_{jk} [r_1(M)]_k e_k} \right).
\end{equation}
\end{result}

\subsection{A \hyph{nearest}{neighbor} model}
\label{subsec:model}

In this subsection we present a \hyph{site}{invariant} \hyph{nearest}{neighbor} communication model which predicts correctly the global outcome of the measurement of any Pauli product. This communication model will supplement one particular choice of LHV table, the one with $v=0$ and $c=0$. The outcome of any definite measurement under this table can be calculated using Eq.~(\ref{eq:lhvvalue}), which looks rather complicated until it is realized that $v$ is the zero vector and $C$ is the zero matrix. Thus we have that $v_g = 0$ for all definite measurements $M_g$; this table assigns the value $+1$ to all definite measurements. Obviously, then, this table correctly predicts all definite measurements that are $+1$ according to quantum mechanics and incorrectly predicts those assigned a $-1$ outcome. 

This table is one we will use throughout this chapter, and so it deserves a moderately involved discussion. Since we only consider graph states, so that $G = \left( \begin{array}{cc} I & \Gamma \end{array} \right)$ with $\Gamma$ being the adjacency matrix of the graph, we find that the dual generator matrix $H$ is $\left( \begin{array}{cc} 0 & I \end{array} \right)$. For these generator matrices, and setting $v=c=0$, Eq.~(\ref{eq:tablecons}) reduces to
\begin{equation} \label{eq:graphtablecons}
\left( \begin{array}{cc} x & z \end{array} \right) \eqmod{2} \left( \begin{array}{cc} 0 & r \end{array} \right) \left( \begin{array}{cc} I & 0 \\ \Gamma & I \end{array} \right) \Rightarrow \begin{array}{ccc} z = r & \mbox{and} & x \eqmod{2} r \Gamma \end{array}.
\end{equation}
Hence we find that the values of $z$ are $0$ and $1$ with equal probability, and for this reason we drop $r$ in favor of $z$ and write $x \eqmod{2} z \Gamma$.

In summary, our LHV table uses $n$ binary random variables, $z_1,\ldots,z_n$, each taking on values $0$ and $1$ with equal probability. The random variable $(-1)^{z_j}$ is thought of as value for $Z_j$ under the LHV table, i.e., the value of the $z$ spin components of the $j$th qubit. For the corresponding values of the $x$ and $y$ spin components, we have that
\begin{align}
\label{eq:lhv}
\begin{split}
&(-1)^{x_j}=\prod_{k\in\nghb(j)}(-1)^{z_{k}},\\
&(-1)^{y_j}=(-1)^{z_j}\prod_{k\in\nghb(j)}(-1)^{z_{k}}.
\end{split}
\end{align}
The values $x_j$ assure that $+1$ values are associated with the generators $g_j$, i.e.,
\begin{equation}
(-1)^{x_j}\prod_{k\in\nghb(j)}(-1)^{z_k}=+1\;,
\end{equation}
in analogy to Eq.~(\ref{eq:graphgens}). The values
\begin{equation}\label{yhv}
(-1)^{x_j} (-1)^{y_j} (-1)^{z_j} = 1
\end{equation}
give instructions for how to multiply LHV entries, as does $X Y Z = i$ for Pauli matrices.

We assume now that each party is given a measurement $M_j$ to perform, chosen from $I$, meaning no measurement, and $X$, $Y$, and $Z$.  After the measurement, there is a round of communication between neighboring sites, and then each party outputs a value $+1$ or $-1$ as the result of the
measurement. When no measurement is performed at a site, the output can be regarded as $+1$.

During the round of communication, site~$j$ sends a bit $b_j$ to each site $k\in\nghb(j)$, where $b_j=0$ if $M_j=I,Z$ and $b_j=1$ if $M_j=X,Y$.  The value $m_j$, for the measurement of $M_j$, output at site~$j$ is determined by the LHV table result for the measurement at that site and by the quantity
\begin{equation}
t_j = \sum_{k\in\nghb(j)}b_k = \sum_{k=1}^n \Gamma_{jk} b_k \;,
\label{eq:nj}
\end{equation}
which is computed from the bits sent to site~$j$ from neighboring sites in a graph with adjacency matrix $\Gamma$ and which is equal to the number of neighboring sites that make an $X$ or $Y$ measurement. The output $m_j$ is determined by rules that decide whether to flip the sign of the LHV entry associated with the measurement at site~$j$:
\begin{enumerate}
\item If $M_j=I$, $m_j=1$.
\item If $M_j=Z$, $m_j=(-1)^{z_j}$.
\item If $M_j=X$, $m_j=\left\{\begin{array}{rl}(-1)^{x_j} &\mathrm{if} \quad t_j \eqmod{4} 0,1,\\
-(-1)^{x_j} &\mathrm{if} \quad t_j \eqmod{4} 2,3.\end{array}\right.$
\item If $M_j=Y$, $m_j=\left\{\begin{array}{rl}(-1)^{y_j} &\mathrm{if} \quad t_j \eqmod{4} 0,1,\\
-(-1)^{y_j} &\mathrm{if} \quad t_j \eqmod{4} 2,3.\end{array}\right.$
\end{enumerate}
If we take the product of the outputs from all the sites, the result, written as $(-1)^{v_M}$, corresponds to a prediction for the measurement of the operator $M=\bigotimes_{j=1}^n M_j$. The number of bits communicated in this model is twice the number of edges in the graph.


\subsection{Proof that the model works for global measurements}

The communication model introduced in the last subsection reproduces the quantum predictions for any global Pauli measurement on graph states. The proof of this fact proceeds as follows. First we will determine the result $(-1)^{v_M}$ of the measurement $M$ according to the model. Once this is done we will determine the result of the measurement according to quantum mechanics, denoted as ${(-1)}^{s_M}$. Finally, we compare the two quantities and see that $s_M \eqmod{2} v_M$ for definite measurements. We will also see that quantum mechanics and the model agree for random measurements.

Let us first determine the result of a measurement according to our model. If we write the result of the measurement according to the LHV table as ${(-1)}^{v'_M}$, then the model says we flip this value under certain circumstances. According to the rules of the model only measurement results of $X$ and $Y$ can change the value of ${(-1)}^{v'_M}$, and they do so only when $t_j \eqmod{4} 2,3$. This can be summarized as
\begin{equation}
{(-1)}^{v_M} = {(-1)}^{v'_M} \prod_{\{ j | b_j = 1 \}} i^{\phantom{.} t_j - \overline{t_j}},
\end{equation}
where, once again, we have the notation that quantities under an overbar are to be taken modulo $2$. Notice that $t_j - \overline{t_j}$ is even, so we are assured that the result will either be $+1$ or $-1$. For random measurements, ${(-1)}^{v'_M}$ is random according to Lemma~\ref{lma:standpp} and therefore so is ${(-1)}^{v_M}$. For definite measurements, ${(-1)}^{v'_M} = +1$ and so our model gives the result of a definite measurement to be
\begin{equation} \label{eq:modelresult}
{(-1)}^{v_M} = \prod_{\{ j | b_j = 1 \}} i^{\phantom{.} t_j - \overline{t_j}}.
\end{equation}
These values for definite measurements are the ones we wish to compare to quantum mechanics.

We want to show that the quantum mechanical result of a global measurement $M$ is equal to the result predicted by our model. The key to relating the quantum mechanical result to our model is by recognizing the relationship between $r(M)$ and values of $b_j$ and $t_j$. First, recall that $[r_1(M)]_j = 1$ when $M_j = X$ or $Y$, which tells us that $[r_1(M)]_j = b_j$, the quantity communicated in the model. Second, if $M$ is a definite measurement then Eq.~(\ref{eq:definite}) must be satisfied, which is to say
\begin{align} \label{eq:evenodd}
\begin{split}
&r_2(M) \eqmod{2} r_1(M) \Gamma \\
\Leftrightarrow \hspace{.5em}& [r_2(M)]_j \eqmod{2} \sum_{k=1}^n \Gamma_{jk} b_k = t_j.
\end{split}
\end{align}

So now we can write $s_M$ from Eq.~(\ref{eq:qmsubsign}) in terms of quantities used in the communication model. Doing so gives the value of $2 s_M$ to be
\begin{equation} \label{eq:qmbres}
\sum_{j=1}^n [r_1(M)]_j \left( \sum_{k=1}^n \Gamma_{jk} [r_1(M)]_k - \overline{\sum_{k=1}^n \Gamma_{jk} [r_1(M)]_k} \right) \eqmod{4} \sum_{j=1}^n b_j \left( t_j - \overline{t_j} \right).
\end{equation}
So we find that ${(-1)}^{s_M}$, the quantum mechanical outcome for a definite measurement $M$, can be written as
\begin{equation} \label{eq:qmpredict}
{(-1)}^{s_M} = i^{\sum_{j=1}^n b_j \left( t_j - \overline{t_j} \right)} = \prod_{\{j|b_j=1\}} i^{\phantom{.} t_j - \overline{t_j}},
\end{equation}
which is exactly the result of definite measurements predicted by our model according to Eq.~(\ref{eq:modelresult}).

Even though we have now completed our proof, we make a few comments here. According to Eq.~(\ref{eq:evenodd}), definite measurements have the property that $t_j$ is even when $M_j = X$, and $t_j$ is odd when $M_j = Y$. This means that we can omit the case when $t_j$ is odd and $M_j = X$ or $t_j$ is even and $M_j = Y$, because in these cases the measurement result is random. The only necessary part of our model is that $X$ measurements change their LHV result when $t_j \eqmod{4} 2$ and $Y$ measurements change their LHV result when $t_j \eqmod{4} 3$. This means that there are variants of our model that also predict correctly the result of all global measurements. One special such variant~\cite{barrett:graphmodel} is that $X$ change its LHV result when $t_j \eqmod{4} 1,2$ and $Y$ change its result when $t_j \eqmod{4} 0,3$. Of course, this change means nothing for global measurements, but when considering submeasurements of a global measurement such variations can make a difference.

\subsection{\hyph{Site}{invariant} communication models}
\label{subsec:invariant}

Both the numbering and the arrangement of nodes in a graph are arbitrary, so it seems reasonable to suppose that a communication model should be insensitive to these things. We refer to this property as site invariance and define it formally as follows. Given a graph, each of whose nodes has been assigned a measurement, a permutation that leaves the graph invariant is one that interchanges nodes and their measurements, letting edges move with the nodes, such that the new graph is identical to the old graph in the sense that they could be placed on top of each other with all nodes, measurements, and edges overlapping. A \hyph{site}{invariant} model is one for which nodes in identical situations, as defined by permutations that leave the graph invariant, make the same sign flipping decision. Surprisingly, we find this trait to be at odds with the modeling of submeasurements.

\begin{figure}
\center
\includegraphics[width=4cm]{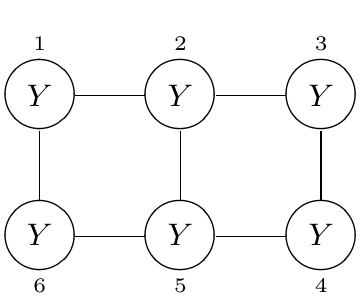}
\caption[Deficiency of \hyph{site}{invariant} models]{Example demonstrating that any communication model based on the LHV table from Eq.~(\ref{eq:lhv}) which is site invariant fails to reproduce some submeasurements. The global measurement $M=Y_1Y_2 Y_3Y_4Y_5Y_6$ has a random outcome, but contains a submeasurement $\tilde M=Y_1Y_2Y_3 I_4Y_5I_6$ such that $-\tilde{M}$ is an element of the stabilizer group. This means that an overall sign flip is required to correct the $+1$ prediction of the LHV table for a measurement of $\tilde M$. The two qubits measuring $Y$ at nodes $1$ and $3$ are in symmetric situations, as are the qubits at nodes $2$ and $5$. Thus, under a \hyph{site}{invariant} model, $1$ and $3$ must make the same sign flipping decision, as must $2$ and $5$. For each pair, the sign flipping decisions cancel one another, producing no overall sign flip and thus giving an incorrect result of $+1$ for the measurement of $\tilde M$. \label{fig:siteinvariant}}
\end{figure}

We demonstrate the limitations imposed by site invariance using the example of a $2\times3$ cluster state, which is depicted in Fig.~\ref{fig:siteinvariant}. The two relevant measurements for this example are $M=Y_1Y_2Y_3Y_4Y_5Y_6$, which has a random outcome, and $\tilde{M}=Y_1Y_2Y_3 I_4Y_5I_6$, which has the certain outcome $-1$.  When either of these is considered as a global measurement, our model yields the correct prediction, as we have already shown in general, but when the second is considered as a submeasurement of the first, the model fails. In this second case, rules~$1$--$4$ say that the two qubits measuring $Y$ at nodes $2$ and $5$ should introduce a sign flip, but the two qubits measuring $Y$ at nodes $1$ and $3$ should not.  The result is no overall sign flip and an outcome $+1$, showing that the model gets the submeasurement outcome wrong.  In contrast, when $\tilde M$ is considered as a global measurement, rules~$1$--$4$ dictate a sign flip for qubit $2$, but no other qubit, thus giving the correct, certain outcome $-1$.  The same measurement $\tilde M$ can lead to different sign-flipping decisions in the two situations because the \hyph{nearest}{neighbor} environments of the qubits differ depending on whether a submeasurement or a global measurement is under consideration.  As is shown in Figure~\ref{fig:siteinvariant}, the counterexample is not limited to the communication model used in this paper.  In fact, any \hyph{site}{invariant} model based on our LHV table yields an incorrect result for the submeasurement $\tilde M$.

This example can easily be generalized by adding $p$ rows and $q$ columns to opposite sides of the $2 \times 3$ cluster state. Doing this results in a class of $(2+2p) \times (3+2q)$ cluster states for which LHV models based on the LHV table of Eq.~(\ref{eq:lhv}) and assisted by a \hyph{site}{invariant} communication model fail for some submeasurements.

It may be possible to find a \hyph{site}{invariant} model for a different LHV table that could reproduce all submeasurement correlations. In this case the symmetry would not be broken by the communication, but rather the table. The point is, nodes in symmetric situation still must behave differently. It is interesting to point out that in this example, the amount of communication and the set of qubits that communicate are both irrelevant. Regardless of how much communication we allow, a \hyph{site}{invariant} model cannot reproduce all submeasurements.

\subsection{Graphs for which the model works}

We have shown that our communication model agrees with quantum mechanics for correlations involving global measurements. We have also demonstrated that because it is site invariant, it will generally fail on submeasurements. In this section we demonstrate that there are some classes of graphs for which the model correctly predicts not only all global measurement correlations, but all correlations present in submeasurements of those global measurements.

The submeasurement results predicted by our model come from a generalization of Eq.~(\ref{eq:modelresult}). That equation gave us the result of a global measurement to be $i$ with an exponent
\begin{equation}
\sum_{j=1}^n b_j \left( \sum_{k=1}^n \Gamma_{jk} b_k - \overline{\sum_{k=1}^n \Gamma_{jk} b_k} \right).
\end{equation}
Generalizing this to a submeasurement, described by a binary vector $e$, is not as simple as replacing all $b$ values with $b$ times $e$; if it were so simple our model would predict all submeasurements correctly. In the above equation, $b_j$ indicates which measurements were present in the measured Pauli product, and $b_k$ indicates which nodes sent their measurement information. Therefore, we need to replace $b_j$ by $b_j e_j$, since we are only multiplying together those results with $e_j = 1$, but we leave $b_k$ alone because everyone still sends a bit $b_k$ if they made an $X$ or $Y$ measurement regardless of whether their measurement is chosen as part of the submeasurement. Thus we get that the prediction of a definite submeasurement by our communication model is $i$ to the power
\begin{equation} \label{eq:modelsubres}
\sum_{j=1}^n b_j e_j \left( \sum_{k=1}^n \Gamma_{jk} b_k - \overline{\sum_{k=1}^n \Gamma_{jk} b_k} \right).
\end{equation}

Since we know our model correctly predicts random submeasurements, our task is to find graphs $\Gamma$ such that Eq.~(\ref{eq:qmsubsign}) and Eq.~(\ref{eq:modelsubres}) agree for all binary vectors $e$ and definite submeasurements satisfying Eq.~(\ref{eq:defsubcon}). We find two such classes of graphs below.

\begin{lemma}
Our \hyph{nearest}{neighbor} communication model correctly predicts all submeasurements for complete bipartite graphs.
\end{lemma}

\begin{proof}

We know our model predicts correctly random submeasurements, so we only need consider definite submeasurements.

The nodes in a complete bipartite graph can be partitioned into two sets, $A$ and $B$, such that each node in $A$ is connected to every node in $B$, yet no node in $A$ is connected to another node in $A$ nor is any node in $B$ connected to another node in $B$. The adjacency matrix for such a graph looks like
\begin{equation}
\Gamma = \left( \begin{array}{cc} 0 & 1 \\ 1 & 0 \end{array} \right),
\end{equation}
where $1$ denotes a matrix with all entries equal to $1$. The nodes in the sets $A$ and $B$ are clumped together to obtain this nice form.

The quantum mechanical result of any submeasurement of a global measurement $M$ is, from Eq.~(\ref{eq:qmsubsign}),
\begin{equation} \label{eq:cbqmsubres}
\sum_{j \in A} b_j e_j \left( \sum_{k \in B} b_k e_k - \overline{\sum_{k \in B} b_k e_k} \right) + \sum_{j \in B} b_j e_j \left( \sum_{k \in A} b_k e_k - \overline{\sum_{k \in A} b_k e_k} \right).
\end{equation}
The result according to our model of any submeasurement is, from Eq.~(\ref{eq:modelsubres}),
\begin{equation} \label{eq:cbmodelsubres}
\sum_{j \in A} b_j e_j \left( \sum_{k \in B} b_k - \overline{\sum_{k \in B} b_k} \right) + \sum_{j \in B} b_j e_j \left( \sum_{k \in A} b_k - \overline{\sum_{k \in A} b_k} \right).
\end{equation}
Notice that for both the quantum mechanical result and the communication model result, the quantities in parentheses are even. Finally, the constraint to be a definite submeasurement, Eq.~(\ref{eq:defsubcon}), becomes for this graph,
\begin{align} \label{eq:cbdefsubcon}
\begin{split}
&[r_2(M)]_{j \in A} e_{j \in A} \eqmod{2} \sum_{k \in B} b_k e_k \\
&[r_2(M)]_{j \in B} e_{j \in B} \eqmod{2} \sum_{k \in A} b_k e_k.
\end{split}
\end{align}

Now we are ready to prove that our model correctly predicts all definite submeasurements. First, since we already know the model predicts the correct result for the global measurement, we can assume that some $e_j = 0$. Without loss of generality, let us suppose this occurs for some $j \in A$. Then Eq.~(\ref{eq:cbdefsubcon}) says that $\sum_{k \in B} b_k e_k$ is even. Since the product of two even numbers is zero modulo $4$, this eliminates the second term from Eqs.~(\ref{eq:cbqmsubres}) and~(\ref{eq:cbmodelsubres}). Now there are two cases to consider, depending on whether $e_j = 0$ for some $j \in B$ or $e_j = 1$ for all $j \in B$. In the first case, Eq.~(\ref{eq:cbdefsubcon}) implies that $\sum_{k \in A} b_k e_k$ is even and both quantum mechanics and our model predict the outcome to be $+1$. In the second case, setting $e_k = 1$ for all $k \in B$ in Eq.~(\ref{eq:cbqmsubres}) gives us Eq.~(\ref{eq:cbmodelsubres}), and so our model agrees with quantum mechanics in this case as well.

\end{proof}

Complete bipartite graphs are interesting because they include the star graphs of GHZ states~\cite{tessier:ghzmodel}. There is also another class of graph for which our model works.

\begin{lemma}
Our \hyph{nearest}{neighbor} communication model correctly predicts all submeasurements for the symmetric difference of two complete graphs.
\end{lemma}

\begin{proof}

Once again, we know our model correctly predicts random submeasurements so we only consider the definite submeasurements.

The symmetric difference of two complete graphs~\cite{diestel:graphtheory} is constructed by beginning with a complete graph, meaning a graph in which every node is connected to every other node, and removing from it a smaller complete graph on a subset $B$ of nodes. The adjacency matrix for this graph looks like
\begin{equation}
\Gamma = \left( \begin{array}{cc} \begin{array}{ccc} 0 & & 1 \\ & \ddots & \\ 1 & & 0 \end{array} & \begin{array}{ccc} 1 & \hdots & 1 \\ \vdots & & \vdots \\ 1 & \hdots & 1 \end{array} \\ \begin{array}{ccc} 1 & \hdots & 1 \\ \vdots & & \vdots \\ 1 & \hdots & 1 \end{array} & \begin{array}{ccc} 0 & \hdots & 0 \\ \vdots & & \vdots \\ 0 & \hdots & 0 \end{array} \end{array} \right),
\end{equation}
The nodes in the graph can be partitioned into sets $A$ and $B$, where $A$ includes all of the remaining nodes not in $B$.

The quantum mechanical result of any submeasurement of a global measurement $M$ is, from Eq.~(\ref{eq:qmsubsign}),
\begin{equation} \label{eq:sdqmsubres}
\sum_{j \in A} b_j e_j \left( \sum_{k =1}^n b_k e_k -1 - \overline{\sum_{k =1}^n b_k e_k -1} \right) + \sum_{j \in B} b_j e_j \left( \sum_{k \in A} b_k e_k - \overline{\sum_{k \in A} b_k e_k} \right).
\end{equation}
The result according to our model of any submeasurement is, from Eq.~(\ref{eq:modelsubres}),
\begin{equation} \label{eq:sdmodelsubres}
\sum_{j \in A} b_j e_j \left( \sum_{k =1}^n b_k -1 - \overline{\sum_{k =1}^n b_k -1} \right) + \sum_{j \in B} b_j e_j \left( \sum_{k \in A} b_k - \overline{\sum_{k \in A} b_k} \right).
\end{equation}
The constraint to be a definite submeasurement, Eq.~(\ref{eq:defsubcon}), becomes for this graph,
\begin{align} \label{eq:sddefsubcon}
\begin{split}
&[r_2(M)]_{j \in A} e_{j \in A} \eqmod{2} \sum_{k =1}^n b_k e_k \\
&[r_2(M)]_{j \in B} e_{j \in B} \eqmod{2} \sum_{k \in A} b_k e_k.
\end{split}
\end{align}

Now, as before, there must be some $e_j = 0$ and there are two possibilities, $j \in A$ and $j \in B$. If $e_j=0$ for some $j \in B$, then Eq.~(\ref{eq:sddefsubcon}) says that $\sum_{k \in A} b_k e_k$ is even. This makes the first term in Eqs.~(\ref{eq:sdqmsubres}) and~(\ref{eq:sdmodelsubres}) zero modulo $4$, since they are the product of two even numbers. Now we look at the second term and consider whether there is some $j \in A$ such that $e_j = 0$. If there is, then Eq.~(\ref{eq:sddefsubcon}) implies that $\sum_{k =1}^n b_k e_k$ is even, and since we know that $\sum_{k \in A} b_k e_k$ is even that means $\sum_{k \in B} b_k e_k$ must also be even. This makes the last term zero and makes Eqs.~(\ref{eq:sdqmsubres}) and~(\ref{eq:sdmodelsubres}) equal. If $e_j = 1$ for all $j \in A$, then $e_k$ can be set to $1$ in Eq.~(\ref{eq:sdqmsubres}) and the two results are still the same.

The last case is when $e_j=1$ for all $j \in B$, but there exists a $j \in A$ such that $e_j = 0$. In this case, Eq.~(\ref{eq:sddefsubcon}) tells us that $\sum_{k =1}^n b_k e_k$ is even. There are two ways this can happen. Either both $\sum_{k \in A} b_k e_k$ and $\sum_{k \in B} b_k e_k$ are even, in which case both Eq.~(\ref{eq:sdqmsubres}) and Eq.~(\ref{eq:sdmodelsubres}) are always zero, or they are both odd. Assuming $\sum_{k \in A} b_k e_k$ and $\sum_{k \in B} b_k e_k = \sum_{k \in B} b_k$ are both odd, Eqs.~(\ref{eq:sdqmsubres}) and~(\ref{eq:sdmodelsubres}) reduce to
\begin{align}
\begin{split}
&\sum_{k =1}^n b_k e_k -1 - \overline{\sum_{k =1}^n b_k e_k -1} + \sum_{k \in A} b_k e_k - \overline{\sum_{k \in A} b_k e_k} \\
&\mbox{and }\sum_{k =1}^n b_k -1 - \overline{\sum_{k =1}^n b_k -1} + \sum_{k \in A} b_k - \overline{\sum_{k \in A} b_k},
\end{split}
\end{align}
because the product of an even and an odd number modulo $4$ is just that even number modulo $4$. Now we use that $\sum_{k =1}^n b_k e_k$ is even and that $\sum_{k \in B} b_k e_k$ and $\sum_{k \in A} b_k e_k$ are both odd to reduce the quantum result further,
\begin{align}
\begin{split}
&\sum_{k =1}^n b_k e_k -1 - \overline{\sum_{k =1}^n b_k e_k -1} + \sum_{k \in A} b_k e_k - \overline{\sum_{k \in A} b_k e_k} \\
=&\sum_{k =1}^n b_k e_k -1 - 1 + \sum_{k \in A} b_k e_k - 1 \\
=&\sum_{k \in B} b_k e_k + 2 \sum_{k \in A} b_k e_k - 3 \eqmod{4} \sum_{k \in B} b_k e_k -1 = \sum_{k \in B} b_k -1,
\end{split}
\end{align}
where once again, in the second to last step, we had the product of an even and an odd number giving just the even number.
As for the result of our model, we use that $\sum_{k \in B} b_k$ is odd to reduce it to
\begin{align}
\begin{split}
&\sum_{k =1}^n b_k -1 - \overline{\sum_{k =1}^n b_k -1} + \sum_{k \in A} b_k - \overline{\sum_{k \in A} b_k} \\
=&\sum_{k \in B} b_k + \sum_{k \in A} b_k -1 - \overline{\sum_{k \in A} b_k} + \sum_{k \in A} b_k - \overline{\sum_{k \in A} b_k} \\
=&\sum_{k \in B} b_k -1 + 2 \left( \sum_{k \in A} b_k - \overline{\sum_{k \in A} b_k} \right) \eqmod{4} \sum_{k \in B} b_k -1,
\end{split}
\end{align}
and so our model and quantum mechanics agree in all cases.

\end{proof}

The symmetric difference of complete graphs includes complete graphs themselves. The symmetric difference of two complete graphs and complete bipartite graphs actually coincide in the case of star graphs.

In all the classes of graphs above, it is possible to get from any node to any other node by traversing no more than two edges. So, in a sense a \hyph{nearest}{neighbor} model communicates through half the graph. This may be why our model works, and so it would be interesting to find graphs without this property for which our \hyph{site}{invariant} \hyph{nearest}{neighbor} communication model works.

\section{Other communication models \label{sec:universalscheme}}

We have looked at one of the simplest kinds of communication models imaginable, this being a \hyph{site}{invariant} model that limits communication to nearest neighbors in the graph. In this section we will go beyond this. We will consider \hyph{site}{invariant} models with unlimited communication, models whose communication distance is limited but whose decision algorithm can vary from site to site, and even a model with no restrictions at all.

\hyph{Site}{invariant} models will always use the LHV table in the last section. The reason is that allowing the LHV table to break the symmetry of the graph defeats the purpose of considering a \hyph{site}{invariant} model. We proved earlier that a \hyph{site}{invariant} model is incapable of reproducing all submeasurement results on a \hyph{two}{dimensional} cluster state. Motivated by results from \hyph{measurement}{based} quantum computation~\cite{vandennest:resources}, we compare this result to \hyph{one}{dimensional} cluster states, i.e., linear chains. We find that a \hyph{site}{invariant} model is indeed capable of reproducing all submeasurement results on a linear chain, although more than \hyph{nearest}{neighbor} communication is needed.

We also consider models that are not site invariant, and for this we must account for an arbitrary underlying LHV table. We analyze such models based on \emph{communication distance\/}, which we define to be the number of successive edges through which information can be sent. For the \hyph{one}{dimensional} cluster state, we find that no communication distance $1$, meaning \hyph{nearest}{neighbor}, model is capable of reproducing all submeasurement results even if it is not site invariant. In general, we prove that any model that reproduces all submeasurement results for all graph states must have a communication distance that scales linearly with the size of the graph.

\subsection{\hyph{Site}{invariant} model for $1$D cluster states}
\label{subsec:cluster}

In Subsec.~\ref{subsec:invariant} it was shown that for certain graph states, no model based on the LHV table of Eq.~(\ref{eq:lhv}) and assisted by a \hyph{site}{invariant} communication model is capable of reproducing all submeasurement correlations. This result holds even if the model allows unlimited distance communication.

This subsection further explores the symmetry of site invariance by considering it in the context of \hyph{one}{dimensional} cluster states, showing that linear chains do indeed permit successful \hyph{site}{invariant} models. The model we describe involves communication over a distance equal to the number of edges in the \hyph{one}{dimensional} cluster state, i.e unlimited communication. At present it is unknown whether the submeasurement correlations of a linear chain could be reproduced by a model with limited distance communication, although this is unlikely since in the next subsection we show that \hyph{nearest}{neighbor} communication is insufficient.

The fact that unlimited communication is allowed is in the same spirit as our counterexample of Fig.~\ref{fig:siteinvariant}, where communication spans the entire graph and the only restriction is site invariance.  The key simplification in the case of \hyph{one}{dimensional} cluster states is that all qubits, except those at the ends of the chain, have exactly two neighbors. As a consequence, the form of stabilizer elements whose LHV result from Eq.~(\ref{eq:lhv}) requires correction is constrained so that the correction can be effected by a \hyph{site}{invariant} model.

Before explaining our \hyph{site}{invariant} model for linear chains we should learn more about them, introducing terminology as appropriate. For an \hyph{$n$}{qubit} chain, the $n$ stabilizer generators are given by $g_1=X_1Z_2$, $g_j=Z_{j-1}X_jZ_{j+1}$ for $j=2,\ldots,n-1$, and $g_n=Z_{n-1}X_n$. Any stabilizer element is a product of generators. An arbitrary product of generators can be decomposed into a product of terms each of which is a product of successive generators. We call these terms \emph{primitive stabilizers\/} or just \emph{primitives\/}. The primitive stabilizers are separated by the omission of one or more generators in the product of generators. An example of a stabilizer element for $n=10$ qubits is $g_1g_2g_3g_5g_6g_9=- Y_1X_2Y_3I_4Y_5Y_6Z_7Z_8X_9Z_{10}$. The primitives in this example are $g_1g_2g_3$, $g_5g_6$, and $g_9$.

Associated with each primitive is a Pauli product, with the sign omitted, for the qubits corresponding to the generators in the primitive. We call these Pauli products \emph{words\/}. For the 10-qubit example above, the words are $Y_1X_2Y_3$, $Y_5Y_6$, and $X_9$. At each end of a word, there is an $I$ if one generator is omitted and a $Z$ if two or more generators are omitted. We can make these word boundaries apply even at the end of the linear chain by embedding our cluster state in an infinite linear chain. The generators for the qubits to the left of $j=1$ and to the right of $j=n$ are always omitted, and we redefine $g_1=Z_0X_1Z_2$ and $g_n=Z_{n-1}X_nZ_{n+1}$. We can list the entire set of words by considering all possible primitives:
\begin{enumerate}
\item $X$ for a primitive with one Pauli operator.
\item $Y\otimes Y$ for a primitive with two Pauli operators.
\item $Y\otimes X^{\otimes(j-2)}\otimes Y$ for a primitive with $j\ge3$ Pauli operators.
\end{enumerate}

If a word is bounded by an $I$, there must be another word immediately on the other side of the $I$. A \emph{sentence\/} is a Pauli product consisting of a set of words separated by singleton $I$'s and bracketed by $Z$'s at both ends. Words are not stabilizer elements, but sentences are. The example above contains two sentences, $Z_0Y_1X_2Y_3I_4Y_5Y_6Z_7$, including the zeroth qubit, and $Z_8X_9Z_{10}$. The $Z$ bookends on a sentence separate it from other, nonoverlapping sentences in the same overall stabilizer element. Between the $Z$'s in successive sentences, there can be an arbitrary number of $I$'s. Any stabilizer element is a product of nonoverlapping sentences, and for these elements $I$'s occur only between sentences or as singletons between words, $X$'s and $Y$'s occur only in words, and $Z$'s occur only as the boundaries of sentences.

Recall that the goal of the communication model is to introduce a sign flip into the product of LHV table entries for those Pauli products that are the negative of a stabilizer element. The only words that introduce a minus sign into the corresponding product of generators are those of the form $Y\otimes X^{\otimes(j-2)}\otimes Y$ with $j$ odd.  Thus a candidate for a \hyph{site}{invariant} communication model is the following.
\begin{enumerate}
\item Each site at which an $X$ or a $Z$ is measured broadcasts the measurement performed upon it.
\item Each site that measures $X$ determines if it is the middle, implying an odd number of $X$'s, qubit
in a word of the form $Y\otimes X^{\otimes(j-2)}\otimes Y$ in a submeasurement sentence, and if so, flips its LHV entry, i.e., changes $(-1)^x$ to $-(-1)^x$.
\end{enumerate}
This clearly gets any stabilizer right and thus all global measurement correlations right.

The only question remaining is whether this model works for submeasurements. We answer this question by showing the following: two sentences, $S_1$ and $S_2$, that are submeasurements of the same global measurement, generally not a stabilizer element, must be identical on the region where they overlap, except possibly at bracketing $Z$'s. This property implies that $S_1$ and $S_2$ have exactly the same words in the region of overlap. Thus, for any pair of submeasurements of the global measurement, a sign flip arising from a word of the form $Y\otimes X^{\otimes(j-2)}\otimes Y$ in the overlap region is common to both submeasurements. Since both the word and the sign flip occur in both submeasurements, our model correctly predicts both outcomes. Therefore, proving this property will complete the proof that our \hyph{site}{invariant} model correctly predicts all submeasurements.

To prove this property, notice first that if $S_1$ and $S_2$ overlap, there are two cases: the region of overlap coincides with one of the sentences, or it does not. In the former case, we choose $S_2$ to be the sentence that coincides with the region of overlap, and in the latter case, we choose $S_1$ to be the sentence on the left and $S_2$ to be the one on the right. With these conventions, the left boundary of the overlap region coincides with the $Z$ that bounds the left end of $S_2$, and the right boundary of the overlap region coincides in the former case with the $Z$ that bounds the right end of $S_2$ and in the latter case it coincides with the $Z$ that bounds the right end of $S_1$.

To be submeasurements of the same global measurement, the two sentences must satisfy the following basic rule: in the overlap region, sites within a word of one sentence must be occupied in the other sentence by the same Pauli operator or by an $I$. Since $Z$'s do not occur in words, this rule implies that the $Z$'s that bound the overlap region at either end in one of the two sentences cannot occupy a site within a word in the other sentence and thus must be a bounding $Z$ or a singleton $I$ in the other sentence. The submeasurement requirement, by itself, implies that in the overlap region, the site of a singleton $I$ in one sentence can be occupied by anything in the other sentence, but the available words impose a much stronger constraint, as we now show.

Consider the left boundary of the overlap region, which is occupied by the leftmost $Z$ in $S_2$ and by a $Z$ or a singleton $I$ in $S_1$. Immediately to the right in both $S_1$ and $S_2$ is a word. When one of
these words is shorter than the other, the basic rule implies that the shorter word must be a prefix of the longer one.  A glance at the allowed words $1$--$3$ shows, however, that none is a prefix of another. Thus $S_1$ and $S_2$ must have the same word in this first overlap position, which is followed by a singleton~$I$ in both sentences. Applying the same logic to this and subsequent singleton $I$'s shows, as promised, that $S_1$ and $S_2$ are identical in the overlap region, except possibly at the boundaries.

\subsection{\hyph{Nearest}{neighbor} models for $1$D cluster states}
\label{subsec:onedcluster}

Two properties that make for a simple communication model are site invariance and \hyph{nearest}{neighbor} communication. We showed that a \hyph{site}{invariant} model can be made to work for linear chains if unlimited communication is allowed. In this subsection we prove that \hyph{nearest}{neighbor} communication models are incapable of reproducing all submeasurement correlations for a linear chain.

\begin{table}[h!]
\center
\begin{tabular}{ll}
\underline{Measurement} & \underline{Outcome of the Model} \\
$Z_1X_2{\color{gray}Y_3}X_4{\color{gray}X_5}X_6Z_7$ & $z_1^{X} x_2^{ZY} x_4^{YX} x_6^{XZ} z_7^{XI} = 1$ \\
$Z_1X_2{\color{gray}Y_3}X_4{\color{gray}Y_5}X_6Z_7$ & $z_1^{X} x_2^{ZY} x_4^{YY} x_6^{YZ} z_7^{XI} = 1$ \\
$Z_2Y_3X_4X_5X_6Y_7Z_8$ & $z_2^{IY} y_3^{ZX} x_4^{YX} x_5^{XX} x_6^{XY} y_7^{XZ} z_8^{YI} = -1$ \\
$Z_2X_3{\color{gray}X_4}X_5{\color{gray}X_6}X_7Z_8$ & $z_2^{IX} x_3^{ZX} x_5^{XX} x_7^{XZ} z_8^{XI} = 1$ \\
$Z_2Y_3X_4Y_5{\color{gray}X_6}Y_7X_8Y_9Z_{10}$ & $z_2^{IY} y_3^{ZX} x_4^{YY} y_5^{XX} y_7^{XX} x_8^{YY} y_9^{XZ} z_{10}^{YI} = 1$ \\
$Z_2X_3{\color{gray}X_4}Y_5X_6Y_7Z_8$ & $z_2^{IX} x_3^{ZX} y_5^{XX} x_6^{YY} y_7^{XZ} z_8^{YI} = -1$ \\
$Z_3X_4{\color{gray}X_5}X_6{\color{gray}Y_7}X_8Z_9$ & $z_3^{IX} x_4^{ZX} x_6^{XY} x_8^{YZ} z_9^{XI} = 1$ \\
$Z_3X_4{\color{gray}X_5}X_6Z_7$ & $z_3^{IX} x_4^{ZX} x_6^{XZ} z_7^{XI} = 1$ \\
$Z_3X_4{\color{gray}Y_5}X_6{\color{gray}Y_7}X_8Z_9$ & $z_3^{IX} x_4^{ZY} x_6^{YY} x_8^{YZ} z_9^{XI} = 1$ \\
$Z_3X_4{\color{gray}Y_5}X_6Z_7$ & $z_3^{IX} x_4^{ZY} x_6^{YZ} z_7^{XI} = 1$ \\
$Z_4X_5{\color{gray}X_6}Y_7X_8Y_9Z_{10}$ & $z_4^{IX} x_5^{ZX} y_7^{XX} x_8^{YY} y_9^{XZ} z_{10}^{YI} = -1$ \\
$Z_4X_5{\color{gray}X_6}X_7Z_8$ & $z_4^{IX} x_5^{ZX} x_7^{XZ} z_8^{XI} = 1$ \\
$Z_5X_6{\color{gray}Y_7}X_8{\color{gray}Y_9}X_{10}Z_{11}$ & $z_5^{IX} x_6^{ZY} x_8^{YY} x_{10}^{YZ} z_{11}^{X} = 1$ \\
$Z_5X_6{\color{gray}Y_7}X_8Z_9$ & $z_5^{IX} x_6^{ZY} x_8^{YZ} z_9^{XI} = 1$ \\
$Z_5X_6{\color{gray}Y_7}X_8{\color{gray}Y_9}X_{10}Z_{11}$ & $z_5^{IX} x_6^{ZY} x_8^{YY} x_{10}^{YZ}  z_{11}^{X} = 1$ \\
$Z_5X_6{\color{gray}Y_7}X_8Z_9$ & $z_5^{IX} x_6^{ZY} x_8^{YZ} z_9^{XI} = 1$
\end{tabular}
\caption[Nonexistence of a \hyph{nearest}{neighbor} model for the linear chain]{In the left column is listed Pauli measurements on a linear chain with $11$ qubits. For each global measurement, a definite submeasurement is shown in black. These definite submeasurements lead to the contraints on the outcome of a \hyph{nearest}{neighbor} communication model listed in the right column. If these $16$ equations in the right column are carefully multiplied together, all the outcomes will cancel and what will remain is the equation $1=-1$. Since this is impossible, no \hyph{nearest}{neighbor} communication model can reproduce all $16$ of these correlations. Note that this argument is independent of the underlying LHV table and assumes nothing about the details of the communication strategy other than communication being only between nearest neighbors. \label{tab:contra}}
\end{table}

In this subsection, for the first time, we need to consider communication models that are not site invariant. This means that we also need to incorporate arbitrary LHV tables on which the model is based. We do this  by considering the outcome, according to a \hyph{nearest}{neighbor} communication model, of a measurement of a Pauli matrix $\sigma$ on qubit $j$. The output of this measurement in such a model depends on both the qubit in question and the measurements made on neighboring qubits. We write this in the form ${s_{\!j}}^\alpha$ where $j$ is the qubit being measured, $s$ is the lower case version of the Pauli operator measured upon it, and $\alpha$ consists of the Pauli matrices being measured on the neighboring qubits. For instance, $x_j^{YZ}$ is the output for the model of an $X$ measured on the $j$th qubit when the left neighbor measures $Y$ and the right neighbor measures $Z$, and $z_j^{X}$ is the outcome of a $Z$ measurement on a qubit $j$ who is at the end of the chain and whose neighbor chooses to measure $X$.


We construct our counterexample on a chain of $11$ qubits. By considering appropriate measurements, we arrive at constraints on the communication model that yield a contradiction. It turns out that we can get a contradiction after considering the $16$ measurements listed in Table \ref{tab:contra}. This proves that no \hyph{nearest}{neighbor} communication model can reproduce all submeasurement correlations arising from Pauli measurements on a linear chain. This result is independent of the underlying LHV table and does not assume a \hyph{site}{invariant} model.

We can most likely generalize this result to say that no limited distance communication model can reproduce all submeasurements on a linear chain. The reasoning would be similar to the strategy of the next section. However, we have not yet done this and so the question of whether a limited distance communication model can reproduce Pauli measurement correlations on a $1$D cluster state still remains open.

\subsection{Non-nearest-neighbor communication models}
\label{subsec:triangle}

We have shown that \hyph{site}{invariant} communication models are incapable, for at least some measurements, of reproducing all submeasurement correlations. Therefore any successful communication model must reject this property. In this section we show that any model whose communication distance is constant or scales less than linearly with the number of qubits fails as well to predict some submeasurements correctly. As a result we have that any model reproducing all quantum correlations must have communication that spans nearly the entire graph, as well as not being site invariant.

In a communication model with communication distance $d$, nodes $j$ and $k$ can signal to each other if there exists within the graph a path from $j$ to $k$ that traverses $d$ or fewer edges~\cite{diestel:graphtheory}. Put another way, this is the statement that information can only be transmitted along edges and that the number of successive edges through which some piece of information can be sent is at most $d$. In this section we prove, via contradiction, that no \hyph{communication}{assisted} LHV model for which the communication distance satisfies
\begin{equation}
  d\leq4\left\lfloor \frac{n}{24}-\frac{1}{2}\right\rfloor+1
\end{equation}
correctly reproduces the predictions of quantum mechanics for all submeasurements on all graph states of $n$ qubits.

The proof relies on an infinite class of graph states for which a set of five global measurements can be chosen that are not locally distinguishable. Each of these global measurements includes a submeasurement that can be written in terms of stabilizer elements and is thus definite. The output of each qubit, however, must be such that the correct values are obtained for all submeasurements that are consistent with its observable surroundings. This requirement, for the particular states and measurements chosen, yields a contradiction.

\begin{figure}
\center
\includegraphics[width=7cm]{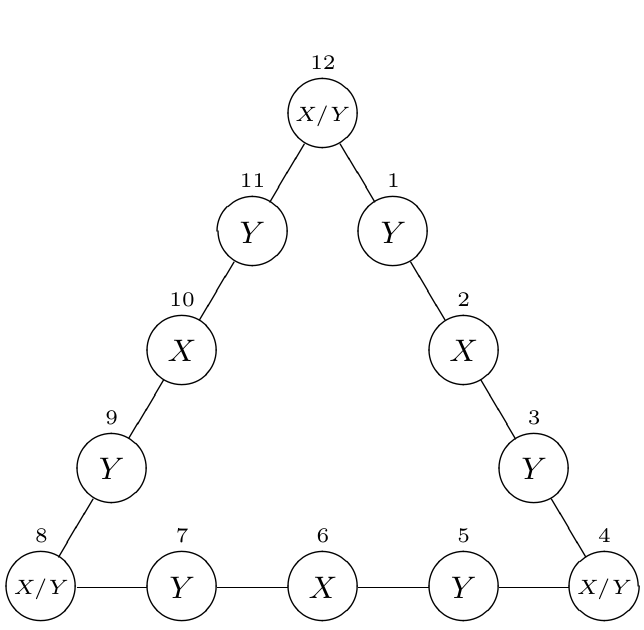}
\caption[Deficiency of \hyph{nearest}{neighbor} models]{Example demonstrating that any communication-assisted LHV model with communication distance $d \leq 1$ fails to reproduce some submeasurements for the ring with $n = 12$ nodes. Five of the global measurements shown on the \vspace{.4em}ring,
\vspace{.4em}\centerline{${\color{gray}Y_1}X_2{\color{gray}Y_3}X_4{\color{gray}Y_5}X_6{\color{gray}Y_7}X_8{\color{gray}Y_9}X_{10}{\color{gray}Y_{11}}X_{12}$\;,}
\vspace{.4em}\centerline{$Y_1X_2Y_3{\color{gray}Y_4}Y_5X_6Y_7{\color{gray}Y_8}Y_9X_{10}Y_{11}{\color{gray}Y_{12}}$\;,}
\vspace{.4em}\centerline{$Y_1{\color{gray}X_2}Y_3Y_4{\color{gray}Y_5}X_6{\color{gray}Y_7}X_8{\color{gray}Y_9}X_{10}{\color{gray}Y_{11}}Y_{12}$\;,} 
\vspace{.4em}\centerline{${\color{gray}Y_1}X_2{\color{gray}Y_3}Y_4Y_5{\color{gray}X_6}Y_7Y_8{\color{gray}Y_9}X_{10}{\color{gray}Y_{11}}X_{12}$\;,}
\vspace{.4em}\centerline{${\color{gray}Y_1}X_2{\color{gray}Y_3}X_4{\color{gray}Y_5}X_6{\color{gray}Y_7}Y_8Y_9{\color{gray}X_{10}}Y_{11}Y_{12}$\;,}
contain submeasurements shown in black that are useful for exhibiting a contradiction. These
submeasurements imply the following constraints on a \hyph{nearest}{neighbor} communication \vspace{.2em}model,
\vspace{.4em}\centerline{\hspace{3.2em}$x_2^{}x_4^{}x_6^{}x_8^{}x_{10}^{}x_{12}^{}=1$\;,\hspace{1em}}
\vspace{.4em}\centerline{$y_1^Yx_2^{}y_3^Yy_5^Yx_6^{}y_7^Yy_9^Yx_{10}^{}y_{11}^Y=-1$\;,\hspace{1em}}
\vspace{.4em}\centerline{\hspace{1.9em}$y_1^Yy_3^Yy_4^{}x_6^{}x_8^{}x_{10}^{}y_{12}^{}=1$\;,\hspace{1em}}
\vspace{.4em}\centerline{\hspace{1.9em}$x_2^{}y_4^{}y_5^Yy_7^Yy_8^{}x_{10}^{}x_{12}^{}=1$\;,\hspace{1em}}
\vspace{.4em}\centerline{\hspace{2.5em}$x_2^{}x_4^{}x_6^{}y_8^{}y_9^Yy_{11}^Yy_{12}^{}=1\;,$\hspace{1em}}
which when multiplied together yield a contradiction.
\label{fig:triangle}}
\end{figure}

To begin, consider the graph state corresponding to an $n$ node ring where $n=12f$ and $f$ is an odd positive integer.  Let the qubits be numbered sequentially, starting with $1$ at an arbitrary point on the ring and moving clockwise along it. Additionally, define the following subsets of the $n$ labels:
\begin{align}
\begin{split}
  \mathcal{V}&=\{4f,8f,12f\}\;,\\
  \mathcal{M}&=\{2f,6f,10f\}\;,\\
  \mathcal{Y}&=\{j|j\equiv1\text{ mod }2\}\;,\\
  \mathcal{L}&=\{j|j\not\in\mathcal{V},\mathcal{M}\text{ and }j\equiv2\text{ mod }4\}\;,\\
  \mathcal{R}&=\{j|j\not\in\mathcal{V},\mathcal{M}\text{ and }j\equiv0\text{ mod }4\}\;,\\
  \mathcal{S}_k&=\{j|2f(k-1)<j<2f k\}\;.
\end{split}
\end{align}
For our purposes, it is useful to think of the ring as arranged in an equilateral triangle with vertices specified by the subset $\mathcal{V}$ as shown in figure \ref{fig:triangle}. The midpoints of the legs of the triangle are then given by the subset $\mathcal{M}$, and the segments between adjacent vertices and midpoints are given by the $\mathcal{S}_j$'s. We use the notation $\mathcal{S}_{j,k}$ as shorthand for $\mathcal{S}_j\cup\mathcal{S}_k$, and $\mathcal{A}\backslash\mathcal{B}$ is used to denote the set consisting of the elements of $\mathcal{A}$ that are not in $\mathcal{B}$.

Now consider global measurements of the form $\bigotimes_{j=1}^{12f} M_j$, with
\begin{equation}
M_j =
  \begin{cases}
    X\text{ or }Y & \text{if } j\in\mathcal{V},\\
    Y & \text{if } j\in\mathcal{Y},\\
    X & \text{otherwise,}
  \end{cases}\label{eq:globalmeas}
\end{equation}
such that the number of vertices measuring $Y$ is not one. These global measurements include the following submeasurements, for which quantum mechanics predicts an outcome with certainty.\\

For $M_{4f}=X$, $M_{8f}=X$, $M_{12f}=X$,
\begin{align} \label{eq:XXX}
X_{2f}X_{4f}X_{6f}X_{8f}X_{10f}X_{12f}\prod_{j\in\mathcal{L}\cup\mathcal{R}} X_{j} = \prod_{j=1}^{6f} g_{2j}\;,
\end{align}
implying a measurement outcome of $+1$.\\

For $M_{4f}=Y$, $M_{8f}=Y$, $M_{12f}=Y$,
\begin{align} \label{eq:YYY}
\begin{split}
X_{2f}&X_{6f}X_{10f}\prod_{j\in\mathcal{Y}} Y_{j} \prod_{k\in\mathcal{L}} X_{k} \hspace{8em} \\
&= \prod_{j=1}^{3f} \Bigl(-g_{4j-3}g_{4j-2}g_{4j-1}\Bigr)\;,
\end{split}
\end{align}
implying a measurement outcome of $(-1)^{3f} = -1$.\\

For $M_{4f}=Y$, $M_{8f}=X$, $M_{12f}=Y$,
\begin{align} \label{eq:YXY}
\begin{split}
Y_{4f}X_{6f}X_{8f}X_{10f}Y_{12f}\prod_{j\in\mathcal{Y}\cap\mathcal{S}_{1,2}} Y_j
\prod_{k\in\mathcal{R}\cup(\mathcal{L}\backslash\mathcal{S}_{1,2})} X_k \\
= g_1 \prod_{j=1}^{f-1} \Bigl(-g_{4j-1}g_{4j}g_{4j+1}\Bigr) g_{4f-1} \prod_{k=2f}^{6f} g_{2k}\;,
\end{split}
\end{align}
implying a measurement outcome of $(-1)^{f-1} = +1$.\\

Cyclic permutation of this last measurement yields two more with $+1$ outcomes.\\

For $M_{4f}=Y$, $M_{8f}=Y$, $M_{12f}=X$, we have
\begin{align} \label{eq:YYX}
X_{2f}Y_{4f}Y_{8f}X_{10f}X_{12f}\prod_{j\in\mathcal{Y}\cap\mathcal{S}_{3,4}} Y_j
\prod_{k\in\mathcal{R}\cup(\mathcal{L}\backslash\mathcal{S}_{3,4})} X_k\;,
\end{align}
and for $M_{4f}=X$, $M_{8f}=Y$, $M_{12f}=Y$, we have
\begin{align} \label{eq:XYY}
X_{2f}X_{4f}X_{6f}Y_{8f}Y_{12f}\prod_{j\in\mathcal{Y}\cap\mathcal{S}_{5,6}} Y_j
\prod_{k\in\mathcal{R}\cup(\mathcal{L}\backslash\mathcal{S}_{5,6})} X_k\;.
\end{align}

Now assume there exists a distance $d=2f-1$ \hyph{communication}{assisted} LHV model that correctly replicates the predictions of quantum mechanics for all Pauli measurements on $n$ qubits. As in the previous subsection, the output of such a model can be fully described in terms of \hyph{single}{qubit} LHV entries whose value depends both on the qubit in question and on the measurements made by other qubits within its communication range. We write these hidden variables in the form ${s_{\!j}}^\alpha$, where in this case $\alpha$ indicates the measurements made on all qubits within its communication range. The global measurements utilized for Eqs.~(\ref{eq:XXX}) through~(\ref{eq:XYY}) have the virtue that the communication range of each qubit includes at most one other qubit whose measurement is changeable, and that is the qubit at the nearest vertex. Thus, in comparisons between them, the measurement performed on, at most, a single qubit need be included in $\alpha$. Moreover, the qubits at the center of each side of the triangle cannot see the changes at the vertices. Consequently, the constraints implied by Eqs.~(\ref{eq:XXX}) through~(\ref{eq:XYY}) on a hidden variable model with communication range $d$ can be expressed as follows:

\begin{align} \label{eq:xxx}
\begin{split}
1=&x_{2f}x_{4f}x_{6f}x_{8f}x_{10f}x_{12f}\prod_{j\in\mathcal{L}\cup\mathcal{R}} x_{j}^X\;, \\
-1=&x_{2f}x_{6f}x_{10f} \prod_{j\in\mathcal{Y}} y_{j}^Y \prod_{k\in\mathcal{L}} x_{k}^Y\;, \\
1=&y_{4f}x_{6f}x_{8f}x_{10f}y_{12f}
\prod_{j\in\mathcal{Y}\cap\mathcal{S}_{1,2}} y_j^Y \\
&\hspace{2em} \prod_{k\in(\mathcal{L}\cup\mathcal{R})\cap\mathcal{S}_{4,5}} x_k^X
\prod_{l\in(\mathcal{L}\cap\mathcal{S}_{3,6})\cup(\mathcal{R}\backslash\mathcal{S}_{4,5})} x_l^Y\;, \\
1=&x_{2f}y_{4f}y_{8f}x_{10f}x_{12f}
\prod_{j\in\mathcal{Y}\cap\mathcal{S}_{3,4}} y_j^Y \\
&\hspace{2em}\prod_{k\in(\mathcal{L}\cup\mathcal{R})\cap\mathcal{S}_{6,1}} x_k^X
\prod_{l\in(\mathcal{L}\cap\mathcal{S}_{5,2})\cup(\mathcal{R}\backslash\mathcal{S}_{6,1})} x_l^Y\;, \\
1=&x_{2f}x_{4f}x_{6f}y_{8f}y_{12f}
\prod_{j\in\mathcal{Y}\cap\mathcal{S}_{5,6}} y_j^Y \\
&\hspace{2em}\prod_{k\in(\mathcal{L}\cup\mathcal{R})\cap\mathcal{S}_{2,3}} x_k^X
\prod_{l\in(\mathcal{L}\cap\mathcal{S}_{1,4})\cup(\mathcal{R}\backslash\mathcal{S}_{2,3})} x_l^Y\;.
\end{split}
\end{align}

Using the identity $\mathcal{A}=\mathcal{A}\cap(\mathcal{S}_1\cup\mathcal{S}_2\cup \mathcal{S}_3\cup\mathcal{S}_4\cup\mathcal{S}_5\cup\mathcal{S}_6)$ for $\mathcal{A}=\mathcal{Y}\text{, }\mathcal{L}\text{, or }\mathcal{R}$ and the fact that all variables square to $1$, it can be shown that the right hand side of the first equation in Eq.~(\ref{eq:xxx}) is equal to the product of the right hand sides of the other four equations. Thus, we have the contradiction $1=-1$, showing that no distance $d$ \hyph{communication}{assisted} LHV model reproduces the predictions of quantum mechanics in this instance.

For other values of $n \neq 12f$, with $f$ odd, an identical contradiction applies to a graph consisting of $r=(n-12) \pmod{24} < 24$ unconnected nodes and a ring of size $n-r$. So for any value of $n$ we have a ring of size $n-r = 12(2k+1)$ for some integer $k$, and we have proven that in order to reproduce all quantum correlations we must have
\begin{equation}
d > 2(2k+1)-1 = 4k+1 = 4 \left\lfloor \frac{n-12}{24} \right\rfloor+1,
\end{equation}
as desired.

\subsection{Universal model}

To end this chapter, we lay out a communication model that actually does reproduce all quantum correlations for Pauli measurements on all graph states. We know that this model cannot be site invariant and that it must also have a communication distance that scales linearly with the number of qubits. For these reasons we impart our model with two properties. One property, motivated by our understanding of site invariance, is that there is a special node capable of breaking symmetry. The second property, motivated by what we know about communication distance, is that we allow communication to span the entire graph.

Our universal model involves two rounds of communication. During the first round, everyone sends their measurement choice to a single party. During the second round of communication, this special party sends a classical bit to each qubit needing to change their LHV output. The algorithm allowing this special party to decide which qubits must change their LHV result is explained below.

There are $2^n$ possibly \hyph{non}{unique} submeasurements, for an \hyph{$n$}{qubit} graph state, which need to be correct in the end. These correspond to the $2^n$ binary vectors $e$ that specify which measurement results of the global measurement, $M$, to include in the submeasurement. However, we need not worry about random submeasurements and only focus on those definite submeasurements satisfying Eq.~(\ref{eq:qmsubsign}). In fact, Eq.~(\ref{eq:qmsubsign}) can be used in a simply way to find all definite submeasurements. To see this, rewrite the condition to be a definite submeasurement as
\begin{align}
\begin{split}
& [r_2(M)]_j e_j \eqmod{2} \sum_{k=1}^n \Gamma_{jk} [r_1(M)]_k e_k \\
\Leftrightarrow \hspace{.5em} & \sum_{k=1}^n \Bigl( \Gamma_{jk} [r_1(M)]_k - [r_2(M)]_j \delta_{jk} \Bigr) e_k \eqmod{2} 0 \\
\Leftrightarrow \hspace{.5em} & \hspace{.5em} \tilde{\Gamma} e^T \eqmod{2} 0,
\end{split}
\end{align}
where we have defined $\tilde{\Gamma}_{jk} = \Gamma_{jk} [r_1(M)]_k - [r_2(M)]_j \delta_{jk}$. We also have, by multiplying these equations by $[r_1(M)]_j$, that definite submeasurements must satisfy
\begin{align} \label{eq:symeven}
\begin{split}
& \sum_{k=1}^n \Bigl( [r_1(M)]_j \Gamma_{jk} [r_1(M)]_k - [r_1(M)]_j [r_2(M)]_j \delta_{jk} \Bigr) e_k \eqmod{2} 0 \\
\Leftrightarrow \hspace{.5em} & \hspace{.5em} \tilde{\Gamma}^s e^T \eqmod{2} 0,
\end{split}
\end{align}
where $\tilde{\Gamma}^s_{jk} = [r_1(M)]_j \Gamma_{jk} [r_1(M)]_k - [r_1(M)]_j [r_2(M)]_j \delta_{jk}$ is a symmetrized version of $\tilde{\Gamma}$, i.e., $\tilde{\Gamma}^s_{jk} = \tilde{\Gamma}^s_{kj}$.

By rewriting the condition to be a definite submeasurement in this way what we find is that definite submeasurements are binary vectors in the modulo $2$ nullspace of some matrix $\tilde{\Gamma}$. This means that definite submeasurements form a subspace. If we can arrange so that all definite submeasurements corresponding to a basis of this nullspace are correct, then we claim that all definite submeasurements will be correct. For, consider two binary vectors in a basis for the modulo $2$ nullspace of $\tilde{\Gamma}$, $e$ and $f$. The quantum mechanical prediction for one of these submeasurements, say the one corresponding to $e$, is denoted by ${(-1)}^{s_e}$ and can be written from Eqs.~(\ref{eq:qmsubsign}) and~(\ref{eq:defsubcon}) as $i$ to the power
\begin{align}
\begin{split}
&\sum_{j=1}^n [r_1(M)]_j e_j \Bigl( \Gamma_{jk} [r_1(M)]_k e_k - [r_2(M)]_j e_j \Bigr) \\
=&\sum_{j=1}^n e_j \sum_{k=1}^n \Bigl( [r_1(M)]_j \Gamma_{jk} [r_1(M)]_k - [r_1(M)]_j [r_2(M)]_j \delta_{jk} \Bigr) e_k \\
=& \sum_{j,k=1}^n e_j \tilde{\Gamma}_{jk}^s e_k= e \tilde{\Gamma}^s e^T.
\end{split}
\end{align}
From this, we find that the quantum mechanical outcome for the submeasurement given by $\overline{e+f}$ is $i$ to the power
\begin{equation} \label{eq:undonespace}
\Bigl(\overline{e+f}\Bigr) \tilde{\Gamma}^s \Bigl(\overline{e+f}\Bigr)^T.
\end{equation}

We want to know the value of Eq.~(\ref{eq:undonespace}) modulo $4$. We can do this with the help of a basic fact of modular arithmetic. That is, a coefficient times an even number modulo $4$ is the same as that coefficient modulo $2$ times the even number, i.e.,
\begin{equation}
c\cdot \mbox{(even number)} \eqmod{4} \overline{c}\cdot\mbox{(even number)}.
\end{equation}
The reason this is relevant is because, since $\overline{e+f}$ describes a definite submeasurement, the entries of $\tilde{\Gamma}_{jk}^s \Bigl( \overline{e+f} \Bigr)^T$ are even by Eq.~(\ref{eq:symeven}). Hence we have that this submeasurement result is $i$ to the power
\begin{align}
\begin{split}
(e+f) \tilde{\Gamma}^s \Bigl(\overline{e+f}\Bigr)^T &= e \tilde{\Gamma}^s \Bigl(\overline{e+f}\Bigr)^T + f \tilde{\Gamma}^s \Bigl(\overline{e+f}\Bigr)^T.
\end{split}
\end{align}
Now, once again, $e \tilde{\Gamma}^s$ has even entries since $e$ corresponds to a definite submeasurement and $\tilde{\Gamma}^s_{jk} = \tilde{\Gamma}^s_{kj}$. Therefore the exponent of $i$ is really,
\begin{align}
\begin{split}
e \tilde{\Gamma}^s \Bigl( e+f \Bigr)^T + f \tilde{\Gamma}^s \Bigl( e+f \Bigr)^T &= e \tilde{\Gamma}^s e^T + f \tilde{\Gamma}^s f^T + e ( \tilde{\Gamma}^s + (\tilde{\Gamma}^s)^T ) f^T \\
&= s_e + s_f + 2 e \tilde{\Gamma}^s f^T \eqmod{4} s_e + s_f,
\end{split}
\end{align}
where we used the even entries of $\tilde{\Gamma}^s f^T$ to say that $2 e \tilde{\Gamma}^s f^T \eqmod{4} 0$. Thus if our model correctly predicts $s_e$ and $s_f$, it also correctly predicts $s_{e+f}$.

So it only remains to fix the LHV table for those submeasurements corresponding to a basis for the modulo $2$ nullspace of $\tilde{\Gamma}$. Any basis will do, so we choose a basis with the property that each basis vector contains a measurement that is not contained in any other basis vector. Such a basis exists and can be obtained by writing the vectors of any basis as the rows of a matrix, and then row reducing that matrix to obtain a basis with the desired property. Having such a basis is nice because the submeasurement corresponding to each basis vector can be corrected, if need be, by having the measurement unique to that basis vector change its LHV result.

We now have a universal communication model for Pauli measurements on stabilizer states. First, each party should communicate their Pauli measurement to a special party. This party then calculates a basis for the modulo $2$ nullspace of $\tilde{\Gamma}$ as described above. For each basis vector whose corresponding submeasurement disagrees with the quantum prediction, this special party sends one bit of classical information to a party who can fix it. With this done, all submeasurements are guaranteed to agree with quantum mechanics.


\section{Concluding remarks}

This chapter explores the degree of nonlocality present in stabilizer states by considering \hyph{communication}{assisted} LHV models. These models can be classified into two types, depending on whether or not they are site invariant, and can be parameterized by the allowed distance of communication, where the distance between two qubits is defined as the minimal number of edges between the corresponding nodes in the \hyph{stabilizer}{state} graph. Interestingly, a simple \hyph{nearest}{neighbor} \hyph{site}{invariant} communication model is capable of yielding the global quantum mechanical correlation for any measurement of Pauli products on any stabilizer state, but the submeasurements of these global measurements are much harder to reproduce. To replicate the predictions of quantum mechanics for all submeasurements, it is necessary to violate site invariance and the communication distance must scale as $n/6$ or faster in the number $n$ of qubits.

We also looked at specific graphs of interest in \hyph{measurement}{based} quantum computation. In particular, for the \hyph{two}{dimensional} cluster state, site invariance must be broken, while for the \hyph{one}{dimensional} cluster state a successful \hyph{site}{invariant} model exists provided communication span the entire graph. \hyph{Nearest}{neighbor} communication is insufficient to reproduce all submeasurement correlations for both types of graphs, suggesting that site invariance, rather than communication distance, is a more relevant quantity in the context of \hyph{measurement}{based} quantum computation. Indeed, a \hyph{site}{invariant} model can be made to reproduce all submeasurement correlations in a Pauli product measurement for the \hyph{one}{dimensional} cluster state, but not for the \hyph{two}{dimensional} cluster state. Finally, a successful model exists for all submeasurements if the restrictions of both site invariance and limited distance communication are relaxed.

The obvious next step with communication models is to go beyond Pauli measurements on stabilizer states. Including one additional measurement is problematic because \hyph{probability}{preserving} tables do exist, and so it is likely that any communication strategy will have to be probabilistic in nature. An alternative is to consider Pauli measurements on a \hyph{non}{stabilizer} state for which Pauli measurements are universal. The first step there would be to see if \hyph{probability}{preserving} tables exist for Pauli measurements on that state. Keeping in mind the results for \hyph{non}{Pauli} measurements on stabilizer states, it seems unlikely that \hyph{probability}{preserving} tables exist for Pauli measurements on these \hyph{non}{stabilizer} states. However, if they do exist, then examining communication strategies for such correlations would hopefully result in a plethora of new and interesting results. If \hyph{probability}{preserving} tables do not exist in either case, then we must begin looking at probabilistic communication models.


\chapter{Conclusion}

Fully harnessing the information processing power inherent in quantum systems requires an understanding of entanglement and its role in quantum information processing tasks. One perspective on the role entanglement plays in empowering quantum computation derives from \hyph{measurement}{based} quantum computation, a unique method by which universal quantum computation can be achieved. A significant implication of this scheme is that the entanglement resource required for the computation is independent of the particular algorithm being implemented; every algorithm begins with the same fiducial state, usually a stabilizer state, so that algorithms differ only in the particular measurements made. The measurement outcomes from the computation are correlated in such a way that is impossible to reproduce with classical correlations alone.

The central theme in this work is the study of correlations, which are usually imagined as resulting from projective measurements on quantum states. According to \hyph{measurement}{based} quantum computation, these quantum correlations somehow capture all the computational power a quantum computer can possess. The difficulty now is to understand how these correlations relate to computational power. We take the point of view that one should ``subtract off'' classical correlations by modeling measurement results with necessarily imperfect local-hidden-variable tables. The focus then shifts to how one should quantify the difference between classical and quantum correlations based off the performance of these LHV tables. There are indeed many ways one could do this.

We quantify the difference between classical and quantum correlations by supplementing LHV tables with classical communication and analyzing the success or failure of a communication model with regards to various properties of the model. The idea of using classical communication to help understand entanglement properties of quantum states is a familiar one~\cite{nielsen:locc,dur:slocc}. In fact, one perspective is that quantum mechanics is a nonlocal theory; that is, measurement results are predetermined, but measurements on one system can effect another system instantaneously. The role of classical communication is to make up for this ``spooky action at a distance.''

The idea of using classical communication and shared randomness to reproduce quantum correlations goes back, in the bipartite scenario, to Toner and Bacon~\cite{toner:bell}, and, in the multipartite scenario, to Tessier \textit{et al.\/}~\cite{tessier:ghzmodel}. One big difference between these two cases arises from the lack of submeasurements in bipartite systems. Results in this work suggest that submeasurements give rise to correlations that are much more difficult to reproduce than global measurements alone. In the bipartite scenario all measurement correlations can be reproduced with just one bit of communication. This includes not only Pauli measurements, but all projective measurements. Therefore, in the attempt to generalize our results beyond Pauli measurements, it is reasonable to suppose that a similarly simple strategy may work for global measurements on multipartite states. Indeed, extending our simple communication strategy beyond Pauli measurements would strengthen the claim that submeasurements should receive more attention than global measurements. Submeasurement correlations, then, would hold the key to gaining more understanding of the power of \hyph{measurement}{based} quantum computation.

Perhaps a reasonable approach to extending our results beyond Pauli measurements is to use Bell inequalities with auxiliary communication~\cite{bacon:bell}. This is essentially the main idea behind our proofs in Subsecs.~\ref{subsec:onedcluster}~and~\ref{subsec:triangle}. Extending the results in Chapter~\ref{chap:localrealism} could help in this endeavor. Since one expects the communication required to reproduce quantum correlations for all submeasurements to grow exponentially, Bell inequalities with auxiliary communication seem more likely than a direct approach to yield interesting results. Perhaps Bell inequalities with auxiliary communication could even lead to an entanglement measure capable of capturing the computational power inherent in a quantum state.


\chapter*{Appendices}
\label{chap:appendix}

\appendix

\chapter[Graphical description of controlled-Z gates]{Graphical description of controlled-Z gates \label{app:czgates}}

The transformation rules given in Subsec.~\ref{subsec:transrules} suffice to describe the effect of any local Clifford operation on a stabilizer state. In order to complete the set of transformation rules for the Clifford group, we include transformation rules for the $\CZ$ gate here. In the interest of brevity, we consider only transformations of reduced stabilizer graphs. Transformation rules for general stabilizer graphs can be derived by first using equivalence rules E$1$ and E$2$ in Sec.~\ref{subsec:equivrules} to convert to an equivalent reduced graph and then applying the transformation rules below.

\begin{figure}[h!]
\center
\includegraphics[width=15cm]{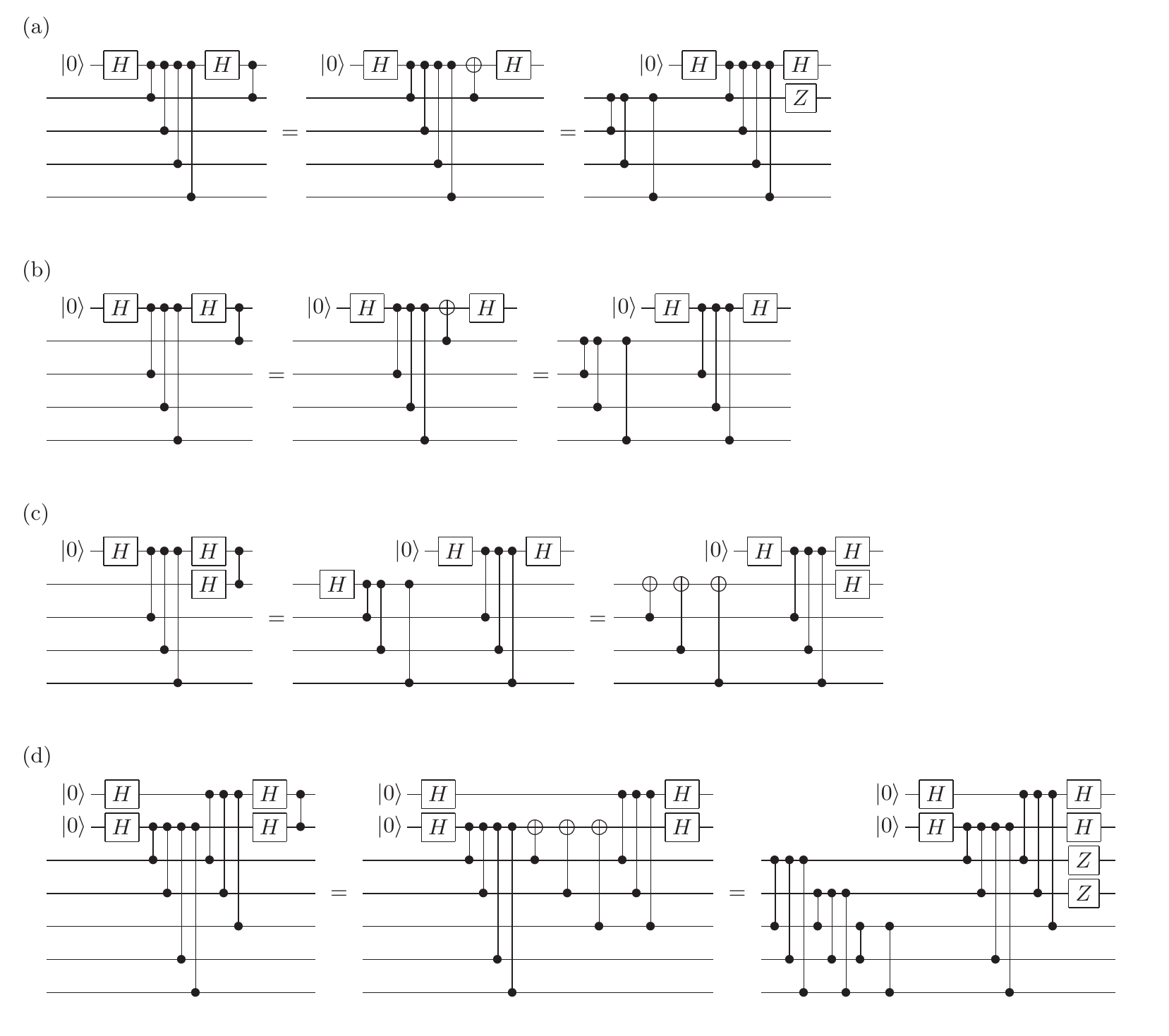}
\caption[Circuit identity for controlled-$Z$]{Circuit identities for transforming a reduced stabilizer graph under $\protect\CZ$ gates. In~(a)--(c) the identity is illustrated for the case of four neighbors connected to a qubit of interest; equivalent expressions hold for other numbers of neighbors. Identities~(a) and (b) give the rule~T(ix). The effect of a $Z$ gate on the hollow qubit in the third layer of~(a) or~(b) is to deposit an additional $Z$ gate on both the hollow and solid qubits in the third layer of the final circuit. The identity in~(c) relies on~(b); this identity is extended in~(d) to a demonstration of the rule~T(x). If there is a $Z$ gate on the lower hollow qubit in the third layer of~(c), the result in the third layer of the final circuit is to put a $Z$ gate on that qubit and on the neighbors of the other hollow qubit. Translated to~(d), this means that a $Z$ gate in the third layer on either hollow qubit leads in the third layer of the final circuit to additional $Z$ gates on that qubit and on the neighbors of the other hollow qubit.\label{fig:CZident}}
\end{figure}

\begin{enumerate}

\item[T(viii).] Applying $\CZ$ between two solid nodes complements the edge between them.

\vspace{0pt}

\item[T(ix).] Applying $\CZ$ between a hollow node and a solid node complements the edges between the solid node and the neighbors of the hollow node.\\
\rule{0em}{0em}\vspace{-.5em}\\
Flip the solid node's sign if the two nodes were initially connected and the hollow node did not have a sign or if the two nodes were not connected and the hollow did have a sign.

\vspace{0pt}

\item[T(x).] Applying $\CZ$ between two hollow application nodes complements an edge provided its endpoints neighbor distinct application nodes.

Nodes that neighbor both application nodes flip their signs. If an application node initially had a sign, flip the signs of nodes connected to the other application node.
\end{enumerate}

Transformation rule~T(viii) is trivial since the $\CZ$ gate simply commutes into layer two of the reduced-graph-form circuit. The circuit identities needed to prove rules~T(ix) and~T(x) are given in Fig.~\ref{fig:CZident}.

\chapter[Proof that the equivalence rules are complete]{Proof that the equivalence rules are complete \label{app:equivproof}}

Having described a set of rules in Subsec.~\ref{subsec:equivrules} for converting between equivalent stabilizer graphs, we show in this appendix that, in each case, the aforementioned rules generate the entire equivalence class of stabilizer graphs. The proof is divided into three parts. The first part shows how to use rules~E$1$ and E$2$ to convert an arbitrary stabilizer graph to an equivalent graph in reduced form. The second part explains how testing the equivalence of a pair of reduced stabilizer graphs can be simplified, using rules~E(i) and~E(ii), to a special form. Finally, the third part proves that the graphs on the two sides of such a simplified equivalence test are equivalent only if they are trivially identical. Taken as a whole, this proves that the set of equivalence rules given in Subsec.~\ref{subsec:equivrules} is sufficient to convert, reversibly, between any two equivalent stabilizer graphs and thus that they generate all stabilizer graphs that are equivalent to the same stabilizer state. Similarly, considering only the final two parts of the proof shows that rules~E(i) and~E(ii) are sufficient to generate all reduced stabilizer graphs.

\section{Converting stabilizer graphs to reduced form}
\label{sec:convert}

Two features identify a stabilizer graph as reduced. In a reduced graph, hollow nodes never have loops, and hollow nodes are never connected to each other. Equivalence rule~E$1$ can be used to convert looped nodes from hollow to solid. Applying rule E$1$ in this sort of situation can cause hollow nodes to acquire loops, but each application fills one hollow node of the graph, so the procedure will terminate in at most a number of repetitions equal to the number of hollow nodes in the graph. Similarly, connected hollow nodes in the resulting graph can be converted to solid nodes using the appropriate case of rule E$2$. Once again, the process is guaranteed to terminate because the number of hollow nodes to which the rule might be applied decreases by two with each application of the rule. Concomitantly, the conversion of a stabilizer graph to an equivalent reduced graph never increases the number of hollow nodes.

\section{Simplifying \hyph{reduced}{graph} equivalence \\ testing}
\label{sec:simplify}

Equivalence testing for pairs of reduced graphs is facilitated by simplifying the equivalence such that nodes that are hollow only in the first graph never connect to nodes that are hollow only in the second. This simplification can be accomplished by iterating the following process. Choose a pair of nodes, $a$ and $b$, that are connected in either graph and such that $a$ is hollow in one graph and $b$ is hollow in the other. To the graph in which they are connected, apply the relevant reduced equivalence rule to the selected nodes. Among other things, the equivalence operation reverses the fill of the two nodes it is applied to. Since one node is hollow and the other solid, this preserves the total number of hollow nodes while yielding a node that is hollow in both graphs. Because it is applied only to unpaired hollow nodes, subsequent uses of this equivalence rule do not disturb the newly paired hollow node. Consequently, this process also terminates in at most a number of repetitions equal to the number of hollow nodes in each of the graphs.

\section{Trivial evaluation of simplified \hyph{reduced}{graph} equivalence tests}
\label{sec:evaluate}

The two reduced graphs composing a simplified \hyph{reduced}{graph} equivalence test are equivalent, i.e., correspond to the same state, if and only if the graphs are identical. To see why this is so, we return to \hyph{graph}{form} quantum circuits. In addition to the standard restrictions for reduced graphs, the circuits corresponding to the graphs in a simplified equivalence test have the following property: if in one of the circuits, a qubit with a terminal $H$ participates in a $\CZ$ gate with a second qubit, which cannot have a terminal $H$, then in the other circuit, it cannot be true that the second qubit has a terminal $H$ and the first does not. We prove the triviality of simplified \hyph{reduced}{graph} equivalence testing by considering an arbitrary simplified \hyph{reduced}{graph} equivalence test and showing that the two graphs must be identical if they are to correspond to the same state.

In terms of unitaries, an arbitrary \hyph{graph}{form} circuit equality can be written as
\begin{align}
  \begin{split}
\prod_{g\in\set{H}_l} H_g
\prod_{j\in\set{S}_l} S_j
\prod_{h\in\set{Z}_l} &Z_h
\prod_{\gamma\in\set{C}_l} \CZ_\gamma
\prod_{k} H_k \ket{0}^{\otimes n} \\
=&
\prod_{g\in\set{H}_r} H_g
\prod_{j\in\set{S}_r} S_j
\prod_{h\in\set{Z}_r} Z_h
\prod_{\gamma\in\set{C}_r} \CZ_\gamma
\prod_{k} H_k \ket{0}^{\otimes n}\;,
  \end{split}
  \label{eq:arbSimpGraphFormCircEqual}
\end{align}
where $\set{C}$ lists the pairs of qubits participating in $\CZ$ gates and $\set{H}$, $\set{Z}$, and $\set{S}$ are sets enumerating the qubits to which $H$, $Z$, and $S$ gates are applied respectively. The total number of qubits is denoted by $n$ and the subscripts $l$ and $r$ discriminate between the circuits on the left and right hand sides of the equation. In terms of these sets, a reduced-graph-form circuit satisfies the constraints $\set{H}\cap\set{S}=\varnothing$ and $\{a,b\}\not\in\set{C}$ for all $a,b\in\set{H}$. The circuits in simplified tests also satisfy $\{a,b\}\not\in\set{C}_l,\set{C}_r$ for all $a\in\hhbar$ and $b\in\hbarh$, where $\bar{\set{H}}$ denotes the complement of $\set{H}$, i.e., the set of qubits to which $H$ is not applied.

Suppose now that the two graphs have hollow nodes at different locations; i.e., at least one of the sets, $\hhbar$ and $\hbarh$, is not empty. For specificity, let's say that $\hhbar$ is not empty. In the language of circuits, this means that there exists a qubit $a$ that has a terminal $H$ on the left side of Eq.~(\ref{eq:arbSimpGraphFormCircEqual}), but not on the right side. Since qubit~$a$ is part of a circuit for a reduced graph, it does not participate in $\CZ$ gates with other qubits that possess terminal $H$ gates. Consequently, on the left side of Eq.~(\ref{eq:arbSimpGraphFormCircEqual}), the $\CZ$ gates involving qubit $a$ can all be moved to the end of the circuit where they become $\CX$ gates with $a$ as the target. Doing this and transferring the $\CX$ gates to the other side yields,
\begin{align}
  \begin{split}
\prod_{g\in\set{H}_l} H_g
\prod_{j\in\set{S}_l} S_j&
\prod_{h\in\set{Z}_l} Z_h
\prod_{\gamma\in\set{C}_l\mbox{\scriptsize{\ s.t.\ }}a\not\in\gamma}\!\!\!\!\!\CZ_\gamma\;\;\prod_{k} H_k
\ket{0}^{\otimes n} \\
=&
\prod_{b\in\set{N}_l(a)} \C{X}_{ba}
\prod_{g\in\set{H}_r} H_g
\prod_{j\in\set{S}_r} S_j
\prod_{h\in\set{Z}_r} Z_h
\prod_{\gamma\in\set{C}_r} \CZ_\gamma
\prod_{k} H_k \ket{0}^{\otimes n}\;,
  \end{split}\label{eq:modCircEqual}
\end{align}
where $\set{N}_l(a)$ denotes the set of qubits that participate in $\CZ$ gates with qubit $a$ on the left hand side of Eq.~(\ref{eq:arbSimpGraphFormCircEqual}).

Because the original graph equality, Eq.~(\ref{eq:arbSimpGraphFormCircEqual}), was simplified, $\set{N}_l(a)\cap\set{H}_r=\varnothing$ and, by assumption, $a\not\in\set{H}_r$, so the $\CX$ gates and the terminal Hadamards on the right side of Eq.~(\ref{eq:modCircEqual}) do not act on the same qubits. Thus we can commute the $\CX$ gates past the terminal Hadamards. Moreover, we can then move the $\CX$ gates to the beginning of the circuit where they have no effect and can therefore be dropped. During this migration, however, they generate a complicated menagerie of phases. The resulting expression for the right side of Eq.~(\ref{eq:modCircEqual}) is
\begin{align}
  \begin{split}
\prod_{g\in\set{H}_r} H_g
\prod_{j\in\set{S}_r} S_j
\raisebox{-.35em}{\Bigg(\Bigg.}
\prod&_{d\in\set{N}_{l}(a)}S_d \CZ_{da}
\raisebox{-.35em}{\Bigg.\Bigg)}^{\!\!\mathbf{1}_{\set{S}_r}(a)}
\prod_{h\in\set{Z}_r} Z_h
\raisebox{-.35em}{\Bigg(\Bigg.}
\prod_{f\in\set{N}_{l}(a)} Z_f
\raisebox{-.35em}{\Bigg.\Bigg)}^{\!\!\mathbf{1}_{\set{Z}_r}(a)} \\
&\prod_{\gamma\in\set{C}_r}\CZ_\gamma
\prod_{\delta\in\set{N}_{lr}(a)}\CZ_\delta
\prod_{c\in\set{N}_l(a)\cap\set{N}_r(a)}\!\!\!Z_c\;\;\prod_{k} H_k\ket{0}^{\otimes n}\;,
  \end{split}\label{eq:modCircEqualRight}
\end{align}
where $\set{N}_r(a)$ is defined similarly to $\set{N}_l(a)$,
\begin{equation}
\set{N}_{lr}(a) = \{ \{ p, q \} | p\in\set{N}_l(a), q\in\set{N}_r(a),
p\neq q \}\;,
\end{equation}
and $\mathbf{1}$ represents an indicator function, i.e., $\mathbf{1}_{\set{Z}_r}(a)$ equals $1$ if $a\in\set{Z}_r$ and $0$ otherwise.

It might appear that we have made things substantially worse by this rearrangement, but in an important sense, Eq.~(\ref{eq:modCircEqualRight}) is now very simple with regard to qubit~$a$: $H$ is applied to qubit~$a$ followed by a sequence of unitary gates all of which are diagonal in the standard basis. This implies that there are an equal number of terms in the resultant state where qubit~$a$ is in the state $\ket{0}$ and the state $\ket{1}$. On the left side of Eq.~(\ref{eq:modCircEqual}), however, the only gate remaining on qubit~$a$ is the identity or an $X$, depending on whether $a\in\set{Z}_l$; thus qubit~$a$ is separable and is either in state $\ket{0}$ or $\ket{1}$ depending on whether $a\in\set{Z}_l$. Consequently, our initial assumption that $\set{H}_l \neq \set{H}_r$ is incompatible with satisfying the equality.

\begin{figure}[h!]
\center
\includegraphics[width=15cm]{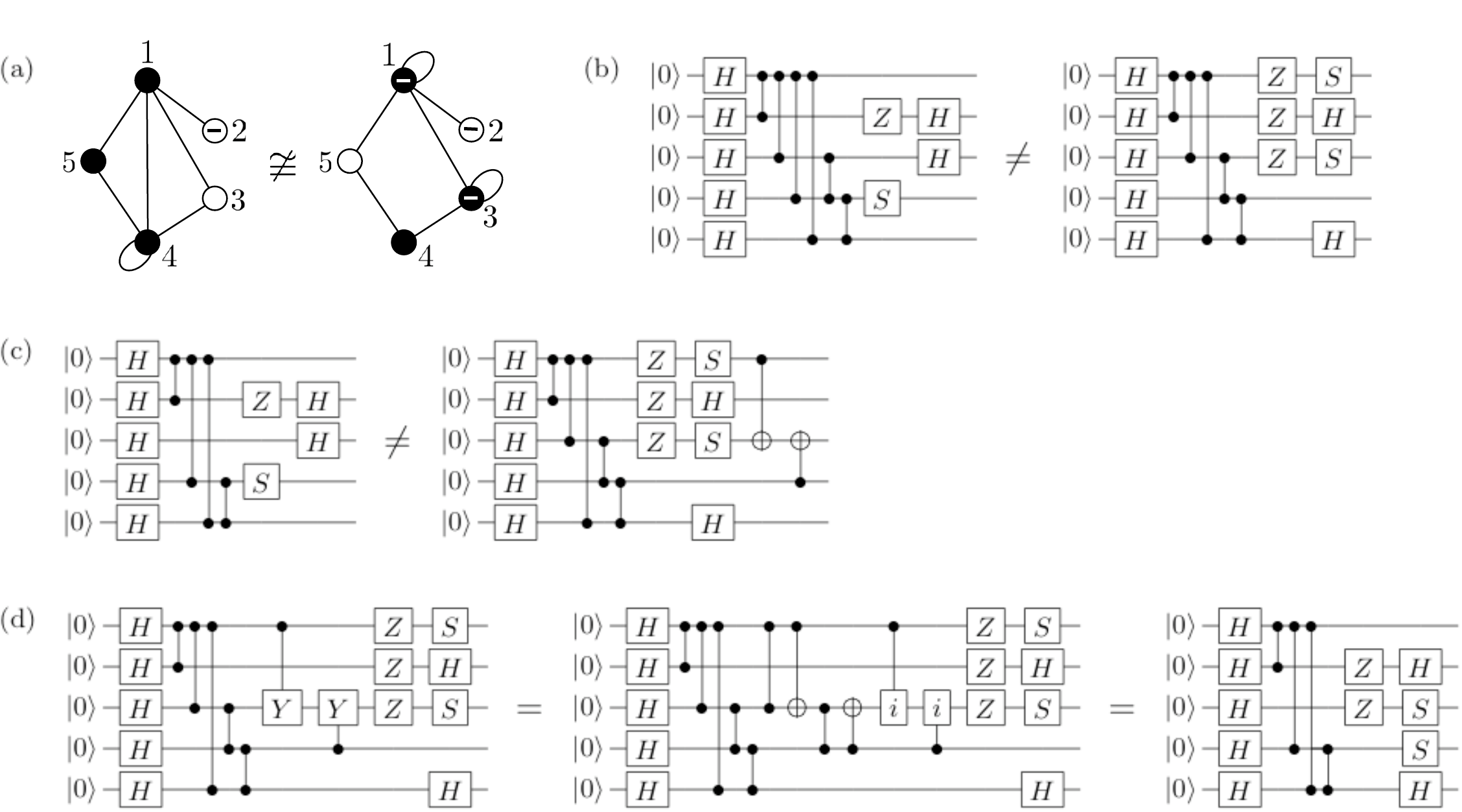}
\caption[Example of a simplified \hyph{reduced}{graph} equivalence test]{An example of the process described in Appendix~\ref{sec:evaluate}. A simplified \hyph{reduced}{graph} equivalence test is shown in (a). The test is simplified because a node that is hollow in only one graph is never connected to a node that is hollow only in the other. The test is also false, since such an equivalence is satisfied if and only if the two graphs are identical. By switching to circuit notation, it is possible to see why these graphs correspond to different states. Translating the graph equality in~(a) into circuit notation yields the circuit equality in~(b). The fact that the graphs are reduced implies that qubits with terminal $H$ gates, such as qubit $3$ in the left hand circuit, are not connected to other qubits with terminal $H$ gates. This allows us to pull out the $\protect\CZ$ gates acting on qubit $3$ on the left side and transfer the resulting $\protect\CX$ gates to the other side of the equation, yielding the equality in~(c). That the graph equality was simplified guarantees that $H$ gates do not prevent us from then pushing the $\protect\CX$ gates to the beginning of the right hand circuit in~(c), as shown in~(d). Though this is not generally the case, qubit $3$ winds up separable on both sides of the equation, allowing us to verify the inequivalence of the prepared states since $\ket{0}\ne(\ket{0}-i\ket{1})/\sqrt2$. \label{fig:partthreeexample}}
\end{figure}

The preceding discussion shows that two graphs composing a simplified equivalence test are equivalent only if they have hollow nodes in exactly the same locations. Given this constraint, the terminal Hadamards can be canceled from both sides of Eq.~(\ref{eq:arbSimpGraphFormCircEqual}), giving
\begin{align}
\begin{split}
\prod_{h\in\set{Z}_l} Z_h \prod_{j\in\set{S}_l} S_j \prod_{\gamma\in\set{C}_l} \CZ_\gamma
\prod_{k} &H_k \ket{0}^{\otimes n} \\
=&\prod_{h\in\set{Z}_r} Z_h \prod_{j\in\set{S}_r} S_j \prod_{\gamma\in\set{C}_r} \CZ_\gamma
\prod_{k} H_k \ket{0}^{\otimes n}\;.\
\end{split}\label{eq:paredSimpGraphFormCircEqual}
\end{align}
The state after the initial Hadamards is an equally weighted superposition of all the states in the standard basis. The subsequent unitaries are diagonal in the standard basis, so they put various phases in front of the terms in the equal superposition. Since a unitary is fully described by its action on a complete set of basis states, demanding equality term by term in Eq.~(\ref{eq:paredSimpGraphFormCircEqual}) amounts to requiring that
\begin{align}
\prod_{h\in\set{Z}_l}& Z_h
\prod_{j\in\set{S}_l} S_j
\prod_{\gamma\in\set{C}_l} \CZ_\gamma=
\prod_{h\in\set{Z}_r} Z_h
\prod_{j\in\set{S}_r} S_j
\prod_{\gamma\in\set{C}_r} \CZ_\gamma\;,
\end{align}
which is only satisfied when $\set{Z}_l=\set{Z}_r$, $\set{S}_l=\set{S}_r$, and $\set{C}_l=\set{C}_r$. Thus, after simplification, the equivalence of pairs of reduced graphs is trivial to evaluate, since equivalence requires that the two graphs be identical.

\begin{figure}
\center
\includegraphics[width=15cm]{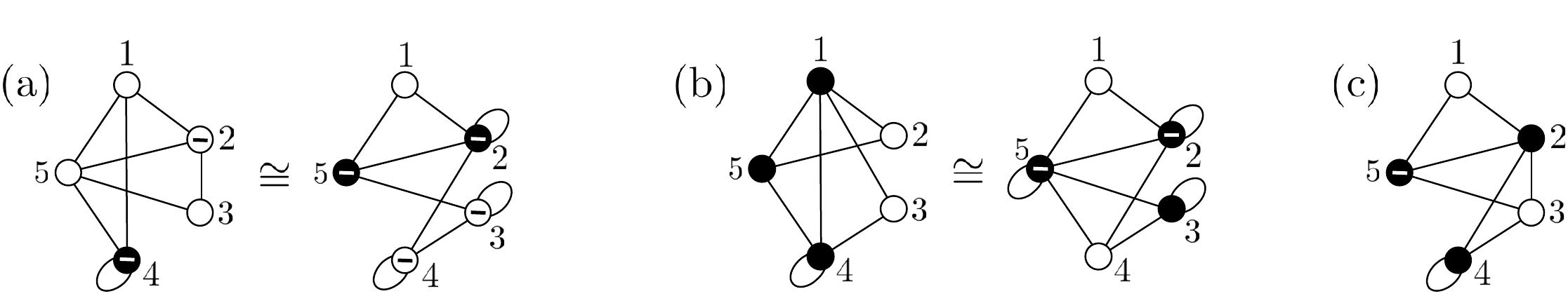}
\caption[Example of testing the equivalence of two graphs]{An example sequence of graphs representing parts~$1$ through $3$ of the equivalence checking procedure in Appendix~\ref{app:equivproof}. The graph on the left hand side of~(a) is equivalent to the reduced graph on the left hand side of~(b) by the application of~E$2$ to the pair of nodes~$\{1,5\}$. Likewise, an application of~E$1$ to node~$3$ in the graph on the right hand side of~(a) yields the reduced graph on the right hand side of~(b). The graph of~(c) results from the application of~E(ii) to the pair of nodes~$\{1,2\}$ in the graph on the left of~(b). An application of~E(i) to the pair of nodes~$\{3,4\}$ in the graph on the right hand side of~(b) also yields the graph in~(c), verifying, as per Appendix~\ref{sec:evaluate}, that the graphs in~(a) represent the same state.\label{fig:equivalenceexample}}
\end{figure}

An example which illustrates the circuit manipulations described algebraically in the text of this section is given in Fig.~\ref{fig:partthreeexample}. An example of the entire process of testing graph equivalence is given in Fig.~\ref{fig:equivalenceexample}.

As mentioned above, our proof provides a constructive procedure for testing the equivalence of stabilizer graphs. Moreover, it shows that equivalence rules E$1$ and E$2$ given in Subsec.~\ref{subsec:equivrules} are sufficient to convert between any equivalent stabilizer graphs, and that the rules, E(i) and E(ii), for reduced graphs are sufficient to convert between any two equivalent reduced stabilizer graphs. Since the conversion of an arbitrary stabilizer graph to reduced form never increases the number of hollow nodes and the rules~E(i) and E(ii) preserve the number of hollow nodes, we conclude that the reduced stabilizer graphs for a stabilizer state are those that have the least number of hollow nodes.

\chapter[Proof of the \hyph{Pauli}{product} measurement rule]{Proof of the \hyph{Pauli}{product} measurement rule}
\label{app:measproof}

In this Appendix we derive the measurement rule given in Subsec.~\ref{subsec:graphmeas} for transforming a stabilizer graph under a measurement $M$, where $M$ is any tensor product of $I$ and $Z$ Pauli operators.  The proof proceeds in three stages.  In the first, we determine the effect of $M$, considered as a Clifford unitary operation, on a stabilizer state $\ket\psi$.  This enables us, in the second stage, to find the action of the measurement projector $P_m =[I+{(-1)}^m M]/2$ on $\ket\psi$ and thus to determine whether the measurement is certain or random.  The \hyph{post}{measurement} state is then found via a simple circuit identity in the last stage.  Notice that, in determining the effect of $M$ on $\ket\psi$, we must retain the overall phase, since, in the second stage, the overall phase becomes a relative phase in the superposition $[\ket\psi+(-1)^m M\ket\psi]/2$.

\section{Action of $Z$ on a stabilizer state}

\begin{figure}
\center
\includegraphics[width=15cm]{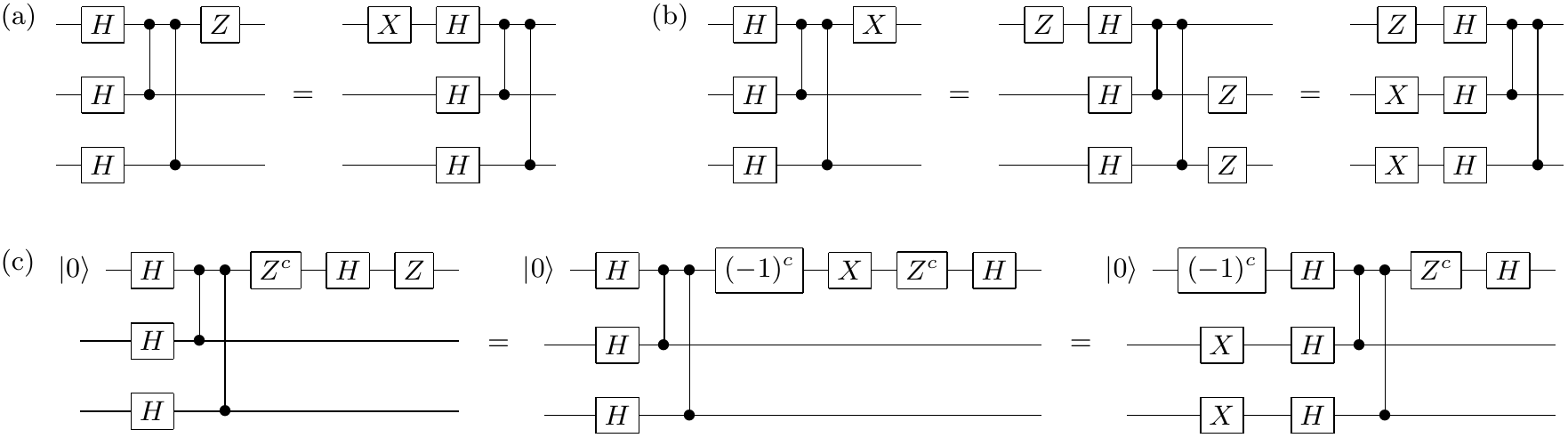}
\caption[Effect of $Z$ operators on a stabilizer state]{Circuit identities used to determine the effect of a $Z$ operator on a stabilizer state.  Identity~(a) follows from the fact
that $Z$ and controlled-sign gates commute and from the identity $HZH
= X$.  A similar identity holds if the node has a loop or a sign
since $Z$ commutes with itself and with $S$.  The first equality in
identity~(b) is easily verified in the standard basis, and the second
equality is an application of~(a).  Identity~(c) follows from the
equalities shown; the second equality uses the identity in~(b).
\label{fig:Zoperators}}
\end{figure}

To begin, consider the effect of a unitary $M$ on the \hyph{$n$}{qubit} stabilizer state $\ket{\psi}$, where $M$ is known to be a tensor product consisting of only the operators $I$ and $Z$.  The action of $I$ is trivial, so we can focus on determining the action of $Z$.  As illustrated in Fig.~\ref{fig:Zoperators}(a), applying $Z$ to a solid node is equivalent to applying an $X$ operator to the same node prior to all other Clifford gates in the circuit.  Similarly, the action of $Z$ on a hollow node can be reexpressed using the circuit identity in
Fig.~\ref{fig:Zoperators}(c).  This identity shows that an identical state is obtained by adding a leading $X$ operator to each neighbor of the hollow node and, if the hollow node has a sign, introducing an overall phase of $-1$. For properly simplified measurements, hollow measured nodes are only neighbored by solid measured nodes, so only members of $\setMS$, the measured solid nodes, collect leading $X$ operators.  The number of $X$ operators collected by each member of $\setMS$ is $1$ plus the number of neighbors it has in the set $\setMH$.  Since $X^2=I$, the net result is that $X$ operators are added only to members of $\setMSE$, the set of solid measured nodes with an even number of hollow measured neighbors.

Summarizing, the stabilizer state $M\ket{\psi}$ can be obtained from the state $\ket{0}^{\otimes n}$ by first applying an $X$ to each member of $\setMSE$ and then applying the Clifford gates needed to obtain $\ket{\psi}$ from the initial state $\ket{0}^{\otimes n}$.  In addition, an overall phase of $(-1)^b$ must be applied where $b=|\setMH\cap\setZ|$ is the number of measured hollow nodes with a sign.  That is,
\begin{equation}
M\ket\psi=
MU\ket{0}^{\otimes n}=(-1)^b
U\Biggl(\prod_{j\in\smallsetMSE}X_j\Biggr)\ket{0}^{\otimes n}\;,
\end{equation}
where $U$ is shorthand for the sequence of Clifford gates for preparing $\ket{\psi}=U\ket{0}^{\otimes n}$ as indicated by the stabilizer graph.

\section{Certain and random outcomes}

The second stage of our proof applies this result to find the action of $P_m$ on $\ket\psi$ and the probability of obtaining measurement outcome ${(-1)}^m$, which is given by $\bra{\psi} P_m \ket{\psi}$.
We have immediately that
\begin{equation}
P_m \ket{\psi} = \frac{1}{2}[I+{(-1)}^m M]\ket\psi =U\frac{1}{2}\Biggl(I+(-1)^{m+b}
\prod_{j\in\smallsetMSE}X_j\Biggr)\ket{0}^{\otimes n}\;,
\end{equation}
which gives
\begin{align}
\bra{\psi} P_m &\ket{\psi}
= \frac{1}{2}\!\left(1 + (-1)^{m+b}\bra{0}^{\otimes n}
\Biggl(\prod_{j\in\smallsetMSE}X_j\Biggr)\ket{0}^{\otimes n}\right)\;.
\end{align}
Since $\bra{0} X \ket{0} = 0$, this means that measurements are classified into two types: if $\setMSE = \varnothing$, the outcome probabilities are $1$, for $m \eqmod{2} b$, and $0$, for $m \ne_{\scriptscriptstyle{2}} b$, but if $\setMSE \ne \varnothing$, $m$ has a $50\%$ chance of being either $0$ or $1$.

\begin{figure*}
\center
\includegraphics[width=15cm]{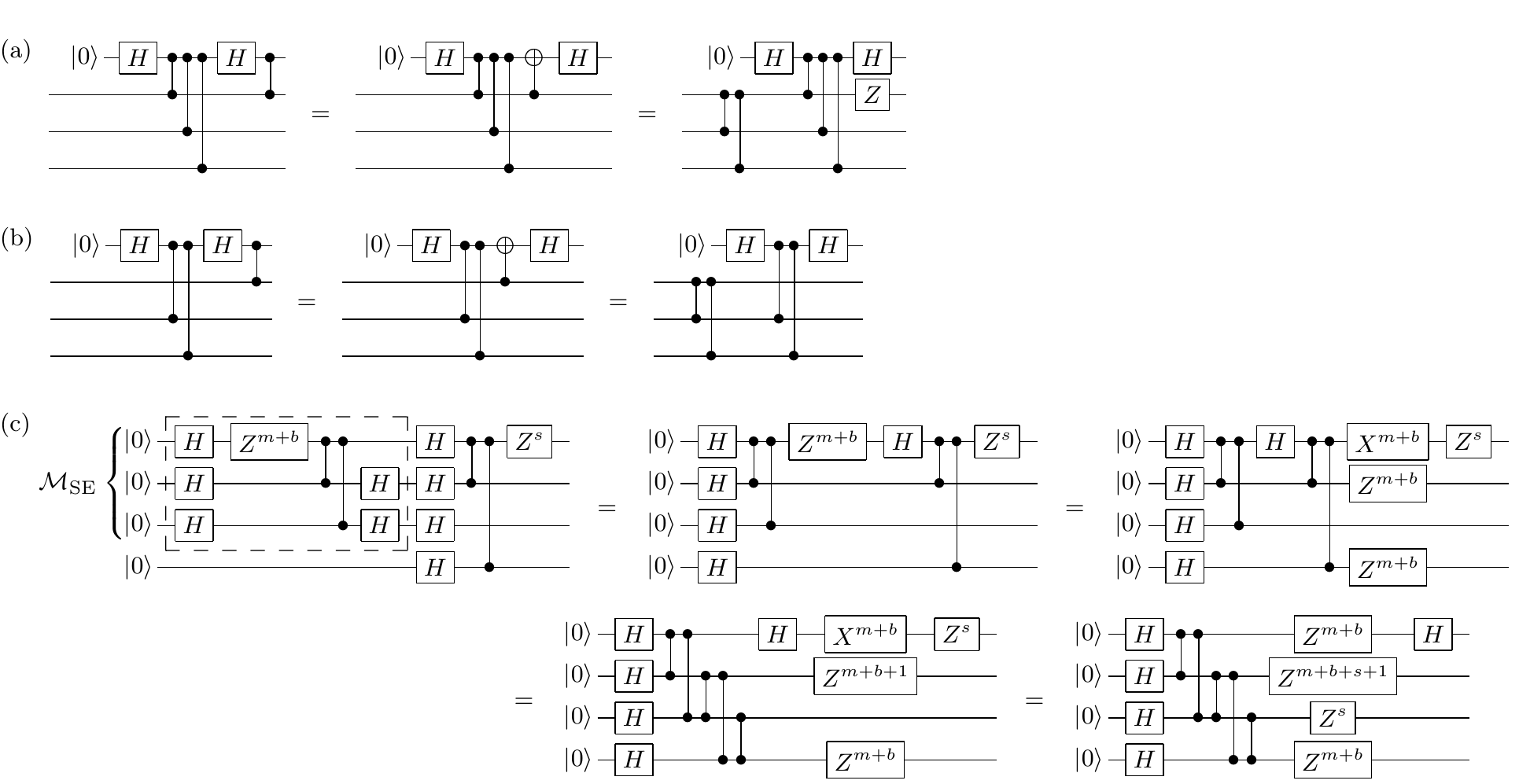}
\caption[Circuit yielding the \hyph{post}{measurement} state]{Circuit identities used to determine the \hyph{post}{measurement} state. Identities~(a) and~(b) utilize basic circuit identities for pushing a controlled-$X$ gate past a controlled-$Z$ gate.  For the identity in~(c), the first three qubits represent the elements of $\setMSE$, with the chosen qubit $p$ placed on top.  The dashed box delineates the Clifford operations that put the qubits in $\setMSE$ into an appropriate cat state, i.e., those operations in Eq.~(\ref{eq:poststate}) that are applied before $U$, whereas the
gates after the dashed box are the relevant portion of the gates in $U$, i.e., the gates that create the original stabilizer state. For illustration, $U$ contains no $S$ or $Z$ gates except that the sign of the chosen node is indicated by $Z^s$. The first equality in~(c) is trivial.  The second uses the identity in Fig.~\ref{fig:Zoperators}(b) to push $Z^{m+b}$ to the right, turning it into $X^{m+b}$ and depositing $Z^{m+b}$ on each neighbor of $p$.  The next equality eliminates the controlled-$Z$ gates that initially connected $p$ to other nodes using the identity in~(a) for connections to nodes in $\setMSE$
and using~(b) for connections to nodes outside of $\setMSE$. The final simple equality returns the circuit to graph form and produces a possible overall phase which is ignored.
\label{fig:catcircuit}}
\end{figure*}

\section{Post-measurement state}

When the measurement outcome is ${(-1)}^m$, the post-measurement
state is $\ket{\psi'}=P_m\ket{\psi}/\sqrt{\bra{\psi}P_m\ket{\psi}}$.
When $\setMSE = \varnothing$, the outcome $m \eqmod{2} b$ occurs with
certainty, and the post-measurement state is the same as the original
stabilizer state $\ket\psi$.

When $\setMSE \ne \varnothing$, the situation is more complicated. In
this case $\bra{\psi}P_m\ket{\psi}=1/2$, so
\begin{equation}
\ket{\psi'} =
U\frac{1}{\sqrt{2}}\Biggl(I+{(-1)}^{m+b}
\prod_{j\in\smallsetMSE}X_j\Biggr)\ket{0}^{\otimes n}\;.
\end{equation}
Thus the post-measurement state is obtained using the Clifford
circuit that created the original stabilizer state, but applied to an
initial state that, for the qubits in $\setMSE$, is changed to a cat
state, i.e., an equal superposition of all 0s and all 1s, with the
sign between the two terms in the superposition given by
$(-1)^{m+b}$.  To construct a graph for the post-measurement state,
we introduce a standard quantum circuit for making the cat state from
$\ket{0}^{\otimes n}$ and use circuit identities to put the overall
quantum circuit into graph form. Thus we write
\begin{equation} \label{eq:poststate}
\ket{\psi'}= U\Biggl(\prod_{l\in\smallsetMSE\backslash \{p\}} H_l\Biggr)
\Biggl(\prod_{k \in\smallsetMSE\backslash \{p\}}\CZ_{pk}\Biggr) Z_p^{m+b}\prod_{j\in\smallsetMSE}H_j\ket{0}^{\otimes n}\;,
\end{equation}
where $p$ denotes the chosen node.  Figure~\ref{fig:catcircuit}
translates these Clifford operations into circuit notation and
develops the identities needed to determine the post-measurement
state $\ket{\psi'}$.

Steps~$1$--$3$ in Sec.~\ref{subsec:graphmeas} follow directly from Fig.~\ref{fig:catcircuit}(c), while step $4$ follows from applying an $S$ gate to both sides of Fig.~\ref{fig:catcircuit}(c) and using T$3$. Step~$1$ describes the edge complementation that occurs in the third equality in the figure. Step~$2$ correctly describes the node signs indicated in the final circuit. Finally, step~$3$ expresses the fact that neighbors of the chosen node are now those nodes in $\setMSE$, and the chosen node is hollow.

\chapter[Proof of Theorem~1]{Proof of Theorem~1}
\label{app:proofnvars}

Before we can proceed with the proof of Theorem~\ref{thm:nvars} there is one definition to get out of the way. The \emph{correlation range\/} of $n$ random variables $R(a_1)$, \ldots, $R(a_n)$ is the range of possible values for the expectation value of their product. This interval is denoted as $\mathcal{C}(a_1, \ldots, a_n)$ so that $R(a_1) \cdots R(a_n) = R(p) \Leftrightarrow p \in \mathcal{C}(a_1, \ldots, a_n)$. The idea of correlation ranges is crucial for our proof.

We prove Theorem~\ref{thm:nvars} by induction on the number of random variables. It is convenient that the case $n=2$ is already covered. Thus we assume that the theorem is true for $n$ variables and we wish to find all the possible $p$ such that
\begin{equation}
\label{eq:rtocorr}
\begin{array}{l}
R(a_1) \cdots R(a_n) R(a_{n+1}) = R(p) \\ \\
\Leftrightarrow \forall j \hspace{.2cm} R(p) R(a_j) = \prod_{k \ne j} R(a_k) \\ \\
\Leftrightarrow \mathcal{C}(p,a_j) \cap \mathcal{C}(a_1, \ldots, a_{j-1}, a_{j+1}, \ldots, a_{n+1}) \ne \emptyset.
\end{array}
\end{equation}
The inductive hypothesis now lets us find both these correlation ranges to see if they intersect. We can immediately write down the second correlation range.
\begin{equation}
\begin{array}{l}
\mathcal{C}(a_1, \ldots, a_{j-1}, a_{j+1}, \ldots, a_{n+1}) = \\ \\ \left[ \sum_{k \ne j} a_k  - (n-1), (n-1) - \sum_{k \ne j}^{n} a_k + a_{n+1} \right]
\end{array}
\end{equation}
Since the sequence in Eq.~(\ref{eq:rtocorr}) is valid for any $j$, we go ahead and find the correlation ranges for some $j \ne n+1$.

\underline{Case 1:} $p<0$.

In this case, $\mathcal{C}(p,a_j) = \Bigl[ \left| p+a_j \right| -1,1-(a_j-p) \Bigr]$. There are now three conditions which, taken together, are equivalent to the correlation ranges overlapping. The first is that the correlation ranges be increasing, i.e., that they are actually ranges, which is really just a technical requirement equivalent to $p \in [-1,1]$. The second is that the lowest value in $\mathcal{C}(p,a_j)$ be below the highest value in $\mathcal{C}(\{ a_k \}_{k \ne j})$,
\begin{equation}
\label{eq:neglower}
\left| p + a_j \right| -1 \leq (n-1) - \sum_{k \ne j}^{n} a_k + a_{n+1}.
\end{equation}
The third is that the highest value in $\mathcal{C}(p,a_j)$ be above the lowest value in $\mathcal{C}(\{ a_k \}_{k \ne j})$,
\begin{equation}
\label{eq:negupper}
1-a_j + p \geq \sum_{k \ne j} a_k  - (n-1).
\end{equation}
We want to show that these are equivalent to $p \in [-1,1]$ and
\begin{equation}
\label{eq:result}
\sum_k a_k - n \leq p \leq n - \sum_{k=1}^n a_k + a_{n+1}.
\end{equation}

$\left[ \Rightarrow \right]$ First we show that $\mathcal{C}(p,a_j)$ being a range gives us that $p \in [-1,1]$. For,
\begin{equation}
\label{eq:incrangeneg}
\begin{array}{l}
\begin{array}{r}
-p-a_j -1 \leq 1-(a_j-p) \hspace{.5em} \mbox{ or} \\ \\
a_j + p -1 \leq 1-(a_j-p)
\end{array} \\ \\
\Leftrightarrow p \geq -1.
\end{array}
\end{equation}
Furthermore, $p \leq 1$ since it is negative. Now we show that Eqs.~(\ref{eq:neglower}) and (\ref{eq:negupper}) imply Eq.~(\ref{eq:result}). From Eq.~(\ref{eq:negupper}) we get
\begin{equation}
1-a_j + p \geq \sum_{k \ne j} a_k - (n-1) \Leftrightarrow p \geq \sum_k a_k - n,
\end{equation}
which gives us the lower bound in Eq.~(\ref{eq:result}). The upper bound in Eq.~(\ref{eq:result}) is positive by the fact that $a_k \in [0,1]$. Therefore, since $p$ is negative in this case, we get the entirety of Eq.~(\ref{eq:result}).

$\left[ \Leftarrow \right]$ First, it is easy to see that $p \in [-1,1]$ implies that the correlation ranges are increasing from the backwards arrow in Eq.~(\ref{eq:incrangeneg}). Now, as we have seen, Eq.~(\ref{eq:negupper}) is equivalent to $p \geq \sum_k a_k - n$. To show Eq.~(\ref{eq:neglower}), we have to consider two cases.

If $p + a_j \geq 0$, consider the upper bound in Eq.~(\ref{eq:result}).
\begin{equation}
p \leq n - \sum_{k=1}^n a_k + a_{n+1} \Rightarrow p + a_j - 1 \leq (n-1) - \sum_{k \ne j}^n a_k + a_{n+1}.
\end{equation}

If $p + a_j < 0$, consider the lower bound.
\begin{equation}
\begin{array}{l}
p \geq \sum_k a_k - n \geq \sum_k a_k - 2(a_{n+1} + a_j) - n \\ \\
\Rightarrow -(p+a_j)-1 \leq (n-1) - \sum_{k \ne j}^{n} a_k + a_{n+1}.
\end{array}
\end{equation}
And we have the result.

\underline{Case 2:}  $p \geq 0$.

Now, $\mathcal{C}(p,a_j) = \Bigl[p + a_j - 1,1 - |a_j-p|\Bigr]$, and the ranges overlapping is equivalent to
\begin{equation}
\label{eq:poslower}
p + a_j -1 \leq (n-1) - \sum_{k \ne j}^{n} a_k + a_{n+1}
\end{equation}
\begin{equation}
\label{eq:posupper}
1- \left| a_j - p \right| \geq \sum_{k \ne j} a_k  - (n-1).
\end{equation}
We want to show that these equations, along with the correlation range being a valid range, is equivalent to Eq.~(\ref{eq:result}) when supplemented with $p \in [-1,1]$.

$\left[ \Rightarrow \right]$
As before, we begin by showing that $\mathcal{C}(p,a_j)$ being a range gives us $p \in [-1,1]$. For,
\begin{equation}
\label{eq:incrangepos}
\begin{array}{l}
\begin{array}{r}
p+a_j -1 \leq 1-(a_j-p) \hspace{.5em} \mbox{ or} \\ \\
p+a_j -1 \leq 1-(p-a_j)
\end{array} \\ \\
\Leftrightarrow p \leq 1.
\end{array}
\end{equation}
Again, $p$ being \hyph{non}{negative} takes care of the lower bound. There are now two cases to consider.

If  $p < a_{n+1}$, then from Eq.~(\ref{eq:posupper}) we get
\begin{equation}
1-a_j+p \geq \sum_{k \ne j} a_k  - (n-1) \Rightarrow p \geq \sum_{k} a_k  - n.
\end{equation}
But then,
\begin{equation}
\begin{array}{l}
0 \leq n - \sum_{k} a_k + p = n - \sum_{k = 1}^n a_k + p - a_{n+1} \\ \\
\leq n - \sum_{k = 1}^n a_k - p + a_{n+1}.
\end{array}
\end{equation}
This means that $p \leq n - \sum_{k = 1}^n a_k + a_{n+1}$.

If $p \geq a_{n+1}$, then we want to consider Eq.~(\ref{eq:poslower}).
\begin{equation}
p+a_j - 1 \leq (n-1) - \sum_{k \ne j}^{n} a_k + a_{n+1} \Leftrightarrow p \leq n - \sum_{k = 1}^n a_k + a_{n+1}.
\end{equation}
Now the other inequality follows from this.
\begin{equation}
0 \geq \sum_{k = 1}^n a_k - a_{n+1} + p - n \geq \sum_k a_k - n - p \Rightarrow p \geq \sum_k a_k - n.
\end{equation}

$\left[ \Leftarrow \right]$ The correlation ranges that yield the overlapping conditions in Eqs.~(\ref{eq:poslower}) and (\ref{eq:posupper}) are valid ranges by Eq.~(\ref{eq:incrangepos}). Also, we have already seen that Eq.~(\ref{eq:poslower}) is equivalent to $p \leq n - \sum_{k = 1}^n a_k + a_{n+1}$. So, to show Eq.~(\ref{eq:posupper}), consider the two cases.

If $p < a_j$, consider the lower bound in Eq.~(\ref{eq:result}).
\begin{equation}
p \geq \sum_k a_k - n \Rightarrow 1+p-a_j \geq \sum_{k \ne j} a_k - (n-1).
\end{equation}

If $p \geq a_j$, then consider the upper bound.
\begin{equation}
\begin{array}{l}
p \leq n - \sum_{k \ne j}^n a_k - a_j + a_{n+1} \\ \\
\leq n - \sum_{k \ne j}^n a_k - a_{n+1} + a_j \\ \\
\Rightarrow 1-p+a_j \geq \sum_{k \ne j} a_k - (n-1).
\end{array}
\end{equation}
Thus we have proven the result in this case as well.

\bibliographystyle{plain}
\bibliography{Bibliography}

\end{document}